Université
Paris Cité

UNIVERSITÉ PARIS CITÉ

École doctorale des Sciences de la Terre et de l'Environnement et Physique de l'Univers — ED 560
Laboratoire AstroParticules et Cosmologie (APC) — Groupe Théorie

# BLACK HOLE PERTURBATIONS IN MODIFIED GRAVITY THEORIES

Par Hugo ROUSSILLE

Thèse de doctorat de Physique de l'Univers

Dirigée par David LANGLOIS
et Karim NOUI

*présentée et soutenue publiquement le 17/06/2022*
*devant le jury composé de :*

| | | |
|---|---|---|
| Laura Bernard | Chargée de recherche (Observatoire de Paris) | Examinatrice |
| Emanuele Berti | Professor (John Hopkins University) | Rapporteur |
| Éric Gourgoulhon | Directeur de recherche (Observatoire de Paris) | Rapporteur |
| Panagiota Kanti | Professor (Ioannina University) | Examinatrice |
| David Langlois | Directeur de recherche (Université Paris Cité) | Directeur |
| Karim Noui | Professeur d'université (Université Paris-Saclay) | Directeur |
| Danièle Steer | Professeure d'université (Université Paris Cité) | Examinatrice |

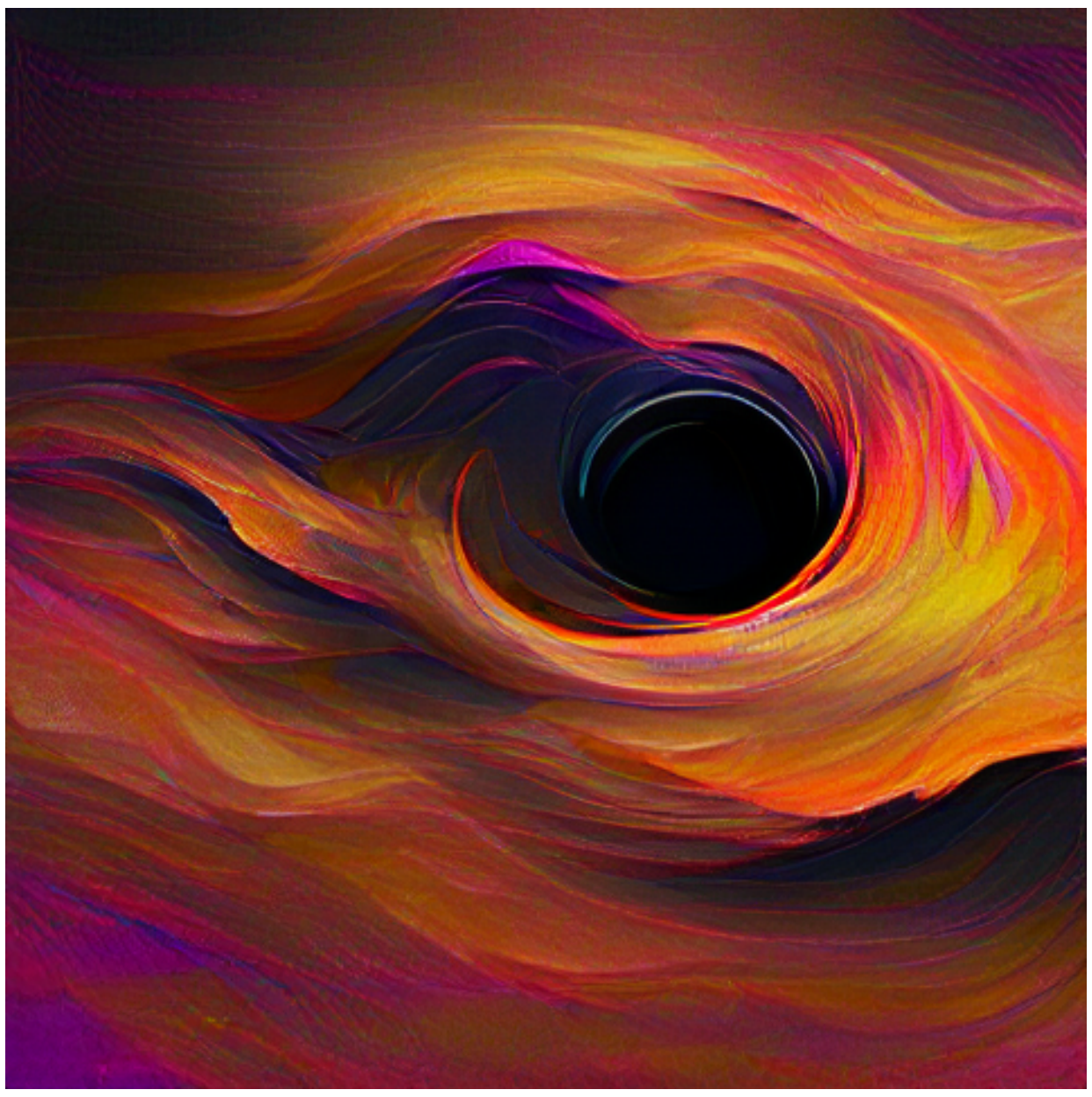

*Perturbations around a black hole*, Artist's impression

## Abstract

The recent first detection of gravitational waves from binary black hole mergers has opened a new field in gravitational physics: gravitational wave astronomy, the study of the Universe through the gravitational waves we can detect from various sources. It has also spurred a renewed interest in possible deviations from General Relativity, since they could be detected in the gravitational waves emitted by such compact binaries.

Of particular interest is the ringdown phase of a binary black hole merger, which can be described by linear perturbations about a background stationary black hole solution. These perturbations mainly correspond to a superposition of modes called quasi-normal modes, whose frequencies form a discrete set. One expects that modified gravity models could predict quasi-normal modes that differ from their General Relativity counterpart and the detailed analysis of the gravitational wave signal, commonly called "black hole spectroscopy", represents an invaluable window to test General Relativity and to look for specific signatures of modified gravity.

The work done in this thesis takes place in the context of scalar-tensor theories of gravity, and more particularly the Degenerate Higher-Order Scalar-Tensor theories which are the most general scalar-tensor theories that do not exhibit Ostrogradsky instabilities. We start by a review of these theories and their properties, and describe a way to reformulate them in a framework with a clear geometrical interpretation, providing us with a better physical understanding of their construction.

We then study linear perturbations about several existing nonrotating black hole solutions of such theories, and show why the perturbation equations obtained are very hard to decouple in general. When it is possible, in the case of odd parity perturbations, we describe the propagation of waves and relate it to the stability of the underlying spacetime. When it is not, we circumvent the difficulty by making use of an algorithm proposed recently in the mathematical literature that allows us to decouple the equations both at the black hole horizon and at infinity. This allows us to get the asymptotic behaviour of waves on such spacetimes, yielding valuable information that can allow us to rule some of them out. Finally, we use the asymptotic behaviours obtained to compute quasi-normal modes numerically.

**Keywords**: modified gravity, scalar-tensor, black hole, quasinormal modes, stability

## Court résumé de la thèse

La première détection d'ondes gravitationnelles issues de binaires de trous noirs a permis l'émergence d'une nouvelle branche de l'astronomie : l'astronomie gravitationnelle. Celle-ci a également renouvelé l'intérêt de la communauté scientifique pour les théories alternatives de la gravité, car des écarts à la Relativité Générale pourraient être détectés par de tels binaires d'objets compacts.

Parmi les différentes phases constituant le signal gravitationnel émis par une binaire de trous noirs, la phase de « vibration atténuée » (ou *ringdown*) est particulièrement intéressante ; en effet, lors de celle-ci, le système se comporte dans l'approximation linéaire comme un trou noir stationnaire autour duquel évoluent des perturbations. Ces perturbations correspondent principalement à une superposition de modes appelés modes quasinormaux, dont les fréquences sont discrètes. On peut s'attendre en général à ce que les théories alternatives de gravité prédisent des modes quasinormaux différents de ceux prédits par la Relativité Générale ; l'étude du signal gravitationnel pour remonter aux fréquences de ces modes, appelée « spectroscopie des trous noirs », est un outil précieux pour tester cette dernière et rechercher des signes de l'existence de théories modifiées.

Le travail réalisé au cours de cette thèse s'inscrit dans le cadre des théories Degenerate Higher-Order Scalar-Tensor qui constituent la classe la plus générale de théories scalaire-tenseur dépourvues d'instabilités d'Ostrogradsky. Nous commençons par une revue de ces théories et de leurs propriétés et présentons une nouvelle façon de formuler leur action qui permet de mieux comprendre ces propriétés à l'aide d'une interprétation géométrique.

Nous nous tournons ensuite vers l'étude des perturbations linéaires de trous noirs sans rotation dans ces théories. Nous montrons que les équations obtenues sont difficiles à découpler en général. Quand cela est possible, dans le cas des perturbations de parité impaire, nous étudions la propagation des ondes gravitationnelles et son lien avec la stabilité de la solution. Lorsque cela n'est pas possible, nous faisons appel à un algorithme proposé récemment dans la littérature mathématique qui nous permet de découpler les équations des perturbations à l'horizon et à l'infini, nous permettant d'obtenir le comportement asymptotique des ondes gravitationnelles et d'en déduire si le trou noir est viable ou non. Nous utilisons enfin ce comportement asymptotique pour calculer les modes quasinormaux numériquement.

**Mots-clés** : gravité modifiée, scalaire-tenseur, trou noir, modes quasinormaux, stabilité

## Résumé de la thèse

La gravité est la première force fondamentale à avoir été étudiée par les scientifiques, depuis les travaux d'Aristote au IV$^{\text{ème}}$ siècle av. J-C. Sa compréhension fut au centre des préoccupations de nombreux scientifiques au cours de l'histoire, et progressa de pair avec le concept de méthode scientifique du Moyen-Âge jusqu'aux Temps Modernes. La première théorie en proposant une modélisation fut développée par Newton au XVII$^{\text{ème}}$ siècle. Cette théorie, appelée « théorie newtonienne de la gravitation », permit l'explication de tous les phénomènes gravitationnels connus jusqu'alors, de la chute des corps jusqu'aux mouvements des astres. Notamment, des calculs précis de trajectoires des planètes dans le Système Solaire amenèrent les scientifiques à prédire l'existence de Neptune et ses propriétés avant même son observation. Ainsi, au XIX$^{\text{ème}}$ siècle, la théorie de la gravitation semblait complète et bien comprise, le seul problème restant étant une légère déviation de la prédiction newtonienne dans le cas de la trajectoire de Mercure.

Cependant, au début du XX$^{\text{ème}}$ siècle, Einstein développa sa théorie de la relativité restreinte, dont l'un des fondements est qu'aucun signal ne peut se propager à une vitesse supérieure à celle de la lumière dans le vide. La gravitation newtonienne reposant sur une force d'attraction immédiate et à portée infinie, elle était en claire contradiction avec ce principe et il devint nécessaire de modifier cette théorie pour la rendre compatible avec la relativité restreinte. Plusieurs scientifiques se mesurèrent à l'exercice, mais un seul proposa une théorie satisfaisante : Einstein à nouveau, qui fonda la théorie de la Relativité Générale en 1915.

Cette théorie fut testée à de nombreuses reprises au cours du XX$^{\text{ème}}$ siècle, et se vit confirmée à chaque fois. Il s'agit aujourd'hui, avec la physique quantique, de la théorie scientifique la mieux testée, avec des confirmations expérimentales précises et indépendantes provenant de domaines variés. Récemment, l'observation d'ondes gravitationnelles émises lors de la fusion de deux trous noirs, récompensée par le prix Nobel de physique 2017, a été un nouveau succès prédictif de la Relativité Générale dans un nouveau régime.

Cependant, de légers problèmes restent présents dans la description de la gravitation proposée par la Relativité Générale, notamment en ce qui concerne le domaine de validité et la compatibilité avec l'autre grande théorie physique du XX$^{\text{ème}}$ siècle, la physique quantique. En effet, la Relativité Générale n'est plus valide aux hautes énergies, par exemple proche du Big Bang ou du centre d'un trou noir. De plus, l'origine de l'expansion accélérée de l'Univers n'est pas certaine : bien que celle-ci soit modélisable par l'ajout d'un terme *ad hoc* dans les équations d'Einstein, la justification de l'origine de ce terme par un raisonnement de physique quantique échoue totalement.

Ainsi, nous nous trouvons présentement dans une situation assez similaire à celle d'il y a 120 ans : la théorie existante de la gravitation est très puissante, mais les quelques limites de celle-ci sont autant d'indices nous laissant supposer l'existence d'une théorie plus générale. C'est à partir de ce constat que de nombreux et nombreuses scientifiques ont cherché à développer, depuis les années 1960, de nouvelles théories de la gravitation.

Il existe plusieurs façons de développer de telles théories. Par exemple, on peut chercher à transformer fondamentalement la Relativité Générale en proposant une théorie complètement différente mais qui permet de retrouver, dans une certaine limite, la première ; une telle approche est appelée *top-down*, et a été à la base de la conception de la théorie des cordes et de la gravitation quantique à boucles. On peut aussi chercher à créer une légère déviation de la théorie existante, afin de comprendre dans quelle mesure il est possible de s'éloigner de la Relativité Générale tout en conservant l'essentiel de ses propriétés remarquables. Ce type d'approche est nommé *bottom-up*, et on parle alors de théories de gravité modifiée.

Cette thèse se concentre sur l'étude d'une classe de théories construite en suivant la seconde approche : les théories scalaire-tenseur. Dans ces théories, la gravitation n'est plus seulement décrite par la géométrie de l'espace-temps, comme en Relativité Générale, mais également par un champ scalaire défini en tout point de l'espace-temps, que l'on peut interpréter comme une cinquième force fondamentale. Ainsi, l'image habituelle d'un espace-temps courbé par la matière doit être corrigée en ajoutant, en tout point, une « couleur » qui décrit la valeur du champ scalaire, comme représenté sur la figure 1. Ces théories sont d'un intérêt particulier car ajouter un champ scalaire à la Relativité Générale usuelle est la façon la plus simple de modifier cette théorie. De plus, de nombreuses théories de gravité modifiée plus complexes aboutissent, dans une certaine limite, à la dynamique d'une théorie scalaire-tenseur.

Nous nous concentrerons dans ce manuscrit plus précisément sur l'étude des théories Degenerate Higher-Order Scalar-Tensor (DHOST), dont la construction et classification ont été réalisées par David Langlois, Karim Noui et Jibril Ben Achour. Ces théories décrivent un scalaire couplé à la métrique de façon non minimale, et sont telles que les équations du mouvement sont d'ordre supérieur à 2. Cette dernière propriété rendant la théorie généralement instable, il est nécessaire d'imposer certaines conditions de dégénérescence pour se débarrasser des instabilités.

Les théories DHOST étant plus riches que la Relativité Générale, il est possible de trouver de nouvelles solutions de trous noirs différant de la solution de Kerr-Newman. La physique des trous noirs étant récemment entrée dans un stade de confirmation expérimentale, notamment au travers des observations d'ondes gravitationnelles par la collaboration LIGO/Virgo, il est important de réaliser des prédictions théoriques dans le contexte des théories de gravité modifiée et de les comparer aux observations futures afin de tester quantitativement la Relativité Générale. Pour ce faire, nous nous intéressons dans ce manuscrit à l'étude des modes quasinormaux des trous noirs.

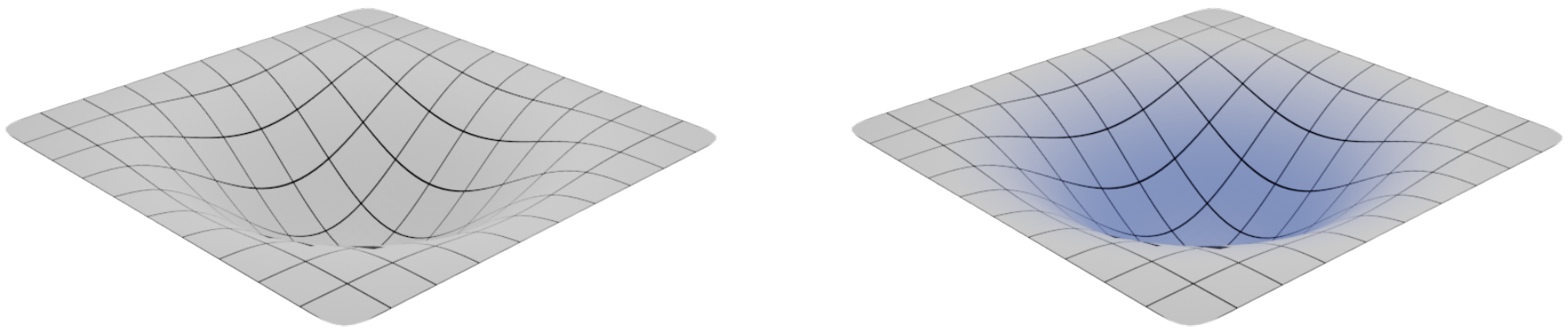

(a) Relativité Générale (b) Théories tenseur-scalaire

Figure 1. – En Relativité Générale, l'espace-temps est courbé par la présence de matière. Dans les théories tenseur-scalaire, la présence d'un objet courbe l'espace-temps et lui fournit aussi une valeur non nulle de champ scalaire en tout point, représentée dans (b) par une teinte de couleur.

Les modes quasinormaux d'un trou noir sont les fréquences des ondes gravitationnelles émises par celui-ci lorsqu'il est légèrement perturbé hors de son équilibre. Ils constituent l'analogue gravitationnel des modes propres d'une corde de guitare. En effet, ces derniers forment un ensemble discret de fréquences, dépendent de la longueur de la corde de guitare ainsi que de la façon dont l'onde mécanique se propage le long de la corde. De la même façon, les modes quasinormaux de trous noirs forment un ensemble discret, dépendent de la solution de trou noir considérée et de la façon dont les ondes gravitationnelles se propagent dans la théorie étudiée. Ainsi, ils constituent un test très puissant de la Relativité Générale, car ils dépendent à la fois de la solution et de la théorie elle-même et permettent donc de tester les deux.

Ces modes sont observables dans la dernière phase de la fusion de deux trous noirs, appelée « vibration atténuée » (ou *ringdown*). Lors de cette phase, le signal gravitationnel émis est constitué notamment d'une somme d'ondes dont les fréquences sont celles des modes quasinormaux. Il est d'ores et déjà possible de mesurer ces fréquences expérimentalement et de les comparer aux prédictions issues de la Relativité Générale ; cependant, la précision des mesures étant faible, la mesure est pour le moment imprécise et aucun écart à la théorie d'Einstein n'a été mesuré. Une amélioration de la précision sera rendue possible par l'exploitation des résultats issus des détecteurs d'ondes gravitationnelles de génération ultérieure, à savoir LISA et Einstein Telescope.

Afin de calculer les modes quasinormaux de trous noirs dans les théories DHOST, nous suivons une procédure semblable à celle utilisée en Relativité Générale : nous perturbons la métrique en-dehors du trou noir, calculons les équations du mouvement des perturbations, et cherchons à les mettre sous une forme similaire à une équation de Schrödinger, facilement interprétable physiquement. Cependant, les théories scalaire-tenseur possèdent un degré de liberté supplémentaire par rapport à la Relativité Générale : des ondes scalaires peuvent se propager, et celles-ci seront en général couplées aux ondes gravitationnelles. Le découplage ne sera pas toujours possible : il est ainsi nécessaire de séparer deux cas, selon que les perturbations soient de parité paire ou impaire.

Dans ce manuscrit, nous commençons par l'étude des perturbations de parité impaire, pour lesquelles nous trouvons une forme d'équation de Schrödinger. Cela nous permet de conclure quant au comportement des ondes autour des différentes solutions trous noirs présentées. Notamment, nous trouvons l'expression de la métrique effective vue par ces ondes et en déduisons des critères de stabilité pour ces solutions.

Nous nous tournons ensuite vers l'étude des perturbations paires, qui contiennent un degré de liberté gravitationnel et un degré de liberté scalaire. Le découplage de ces deux degrés de liberté n'étant pas possible en général, nous faisons appel à un algorithme proposé récemment dans la littérature mathématique nous permettant de le réaliser asymptotiquement, à l'horizon du trou noir et à l'infini. Cela nous permet d'obtenir des informations précieuses sur le comportement propagatif des ondes dans ces deux zones, et d'obtenir des conditions aux limites pour le calcul numérique des modes quasinormaux.

Dans le chapitre 1, nous réalisons une revue des théories de gravité modifiée et notamment des théories DHOST. Nous présentons ensuite au chapitre 2 quatre nouvelles solutions de trous noirs dans ces théories. Au chapitre 3, nous proposons une reformulation des théories DHOST dans un cadre possédant une simple interprétation géométrique, nous permettant de comprendre plus intuitivement les propriétés de ces théories. Ensuite, nous présentons au chapitre 4 le formalisme des perturbations des trous noirs en Relativité Générale ; au chapitre 5, nous décrivons l'algorithme mathématique qui nous servira pour l'étude de ces perturbations dans le cadre des DHOST. Enfin, nous réalisons l'étude des perturbations respectivement impaires et paires de trous noirs solutions de DHOST dans les chapitres 6 et 7. Pour finir, nous calculons numériquement des modes quasinormaux à l'aide des résultats obtenus au chapitre 7 dans le chapitre 8.

## Remerciements

CES trois dernières années ont été pour moi extrêmement riches en rencontres et en échanges. J'aimerais ici remercier autant que quelques pages me le permettent les personnes qui m'ont soutenu tout au long de ma thèse et sans lesquelles ce travail n'aurait pas été possible.

Le doctorat marque, dans le monde académique, le passage entre le statut d'étudiant et le statut de chercheur. Je remercie infiniment mes directeurs de thèse, David Langlois et Karim Noui, pour avoir été les meilleurs guides que j'aurais pu souhaiter avoir lors de cette transition. Vous m'avez accueilli dans le monde de la recherche avec bienveillance, me guidant lorsque cela était nécessaire, me laissant autonome lorsque cela était préférable. Bien plus que d'être des encadrants, j'ai eu l'impression que vous étiez des collègues, avec lesquels je pouvais échanger sans hésiter. Les nombreux échanges que nous avons eus, devant un tableau, dans un RER ou à travers un écran d'ordinateur ont toujours été très enrichissants et je suis heureux de dire que vous m'avez transmis votre passion pour l'étude de la gravitation. Vous avez également su occuper la fonction de directeurs de thèse à distance durant la pandémie de coronavirus : bien que cela fût nouveau pour nous tous, nous sommes parvenus à garder le contact, malgré les passages occasionnels de chats ou de membres de la famille. Merci pour cela !

Je souhaiterais remercier Emanuele Berti ainsi que Éric Gourgoulhon pour avoir accepté de relire en détail mon manuscrit. Leurs commentaires ont été précieux pour améliorer la qualité de ce qui suit. Merci également à Laura Bernard, Panagiota Kanti et Danièle Steer d'avoir accepté de faire partie du jury de thèse. Merci aux membres de l'école doctorale, Alessandra Tonazzo, Alissa Marteau et Irena Nikolic pour leurs réponses rapides à toutes mes questions sur le dédale administratif qu'est le doctorat. Merci à Eugeny Babichev d'avoir fait partie de mon comité de thèse.

Les années de thèse ne seraient rien sans le soutien des autres doctorants et doctorantes, plus jeunes ou plus anciens. Edwan, Konstantin, Pierre, c'est grâce à vous que notre bureau est le meilleur bureau de APC : merci pour les conseils de jeux de société, les trouvailles informatiques farfelues, les énigmes et les tableaux remplis ! Merci à Jani, Thomas et Valentin pour les repas et les pauses cafés, sans lesquels la recherche serait bien moins intéressante... et merci à David et Konstantin pour m'avoir montré que les discussions physiques les plus intéressantes ont souvent lieu en altitude. Merci enfin à tout le monde, Alexandre, Aurélien, Baptiste, Bastien, Calum, Clara, Gabriel, Hamza, Léna, Louise, Magdy, Makarim, Marc, Marion, Mikel, Moana, Nicolas, Pilar, Raphaël, Samuël, Sruthi, Thomas, Thibaut pour ces années passées avec vous, les Pub Quizz du confinement, les pots, les séminaires. Ce sont ces moments qui font que les années de doctorat sont bien plus que 3 années de recherche !

J'ai aussi eu la chance d'être entouré par des amis et amies extraordinaires pendant ma thèse (et bien avant !). Les différentes vagues de covid ont été l'occasion de se voir différemment mais de créer ainsi des liens plus forts : merci pour les pizzas du couvre-feu, les soirées sur Zoom et les retrouvailles du déconfinement. Merci pour les essais de cocktails plus ou moins fructueux, les randos, l'Eurovision, les découvertes de restaurants, les Drive de planification de repas 4 mois à l'avance, les plages, les calanques, la Normandie, la Dordogne, les traversées de la France en voiture ou en train... Merci aussi pour les discussions inarrêtables, les partages de mèmes à 2h du matin, les soirées où l'on refait le monde. Je suis impatient de découvrir les moments que nous partagerons dans les années à venir !

Merci également à Jules et Julien pour leur engouement dans la rédaction de notre manuel de physique. Les heures de discussion (sur la physique souvent, sur la meilleure trajectoire à suivre dans un circuit Mario Kart parfois) m'auront certainement aidé à traverser les différentes vagues de covid, et j'espère qu'elles porteront bientôt leurs fruits !

J'aimerais enfin remercier ma famille. Vous avez toujours été un soutien extraordinaire au cours de ma vie ; cela n'a bien sûr pas changé durant mon doctorat. Merci à mes parents pour tous ces week-ends précieux de retour à Nantes, pour les concours de pétrissage de pain ou de Wordle, pour les jeux de rôle et les Gartic Phone en famille ! Colin, Éliane, merci pour les soirées films ou concerts, pour les vacances et pour toutes les soirées couvre-feu. J'ai beaucoup de chance de tous vous avoir dans ma vie et j'espère que vous le savez.

Pour finir, je souhaiterais remercier Iris, la personne la plus importante dans ma vie. Tu as toujours été là ces dernières années, que ce soit pour rire avec moi pendant les hauts ou me soutenir pendant les bas. Si ce travail de thèse a été réalisé, c'est bien grâce à toi ; mais bien plus, si je suis la personne que je suis aujourd'hui, c'est à toi que je le dois. Merci de rendre chaque instant joyeux et unique ! Merci pour les petits moments du quotidien, partagés avec Yoko, Kira et Calsi, merci pour les grands projets, et merci tout simplement d'être là.

# CONTENTS

# LIST OF ABBREVIATIONS

**4dEGB** 4-Dimensional-Einstein-Gauss-Bonnet.

**ADM** Arnowitt-Deser-Misner.

**BBH** Binary Black Hole.

**BCL** Babichev-Charmousis-Lehébel.

**BH** Black Hole.

**CMB** Cosmic Microwave Background.

**DGP** Dvali-Gabadadze-Porrati.

**DHOST** Degenerate Higher-Order Scalar-Tensor.

**EEP** Einstein's Equivalence Principle.

**EFT** Effective Field Theory.

**EGB** Einstein-Gauss-Bonnet.

**EsGB** Einstein-Scalar-Gauss-Bonnet.

**ET** Einstein Telescope.

**FLRW** Friedmann-Lemaître-Robertson-Walker.

**GR** General Relativity.

**GW** Gravitational Wave.

**IR** Infrared.

**LISA** Laser Interferometer Space Antenna.

**MTMG** Minimal Theory of Massive Gravity.

**ODE** Ordinary Differential Equation.

**QNM** Quasi-Normal Mode.

**RN** Reissner-Nordström.

**RW** Regge-Wheeler.

**SM** Standard Model.

**UV** Ultraviolet.

**WKB** Wentzel–Kramers–Brillouin.

# LIST OF SYMBOLS

$f^{(n)}(\phi)$ $n$-th derivative of the function $f(\phi)$.

$F_\phi$ Derivative of the function $F(\phi,X)$ with respect to $\phi$.

$F_X$ Derivative of the function $F(\phi,X)$ with respect to $X$.

$A_i$ Functions describing quadratic DHOST theories ($i \in [\![1,5]\!]$).

$B_i$ Functions describing cubic DHOST theories ($i \in [\![1,10]\!]$).

$E_{\mu\nu}$ 4-dimensional Einstein tensor.

$\varepsilon_{\mu\nu\rho\sigma}$ Fully antisymmetric tensor density.

$\approx$ Equality up to subdominant terms in a given variable ($r$ or $z$).

$g_{\mu\nu}$ 4-dimensional metric tensor.

$\lambda$ Quantity related to the angular momentum $\ell$ by $2\lambda = \ell(\ell+1) - 2$..

$A(r)$ $\mathrm{d}t^2$ term for a spherically symmetric line element.

$B(r)$ Inverse of the $\mathrm{d}r^2$ term for a spherically symmetric line element.

$C(r)$ $\mathrm{d}\Omega^2$ term for a spherically symmetric line element.

$\mathrm{d}s^2$ Line element for a given metric.

$\mathfrak{a}$ Asymptotically decoupled axial modes.

$\mathfrak{p}$ Asymptotically decoupled polar modes.

$\mathfrak{g}$ Asymptotically decoupled polar gravitational modes.

$\mathfrak{s}$ Asymptotically decoupled polar scalar modes.

$\phi$ Scalar field.

$\phi_\mu$ First derivative of the scalar field: $\phi_\mu = \nabla_\mu\phi$.

$\phi_{\mu\nu}$ Second derivative of the scalar field: $\phi_{\mu\nu} = \nabla_\mu\nabla_\nu\phi$.

$X$ Kinetic energy density of the scalar field: $X = \phi_\mu\phi^\mu$.

$M_{\mathrm{P}}$ Reduced Planck mass.

${}^{(3)}R$ 3-dimensional Ricci tensor.

$R_{\mu\nu\rho\sigma}$ 4-dimensional Riemann tensor.

$R_{\mu\nu}$ 4-dimensional Ricci tensor.

$R$ 4-dimensional Ricci scalar, sometimes written ${}^{(4)}R$.

$\mathrm{sgn}(x)$ Sign function: for $x \in \mathbf{R}$, $\mathrm{sgn}(x) = x/|x|$.

$r_*$ Tortoise coordinate defined from $r$ by $\mathrm{d}r/\mathrm{d}r_* = n(r)$.

# INTRODUCTION

## The search for a complete theory of gravity

The first mathematical description of gravity was proposed by Newton in the 17$^{\text{th}}$ century [1]. This description, called today "Newtonian gravity", explained perfectly at the time the motion of bodies, whether they were on the surface of the Earth or in the Solar System. However, increased precision in the measurements of the movement of planets [2, 3] led astronomers to discover some discrepancies in Newton's theory. Furthermore, the special theory of relativity proposed by Einstein in 1905 [4] shook the fundations of classical physics and created a need to develop a new theory of gravity that would be compatible with relativity.

The search for this new theory was an intense race at the beginning of the 20$^{\text{th}}$ century, and while several theories were proposed, only the one proposed by Einstein in 1915 [5] satisfied the requirements of coherence with special relativity and present observations. This theory, called General Relativity (GR), relied on the description of gravity as curvature of spacetime. Up to now, this theory has successfully explained all experimental observations, on many different scales, and can be said to be the most thoroughly verified physical theory along with quantum mechanics.

However, small issues are still present on the horizon, and looking at them carefully exhibits limits of both the domain of validity of GR and its compatibility with other theories (such as the ones describing particle physics). In particular, the lack of a quantum completion of GR and the lack of a clear understanding of origin of the accelerated expansion of the Universe are two of the main hints that a more general theory must be proposed. In a way, the present situation is comparable with the situation of 110 years ago: the defects of the present theory of gravity are small, but present nonetheless, and curing them requires proposing a whole new theory.

To this end, many modified theories of gravity have been proposed in the literature in the past 60 years. The search for a more complete theory of gravity is much more organised now than a century ago; indeed, unicity theorems concerning the possible theories and their solutions have greatly constrained deviations from GR and provided physicists with ideas about how to realise such deviations.

Generally, extending a theory can be done in one of two ways: one can either propose small deviations, with a clear limit through which one recovers the former theory; one can also propose a whole new theory based on completely different mathematical principles. The first way is usually technical and can lead to many dead-ends, but the second way is very hard to perform since it requires entirely new ideas — the kind of which Einstein had 120 years ago. In that sense, much of nowadays research is focused on the former way of extending GR: taking the same phenomenology, adding a small freedom, and seeing what are the consequences. The latter way has been also tried, through the development of string theory and loop quantum gravity, but it has been limited by the lack of confirmation of existing predictions and hypotheses in laboratory-based experiments for now.

This manuscript focuses on one particular kind of modified theory of gravity, that is useful in many different ways: the scalar-tensor theories of gravity, that describe gravity using the usual metric field plus an additionnal scalar field. Such theories have the very interesting feature that they constitute the easiest way to add some freedom to GR while keeping a simple GR limit (taking a vanishing scalar field). Such theories have been proposed initially by Brans and Dicke [6], and have been extremely well studied and generalised in the next decades: adding more freedom in the kinetic term led to the Bergmann-Wagoner theory [7], proposing that the action contain higher derivatives led to the Galileons theory [8], generalizing the Galileons onto curved spacetime led to Horndeski theories [9, 10, 11]... Finally, the most general scalar-tensor theory with higher derivatives, called Degenerate Higher-Order Scalar-Tensor (DHOST), was proposed recently in [12] by David Langlois and Karim Noui. These theories have been linked to many other theories of gravity in some specific cases.

## Tests of gravity using quasi-normal modes

While modifications of gravity have been proposed since the 1960s, they have known an incredible renewal of interest in the past 20 years, thanks to ever increasing precision in the experimental checks of GR. Indeed, two milestones have been reached respectively by the Planck and LIGO/Virgo collaborations: the measurement of the cosmic microwave background [13], that opens a window on the primordial Universe, and the observations of gravitational waves emitted from a binary black hole merger [14]. With these observations, GR enters a new phase of experimental evidence, that will lead to a great improvement of existing constraints in the upcoming years. It is therefore crucial for theoretical physicists that develop modified theories of gravity to compute physical predictions of their theories in order to compare them to what has and will soon be obtained.

A system of particular interest is a binary black hole (BBH) merger, since its dynamics is described by strong-regime gravity. As a consequence, a small modification of gravity would lead to different behaviour for such mergers, while it would not impede with the existing checks that have been performed at a Solar System scale. The time evolution of these systems can be split into three parts: the inspiral phase, during which the two black holes (BHs) orbit each other and get closer and closer; the merger, during which both BHs merge into a single one; the ringdown, during which the newly-formed BH settles down into a BH solution of the theory in vacuum.

In this manuscript, we focus on the last part, during which standard perturbation theory can be applied in order to compute the dynamics of the emitted waves. Such waves will have a discrete set of frequencies, called the quasi-normal modes (QNMs): these modes are expected to vary if the theory of gravity is not GR. The main goal of this thesis is to find a way to compute these modes in modified theories of gravity, such as the DHOST theories.

## Thesis outline

The outline of this manuscript is the following. In chapter 1, we give a review of the existing theories of modified gravity, explain in details how Horndeski and DHOST theories are constructed and explain their properties. In chapter 2, we write some nonrotating BH solutions that exist in these theories; these solutions are the ones we will study in the rest of the manuscript. In chapter 3, we describe a new way to construct DHOST theories that arises from a simple geometrical argument and allows us to understand physically the different stability results that were obtained for these theories. In chapter 4, we apply perturbation theory to BH spacetimes in GR in order to define the relevant quantities. The study of perturbations in modified gravity is then performed in chapters 6 and 7, using a specific algorithm that we discovered in the mathematical literature and summarised in chapter 5. Finally, a numerical computation of QNMs is proposed in chapter 8.

## Publications resulting from this thesis

This thesis led to the publication of three articles in peer-reviewed journals, two preprints on arXiv and one conference proceedings [15]. The references are given hereafter.

CHAPTER 1

# MODIFIED GRAVITY AND THE DHOST THEORIES

CONTENTS

## 1.1. Gravity, General Relativity and its limits

### 1.1.1. Einstein's theory of gravity

GRAVITY is the fundamental force that has been studied for the longest time, starting with Aristotle's considerations in 4$^{\text{th}}$ century BC. Its properties were studied by many different scientists throughout the following two millennia; while Al-Khwârizmî was the first to propose that gravity was mediated by an attractive force, its first mathematical description was proposed by Newton in 1687 [1] and featured an inverse square law: the force $\vec{F}$ applied by a mass $m_1$ on a mass $m_2$ was written in this framework as

$$\vec{F} = -\mathscr{G}\frac{m_1 m_2}{r^2}\vec{u}\,, \tag{1.1}$$

with $\vec{u}$ a unitary vector oriented from the mass $m_1$ to the mass $m_2$, $r$ the distance between those two masses and $\mathscr{G}$ Newton's constant. This theory led to many results that were consistently in agreement with experiments. Two of them were of particular importance:

- the proof of Kepler's laws concerning the motion of celestial bodies [16];
- the prediction of the existence of Neptune before its observation [17].

It therefore seemed, at the end of the XIX$^{\text{th}}$ century, that gravity was perfectly understood — as was the case for many other domains of physics.

However, the development of special relativity by Einstein in 1905 [4] changed the game: in this framework, no information can travel at a speed higher than the one of light; yet the force of eq. (1.1) changes immediately in all of space if one changes the mass $m_1$. Therefore, Newton's theory of gravity needed to be changed to account for special relativity.

This was attempted by several physicists between 1905 and 1915: Poincaré, Minkowski, Sommerfeld, Nordström, Abraham, Mie and Einstein. Important features were introduced step by step: the notion of 4-vectors and covariance appeared progressively in the works of Poincaré, Minkowski and Sommerfeld; the principle of equivalence was introduced by Einstein in 1907; the idea of gravity as a consequence of spacetime curvature was proposed by Nordström [18] (but understood as such only later by Einstein and Fokker [19]).

Let us present in more details the reasonings that led to the development of GR. The main ingredient is Einstein's Equivalence Principle (EEP), that can be formulated as follows: "it is impossible to tell the difference between a uniform gravitational field and an acceleration of the frame of reference". In other words, this means that the gravitational mass — the mass appearing in the gravitational interaction — is equal to the inertial mass — the mass appearing in the expression of the momentum. From this principle, one can infer that an object in free fall in a gradient of gravitational field will be redshifted when seen from infinity: this is the *gravitational redshift*, that links the description of time and the description of gravity.

This feature leads one to consider a full 4-dimensional theory of gravity; by consistency with special relativity, such a theory should exhibit Lorentz covariance, and therefore be built from scalars, 4-vectors or higher dimensonal 4-tensors. Several possibilities were investigated: a gravitational vector, a gravitational 6-vector (which was actually a rank 2 antisymmetric tensor), a scalar... All these theories contained specific issues [20], and the only theory that is generally covariant and satisfies the equivalence principle was finally proposed by Einstein in 1915 [5]: gravity was described by a rank 2 symmetric tensor, the metric $g_{\mu\nu}$ corresponding to the geometry of spacetime. The field equations for the metric were

$$E_{\mu\nu} + \Lambda g_{\mu\nu} = 8\pi\mathcal{G} T_{\mu\nu}\,, \tag{1.2}$$

with $E_{\mu\nu}$ the Einstein tensor defined by

$$E_{\mu\nu} = R_{\mu\nu} - \frac{1}{2} R g_{\mu\nu} \tag{1.3}$$

with $R_{\mu\nu}$ the Ricci tensor, $R$ the Ricci scalar, $T_{\mu\nu}$ the stress-energy tensor and $\Lambda$ a constant. One should note that units were chosen such that $c = 1$; such units will be used throughout the whole manuscript. The quantity $8\pi\mathscr{G}$ is such that

$$8\pi\mathscr{G} = \frac{1}{M_{\rm P}^2}\,, \tag{1.4}$$

where $M_{\rm P}$ is the *reduced Planck mass*. The equations of motion (1.2) can be deduced from applying the principle of least action to the action

$$S_{\rm GR}[g_{\mu\nu}] = \frac{M_{\rm P}^2}{2}\int {\rm d}^4x\,\sqrt{-g}\,(R-2\Lambda) + S_{\rm matter}\,, \tag{1.5}$$

with $S_{\rm matter}$ the action of all matter fields.

At the time of its proposal by Einstein, the only available test of the theory described by eq. (1.2) was the prediction of the advance of Mercury's perihelion; this theory was the only one that predicted it successfully, paving the way for a century of immense success in experimental tests. To this day, GR has been verified up to very high precision in various systems, on many scales, making it the most successfully tested theory along with quantum mechanics. More details on the various tests will be given in section 1.5.

### 1.1.2. Motivations for theories beyond GR

We are at present in many ways in a situation similar to 120 years ago: the widely accepted theory of General Relativity has been thoroughly tested, and it performed very successfully on different energy scales (see section 1.5). However, it has inherent flaws that need to be addressed in order to make it compatible with theories in other domains of physics, such as quantum physics. We now review a few of these flaws and show why their existence implies it is necessary to go beyond GR to fully describe gravity.

#### 1.1.2.1. Absence of quantum completion

A concerning issue of GR is that it does not have a coherent quantum field theory description at high energy. This is explained by the fact that if the energy density is higher than a given threshold, GR predicts that black holes are formed. However, these objects behave quite differently when compared to particles, which implies that gravity has a high energy regime that differs a lot from such regimes for quantum field theories [21]. Henceforth, techniques used in the latter context — in that case, renormalization — are no longer working for the former theory [22, 23, 24, 25]. This means that the ultraviolet (UV) limit of GR needs to be changed in order to account for quantum effects.

#### 1.1.2.2. Fine-tuning problem

GR contains another inconsistency with quantum field theory, that appears when one wants to describe cosmology. In cosmology, one assumes that the Universe is homogeneous and isotropic; with these hypotheses, the line element can be written as

$$\mathrm{d}s^2 = g_{\mu\nu}\,\mathrm{d}x^\mu\,\mathrm{d}x^\nu = -\,\mathrm{d}t^2 + a(t)^2\left(\frac{\mathrm{d}r^2}{1-kr^2} + r^2\,\mathrm{d}\Omega^2\right), \tag{1.6}$$

with $k \in \{-1, 0, 1\}$ and $\mathrm{d}\Omega^2$ the line element of a 2-dimensional sphere, while the function $a(t)$ is the *scale factor*. This ansatz is the Friedmann-Lemaître-Robertson-Walker (FLRW) metric. In this framework, it can be shown that the cosmological constant $\Lambda$ leads to accelerated increase of the scale factor $a(t)$. Such an accelerated increase is experimentally observed and allows one to measure $\Lambda$ experimentally. One obtains $\Lambda = (1.11 \pm 0.02) \times 10^{-52}\,\mathrm{m}^{-2}$ [13]. However, this value cannot be understood from a theoretical point of view.

To understand the problem, we can rewrite eq. (1.2):

$$E_{\mu\nu} = \frac{1}{M_\mathrm{P}^2} T_{\mu\nu} - \Lambda g_{\mu\nu}\,. \tag{1.7}$$

Using this form of the equations, we can interpret $\Lambda$ as a stress-energy density in all spacetime. This density has three possible sources when one takes into account the Standard Model (SM):

- a classical contribution $\Lambda_\mathrm{class}$ corresponding to the potential energy of all SM fields that have reached the minimum of their potential [1];
- a quantum contribution $\Lambda_\mathrm{quant}$ corresponding to the vacuum energy of all SM fields;
- a bare contribution $\Lambda_\mathrm{bare}$ that could appear in the theory.

The total cosmological constant is then given by

$$\Lambda = \Lambda_\mathrm{bare} + \Lambda_\mathrm{class} + \Lambda_\mathrm{quant}\,. \tag{1.8}$$

When one takes into account all SM fields, one obtains [26]

$$\Lambda_\mathrm{class} + \Lambda_\mathrm{quant} \sim -10^3\,\mathrm{m}^{-2}\,. \tag{1.9}$$

Therefore, in order to recover the measured value for $\Lambda$, the constant $\Lambda_\mathrm{bare}$ should be such that it cancels exactly the first 55 digits of the sum $\Lambda_\mathrm{class} + \Lambda_\mathrm{quant}$! Such a *fine-tuning problem*, while not technically pathological, is not a good feature of the theory and it is relevant to try to find a mechanism that gives either other contributions to the total $\Lambda$ or a way to "screen" $\Lambda$ in the case of cosmological experiments. Henceforth, an infrared (IR) modification of GR might be needed to solve this issue [27].

1. This energy cannot be set to zero in GR as energy is no longer relative in this theory: any contribution to $T_{\mu\nu}$ gravitates and thus cannot be removed!

#### 1.1.2.3. Singularities

Finally, existing solutions of GR exhibit singular behaviour, in two different ways: either the spacetime curvature becomes infinite somewhere in spacetime, or some worldlines end abruptly at some point of spacetime. This is for instance the case of the Schwarzschild solution which exhibits a singularity of the former type. The line element for a Schwarzschild black hole is

$$\mathrm{d}s^2 = -\left(1 - \frac{2M}{r}\right)\mathrm{d}t^2 + \left(1 - \frac{2M}{r}\right)^{-1}\mathrm{d}r^2 + r^2\,\mathrm{d}\Omega^2\ , \tag{1.10}$$

with $M$ an integration constant that corresponds to the ADM mass. Such a solution has a Kretschmann scalar $K$ that is singular at $r = 0$:

$$K = R^{\mu\nu\rho\sigma}R_{\mu\nu\rho\sigma} = \frac{48M^2}{r^6}\ . \tag{1.11}$$

This means that the solution must be modified when $r$ goes to zero. However, it can be proven that the Schwarzschild solution (1.10) is unique in GR when one makes some simple assumptions (see section 2.1.2); the only way to modify the solution is therefore to look for new theories that have solutions close to eq. (1.10) but are regular everywhere. A very similar problem appears when one considers cosmology: in the Big Bang model, the scale factor $a(t)$ of eq. (1.6) goes to zero at some time $t_{\mathrm{BB}}$. At this point, geodesics stop, which means that spacetime also contains a singularity.

#### 1.1.2.4. Different approaches for the modification of gravity

All-in-all, looking for new theories of gravity is necessary in order to find solutions to these problems. Two different approaches are possible: one can either try to find a different new theory that solves part of these problems — this is a "top-down" approach — or one can try to find all possible extensions of GR and see which ones have interesting features — this is a "bottom-up" approach.

Top-down approaches will be based on adding features that were absent in GR, for example providing the graviton with a mass or considering that the Ricci scalar appears no longer as $R$ in the action but as $f(R)$, where $f$ is an arbitrary function. The resulting theory will therefore differ from GR. Specific examples will be given in sections 1.2.4 and 1.2.5.

Bottom-up approaches can be seen as a search for a simple way to parametrize deviations from GR. As an example, let us assume that a coherent quantum theory of gravity exists. At the endpoint of the low energy limit, this theory must be equivalent to GR. Therefore, before one arrives at the GR limit, this theory must become GR to which some dynamical freedom is added in the form of a *new field*. This Effective Field Theory (EFT) approach has been motivated by the existence of interesting modified gravity limits of string theory [28, 29] and quantum gravity [30, 31, 32]. In this type of modification of gravity, new physics must appear around the Planck mass $M_\mathrm{P}$, which is the limit of validity of GR as an EFT. The goal of the modified gravity theory is then to *extend* the validity of the effective description up to a higher scale $\Lambda_\mathrm{cut}$.

It is also possible to modify GR in the IR regime in order to describe a different behaviour on cosmological scales. This setup leads to a different IR limit for gravity, which can also be described from an effective point of view as a modified gravity theory. Such modifications imply that GR is replaced by a new theory, and one should make sure this new theory has a well-behaved UV limit. Moreover, this EFT will be valid up to a given scale $\Lambda_\mathrm{cut}$, which will in general be lower [2] than $M_\mathrm{P}$. Both approaches are summed up on fig. 1.1.

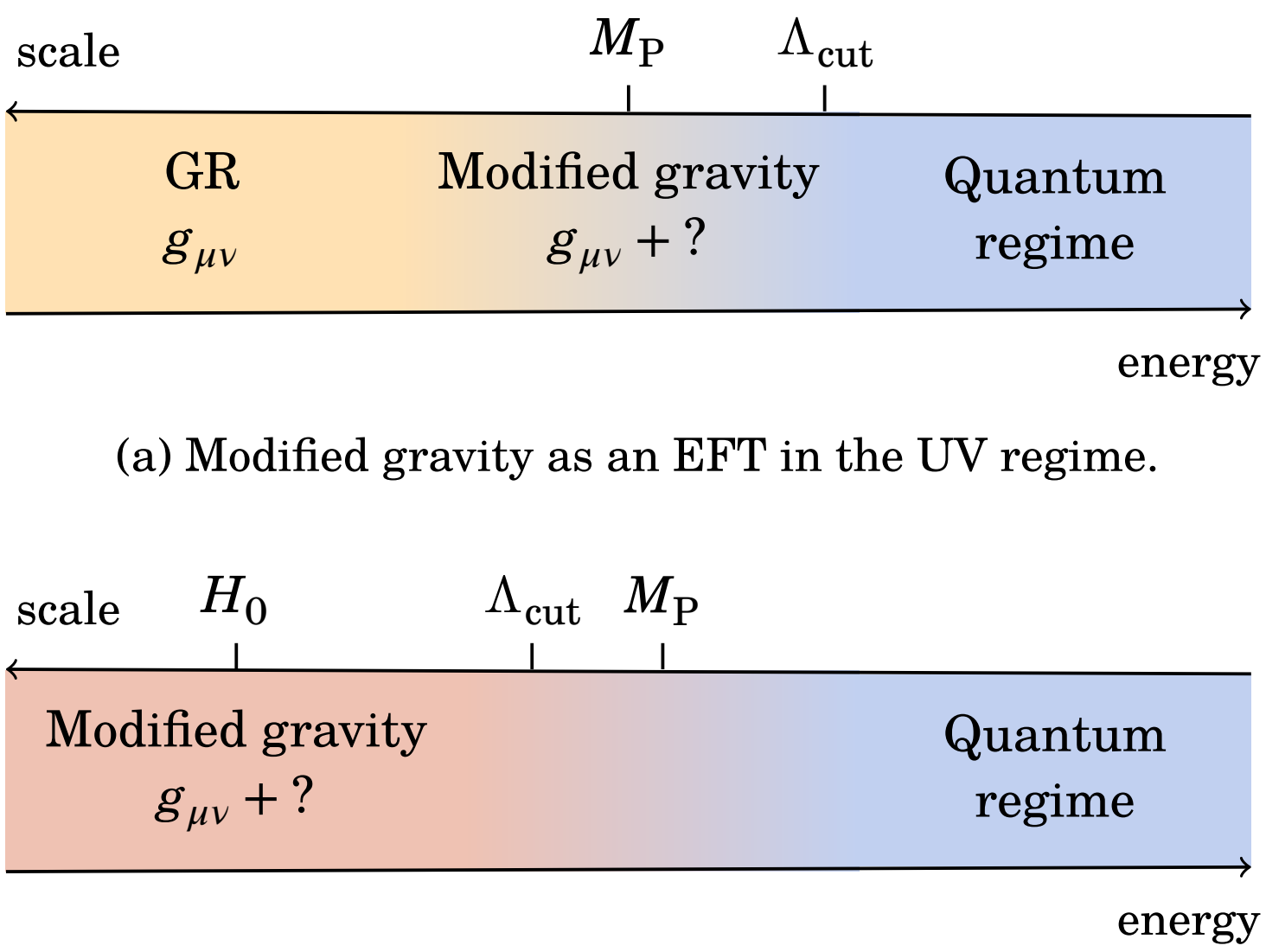

(a) Modified gravity as an EFT in the UV regime.

(b) Modified gravity as an EFT in the IR regime.

Figure 1.1. – GR is a coherent EFT up to the Planck mass scale. One can either use (a) a modified gravity theory in the UV to expand the validity of GR up to a higher scale, or (b) modify gravity in the IR to account for a different cosmology; in that case, one must make sure the theory's UV limit is well-defined. In both cases, $\Lambda_\mathrm{cut}$ is the limit of the modified theory of gravity. In the case of (b), one also wants to mimic GR on Solar System scales if $\Lambda_\mathrm{cut}$ is high enough: this can be done via several mechanisms such as the Vainshtein screening [33, 34].

2. It is possible to look for a modified theory of gravity improving both the IR and the UV limit, but most of the time this is not what will happen.

Finally, bottom-up approaches can be seen heuristically as a way to parametrize deviations from GR in order to develop new tests of this theory. This way of looking at the problem requires no motivation for the modification of GR: it is simply a way to obtain better tests that go beyond a simple null hypothesis check and compute actual deviations. From an experimental point of view, this is sufficient motivation.

## 1.2. Zoology of modified gravity theories

The search for a modified theory of gravity having been motivated, let us now look at the different tools one can use in order to obtain a coherent theory of gravity that is not GR. We then review a few of the theories that have been proposed since Einstein's publication in 1915.

### 1.2.1. Constructing modified gravity theories: Lovelock's theorem

As soon as Einstein proposed his theory of gravity in 1915, the uniqueness of the left-hand side of eq. (1.2) was studied, in order to understand if another theory with the same properties as GR could exist. Vermeil obtained a first result [35] that was then developed by Cartan and Weyl [36, 37]: let us consider a rank-2 tensor $A_{\mu\nu}$, and assume that

1. $A_{\mu\nu}$ is symmetric: $A_{\mu\nu} = A_{\nu\mu}$;
2. $A_{\mu\nu}$ is conserved, or divergence-free: $\nabla^\mu A_{\mu\nu} = 0$;
3. $A_{\mu\nu}$ depends only on the metric and its first and second derivatives;
4. $A_{\mu\nu}$ transforms covariantly under diffeomorphisms;
5. the dependence of $A_{\mu\nu}$ on the second order derivative of the metric is linear;

then in any dimension the tensor $A_{\mu\nu}$ must be of the form

$$A_{\mu\nu} = \alpha g_{\mu\nu} + \beta E_{\mu\nu}\,, \tag{1.12}$$

with $\alpha$ and $\beta$ some constants.

In order to understand how powerful this theorem is, let us motivate its different assumptions. First, item 1 is a very natural assumption to make if one requires that the equations of motion can be obtained from the variation of an action $S$. Indeed, in that case the equations of motion are of the form $\delta S/\delta g^{\mu\nu} = 0$, and are therefore symmetric since the metric itself is. Then, item 3 is coherent with the fact that most equations in physics imply derivatives up to the second order only. Last, the requirement that $A_{\mu\nu}$ preserve diffeomorphism invariance means that no observer should be able to infer their global position in spacetime via a local measurement.

We see that items 1, 3 and 4 are deeply linked to the original postulates made by Einstein when he proposed GR. Item 5 can however look like an oversimplification. Nonetheless, it was proven by Lovelock in 1971 that one can remove this assumption and replace it by the assumption that spacetime is 4-dimensional [38]. Furthermore, one can prove that if the equations of motion derive from an action $S$, item 4 and item 2 are equivalent [39]: indeed, one obtains in that case

$$\delta S = \int \mathrm{d}^4x\,\sqrt{-g}\,\frac{\delta S}{\delta g^{\mu\nu}}\,\delta g^{\mu\nu} = 0\,. \tag{1.13}$$

This would lead to the field equations $A_{\mu\nu} = 0$ by defining $\delta S/\delta g^{\mu\nu} = A_{\mu\nu}$. Imposing diffeomorphism invariance then corresponds to the choice $\delta g^{\mu\nu} = \nabla^\mu \xi^\nu$, with $\xi^\nu$ some vector field. One writes

$$\int \mathrm{d}^4x\,\sqrt{-g} A_{\mu\nu} \nabla^\mu \xi^\nu = 0\,, \tag{1.14}$$

which gives after integration by parts

$$\int \mathrm{d}^4x\,\sqrt{-g}\,(\nabla^\mu A_{\mu\nu})\,\xi^\nu = 0\,. \tag{1.15}$$

This being true for any diffeomorphism, one obtains $\nabla^\mu A_{\mu\nu} = 0$.

It is therefore possible to simplify the hypotheses of the theorem, leading to what is now called Lovelock's theorem:

**Lovelock's theorem.** *In four spacetime dimensions, the only rank-2 tensor constructed solely from the metric $g_{\mu\nu}$ and its derivatives up to second differential order, preserving diffeomorphism invariance and arising from a least action principle is the Einstein tensor plus a cosmological term.*

This theorem is a strong uniqueness statement. However, it can be "turned around" and made into a very convenient toolbox that one can use when one wants to construct new theories of gravity. Indeed, it is sufficient to break one of its hypotheses in order to get a plethora of new theories of gravity. Modified theories of gravity will be classified into categories depending on which hypothesis of Lovelock's theorem they violate [40]:

- some theories will make use of new fields in order to build the left-hand side of eq. (1.2), removing the constraints on its structure as a rank-2 tensor;
- theories that violate diffeomorphism invariance will be classified into several subcategories depending on the way they violate it (massive gravity, Lorentz-violating theories...);

– finally, theories involving higher-dimensional spacetimes will have their own category that requires some compactification procedure in order to recover our usual 4-dimensional spacetime.

In fig. 1.2, we give a summary of all the different ways Lovelock's theorem can be broken, and a few examples of the resulting theories [3]. In the following sections, we give several examples of such theories, with a specific emphasis on the scalar-tensor theories of gravity which we will study for the rest of this manuscript. We also give a few examples of other theories that give a specific scalar-tensor theory in some limit, in order to illustrate how general the former theories are.

### 1.2.2. Horndeski theories

Horndeski theories belong to the class of modified gravity theories that add new fields in order to circumvent the hypotheses of Lovelock's theorem, while keeping all other hypotheses verified. They were proposed by Horndeski in 1974 [9] and unearthed recently [11]. They rely on the addition of the simplest possible field to the dynamics: a scalar field $\phi$. For this reason, they belong to a class of theories dubbed "scalar-tensor" theories. This scalar field will couple nonminimally to the metric, leading to qualitatively different behaviour when compared to GR. These theories are quite interesting as bottom-up approaches to modified gravity since they provide the most general way to add a scalar to GR without adding higher derivatives in the action and assuming anything about the scalar dynamics.

Before describing the Horndeski action, we review the historical construction of scalar-tensor theories. The easiest way to add a scalar field to the Einstein-Hilbert action is to provide it with a kinetic term and a potential $V$:

$$S[g_{\mu\nu}, \phi] = \int \mathrm{d}^4x \sqrt{g}\left(R + \frac{1}{2}X - V(\phi)\right), \tag{1.16}$$

where we have defined

$$X = \phi_\mu \phi^\mu \quad \text{and} \quad \phi_\mu = \nabla_\mu \phi . \tag{1.17}$$

However, this does not yield a modification of gravity: it corresponds only to the presence of a matter field. In order to couple gravity and this new scalar field, one can for example assume that the gravitational constant $\mathcal{G}$ has to be replaced by the scalar field, which would in that case describe a spatial variation of Newton's constant. This was done by Brans and Dicke in [6], building on a theory proposed by Jordan [41], and led to the action

$$S_{\mathrm{BD}}[g_{\mu\nu}, \phi] = \int \mathrm{d}^4x \sqrt{-g}\left(\phi R - \frac{\omega}{\phi}X\right). \tag{1.18}$$

3. One should note that many of the resulting theories are equivalent in some way, as will be proven in section 1.3.

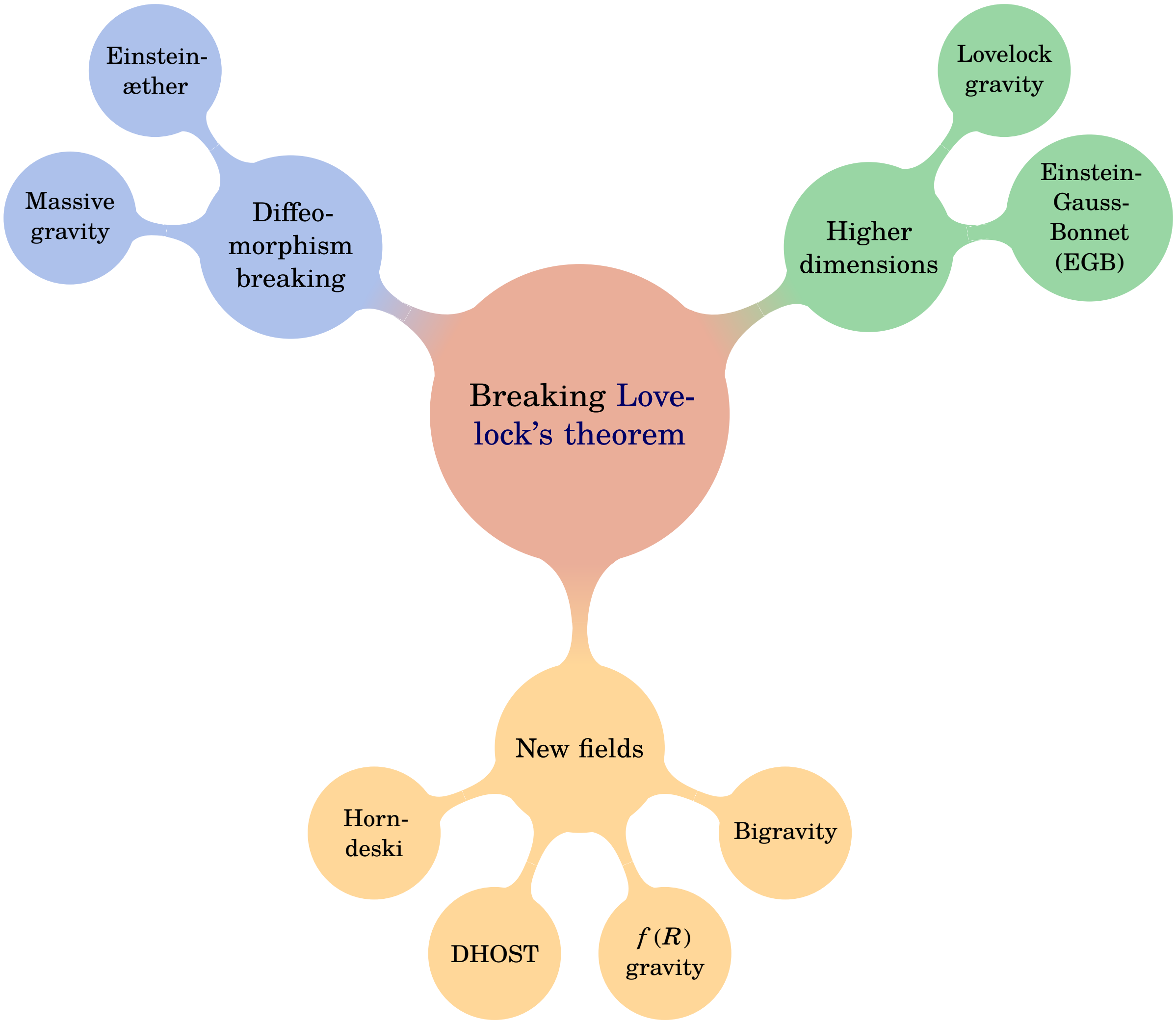

Figure 1.2. – Mindmap of the different ways Lovelock's theorem can be broken, with a few examples of the resulting theories.

This can be further generalized: one can add arbitrary functions of $\phi$ in front of each term of the action (1.16). Removing the extra freedom in the definition of those functions, one obtains the so-called Bergmann-Wagoner theory [7, 42]:

$$S_{\mathrm{BW}}[g_{\mu\nu},\phi] = \int \mathrm{d}^4x\,\sqrt{-g}\left(\phi R - \frac{\omega(\phi)}{\phi}X - V(\phi)\right), \tag{1.19}$$

which is very similar to the Brans-Dicke action in which $\omega$ was upgraded to a function and a potential $V$ was added. This theory was then generalized in [8] in order to reproduce the behaviours obtained in some existing theory of massive gravity (the DGP model [43], see section 1.2.5): called "Galileons" (due to the specific Galilean invariance of its building blocks), it contained higher-derivative terms $\phi_{\mu\nu}$, and was valid only on a flat space background. Despite the presence of higher derivatives, the equations of motion were still only of second order thanks to specific cancellations of the higher derivative terms constructed using the totally antisymmetric tensor $\varepsilon^{\mu\nu\rho\sigma}$ [44]. As a consequence, only terms quadratic and cubic in the second derivatives of $\phi$ could appear in the action.

The Galileons being only defined in flat spacetimes, generalizing their construction in curved spacetimes was necessary. This was done in [10], recovering a result from Horndeski [9]. In order to keep the quadratic and cubic dependency [4] in $\phi_{\mu\nu}$, one poses the Horndeski action:

$$S[g_{\mu\nu},\phi] = \int \mathrm{d}^4x\,\sqrt{-g}\Big(P + Q\Box\phi + L^{(2)}_{\text{Horn}} + L^{(3)}_{\text{Horn}}\Big)\,, \tag{1.20}$$

with

$$\begin{aligned} L^{(2)}_{\text{Horn}} &= FR + 2F_X(\phi_{\mu\nu}\phi^{\mu\nu} - (\Box\phi)^2)\,,\\ L^{(3)}_{\text{Horn}} &= GE_{\mu\nu}\phi^{\mu\nu} + \frac{1}{3}G_X((\Box\phi)^3 - 3\Box\phi\phi_{\mu\nu}\phi^{\mu\nu} + 2\phi_{\mu\rho}\phi^{\rho\nu}\phi_{\nu}{}^{\mu})\,. \end{aligned} \tag{1.21}$$

The quantities $F$, $P$, $Q$ and $G$ are functions of $\phi$ and $X$, and we introduced the notations

$$\phi_{\mu\nu} = \nabla_\mu\nabla_\nu\phi\,, \quad F_X = \frac{\partial F}{\partial X}\,. \tag{1.22}$$

One should note that although second derivatives of the scalar field appear in the action, the equations of motion will still be of second order only thanks to a specific cancellation of the higher-order terms.

We will refer to *quadratic* Horndeski theories when $G(\phi,X) = 0$, and *cubic* Horndeski theories when this is not the case. When one also has $Q = 0$ and $F = 1$, the theory is called K-essence [45]. It is also common to consider only Horndeski theories that are invariant under a constant shift of the scalar field

$$\phi \longrightarrow \phi + c\,, \tag{1.23}$$

with $c$ a constant. In such theories, dubbed *shift-symmetric* theories, the Horndeski functions $F$, $P$, $Q$ and $G$ cannot depend on $\phi$ and are therefore functions of $X$ only. Henceforth, the action for quadratic shift-symmetric Horndeski theories is

$$S[g_{\mu\nu},\phi] = \int \mathrm{d}^4x\,\sqrt{-g}\Big[F(X)R + P(X) + Q(X)\Box\phi$$

4. It is possible in general to look at scalar-tensor theories depending on $(\phi_{\mu\nu})^n$ with any integer $n$. However, when $n > 4$, these theories do not contain physics beyond the Horndeski theories presented here in the flat space limit, which indicates that they might not be physically relevant.

$$+ 2F_X(X)\left(\phi_{\mu\nu}\phi^{\mu\nu} - (\Box\phi)^2\right)\Big] . \quad (1.24)$$

The motivation for such a simplification is a better consistency with observations in some cases [46]. Typically, in a cosmological setting, the field $\phi$ would relax to a constant if the theory explicitly depends on $\phi$. In a shift-symmetric theory depending only on $X$, only this field would relax to a constant, allowing $\phi$ to evolve (driving, for example, an accelerated expansion) [47]. A similar feature is observed in the context of BHs, where scalar field ansätze linearly depending on time are proposed for solutions of shift-symmetric theories (see chapter 2).

### 1.2.3. DHOST: the most general scalar-tensor theories

#### 1.2.3.1. DHOST action

Degenerate Higher-Order Scalar-Tensor theories are a generalization of the Horndeski theories presented in section 1.2.2 [5]. These theories are constructed by relaxing two hypotheses of Lovelock's theorem: gravity is supposed to be described by the metric $g_{\mu\nu}$ plus an additional scalar field $\phi$, while the equations of motion are no longer required to be of second order. The Lagrangian density of the theory is therefore built using all possible combinations of second-order derivatives of the scalar field. In practice, we restrict ourselves to terms with at most a cubic dependency on $\phi_{\mu\nu}$ for simplicity:

$$S[g_{\mu\nu}, \phi] = \int \mathrm{d}^4x\, \sqrt{-g}\left(P + Q\Box\phi + L^{(2)}_{\mathrm{DHOST}} + L^{(3)}_{\mathrm{DHOST}}\right), \quad (1.25)$$

with

$$L^{(2)}_{\mathrm{DHOST}} = F_2 R + \sum_{i=1}^{5} A_i L_i^{(2)}, \quad L^{(3)}_{\mathrm{DHOST}} = F_3 E_{\mu\nu}\phi^{\mu\nu} + \sum_{i=1}^{10} B_i L_i^{(3)} . \quad (1.26)$$

The parameters $F_2, F_3, A_i$ and $B_i$ are functions of $\phi$ and $X$. The 5 Lagrangian densities $L_i^{(2)}$ are quadratic in the second derivatives of the scalar field:

$$L_1^{(2)} = \phi_{\mu\nu}\phi^{\mu\nu}, \quad L_2^{(2)} = (\Box\phi)^2, \quad L_3^{(2)} = (\Box\phi)\phi^\mu\phi_{\mu\nu}\phi^\nu,$$
$$L_4^{(2)} = \phi^\mu\phi_{\mu\rho}\phi^{\rho\nu}\phi_\nu, \quad L_5^{(2)} = (\phi^\mu\phi_{\mu\nu}\phi^\nu)^2, \quad (1.27)$$

while the 10 elementary Lagrangian densities $L_i^{(3)}$ are cubic in the second derivatives of the scalar field and are given by

$$L_1^{(3)} = (\Box\phi)^3, \quad L_2^{(3)} = (\Box\phi)\,\phi_{\mu\nu}\phi^{\mu\nu}, \quad L_3^{(3)} = \phi_{\mu\nu}\phi^{\nu\rho}\phi^\mu_\rho,$$
$$L_4^{(3)} = (\Box\phi)^2\,\phi_\mu\phi^{\mu\nu}\phi_\nu, \quad L_5^{(3)} = \Box\phi\,\phi_\mu\phi^{\mu\nu}\phi_{\nu\rho}\phi^\rho, \quad L_6^{(3)} = \phi_{\mu\nu}\phi^{\mu\nu}\phi_\rho\phi^{\rho\sigma}\phi_\sigma,$$

5. As an intermediate step to the construction of DHOST theories, Horndeski theories were first generalized into Beyond Horndeski (also called GLPV) theories [48, 49].

$$L_7^{(3)} = \phi_\mu \phi^{\mu\nu} \phi_{\nu\rho} \phi^{\rho\sigma} \phi_\sigma \,, \quad L_8^{(3)} = \phi_\mu \phi^{\mu\nu} \phi_{\nu\rho} \phi^\rho \, \phi_\sigma \phi^{\sigma\lambda} \phi_\lambda \,,$$
$$L_9^{(3)} = \Box\phi \left(\phi_\mu \phi^{\mu\nu} \phi_\nu\right)^2 \,, \quad L_{10}^{(3)} = \left(\phi_\mu \phi^{\mu\nu} \phi_\nu\right)^3 \,. \tag{1.28}$$

The functions $F_2$, $F_3$, $A_i$ and $B_i$ cannot be chosen independently: in order to avoid instabilities, they have to satisfy degeneracy conditions. More details will be given in sections 1.2.3.2 and 1.2.3.3.

Similarly to the Horndeski case, we will talk about *shift-symmetric* theories when these functions do not depend explicitly on $\phi$; the subcategory of theories for which $F_3$ and all the $B_i$ are zero is called "quadratic DHOST theories", while the theories are called "cubic" if some of these functions are nonzero. One should note that DHOST theories with quartic $L_i^{(4)}$, quintic $L_i^{(5)}$ or even higher-order Lagrangians can be studied. However, such theories would be extremely complicated [6] (there are 5 elementary Lagrangians at quadratic order in $\phi_{\mu\nu}$, 10 at third order, 20 at fourth order, 37 at fifth order...). Furthermore, as explained in section 1.2.2, such theories would not have a relevant nondegenerate Galileon limit.

#### 1.2.3.2. Evading the Ostrogradsky instability

The addition of higher derivatives in the action will lead in general to higher derivatives in the equations of motion. The presence of such higher derivatives is a problem in general because it will imply the presence of an Ostrogradski "ghost", that will manifest itself as an unstable degree of freedom [50, 51].

One can understand this feature using a simple toy model [51, 52]. Let us consider a Lagrangian depending on some field $\phi(t)$ and its second time derivative:

$$L = \frac{1}{2} a \ddot{\phi}^2 - V(\phi) \,, \tag{1.29}$$

with $V$ some potential and $a$ a nonzero constant. The Euler-Lagrange equation for this Lagrangian is

$$a \ddddot{\phi} = \frac{\mathrm{d}V}{\mathrm{d}\phi} \,. \tag{1.30}$$

This equation is fourth order: it is equivalent to two second-order equations, which means that the system contains two degrees of freedom. To see this more intuitively from eq. (1.29), one can introduce an auxiliary field $\psi$ whose equation leads to $\psi = \ddot{\phi}$:

$$L = a\psi\ddot{\phi} - \frac{1}{2} a \psi^2 - V(\phi) \,. \tag{1.31}$$

6. While such theories are complicated to build from scratch using quartic or quintic Lagrangians, it is possible to build arbitrary-order DHOST theories from mimetic gravity theories as explained in [32].

Finally, one can perform an integration by parts on $L$ in order to find a new form for it (not taking boundary terms into account):

$$L = -a\dot{\psi}\dot{\phi} - \frac{1}{2}a\psi^2 - V(\phi)\,. \tag{1.32}$$

One can then define $q_+ = (\phi+\psi)/\sqrt{2}$ and $q_- = (\phi-\psi)/\sqrt{2}$, yielding

$$L = -\frac{1}{2}a\dot{q}_+^2 + \frac{1}{2}a\dot{q}_-^2 - U(q_+,q_-)\,, \tag{1.33}$$

$$U(q_+,q_-) = -\frac{1}{2}a\left(\frac{q_+-q_-}{\sqrt{2}}\right)^2 - V\left(\frac{q_++q_-}{\sqrt{2}}\right)\,.$$

One can clearly identify two degrees of freedom in this Lagrangian, but these two degrees of freedom have opposite signs for their kinetic terms, which means that one of them is unstable: the associated degree of freedom is an "Ostrogradsky ghost". Even though this proof concerned a simplified case, the result is very general and stays true for third-order equations of motion [53].

However, it was proven in [12] (see also [54]) that when a Lagrangian contains several degrees of freedom with higher derivatives, there are cases where no ghosts are present, making the theory relevant again. This relies on degeneracies in the higher derivatives, leading to constraints that prevent Ostrogradsky ghosts from appearing. Therefore, one can still study higher derivatives theories as long as the higher derivative terms in the theory verify some specific degeneracy relations.

While apparently trivial, the argument for instability is not so easy to make rigorously. Indeed, if no interaction is present ($U = 0$), both fields can oscillate without any problem. The issue appears when interactions are turned on and when the theory is studied from a quantum point of view [55]: one can prove that the Hamiltonian is unbounded from below, authorizing the unending creation of pairs of negative-positive energy particles.

One should nevertheless note that in covariant theories, invariance by time reparametrization imposes that the Hamiltonian be zero on-shell. One might wonder in that case if the presence of an Ostrogradsky ghost is really problematic; a detailed study of this case is given in [55].

In the case of DHOST theories, the degeneracy relations were found in [12, 56]. They separate the theories into several classes that we give in the following section.

#### 1.2.3.3. Classification of quadratic DHOST theories

We start with the expression of the degeneracy conditions for quadratic DHOST theories, for which $F_3$ and all $B_i$ are zero. As shown in [12, 57, 58], the degeneracy conditions for such theories are written

$$D_0(X) = 0\,, \quad D_1(X) = 0\,, \quad D_2(X) = 0\,, \tag{1.34}$$

with

$$\begin{aligned}
D_0(X) &= -4(A_1 + A_2)\left[XF_2(2A_1 + XA_4 + 4F_{2X}) - 2F_2^2 - 8X^2F_{2X}^2\right], \\
D_1(X) &= 4\left[X^2A_1(A_1 + 3A_2) - 2F_2^2 - 4XF_2A_2\right]A_4 + 4X^2F_2(A_1 + A_2)A_5 \\
&\quad + 8XA_1^3 - 4(F_2 + 4XF_{2X} - 6XA_2)A_1^2 - 16(F_2 + 5XF_{2X})A_1A_2 \\
&\quad + 4X(3F_2 - 4XF_{2X})A_1A_3 - X^2F_2A_3^2 + 32F_{2X}(F_2 + 2XF_{2X})A_2 \\
&\quad - 16F_2F_{2X}A_1 - 8F_2(F_2 - XF_{2X})A_3 + 48F_2F_{2X}^2\,, \\
D_2(X) &= 4\left[2F_2^2 + 4XF_2A_2 - X^2A_1(A_1 + 3A_2)\right]A_5 + 4A_1^3 \\
&\quad + 4(2A_2 - XA_3 - 4F_{2X})A_1^2 + 3X^2A_1A_3^2 - 4XF_2A_3^2 \\
&\quad + 8(F_2 + XF_{2X})A_1A_3 - 32F_{2X}A_1A_2 + 16F_{2X}^2A_1 \\
&\quad + 32F_{2X}^2A_2 - 16F_2F_{2X}A_3\,.
\end{aligned} \tag{1.35}$$

One can see from eq. (1.35) that eq. (1.34) can be verified by several different choices of conditions linking the $A_i$ together; each choice will correspond to a given class.

**Class I** All subclasses of this class verify the relation $A_2 = -A_1$.

**Subclass Ia** This subclass is defined by $F_2 \neq XA_1$. One can use the conditions $D_1(X) = 0$ and $D_2(X) = 0$ in order to obtain

$$\begin{aligned}
A_4 &= \frac{1}{8(F_2 - XA_1)^2}\Big[-16XA_1^3 + 4(3F_2 + 16XF_X)A_1^2 - X^2F_2A_3^2 \\
&\quad - (16X^2F_{2X} - 12XF_2)A_3A_1 - 16F_{2X}(3F_2 + 4XF_{2X})A_1 \\
&\qquad + 8F_2(XF_{2X} - F_2)A_3 + 48F_2F_{2X}^2\Big]\,, \\
A_5 &= \frac{1}{8(F_2 - XA_1)^2}(4F_{2X} - 2A_1 + XA_3)(-2A_1^2 - 3XA_1A_3 \\
&\qquad + 4F_{2X}A_1 + 4F_2A_3)\,.
\end{aligned} \tag{1.36}$$

**Subclass Ib** This subclass is defined by $F_2 = XA_1$. One obtains $A_3 = 2(F_2 - 2XF_{2X})/X^2$ in order to satisfy the degeneracy conditions. However, the degeneracy appears in the metric sector: it removes one polarisation of gravitational waves (GWs) and not the Ostrogradsky ghost, which means that this subclass is not relevant [57].

**Class II** This class is characterized by $F_2 \neq 0$ and $A_2 \neq A_1$.

**Subclass IIa** This subclass is described by $F_2 \neq XA_1$. Using the conditions $D_1(X) = 0$ and $D_2(X) = 0$, one obtains

$$
\begin{aligned}
A_3 &= \frac{1}{X^2 F_2}\Big[ -4F_2(F_2 - XF_{2X}) - 2X(F_2 - 2XF_{2X}) \\
&\qquad\qquad\qquad\qquad - 4X^2 F_{2X}(A_1 + 3A_2)\Big], \\
A_4 &= \frac{2}{X^2 F_2}\Big[F_2^2 - 2F_2 X F_{2X} + 4X^2 F_{2X}^2 - XF_2 A_1\Big], \\
A_5 &= \frac{1}{X^2 F_2}\Big[4F_2(F_2^2 - 3F_2 X F_{2X} + 2X^2 F_{2X}^2) \\
&\quad + (3XF_2^2 - 8X^2 F_2 F_{2X} + 6X^3 F_{2X}^2)A_1 + 2X(2F_2 - 3XF_{2X})^2 A_2\Big].
\end{aligned}
\tag{1.37}
$$

**Subclass IIb** In this subclass, one has $F_2 = XA_1$. This implies

$$
\begin{aligned}
A_4 &= 4F_{2X}(2F_{2X}/F_2 - 1/F_2), \\
A_5 &= \frac{1}{4X^3 F_2(F_2 + XA_2)}\Big[8X(4XF_{2X}F_2 - F_2^2 - 4X^2 F_{2X}^2)A_2 \\
&\qquad + XF_2(8X^2 F_{2X} + X^3 A_3 - 4F_2)A_3 \\
&\qquad\quad + 4(XF_{2X}F_2^2 - 2X^3 F_{2X}^3 + 2X^2 F_{2X}^2 F_2 - F_2^3)\Big]
\end{aligned}
\tag{1.38}
$$

Similarly to class Ib, the degeneracy in this subclass concerns the metric sector and not the Ostrogradsky ghost.

**Class III** The last case is defined by $F_2 = 0$, and contains three subclasses.

**Subclass IIIa** This subclass is defined by $A_1 + 3A_2 \neq 0$, and is such that

$$
A_4 = -\frac{2}{X}A_1 \quad \text{and} \quad \frac{4A_1^2 + 8A_1A_2 - 4A_1A_3X + 3A_3^2X^2}{4X^2(A_1 + 3A_2)}. \tag{1.39}
$$

One should note that the intersection of classes Ia and IIIa is not empty.

**Subclass IIIb** This subclass corresponds to the case $A_1 + 3A_2 = 0$, for which one has $A_2 = -XA_3/2$.

**Subclass IIIc** Finally, this subclass corresponds to the choice $A_1 = 0$. Its metric sector is also degenerate.

One can notice that all subclasses depend on five arbitrary functions, except for subclass IIIc that depends on six. Such a classification is more than a simple result of the expression of the degeneracy conditions. Indeed, it was proven in [57] that all the previous subclasses are stable under specific transformations of the metric, called disformal transformations. This property will be studied in section 1.4.1. Moreover, this classification of DHOST theories also allows to simplify the study of such theories since classes II and III can be shown to be physically unviable: either tensor modes have pathological behaviour [59] or gradient instabilities of cosmological perturbations are present [60].

The equations defining each class are quite involved, and one might wonder what physical meaning is hidden behind such complicated combinations of $F_2$ and the $A_i$. It is actually possible to rewrite the action of eq. (1.25) in the case of quadratic DHOST theories ($L^{(3)}_{\rm DHOST} = 0$) in a geometrical way, leading to very simple degeneracy conditions. This work is presented in chapter 3.

#### 1.2.3.4. Classification of cubic DHOST theories

The classification of cubic DHOST theories was done in [56]. It is quite involved, and not useful for the present manuscript, which is why we will not give details about the different classes and their properties. The main result is that when the cubic DHOST Lagrangian is added to a quadratic theory of class Ia, several relations must be verified in order to still have a degenerate theory:

$$
\begin{aligned}
&B_1 \neq 0\,, \quad F_2 \neq 0\,, \quad B_2 = -3B_1\,, \quad B_3 = 2B_1\,, \quad B_6 = -B_4\,,\\
&B_4 = \frac{-A_1 F_{3X} X - 6B_1 F_2 + 6B_1 F_{2X} X + 2F_2 F_{3X}}{F_2 X}\,,\\
&B_5 = \frac{2(F_{3X} - 3B_1)^2 - 2B_4 F_{3X} X}{3B_1 X}\,, \quad B_7 = \frac{2B_4 F_{3X} X - 2(F_{3X} - 3B_1)^2}{3B_1 X}\,,\\
&B_8 = \frac{2}{9B_1^2 X^2}(3B_1 + B_4 X - F_{3X})\left((F_{3X} - 3B_1)^2 - B_4 F_{3X} X\right),\\
&B_9 = \frac{2B_4(3B_1 + B_4 X - F_{3X})}{3B_1 X}\,, \quad B_{10} = \frac{2B_4(3B_1 + B_4 X - F_{3X})^2}{9B_1^2 X^2}\,,\\
&A_3 = \frac{2}{3B_1 F_2 X^2}\Big[B_1(9A_1 F_2 X - 12A_1 F_{2X} X^2 + 6F_2 F_{2X} X - 6F_2^2)\\
&\qquad\qquad + 2F_{3X}(F_2 - A_1 X)^2\Big]\,.
\end{aligned}
\tag{1.40}
$$

### 1.2.4. Higher dimensional theories

As explained previously, a simple way to break the unicity statement of Lovelock's theorem is to consider a spacetime with more than 4 dimensions. In this situation, the most general action fulfilling the other hypotheses in $d$ dimensions can be found as a generalization of eq. (1.5). It was first obtained by Lovelock in 1971 [38] and can be written as

$$
S[g_{\mu\nu}] = \frac{M_{\rm P}^2}{2}\int \mathrm{d}^d x\,\sqrt{-g}\sum_{i=0}^{m-1}\frac{\alpha_i}{2^i}\delta^{\mu_1\cdots\mu_{2i}}_{\nu_1\cdots\nu_{2i}} R_{\mu_1\mu_2}{}^{\nu_1\nu_2}\times\cdots\times R_{\mu_{2i-1}\mu_{2i}}{}^{\nu_{2i-1}\nu_{2i}}\,, \tag{1.41}
$$

where the $\alpha_i$ are constants and $m$ is $d/2$ if $d$ is even and $d+1/2$ if $d$ is odd. The tensor $\delta^{\mu\mu_1\cdots\mu_N}_{\nu\nu_1\cdots\nu_N}$ is defined by

$$\delta^{\mu_1\cdots\mu_N}_{\nu_1\cdots\nu_N} = \det\begin{pmatrix} \delta^{\mu_1}_{\nu_1} & \cdots & \delta^{\mu_1}_{\nu_N} \\ \vdots & \ddots & \vdots \\ \delta^{\mu_N}_{\nu_1} & \cdots & \delta^{\mu_N}_{\nu_N} \end{pmatrix} = -\varepsilon^{\mu_1\cdots\mu_N}\varepsilon_{\nu_1\cdots\nu_N}\,; \tag{1.42}$$

by convention, the case $i=0$ corresponds to a constant contribution $\sqrt{-g}\alpha_0$. One observes that in the case $d=4$, or $m=2$, one recovers the result of eq. (1.5): indeed, in this case, the action of eq. (1.41) is such that

$$\begin{aligned} S[g_{\mu\nu}] &= \frac{M_\mathrm{P}^2}{2}\int \mathrm{d}^4x\,\sqrt{-g}\Big[\alpha_0 + \frac{\alpha_1}{2}\delta^{\mu_1\mu_2}_{\nu_1\nu_2}R_{\mu_1\mu_2}{}^{\nu_1\nu_2}\Big]\,, \\ &= \frac{M_\mathrm{P}^2}{2}\int \mathrm{d}^4x\,\sqrt{-g}\Big[\alpha_0 + \frac{\alpha_1}{2}(\delta^{\mu_1}_{\nu_1}\delta^{\mu_2}_{\nu_2} - \delta^{\mu_2}_{\nu_1}\delta^{\mu_1}_{\nu_2})R_{\mu_1\mu_2}{}^{\nu_1\nu_2}\Big]\,, \\ &= \frac{M_\mathrm{P}^2}{2}\int \mathrm{d}^4x\,\sqrt{-g}\Big[\alpha_0 + \frac{\alpha_1}{2}R_{\mu_1\mu_2}{}^{\mu_1\mu_2} - R_{\mu_1\mu_2}{}^{\mu_2\mu_1}\Big]\,, \\ &= \frac{M_\mathrm{P}^2}{2}\int \mathrm{d}^4x\,\sqrt{-g}\Big[\alpha_0 + \alpha_1 R\Big]\,. \end{aligned} \tag{1.43}$$

The number of terms in the sum of eq. (1.41) is $m$. This implies that as soon as the number of dimensions reaches 5, one new term can be added to the action. This new term will be a Lagrangian density written as

$$\begin{aligned} L &= \frac{\alpha_2}{4}\sqrt{-g}\delta^{\mu_1\mu_2\mu_3\mu_4}_{\nu_1\nu_2\nu_3\nu_4}R_{\mu_1\mu_2}{}^{\nu_1\nu_2}R_{\mu_3\mu_4}{}^{\nu_3\nu_4}\,, \\ &= \alpha_2\sqrt{-g}\big(R_{\mu_1\mu_2\mu_3\mu_4}R^{\mu_1\mu_2\mu_3\mu_4} - 4R_{\mu_1\mu_2}R^{\mu_1\mu_2} + R^2\big)\,, \\ &= \alpha_2\sqrt{-g}\mathscr{G}\,, \end{aligned} \tag{1.44}$$

where $\mathscr{G}$ is called the *Gauss-Bonnet* term.

Such theories of gravity in higher-dimensional spacetimes have an obvious drawback: the spacetime in which we live is 4-dimensional from an experimental point of view, so some process "screening" $d-4$ of the dimensions in order to get 4-dimensional experiments must be at work. In practice, one often uses compactification of the extra dimensions; more details will be given in section 1.3.2.

Furthermore, while the Lagrangian density of eq. (1.44) is relevant only for spacetimes of dimensions 5 or more, it is defined also for $d=4$ and can therefore be added to the Einstein-Hilbert action of eq. (1.5). In that case, it has the remarkable property that its integral yields the Euler characteristic of spacetime: as this is a topological invariant, it does not contribute to the equations of motion. However, it is possible to make use of this term in $d=4$ by coupling it to a scalar field, as will be also illustrated in section 1.3.2.

### 1.2.5. Other theories

Many other theories of modified gravity exist. We give here a few examples of theories that will be shown to reduce to a scalar-tensor theory in some limit in section 1.3.

As a first example, one can consider a very simple modification of the Einstein-Hilbert action given by a modification of its dependency on the Ricci scalar $R$. Theories having this feature are called $f(R)$ theories [61, 62] and their action is given in vacuum by

$$S[g_{\mu\nu}] = \frac{M_{\rm P}^2}{2} \int {\rm d}^4x \, \sqrt{-g} f(R) \,, \tag{1.45}$$

where $f$ is an arbitrary function. Such theories have a very interesting feature: the associated cosmology can be self-accelerating [63, 64], which gives an alternative to the standard description of the accelerated expansion of the Universe using a cosmological constant (see section 1.1.2.2).

A natural generalization of GR, which describes the dynamics of a massless spin-2 particle known as the graviton, would be a theory of a massive spin-2 particle. Such a theory would be a good candidate for the resolution of the fine-tuning problem since massive bosons have exponentially suppressed interactions at long range: gravity would be much weaker at very high distances, leading to corrections to the predicted expansion of the Universe.

Theories describing massive spin-2 particles are called "massive gravity theories": the first one was proposed in 1939 [65], and while the modification might not seem too hard to perform, it requires the addition of a mass term for the graviton which will break invariance under diffeomorphisms [66]; this is how those theories can escape the hypotheses of Lovelock's theorem. It is possible to restore the gauge symmetry using auxiliary fields [67], in a procedure similar to what was proposed by Stückelberg for electromagnetism [68, 69, 70]. However, adding these new fields can lead to the presence of ghosts, and these instabilities must be dealt with. Recently, several theories of massive gravity that fix these issues have been proposed [71, 72].

It is also possible to evade the hypotheses of Lovelock's theorem by introducing a second metric field, the "fiducial metric", to which matter is not coupled: such theories are called bigravity theories, and can also describe a massive graviton. They can in general exhibit ghost instabilities, but it is possible to suppress them in order to get a healthy theory [73]. Recently, such a theory has been proposed in [74], generalizing the construction of Minimal Theory of Massive Gravity (MTMG) originally proposed in [72].

Finally, a theory describing a massive graviton can also be obtained from a higher-dimensional reasoning. Indeed, if one assumes that spacetime is five-dimensional but that matter fields are restricted to a four-dimensional subspace, the theory can be shown to be equivalent to massive gravity, without ghost instabilities[7] [77, 78]. This theory is called Dvali-Gabadadze-Porrati (DGP), from the names of its authors, and it was developed in 2000 [43].

## 1.3. Equivalence between theories

The high number of modified gravity theories might seem discouraging, since studying the properties of all of them would be nearly impossible. Fortunately, many of these theories can be related to each other, sometimes entirely at the level of the action and sometimes in some decoupling limit. We give a few examples here, mainly focused on the links between several theories and the Horndeski theory of gravity.

### 1.3.1. Horndeski as particular case of DHOST

Both Horndeski and DHOST theories are scalar-tensor theories; however, Horndeski is the most general theory with second-order equations of motion while DHOST theories do not have this requirement. It is therefore expected that Horndeski theories can be recovered as some specific case of DHOST. It is indeed the case; one can check that the former theories can be seen as a subclass of the latter with

$$\begin{aligned}
&F_2 = F\,, \quad A_1 = -A_2 = 2F_{2X}\,, \quad A_i = 0 \quad \text{for} \quad i \geq 3\,, \\
&F_3 = G\,, \quad 3B_1 = -B_2 = \frac{3}{2}B_3 = F_{3X}\,, \quad B_i = 0 \quad \text{for} \quad i \geq 4\,.
\end{aligned} \tag{1.46}$$

In the specific case of quadratic Horndeski theories, one furthermore sees that these theories belong to the subclass Ia, since $A_1 = -A_2$. In the general case, one can check that eq. (1.46) is coherent with eq. (1.40).

### 1.3.2. Gauss-Bonnet term as Horndeski theory

The Gauss-Bonnet term was introduced in section 1.2.4:

$$\mathcal{G} = R_{\mu\nu\rho\sigma}R^{\mu\nu\rho\sigma} - 4R_{\mu\nu}R^{\mu\nu} + R^2\,. \tag{1.47}$$

The Lagrangian defined by $\sqrt{-g}\mathcal{G}$ does not modify the equations of motion in 4-dimensional spacetimes since it is the density of a topological invariant, but is relevant as soon as $d \geq 4$. There are therefore two different ways to propose a modified theory of gravity relying on this term. First, one can notice that as this

7. One branch of the theory can however be proven to contain a ghost [75, 76].

term does not appear in the equations of motion, it must be a total derivative: therefore, a term of the form $\sqrt{-g}f(\phi)\mathcal{G}$ would have a nontrivial effect even when $d = 4$. Second, one can choose to assume spacetime has more than 4 dimensions, and compactify some of them using a Kaluza-Klein-like scheme, yielding a new action for the 4-dimensional part of the metric. These two methods are explored in the following sections.

#### 1.3.2.1. Coupling to a scalar field

Let us study the action

$$S_{\rm GB}[g_{\mu\nu},\phi] = \int \mathrm{d}^4x\,\sqrt{-g}\,f(\phi)\,\mathcal{G}\,. \tag{1.48}$$

Since $\sqrt{-g}\mathcal{G}$ must be a total derivative, integration by parts should allow us to recover a scalar-tensor action from eq. (1.48). It is proven in [79] that the Einstein-Gauss-Bonnet (EGB) action completed with a kinetic term for the scalar field contains only one scalar degree of freedom [8]. It is therefore expected that eq. (1.48) can be written as a specific case of eq. (1.20).

In [81], an expression for the Gauss-Bonnet term $\mathcal{G}$ (more precisely an expression for all the Lovelock invariants) as a total derivative was found. This reformulation makes use of an auxiliary scalar field $\pi$, by writing the Riemann tensor as a commutator of covariant derivatives of this scalar field and using the Bianchi identities. One should note that the scalar field $\pi$ does not have to be related to the scalar field $\phi$! The only requirement on $\pi$ is $\pi_\mu\pi^\mu \neq 0$. In order to reformulate the action of eq. (1.48) into a Horndeski theory, we take $\pi = \phi$, but we could actually use any other scalar field, for example the Ricci scalar $R$ itself (as long as $\nabla_\mu R\nabla^\mu R$ is nonzero) as Colléaux writes himself.

We start from the following expression, obtained from eq. (47) in [81]:

$$\mathcal{G} = -2\delta^{\mu\nu\alpha\beta}_{\sigma\rho\lambda\delta}\nabla^\delta\left[\frac{\phi_\alpha{}^\lambda\phi_\beta}{X}\left(R_{\mu\nu}{}^{\sigma\rho} + \frac{4}{3}\frac{\phi_\mu{}^\sigma\phi_\nu{}^\rho}{X}\right)\right]. \tag{1.49}$$

The idea of the proof done in [81] is to generate Riemann terms by using the commutation of covariant derivatives acting on $\phi^\mu$, using the formula

$$\left[\nabla_\mu,\nabla_\nu\right]\phi^\rho = R^\rho{}_{\lambda\mu\nu}\phi^\lambda\,. \tag{1.50}$$

In order to obtain the squared Riemann terms present in $\mathcal{G}$, one searches an expression of $\mathcal{G}$ in the schematic form $\nabla_\mu(\phi_{\nu\rho}\phi_\sigma R_{\lambda\delta\alpha\beta})$. The action of the covariant derivative on $\phi_{\nu\rho}$ will lead to a squared Riemann term as expected.

8. This can also be understood by remarking that the construction of the Lovelock invariants in section 1.2.4 is very similar to the construction of Galileons [44, 80].

Recovering the Gauss-Bonnet will require antisymmetrization, since this is how this term was defined as a Lovelock invariant in eq. (1.44). We will therefore contract the expression with the fully antisymmetric tensor. However, several new terms will be created when the covariant derivative acts on the other parts of the expression: specific tuning of prefactors in front of these terms will be required to make sure only the Gauss-Bonnet invariant is left in the end.

We reproduce the proof of [81] in the specific case of a 4-dimensional spacetime. We start from the generic Lagrangian

$$L = \delta^{\mu\nu\alpha\beta}_{\sigma\rho\lambda\delta}\nabla^{\delta}\left[a_0\frac{\phi_\alpha{}^\lambda\phi_\beta}{X}R_{\mu\nu}{}^{\sigma\rho} + a_1\frac{\phi_\alpha{}^\lambda\phi_\beta}{X^2}\phi_\mu{}^\sigma\phi_\nu{}^\rho\right], \tag{1.51}$$

where $a_0$ and $a_1$ are constants. By expanding the covariant derivative $\nabla^\delta$ in $L$, one obtains

$$\begin{aligned} L = a_0\delta^{\mu\nu\alpha\beta}_{\sigma\rho\lambda\delta}&\left[\frac{\nabla^\delta\phi_\alpha{}^\lambda}{X}\phi_\beta R_{\mu\nu}{}^{\sigma\rho} + \frac{\phi_\alpha{}^\lambda\phi_\beta{}^\delta}{X}R_{\mu\nu}{}^{\sigma\rho} - \frac{2}{X^2}\phi_\alpha{}^\lambda\phi_\kappa{}^\delta\phi^\kappa\phi_\beta R_{\mu\nu}{}^{\sigma\rho}\right] \\ + a_1\delta^{\mu\nu\alpha\beta}_{\sigma\rho\lambda\delta}&\left[\frac{3}{X^2}\nabla^\delta\phi_\alpha{}^\lambda\phi_\beta\phi_\mu{}^\sigma\phi_\nu{}^\rho + \frac{\phi_\alpha{}^\lambda\phi_\beta{}^\delta\phi_\mu{}^\sigma\phi_\nu{}^\rho}{X^2} - \frac{4}{X^3}\phi^\kappa\phi_\beta\phi_\kappa{}^\delta\phi_\alpha{}^\lambda\phi_\mu{}^\sigma\phi_\nu{}^\rho\right], \end{aligned} \tag{1.52}$$

by regrouping terms that are equal under contraction with the totally antisymmetric tensor. One notices that the covariant derivatives of the Riemann tensors contracted with the totally antisymmetric tensor disappear by application of the second Bianchi identities,

$$\nabla_\delta R_{\mu\nu\sigma\rho} + \nabla_\sigma R_{\mu\nu\rho\delta} + \nabla_\rho R_{\mu\nu\delta\sigma} = 0\,. \tag{1.53}$$

By using the first Bianchi identities,

$$R_{\mu\nu\sigma\rho} + R_{\mu\rho\nu\sigma} + R_{\mu\sigma\rho\nu} = 0\,, \tag{1.54}$$

along with eq. (1.50), one obtains

$$2\delta^{\mu\nu\alpha\beta}_{\sigma\rho\lambda\delta}\nabla^\delta\phi_\alpha{}^\lambda = -\delta^{\mu\nu\alpha\beta}_{\sigma\rho\lambda\delta}R^{\lambda\delta}{}_{\alpha\kappa}\phi^\kappa\,. \tag{1.55}$$

This allows us to write

$$L = a_0\left[-\frac{1}{2X}\Omega_{2,0} - \frac{2}{X^2}\Omega_{3,1} + \frac{1}{X}\Omega_{1,1}\right] + a_1\left[-\frac{3}{2X^2}\Omega_{2,1} + \frac{1}{X^2}\Omega_{1,2} - \frac{4}{X^3}\Omega_{3,2}\right], \tag{1.56}$$

where the functions $\Omega_{i,j}$ are defined in [81] as

$$\Omega_{1,0} = \delta^{\mu\nu\alpha\beta}_{\sigma\rho\lambda\delta}R_{\mu\nu}{}^{\sigma\rho}R_{\alpha\beta}{}^{\lambda\delta}\,, \qquad \Omega_{1,2} = \delta^{\mu\nu\alpha\beta}_{\sigma\rho\lambda\delta}\phi_\mu{}^\sigma\phi_\nu{}^\rho\phi_\alpha{}^\lambda\phi_\beta{}^\delta\,,$$

$$\Omega_{1,1} = \delta^{\mu\nu\alpha\beta}_{\sigma\rho\lambda\delta}R_{\mu\nu}{}^{\sigma\rho}\phi_\alpha{}^\lambda\phi_\beta{}^\delta\,, \qquad \Omega_{3,1} = \delta^{\mu\nu\alpha\beta}_{\sigma\rho\lambda\delta}\phi_\kappa\phi^\lambda\phi_\alpha{}^\kappa R_{\mu\nu}{}^{\sigma\rho}\phi_\beta{}^\delta\,,$$

$$\Omega_{2,0} = \delta^{\mu\nu\alpha\beta}_{\sigma\rho\lambda\delta}\phi_\kappa\phi^\rho R_{\mu\nu}{}^{\sigma\kappa}R_{\alpha\beta}{}^{\lambda\delta}\,, \qquad \Omega_{3,2} = \delta^{\mu\nu\alpha\beta}_{\sigma\rho\lambda\delta}\phi_\kappa\phi^\sigma\phi_\mu{}^\kappa\phi_\nu{}^\rho\phi_\alpha{}^\lambda\phi_\beta{}^\delta\,,$$
$$\Omega_{2,1} = \delta^{\mu\nu\alpha\beta}_{\sigma\rho\lambda\delta}\phi_\kappa\phi^\rho R_{\mu\nu}{}^{\sigma\kappa}\phi_\alpha{}^\lambda\phi_\beta{}^\delta\,. \tag{1.57}$$

One can then prove the following identities relating the functions $\Omega_{i,j}$:

$$\begin{aligned}
X\Omega_{1,0} - 4\Omega_{2,0} &= \delta^{\mu\nu\alpha\beta\gamma}_{\sigma\rho\lambda\delta\kappa}\phi_\mu\phi^\sigma R_{\nu\alpha}{}^{\rho\lambda}R^{\delta\kappa}{}_{\beta\gamma} = 0\,,\\
X\Omega_{1,1} - 2\Omega_{2,1} - 2\Omega_{3,1} &= \delta^{\mu\nu\alpha\beta\gamma}_{\sigma\rho\lambda\delta\kappa}\phi_\mu\phi^\sigma\phi_\nu{}^\rho\phi_\alpha{}^\lambda R^{\delta\kappa}{}_{\beta\gamma} = 0\,,\\
X\Omega_{1,2} - 4\Omega_{3,2} &= \delta^{\mu\nu\alpha\beta\gamma}_{\sigma\rho\lambda\delta\kappa}\phi_\mu\phi^\sigma\phi_\nu{}^\rho\phi_\alpha{}^\lambda\phi_\beta{}^\delta\phi_\gamma{}^\kappa = 0\,,
\end{aligned} \tag{1.58}$$

since in 4 dimensions the fully antisymmetric tensor $\delta^{\mu\nu\alpha\beta\gamma}_{\sigma\rho\lambda\delta\kappa}$ is zero (there are more indices than dimensions so two indices have to be repeated). Equation (1.56) then becomes

$$L = -\frac{a_0}{8}\Omega_{1,0} + \frac{\Omega_{2,1}}{X^2}\left(2a_0 - \frac{3}{2}a_1\right). \tag{1.59}$$

One can see from eqs. (1.44) and (1.57) that $\Omega_{1,0} = 4\mathscr{G}$. Therefore, by choosing $a_0 = -2$ and $a_1 = 4a_0/3 = -8/3$, one obtains eq. (1.49).

We can now use the expression of $\mathscr{G}$ as a total derivative to express the action (1.48) as a Horndeski theory. Injecting eq. (1.49) into eq. (1.48) and integrating by parts gives

$$S_{\rm GB}[g_{\mu\nu},\phi] = -\int \mathrm{d}^4x\,\sqrt{-g}\frac{2}{X}\frac{\mathrm{d}f}{\mathrm{d}\phi}\varepsilon^{\mu\nu\alpha\beta}\varepsilon_{\sigma\rho\lambda\delta}\phi_\alpha{}^\lambda\phi_\beta\phi^\delta\left(R_{\mu\nu}{}^{\sigma\rho} + \frac{4}{3}\frac{\phi_\mu{}^\sigma\phi_\nu{}^\rho}{X}\right). \tag{1.60}$$

After expanding the products, one finds that the Lagrangian density $L$ of eq. (1.60) is

$$\begin{aligned}
L = -\frac{\mathrm{d}f}{\mathrm{d}\phi}\Big[&8R^{\mu\nu}\phi_{\mu\nu} + \frac{4}{X}\phi^\mu\phi_{\mu\nu}\phi^\nu - 4R\Box\phi - \frac{16}{X}R_\mu{}^\nu\phi^\mu\phi^\rho\phi_{\nu\rho} - \frac{16}{3X}\phi_\mu{}^\nu\phi^{\mu\rho}\phi_{\rho\nu}\\
&+ \frac{8}{X}\Box\phi\phi_{\mu\nu}\phi^{\mu\nu} + \frac{16}{X^2}\phi^\mu\phi^\nu\phi_\mu{}^\rho\phi_\nu{}^\sigma\phi_{\rho\sigma} + \frac{8}{X}R_{\mu\nu}\phi^\mu\phi^\nu\Box\phi\\
&- \frac{8}{3X}(\Box\phi)^3 - \frac{8}{X}R_{\mu\nu\rho\sigma}\phi^\mu\phi^\nu\phi^{\rho\sigma} - \frac{8}{X^2}\phi^\mu\phi_{\mu\nu}\phi^\nu\phi_{\rho\sigma}\phi^{\rho\sigma}\\
&- \frac{16}{X^2}\phi^\mu\phi^\nu\phi_\mu{}^\rho\phi_{\rho\nu}\Box\phi + \frac{8}{X^2}\phi^\mu\phi_{\mu\nu}\phi^\nu(\Box\phi)^2\Big].
\end{aligned} \tag{1.61}$$

One can recognise several total derivatives:

$$\nabla_\mu\left(\frac{1}{X}\right) = -\frac{2}{X^2}\phi^\nu\phi_{\mu\nu} \quad\text{and}\quad \nabla_\mu(\ln(X)) = \frac{2}{X}\phi^\nu\phi_{\mu\nu}. \tag{1.62}$$

integrating by parts the terms containing these total derivatives and writing contractions of the Riemann tensors as commutators of derivatives, one obtains

$$L = \frac{\mathrm{d}f}{\mathrm{d}\phi}\Big[-E_{\mu\nu}\phi^{\mu\nu}(8+4\ln(X)) - \frac{4}{3X}(L_1^{(3)} - 3L_2^{(3)} + 2L_3^{(3)})\Big] \\ + \frac{\mathrm{d}^2 f}{\mathrm{d}\phi^2}\big[2X\ln(X)R + 4\ln(X)L_1^{(2)} + 4(L_1^{(2)} - L_2^{(2)})\big] \\ + \frac{\mathrm{d}^3 f}{\mathrm{d}\phi^3}\times 2X(1-3\ln(X))\Box\phi - \frac{\mathrm{d}^4 f}{\mathrm{d}\phi^4}\times 2X^2\ln(X)\,, \tag{1.63}$$

where the $L_i^{(j)}$ are the DHOST Lagrangians introduced in section 1.2.3.1. Finally, one can rewrite the term $E_{\mu\nu}\phi^{\mu\nu}$ using $\nabla^\mu E_{\mu\nu} = 0$ and writing contractions of the Ricci as commutators of derivatives, yielding

$$\int \mathrm{d}^4x\,\sqrt{-g}E_{\mu\nu}\phi^{\mu\nu}\frac{\mathrm{d}f}{\mathrm{d}\phi} = \int \mathrm{d}^4x\,\sqrt{-g}\Big[\frac{1}{2}R\frac{\mathrm{d}^2 f}{\mathrm{d}\phi^2} + 2(L_1^{(2)} - L_2^{(2)})\frac{\mathrm{d}^2 f}{\mathrm{d}\phi^2} \\ - 3X\Box\phi\frac{\mathrm{d}^3 f}{\mathrm{d}\phi^3} - 4X^2\frac{\mathrm{d}^4 f}{\mathrm{d}\phi^4}\Big]\,. \tag{1.64}$$

Putting eq. (1.64) into eq. (1.63), one concludes that the action of eq. (1.48) is equivalent to a cubic Horndeski theory with

$$G(\phi,X) = -4\frac{\mathrm{d}f}{\mathrm{d}\phi}\ln(X)\,,\quad F(\phi,X) = -2X(2-\ln(X))\frac{\mathrm{d}^2 f}{\mathrm{d}\phi^2}\,, \\ Q(\phi,X) = 2X(7-3\ln(X))\frac{\mathrm{d}^3 f}{\mathrm{d}\phi^3}\,,\quad P(\phi,X) = 2X^2(3-\ln(X))\frac{\mathrm{d}^4 f}{\mathrm{d}\phi^4}\,. \tag{1.65}$$

This result was already proven in [82], but the reasoning given in this reference is only valid at the level of the equations of motion. On the contrary, the computation presented here is done at the level of the action.

#### 1.3.2.2. Compactification

Let us now consider a $D$-dimensional Einstein-Gauss-Bonnet action, denoted $S_D$, containing the usual Einstein-Hilbert action with an additional Gauss-Bonnet term:

$$S_D[g_{D,\mu\nu}] = \int \mathrm{d}^D x\,\sqrt{-g_D}\,({}^D R + \alpha_D\,{}^D\mathcal{G})\,, \tag{1.66}$$

where $g_{D,\mu\nu}$ is the $D$-dimensional metric field, ${}^D R$ is the Ricci in $D$ dimensions and ${}^D\mathcal{G}$ is the Gauss-Bonnet term in $D$ dimensions. It was proven in [83] that one can compactify the last $D-4$ dimensions of this action in order to recover a 4-dimensional action. To do so, one poses

$$\mathrm{d}s_D^2 = \mathrm{d}s^2 + e^{2\phi}\,\mathrm{d}\Sigma_{D-4}^2\,, \tag{1.67}$$

with $\mathrm{d}s^2$ the line element in 4 dimensions and $\mathrm{d}\Sigma^2_{D-4}$ the infinitesimal element of a sphere of dimension $D-4$. Here, the field $\phi$ depends only on the first 4 coordinates. One defines $\eta = D-4$; the action $S_D$ becomes

$$S_D[g_{D,\mu\nu}] = \int \mathrm{d}^4x\,\sqrt{-g}e^{\eta\phi}(R + \eta(\eta-1)X + \alpha_D\mathcal{G} \\ - \alpha^D\eta(\eta-1)(4G^{\mu\nu}\phi_\mu\phi_\nu + 2(\eta-2)X\Box\phi + (\eta-1)(\eta-2)X^2))\,, \quad (1.68)$$

where all geometrical quantities are evaluated on the noncompactified 4-dimensional submanifold.

In the $\eta = 0$ limit, the action of eq. (1.68) becomes similar to the GR action, since the Gauss-Bonnet invariant $\mathcal{G}$ is topological in 4 dimensions. However, if one defines $\alpha$ via $\alpha_D = \alpha/\eta$, it is possible to obtain a richer action. Indeed, in that case, the action becomes

$$S_D[g_{D,\mu\nu}] = \int \mathrm{d}^4x\,\sqrt{-g}e^{\eta\phi}(R + \eta(\eta-1)X + \frac{\alpha}{\eta}\mathcal{G} \\ - \alpha(\eta-1)(4G^{\mu\nu}\phi_\mu\phi_\nu + 2(\eta-2)X\Box\phi + (\eta-1)(\eta-2)X^2))\,. \quad (1.69)$$

The $\eta = 0$ limit of such an action is trickier, since a term is proportionnal to $\alpha/\eta$. Let us look precisely at what happens for this part of the action by expanding the exponential:

$$\int \mathrm{d}^4x\,\sqrt{-g}e^{\eta\phi}\frac{\alpha}{\eta}\mathcal{G} = \int \mathrm{d}^4x\,\sqrt{-g}\frac{\alpha}{\eta}\mathcal{G} + \int \mathrm{d}^4x\,\sqrt{-g}\alpha\phi\mathcal{G} + \mathcal{O}(\eta^2)\,. \quad (1.70)$$

The first term is zero since $\mathcal{G}$ is topological in 4 dimensions: in the $\eta = 0$ limit, one is left with only the second term. Therefore, a new action $S$ is obtained when taking this limit: it involves the 4-dimensional metric tensor $g_{\mu\nu}$ and the scalar field $\phi$ and reads

$$S[g_{\mu\nu},\phi] = \int \mathrm{d}^4x\,\sqrt{-g}\left(R + \alpha(\phi\mathcal{G} + 4G^{\mu\nu}\phi_\mu\phi_\nu - 4X\Box\phi + 2X^2)\right)\,, \quad (1.71)$$

This theory belongs to the class of scalar-tensor theories. Using the results of eq. (1.65), we see that it can be cast into a cubic Horndeski theory via the relations

$$G(X) = -4\alpha\ln(X)\,,\quad F(X) = 1-2X\alpha,\quad Q(X) = -4\alpha X \quad\text{and}\quad P(X) = 2\alpha X^2\,. \quad (1.72)$$

### 1.3.3. Other equivalences

#### 1.3.3.1. $f(R)$ theories as scalar-tensor

It is possible to cast the $f(R)$ theories defined in section 1.2.5 into a scalar-tensor theory of the Bergman-Wagoner form defined in eq. (1.19) [62]. Indeed, if one considers the action

$$S[g_{\mu\nu}, \chi] = \frac{M_P^2}{2} \int \mathrm{d}^4x \sqrt{-g} [f(\chi) + f'(\chi)(R - \chi)] , \tag{1.73}$$

with $\chi$ a scalar field, then the equations of motion impose

$$f''(\chi)(R - \chi) = 0 , \tag{1.74}$$

and therefore the action of eq. (1.73) is equivalent to the one of eq. (1.45) as soon as $f''(\chi)$ is nonzero. One can then define the scalar field of Horndeski $\phi$ as

$$\phi = f'(\chi) . \tag{1.75}$$

Provided $f'$ is a bijection, one can then express $\chi$ as a function of $\phi$, and write

$$S[g_{\mu\nu}, \chi] = \frac{M_P^2}{2} \int \mathrm{d}^4x \sqrt{-g} [\phi R + f(\chi(\phi)) - \phi\chi(\phi)] . \tag{1.76}$$

This corresponds to a theory of the form (1.19) with

$$V(\phi) = \phi\chi(\phi) - f(\chi(\phi)) \quad \text{and} \quad \omega(\phi) = 0 . \tag{1.77}$$

#### 1.3.3.2. Massive gravity as Horndeski

It is possible to recover a Horndeski theory from the DGP theory of massive gravity given in section 1.2.5. This was initially obtained in [75], in which the authors recovered a term of the form $X\Box\phi$. The Galileons were then introduced as a way to generalize theories of this form (they have been presented in section 1.2.2). It is actually possible to recover a complete Horndeski theory in the decoupling limit of a massive gravity theory [67].

## 1.4. Disformal transformations

### 1.4.1. Definition and effect on DHOST theories

In the case of scalar-tensor theories, the additional field $\phi$ provides a vector field $\phi_\mu$. This allows one to define a generalization of conformal transformations, called *disformal transformations*: for a metric $g_{\mu\nu}$, the disformed metric $\tilde{g}_{\mu\nu}$ is defined by

$$\tilde{g}_{\mu\nu} = A(\phi,X)g_{\mu\nu} + B(\phi,X)\phi_\mu\phi_\nu\,. \tag{1.78}$$

The most interesting feature of such transformations in the case of DHOST theories is that if $S[g_{\mu\nu},\phi]$ is the action of a DHOST theory, then $S[\tilde{g}_{\mu\nu},\phi]$ is also the action of a DHOST theory, with different values of functions $F_2$, $F_3$, $A_i$ and $B_i$ [56, 57], meaning that DHOST theories are *stable* under disformal transformations. Furthermore, in the case of quadratic theories, it was proven in [57] that this stability statement extends to each subclass defined in section 1.2.3.3. The explicit transformation laws for $F_2$ and the $A_i$ are given in section 1.4.2.

Quadratic Horndeski theories being a subset of DHOST theories, one could wonder whether these theories are stable under disformal transformations. In general, such theories belong to the class I of DHOST theories, which means that any transformation of the form (1.78) will lead to another theory of class I. However, not all theories of this subclass are Horndeski theories: this means that Horndeski theories are not stable under general disformal transformations. One can still prove that these theories are stable under restricted disformal transformations, namely those that do not depend on $X$ [84]. As a conclusion, this means that each DHOST theory in class I can be cast into a corresponding Horndeski theory; the study of the former is therefore equivalent to the study of the latter and one can choose a preferred formulation depending on which one is the simplest, provided one ignores matter.

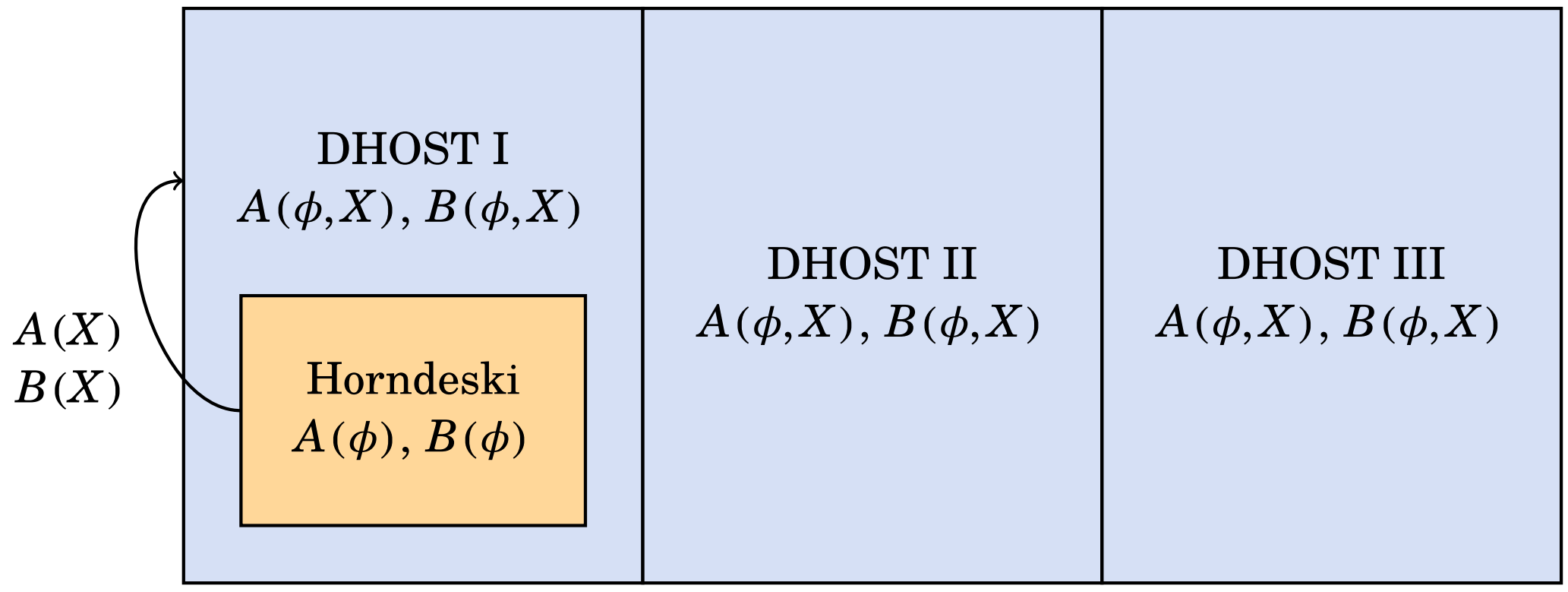

Figure 1.3. – DHOST theories and stability of their classes under disformal transformations (adapted from [54]).

This equivalence becomes indeed more complicated when one wants to couple the DHOST action to matter. Let us consider a theory of the form

$$S[g_{\mu\nu},\phi] = S_{\mathrm{DHOST}}[g_{\mu\nu},\phi] + S_{\mathrm{m}}\,, \tag{1.79}$$

with $S_{\rm m}$ the matter action. The matter fields in $S_{\rm m}$ will be coupled to gravity; usually, the coupling is done with $g_{\mu\nu}$. If one performs a disformal transformation following eq. (1.78), then $S_{\rm DHOST}$ becomes the action of a new DHOST theory, but the matter will now be coupled to a new metric. This means that in general, two disformally-linked DHOST theories are not equivalent in the presence of matter.

### 1.4.2. Formulae for the disformal transformations

In this section, we recall how a quadratic higher-order scalar-tensor action transforms under disformal transformations. The fact that disformal transformations allow one to construct viable scalar-tensor theories that do not belong in the Horndeski class was initially realised in [85].

We consider the quadratic higher-order scalar-tensor action

$$\tilde{S}[\tilde{g}_{\mu\nu},\phi] = \int \mathrm{d}^4x\,\sqrt{-\tilde{g}}\left(\tilde{F}_2(\tilde{X})\,R + \sum_{i=1}^{5}\tilde{A}_i(\tilde{X})\,L_i^{(2)}\right), \tag{1.80}$$

where $\tilde{X} \equiv \tilde{g}^{\mu\nu}\phi_\mu\phi_\nu$. For simplicity, we assume that the theory is shift-symmetric, and consider only shift-symmetric transformations. By the definition of eq. (1.78), we see that such transformations induce a modification of the action according to

$$\tilde{S}[\tilde{g}_{\mu\nu},\phi] \;\longrightarrow\; S[g_{\mu\nu},\phi] = \tilde{S}[A(X)g_{\mu\nu} + B(X)\phi_\mu\phi_\nu]\,, \tag{1.81}$$

where the quadratic part of $S[g_{\mu\nu},\phi]$ is of the form

$$\int \mathrm{d}^4x\,\sqrt{-g}\left(F_2(X)\,R + \sum_{i=1}^{5}A_i(X)\,L_i^{(2)}\right), \tag{1.82}$$

with

$$\tilde{X} = \frac{X}{A+BX}, \tag{1.83}$$

while the functions $F_2$ and $A_i$ can be expressed in terms of $\tilde{F}_2$ and $\tilde{A}_i$ as follows:

$$\begin{aligned}
&F_2 = \frac{J}{A}\tilde{F}_2\,, \quad A_1 = -h + JT_{11}\tilde{A}_1\,, \quad A_2 = h + JT_{22}\tilde{A}_2\,,\\
&A_3 = 2h_X + J\,(\tilde{F}_2\gamma_3 - 2\tilde{F}_{2X}\delta_3 + T_{13}\tilde{A}_1 + T_{23}\tilde{A}_2 + T_{33}\tilde{A}_3)\,,\\
&A_4 = -2h_X + J\,(\tilde{F}_2\gamma_4 - 2\tilde{F}_{2X}\delta_4 + T_{14}\tilde{A}_1 + T_{44}\tilde{A}_4)\,,\\
&A_5 = J\,(\tilde{F}_2\gamma_5 - 2\tilde{F}_{2X}\delta_5 + T_{15}\tilde{A}_1 + T_{25}\tilde{A}_2 + T_{35}\tilde{A}_3 + T_{45}\tilde{A}_4 + T_{55}\tilde{A}_5)\,.
\end{aligned} \tag{1.84}$$

We have introduced the following notations:

$$J = A^{3/2}\sqrt{A+BX}\,, \quad h = -\frac{BJ\tilde{F}_2}{A(A+BX)}\,,$$

$$
\begin{aligned}
T_{11} &= \frac{1}{(A+BX)^2}, \quad T_{13} = \frac{2A_X}{A(A+BX)^2}, \\
T_{14} &= \frac{2\left(X\left(A_X+XB_X\right)^2 - A\left(2\left(A_X+XB_X\right)+B\right)\right)}{A(A+BX)^3}, \\
T_{15} &= \frac{3B^2X^2A_X^2}{A^2(A+BX)^4} - \frac{2B^2XA_X}{A(A+BX)^4} + \frac{B^2}{(A+BX)^4} - \frac{2BX^3B_X^2}{A(A+BX)^4} \\
&\quad - \frac{X^2B_X^2}{(A+BX)^4} - \frac{4BX^2A_XB_X}{A(A+BX)^4} + \frac{4BXA_X^2}{A(A+BX)^4} - \frac{2XA_XB_X}{(A+BX)^4} \\
&\quad + \frac{4BXB_X}{(A+BX)^4} + \frac{2A_X^2}{(A+BX)^4} + \frac{2AB_X}{(A+BX)^4}, \\
T_{22} &= \frac{1}{(A+BX)^2}, \quad T_{23} = -\frac{2\left(A\left(-2A_X+XB_X+B\right) - 3BXA_X\right)}{A(A+BX)^3}, \\
T_{25} &= \frac{\left(A\left(-2A_X+XB_X+B\right) - 3BXA_X\right)^2}{A^2(A+BX)^4}, \\
T_{33} &= \frac{A - X\left(A_X+XB_X\right)}{(A+BX)^4}, \\
T_{35} &= -\frac{\left(A\left(-2A_X+XB_X+B\right) - 3BXA_X\right)\left(A - X\left(A_X+XB_X\right)\right)}{A(A+BX)^5}, \\
T_{44} &= \frac{\left(A - X\left(A_X+XB_X\right)\right)^2}{A(A+BX)^4}, \quad T_{45} = -\frac{B\left(A - X\left(A_X+XB_X\right)\right)^2}{A(A+BX)^5}, \\
T_{55} &= \frac{\left(A - X\left(A_X+XB_X\right)\right)^2}{(A+BX)^6}, \\
\gamma_3 &= -\frac{B\left(BXA_X + A\left(2A_X+XB_X+B\right)\right)}{A^2(A+BX)^2}, \\
\gamma_4 &= \frac{A\left(2BA_X + 4XA_XB_X + 6A_X^2 + B^2 + BXB_X\right)}{A(A+BX)^2} \\
&\quad + \frac{A^2\left(2BA_X + 4XA_XB_X + 6A_X^2 + B^2 + BXB_X\right)}{A^3(A+BX)^2}, \\
\gamma_5 &= -\frac{2A_X\left(BA_X + 2AB_X\right)}{A^3(A+BX)}, \\
\delta_3 &= \frac{B}{A(A+BX)}, \quad \delta_4 = \frac{-4BXA_X - A\left(6A_X + 2XB_X + B\right)}{A^2(A+BX)}, \\
\delta_5 &= \frac{2\left(2BA_X + AB_X\right)}{A^2(A+BX)}.
\end{aligned}
\tag{1.85}
$$

All functions are evaluated in $X$ and not $\tilde{X}$ (and derivatives are with respect to $X$ and not $\tilde{X}$).

## 1.5. Tests and constraints on modified gravity

### 1.5.1. ADM decomposition of the DHOST Lagrangian

In order to obtain experimental predictions from modified gravity theories, and in particular DHOST theories, it is convenient to single out the time coordinate of spacetime in order to describe temporal evolution of spatial quantities. In this framework, it is possible to study the dynamics of a test mass in order to find a modified Poisson equation; one can also understand how the matter content of the Universe impacts the dynamics of its expansion, or study the propagation of waves as small perturbations of the background.

In order to perform such a reformulation, it is relevant to choose a gauge in which the scalar field $\phi$ coincides with the time coordinate:

$$\phi = \phi(t)\,. \tag{1.86}$$

This is always possible, provided that $\phi_\mu$ is timelike, which is the most interesting case in these situations. One then makes use of the Arnowitt-Deser-Misner (ADM) decomposition of spacetime, using a slicing of constant time (and scalar field):

$$\mathrm{d}s^2 = -N^2\,\mathrm{d}t^2 + h_{ij}\,(\mathrm{d}x^i + N^i\,\mathrm{d}t)\,(\mathrm{d}x^j + N^j\,\mathrm{d}t)\,. \tag{1.87}$$

In this decomposition, $N$ is called the *lapse* and $N^i$ is called the *shift*. All quantities with Latin indices $i, j, \ldots$ are covariant when seen from the point of view of the induced 3-dimensional metric $h_{ij}$.

Let us now write the DHOST action of eq. (1.25) using this decomposition, in the case of shift-symmetric theories:

$$\begin{aligned} S[N,N^i,h_{ij}] = \int \mathrm{d}^3x\,\mathrm{d}t\,N\sqrt{h}\Big[&XV(B_2 + XB_6 + F_{3X})K_{ij}K^{ij} \\ &+ (F_2 - XA_1)K_{ij}K^{ij} - (F_2 + XA_2)K^2 \\ &+ F_2\,{}^{(3)}\!R + \ldots\Big]\,. \end{aligned} \tag{1.88}$$

The complete action (containing also the terms that arise in the non-shift-symmetric case) can be found in the appendices of [60]; it is quite involved and not relevant to the study of this manuscript. We introduced the quantities $V = -\dot{X}\big/2N\sqrt{-X}$ and

$$K_{ij} = \frac{1}{2N}\left(\dot{h}_{ij} - {}^{(3)}\nabla_i N_j - {}^{(3)}\nabla_j N_i\right), \tag{1.89}$$

where ${}^{(3)}\nabla$ is the covariant derivative associated to $h_{ij}$. The tensor $K_{ij}$ is the *extrinsic curvature* of each $\phi = \text{cst}$ hypersurface: it is the only geometrical term [9] that contains time derivatives of the metric $h_{ij}$.

9. The trace of $K_{ij}$ does not depend on $\dot{h}_{ij}$ since the trace of $h_{ij}$ is constant.

This expression of the DHOST action can be very useful to understand physical properties of the theory, as we will illustrate in the next sections.

### 1.5.2. Different test scales

#### 1.5.2.1. Cosmological tests

The past 30 years have seen formidable developments in the measurement of cosmological parameters, paving the way for tests of GR on a cosmological scale. The most successful recent result is the measurement of anisotropies in the Cosmic Microwave Background (CMB) by the Planck collaboration [13]. These results give precise measurements of the cosmological parameters of the standard model of cosmology. Naturally, these results can also be used to constrain modified theories of gravity; it is therefore relevant to study cosmology in modified theories of gravity. This was done in the case of DHOST theories in [86, 87]. More extensive computations were performed in the case of Horndeski theories (see for example the review [88]).

We focus here on one specific computation in the framework of DHOST theories, namely the effective action for scalar and tensor perturbations over a cosmological background (such as the one given in eq. (1.6)). These perturbations are very important since their behaviour will allow us to probe the stability of the theory on such a background. The scalar perturbations are parametrized as

$$N = 1 + \delta N\,, \quad N^i = \delta^{ij}\partial_j\psi\,, \quad h_{ij} = a(t)^2 e^{2\zeta}\delta_{ij}\,, \tag{1.90}$$

while the tensor perturbations are parametrized as

$$h_{ij} = a(t)^2(\delta_{ij} + \gamma_{ij})\,, \tag{1.91}$$

with $\gamma_{ij}$ a traceless and divergence-free tensor. As proven in [60] by injecting eqs. (1.90) and (1.91) into eq. (1.88) and performing several further simplifications, one obtains an action of the form [10]

$$S_{\rm pert}[\gamma_{ij},\zeta] = \int \mathrm{d}^3x\,\mathrm{d}t\,a^3\left[A_\zeta\dot\zeta^2 - B_\zeta\frac{(\partial_i\zeta)^2}{a^2} + A_\gamma\dot\gamma_{ij}^2 - B_\gamma\frac{(\partial_k\gamma_{ij})^2}{a^2}\right]. \tag{1.92}$$

One can then prove, by using the explicit degeneracy conditions of eqs. (1.37) and (1.38) for DHOST theories of class II, that the coefficients $A_\zeta$ and $A_\gamma$ have the same sign, while $B_\zeta$ and $B_\gamma$ have opposite signs. This means that one of the modes must exhibit a gradient instability: the DHOST theories of class II are cosmologically unstable, a feature that could not be inferred directly from the expression of the action.

10. The field $\zeta$ in the action (1.92) is not exactly the one defined in eq. (1.90) (see [60] for details).

#### 1.5.2.2. Gravitational wave tests

In a given theory that couples space and time, it is usually expected that perturbations of a solution of the equations of motion leads to the propagation of waves. In the case of gravitation, the existence of "gravitational waves" was therefore quickly postulated by Einstein after he proposed his theory of GR [89]. Such waves can be seen as ripples of curvature propagating through spacetime. However, the very existence of these waves was long debated, because of the coordinate independence of GR: the wave speed could not depend on the coordinates chosen, while quantities computed in a specific gauge could however diverge without causing an issue if the divergence was due to the coordinate system. One must wait until 1957 for the problem to be solved [90]. During this year, arguments were proposed by Bondi [91], Feynman [92] and Pirani [93, 94] in order to justify that GWs are physical quantities. While the clearest and most rigorous treatment was proposed by Pirani, it is easier to understand Feynman's argument: consider a rod with two beads sliding freely with a small amount of friction along it. If a wave passes over the rod, then while the length of the rod is maintained fixed by atomic forces, the proper distance between the beads changes, forcing them to slide along the rod. Due to the friction, this requires work, meaning that GWs are physical quantities.

While the existence of GWs was no longer a debate, a measurement of their effect was still quite hard to imagine. The first physicist to propose an experimental setup was Weber [95]: his experiment relied on the interaction between a passing wave and the natural mode of vibrations of an aluminium bar. Although Weber claimed a discovery in 1969 [96], his results were criticized by his peers; the first accepted proof of the existence of gravitational waves was done in 1974 by Hulse and Taylor [97], for which they received the Nobel Prize in 1993. This observation was indirect: they found that a given binary system exhibited energy loss in a way that could not be explained by the Newtonian motion of celestial bodies, and successfully interpreted this loss by the emission of GWs. The first direct observation of GWs, for which the 2017 Nobel prize was awarded, was done in 2015 by the LIGO/Virgo collaboration [14]. A review of the methods used for the detection as well as other ongoing experiments looking for GWs can be found in [98].

In the case of DHOST theories, two observations had major implications for constraints: the fact that GWs were detected at all, and the observation of an electromagnetic counterpart in the case of GW170817 [99]. First, the fact that GWs emitted one billion lightyears away from the Earth were observed means that such waves have a very small decay rate. However, in the case of scalar-tensor theories, GWs can be expected to decay into scalar waves due to the strong interactions between these two sectors. The observations of [14] therefore places a strict bound on the terms of the DHOST action (1.25) that contribute to this decay. Second, the presence of an electromagnetic counterpart in GW170817 allowed the LIGO/Virgo collaboration to obtain a measurement of the speed of gravitational waves $c_T$ [100]. They obtained

$$-3 \times 10^{-15} \leq c_T - 1 \leq 7 \times 10^{-16} \; . \tag{1.93}$$

This means that only theories that predict a speed of gravitational waves equal to the speed of light should be considered valid.

The speed of gravitational waves in the case of quadratic DHOST theories can easily be deduced from eq. (1.88), if one understand that gradient terms are present only in the ${}^{(3)}R$ term while kinetic terms are present only in the $K_{ij}K^{ij}$ term. The speed is then obtained from the quotient of the coefficients of these terms in eq. (1.88) (without taking the cubic terms into account in the case of quadratic theories):

$$c_T^2 = \frac{F_2}{F_2 - XA_1} \, . \tag{1.94}$$

The requirement that $c_T$ be 1 then imposes $A_1 = 0$. It is proven in [101] that in this case, the most general DHOST Lagrangian with no decay of GWs has also $A_3 = 0$ (one should however note that this is only true if one requires the DHOST theory to describe dark energy, which corresponds to the case (b) in fig. 1.1). Finally, the subclass of DHOST theories that satisfies these requirements is given by

$$A_4 = \frac{6F_{2X}^2}{F_2} \, , \quad (A_i)_{i \neq 4} = 0 \, . \tag{1.95}$$

In the case of cubic DHOST theories, computing the speed of gravitational waves is a bit more involved. It is performed in [60]: it can be read off the coefficient $\alpha_T$ computed in the appendix of this reference using the relation $c_T^2 = 1 + \alpha_T$. Imposing $c_T = 1$ for any background yields $A_1 = 0$, $F_{3\phi} = 0$ and $B_3 = -B_2$; imposing the degeneracy conditions of eq. (1.40) then means that one must have $B_1 = 0$, $B_2 = 0$ and $B_3 = 0$. This is problematic since the requirement for eq. (1.40) is $B_1 \neq 0$. The only solution is then to have all $B_i$ equal to zero [11], which leads back to the quadratic case.

#### 1.5.2.3. Star-scale tests

The previous tests we presented are actually not the ones historically associated with gravitation: first and foremost, the most important prediction of the theory of gravity is the movement of stars. It is therefore natural to look for deviations

11. One could in principle consider cubic DHOST theories with zero quadratic term, in which case the degeneracy conditions are less stringent: cubic-only theories with $c_T = 1$ could exist. However, such theories would belong to a pathological class of DHOST theories.

of GR in astrophysical experiments. Two main categories can be defined: experiments studying the trajectories of astrophysical bodies, and experiments studying gravitationally bound objects such as galaxies, galaxy clusters, neutron stars...

The study of the motion of stars at the center of our Galaxy has led to the discovery of the presence of a supermassive compact object, which is strongly believed to be a black hole, called Sagittarius A*. The trajectories in the strong field created by such an object are a very good natural laboratory for the study of GR; up to now, observation fully agree with Einstein's theory [102, 103]. Half of the Nobel Prize of 2020 was awarded to Reinhard Genzel and Andrea Ghez for this discovery. A BH was also directly observed for the first time in 2019 by the Event Horizon Telescope through the measurement of radio-waves near its event horizon [104].

Experiments based on the motion of objects can also be performed in our Solar System and lead to tight bounds on deviations from GR. One can consider for example the Bepi-Colombo mission that provided constraints on the graviton mass and the parameters of some scalar-tensor theories [105, 106]. Studies of the motion of faraway objects such as pulsars is also a great way to obtain bounds on deviations from GR in the case of strongly self-gravitating bodies [107].

The study of the internal composition of gravitationally bound bodies is booming since neutron stars have been proven to appear in binary mergers observed by the LIGO/Virgo collaboration (the first one being detected was GW170817 [99]). Theoretical predictions are more complicated since the internal composition of stars involved a lot of different phenomena. A review of the work done for neutron stars in the case of Horndeski theories can be found in [108].

An interesting feature of gravitational bodies in scalar-tensor theories is the Vainshtein mechanism. This mechanism allows "screening" of the modification of gravity outside of an astrophysical body, such that Solar System tests stay valid even though the modifications can still be observed on other scales. This phenomenon, initially introduced in the case of massive gravity [33], was studied in the case of Horndeski theories in [109, 110] and for DHOST theories in [111, 112, 113].

#### 1.5.2.4. Laboratory-based tests

The main difficulty of the probes of GR presented up to now is the lack of control over the sources. Indeed, one does not know the exact black hole population of the Universe, or the position of the next binary neutron star merger. This problem disappears in laboratory-based tests of GR, where the experimental setup is fully controled. The price to pay is that the theory can only be tested at relatively small scales, which means that deviations on the cosmological scale will not be detectable. One can cite the MICROSCOPE experiment, that tested the weak equivalence principle up to $10^{-15}$ precision [114]. One can also perform experiments at the quantum level, measuring how particles behave in a gravitational field [115, 116].

### 1.5.3. Why we must go further

The different constraints obtained previously paint a grim landscape of the modified theories of gravity. Indeed, most of these theories are strongly constrained from large or small scale observations. In the remaining theories, few have coherent constraints both in the cosmological and the black hole regime. Therefore, it looks like no theory of modified gravity proposes a coherent generalization of GR.

However, the search for theories of modified gravity is still a very important pillar of research in gravitation. Indeed, although GR has been very successful for many different situations, it has not been tested in several extreme cases, where deviations could be expected. For example, the probe of the event horizon of a black hole has not been realized yet; more information will also come from the increased resolution and longer detection times of GWs emitted by binary mergers thanks to future experiments (Laser Interferometer Space Antenna (LISA), Einstein Telescope (ET)...). On the theoretical side, the freedom of coupling matter to a disformal metric (see section 1.4.1) means that much freedom is still left.

Furthermore, when modified theories of gravity are seen as EFTs, they are not required to verify the all the experimental constraints obtained until now. Indeed, each one of these constraints comes from a probe at given order of magnitude of energy: cosmological probes of dark energy are IR measurements, while BH measurements are much closer to the UV regime. Henceforth, bounds obtained on DHOST functions from cosmology concern only DHOST theories when they are seen as EFTs of dark energy (see fig. 1.1(b)) and they should not be used when one studies BH, and vice-versa.

Finally, even though constraints such as eq. (1.93) are very tight, it can be argued that these are actually only valid for a given frequency regime. Indeed, it is possible that GWs exhibit dispersion, with a speed varying throughout the whole frequency spectrum. This argument was proposed in [117]: observations at different frequencies in the future will give us more insight on this “gravitational rainbow”; in the meantime, the search for extensions of GR must go on.

CHAPTER 2

# BLACK HOLE SOLUTIONS IN DHOST THEORIES

CONTENTS

The need for modified theories of gravity was motivated in chapter 1. Now that we have constructed extensions of GR, we will be interested in the existence of solutions to these new theories. In this manuscript, we focus on a specific class of solutions: black holes, for two main reasons. First, the existence of BHs is a major prediction of GR which was absent in previous theories of gravity, which means that such objects are very effective probes of the theory. Second, understanding black holes can be done without a complex description of the matter within them, contrarily to neutrons stars or other compact objects. In this chapter, we review a few of the existing BH solutions to Horndeski and DHOST theories.

## 2.1. Looking for BH solutions in modified gravity

In this section, we describe the general procedure we use to find BH solutions in a quadratic Horndeski theory. We start by a review of BH solutions in GR, and review the unicity properties of these solutions. We then give the equations of motion that will be solved in the next sections.

### 2.1.1. Assumptions for the background

We consider static spherically symmetric black hole solutions: such solutions have a metric of the form

$$\mathrm{d}s^2 = g_{\mu\nu}\mathrm{d}x^\mu \mathrm{d}x^\nu = -A(r)\,\mathrm{d}t^2 + \frac{1}{B(r)}\,\mathrm{d}r^2 + C(r)\left(\mathrm{d}\theta^2 + \sin^2\theta\,\mathrm{d}\varphi^2\right)\,, \tag{2.1}$$

where $A$ and $B$ are functions of the radial coordinate $r$ only. Although it seems natural to assume a radially dependent scalar field, i.e. of the form $\phi = \phi(r)$, it was realised in [118] that one can adopt in the context of shift-symmetric theories the more general ansatz

$$\phi(t,r) = qt + \psi(r)\,, \tag{2.2}$$

where $q$ is constant, which implies

$$X = -\frac{q^2}{A(r)} + B(r)\psi'(r)^2\,. \tag{2.3}$$

This ansatz was motivated by the link between the Hamilton-Jacobi potential and the scalar field for specific spacetimes [119]. In this case, since only the gradient of the scalar field $\phi_\mu$ is relevant, eq. (2.2) is compatible with a static metric. Note that if $q \neq 0$ the disformal transformation of the metric (2.1) does not conserve the same form, because of the presence of a nonzero $\tilde{g}_{tr} \neq 0$. This implies that, in the case $q \neq 0$, working with the Horndeski action is more restrictive than starting with the general DHOST action even in vacuum.

Black holes described by eq. (2.1) do not correspond to the compact objects that are observed in our Universe. Indeed, in general, such objects are rotating. The metric describing rotating bodies cannot be static: it is stationary, and contains nondiagonal terms mixing angular coordinates and time. An example will be given in section 2.1.2. The study of static and spherically symmetric BHs is still interesting since it is much easier and allows one to grasp the main effects governing BH physics in a given theory of gravity.

### 2.1.2. GR solutions and no-hair theorem

Black hole solutions in GR all depend on three parameters: their mass $M$, their angular momentum $J$ and their charge $Q$. The first solution found by Schwarzschild was static and spherically symmetric [120] and described a chargeless BH with zero charge: it corresponded to eq. (2.1) with the choice

$$A(r) = B(r) = 1 - \frac{\mu}{r} \quad \text{and} \quad C(r) = r^2\,, \tag{2.4}$$

with $\mu = 2M$. This solution was extended to the charged case by Reissner, Weyl and Nordström who all obtained the result independently [1] [121, 122, 123]: the metric element in this case is also of the form eq. (2.1) with

$$A(r) = B(r) = 1 - \frac{\mu}{r} + \frac{Q^2}{r^2} \quad \text{and} \quad C(r) = r^2\,, \tag{2.5}$$

with $\mu = 2M$. This solution is called the *Reissner-Nordström solution*.

Finally, a generalization to the rotating case was obtained by Kerr in a computational tour de force [124], paving the way for comparison between astrophysical observations and GR predictions. This generalization changes completely the shape of the metric, since as explained previously it cannot be static and spherically symmetric. The Kerr metric is given by

$$\mathrm{d}s^2 = -\frac{\Delta}{\rho^2}\left(\mathrm{d}t - a\sin^2(\theta)\,\mathrm{d}\varphi\right)^2 + \frac{\sin^2(\theta)}{\rho^2}\left((r^2+a^2)\,\mathrm{d}\varphi - a\,\mathrm{d}t\right)^2 + \frac{\rho^2}{\Delta}\,\mathrm{d}r^2 + \rho^2\,\mathrm{d}\theta^2\,, \tag{2.6}$$

with

$$\Delta = r^2 - 2Mr + a^2\,, \quad \rho^2 = r^2 + a^2\cos^2(\theta) \quad \text{and} \quad a = \frac{J}{M}\,. \tag{2.7}$$

This solution was extended to the charged rotating BHs by Newman [125]: the only difference with eq. (2.6) is that $\Delta$ becomes $r^2 - 2Mr + a^2 + Q^2$. This most general solution with nonzero mass, charge and angular momentum is called the *Kerr-Newman* solution.

No other BH solutions have been found in GR, and this is due to a very peculiar result: the *no-hair theorem*, which was proven in the 1970s with the combined works of Israel, Carter, Hawking, Robinson and Mazur. This theorem assures the uniqueness of the Kerr-Newman solution [2] under a few simple assumptions:

**No hair theorem.** *The only possible exterior solution for a stationary, rotating, electrovacuum black hole with non-degenerate event horizon and no cosmological constant is the Kerr-Newman solution with no naked singularity.*

This theorem was first proven with the assumptions of a static and spherically symmetric black hole by Israel [126]; the very surprising result was that the only solution with these properties was the Schwarzschild solution of eq. (2.4). It was then extended by Israel himself to charged black hole solutions, proving that Reissner-Nordström (RN) was the only solution [127]. The generalization to rotating solutions was made possible thanks to a result by Hawking [128],

1. The solution was found yet again a few years later by Jeffery.

2. The theorem is only valid for the exterior region, which means outside of the event horizon: inside this horizon, all uniqueness results break down.

who proved that stationary BHs had to be axisymmetric (if rotating) and to have a spherical horizon. The final pieces of the puzzle were obtained by Carter, Robinson and Mazur [129, 130, 131].

The no hair theorem constitutes a very strong constraint when one is looking for modified theories of gravity. Indeed, the statement of uniqueness is so strong in GR that it is reasonable to think that it might still hold in some theories of modified gravity: such theories would therefore exhibit no new BH behaviour when compared to GR. Although such theories would still predict interesting modifications for other compact objects or on cosmological scales, they lose some of their interest if no deviation from GR can be found in the most exotic objects predicted by the latter theory. Therefore, people have searched since the 1970s for ways to generalize the no hair theorem, in order to find which theories contain new BH physics.

Presently, no-hair theorems have been obtained for specific subclasses of Horndeski theories called “galileons” [132], and generalized to some Horndeski theories [133]. Generalizations to the case of stars have also been studied, in the context of Horndeski theories [134]. Such extensions are not always as general as the original no hair theorem; for example, they usually are restricted to static spherically symmetric BHs. One should note that disproving the no-hair theorem for a given theory is much easier than proving it, since finding a single BH solution is sufficient. In the following sections, we will give examples of BH solutions for Horndeski and DHOST theories; we start by giving the explicit formulation of the background equations of motion that will be solved.

### 2.1.3. Background equations of motion

The variation of the quadratic shift-symmetric Horndeski action of eq. (1.24) yields the equations of motion

$$\mathcal{B}_{\mu\nu} \equiv \frac{2}{\sqrt{-g}}\frac{\delta S}{\delta g^{\mu\nu}} = 0\,, \qquad \mathcal{B}_\phi \equiv \frac{1}{\sqrt{-g}}\frac{\delta S}{\delta \phi} = 0\,. \tag{2.8}$$

Due to diffeomorphism invariance, the equation for the scalar field is not independent from the metric equations and therefore can be ignored [52]. Indeed, with the definitions of eq. (2.8), one has

$$\delta S = \int \mathrm{d}^4x\,\sqrt{-g}\left(\frac{1}{2}\mathcal{B}_{\mu\nu}\,\delta g^{\mu\nu} + \mathcal{B}_\phi\,\delta\phi\right). \tag{2.9}$$

Imposing diffeomorphism invariance then requires that $\delta S$ be zero if one chooses $\delta g^{\mu\nu} = \nabla^\mu\xi^\nu + \nabla^\nu\xi^\mu$ and $\delta\phi = \xi^\mu\nabla_\mu\phi$, for some vector field $\xi^\mu$. This implies

$$\int \mathrm{d}^4x\,\sqrt{-g}\big((\nabla^\nu\xi^\mu)\,\mathcal{B}_{\mu\nu} + \xi^\mu(\nabla_\mu\phi)\,\mathcal{B}_\phi\big)\,. \tag{2.10}$$

After an integration by parts and imposing that eq. (2.10) holds for every field $\xi^\mu$, one obtains

$$\nabla^\nu \mathcal{B}_{\mu\nu} = \mathcal{B}_\phi \nabla_\mu \phi \,. \tag{2.11}$$

This means that the scalar equation of motion is always implied by the metric one. It is therefore sufficient to solve only the metric equations of motion to find a BH solution. However, in some specific cases, the scalar equation will still be considered as it can be easier to solve than the metric equations (see section 2.4 for an example).

For a metric of the form (2.1) and a scalar field profile (2.2), one finds that the only non-trivial equations are given, up to a global irrelevant factor, by

$$\begin{aligned}
\mathcal{B}_{tt} \propto\; & Q_X\left(\frac{q^2BA'\psi'}{2A} - \frac{1}{2}AB\psi'X' + \frac{1}{2}q^2B'\psi' + \frac{q^2BC'\psi'}{C} + q^2B\psi''\right)\\
&+ F_X\left(-\frac{q^2BA'C'}{AC} + \frac{AXB'C'}{C} + \frac{2ABXC''}{C} + \frac{ABC'X'}{C} - \frac{ABXC'^2}{2C^2} + \frac{2q^2}{C}\right)\\
&+ F_{XX}\left(-\frac{2q^2BXA'C'}{AC} - \frac{q^4BC'^2}{AC^2} + \frac{2ABXC'X'}{C} - \frac{q^2BXC'^2}{C^2}\right)\\
&+ F\left(-\frac{AB'C'}{2C} - \frac{ABC''}{C} + \frac{ABC'^2}{4C^2} + \frac{A}{C}\right) + \frac{1}{2}AP + q^2P_X\,,\\
\mathcal{B}_{tr} \propto\; & F_{XX}\left(-\frac{2qBXA'C'\psi'}{AC} - \frac{q^3BC'^2\psi'}{AC^2} - \frac{qBXC'^2\psi'}{C^2}\right)\\
&+ F_X\left(-\frac{qBA'C'\psi'}{AC} - \frac{qBC'^2\psi'}{2C^2} + \frac{2q\psi'}{C}\right)\\
&+ Q_X\left(\frac{qXA'}{2A} + \frac{q^3C'}{AC} + \frac{qXC'}{C}\right) + q\psi'P_X\,,\\
\mathcal{B}_{rr} \propto\; & F_X\left(-\frac{q^2A'C'}{A^2C} - \frac{2XA'C'}{AC} + \frac{2q^2}{ABC} - \frac{q^2C'^2}{AC^2} + \frac{2X}{BC} - \frac{XC'^2}{C^2}\right)\\
&+ F_{XX}\left(-\frac{2q^2XA'C'}{A^2C} - \frac{2X^2A'C'}{AC} - \frac{q^4C'^2}{A^2C^2} - \frac{2q^2XC'^2}{AC^2} - \frac{X^2C'^2}{C^2}\right)\\
&+ F\left(\frac{A'C'}{2AC} - \frac{1}{BC} + \frac{C'^2}{4C^2}\right) + Q_X\left(\frac{XA'\psi'}{2A} + \frac{q^2C'\psi'}{AC} + \frac{XC'\psi'}{C}\right)\\
&+ P_X\left(\frac{q^2}{AB} + \frac{X}{B}\right) - \frac{P}{2B}\,,\\
\mathcal{B}_{\theta\theta} \propto\; & F_{XX}\left(-\frac{BCXA'X'}{A} - \frac{q^2BC'X'}{A} - BXC'X'\right)\\
&+ F_X\left(-\frac{BCXA''}{A} - \frac{CXA'B'}{2A} + \frac{q^2BA'C'}{2A^2} - \frac{BXA'C'}{2A} - \frac{BCA'X'}{2A} + \frac{BCXA'^2}{2A^2}\right.\\
&\left.-\frac{q^2B'C'}{2A} - \frac{q^2BC''}{A} + \frac{q^2BC'^2}{2AC} - \frac{1}{2}XB'C' - BXC'' - \frac{1}{2}BC'X' + \frac{BXC'^2}{2C}\right)
\end{aligned}$$

$$+F\left(\frac{BCA''}{2A}+\frac{CA'B'}{4A}+\frac{BA'C'}{4A}-\frac{BCA'^2}{4A^2}+\frac{1}{4}B'C'+\frac{1}{2}BC''-\frac{BC'^2}{4C}\right)$$
$$+\frac{1}{2}BC\psi'X'Q_X-\frac{1}{2}CP\,, \tag{2.12}$$

where a prime denotes a derivative with respect to $r$. One should note that assuming $X$ to be constant drastically simplifies the above metric equations; this fact will be useful in section 2.3. The equation $\mathscr{B}_\phi$ being much more involved, we do not write it explicitly here.

One can see from the complex structure of eq. (2.12) that obtaining a BH solution in Horndeski theories will require assumptions on the functions $F$, $P$ and $Q$. Such assumptions are however constrained by the structure of the equations; a review of possible choices and their implications is provided in [135].

## 2.2. BCL solution

### 2.2.1. Background

The equations of motion for quadratic Horndeski theories presented in section 2.1.3 are quite involved, even for a static and spherically symmetric metric, and contain a lot of freedom when $F$, $P$ and $Q$ are arbitrary. It is therefore interesting to search for solutions that correspond to a given choice of these three functions. We present here a solution obtained in [136] for a subset of quadratic shift-symmetric Horndeski theories (1.20) characterized by the functions

$$F(X)=f_0+f_1\sqrt{X},\quad P(X)=-p_1X,\qquad Q(X)=0\,, \tag{2.13}$$

where $f_0$, $f_1$ and $p_1$ are constants (we take $f_0, p_1>0$) and $X$ is supposed to be positive. For simplicity, we restrict ourselves to the case where the scalar field (2.2) has no time dependence, i.e. $q=0$.

The BH solution found in [136], which we will name Babichev-Charmousis-Lehébel (BCL) after the authors, is described by a metric of the form of eq. (2.1) with

$$A(r)=B(r)=\left(1-\frac{r_+}{r}\right)\left(1+\frac{r_-}{r}\right)\,, \tag{2.14}$$

where $r_-$ and $r_+$ are defined by the relations

$$r_+r_-=\frac{f_1^2}{2f_0p_1}\,,\quad r_+-r_-=\mu\equiv 2m\,,\quad r_+>r_->0\,. \tag{2.15}$$

Using these definitions, the metric function $A$ can be rewritten as

$$A(r)=1-\frac{\mu}{r}-\frac{\xi\mu^2}{2r^2}\,, \tag{2.16}$$

where we defined

$$\xi = 2\frac{r_+ r_-}{\mu^2} = \frac{f_1^2}{f_0 p_1 \mu^2}\,. \tag{2.17}$$

Note that the expression for $A(r)$ is reminiscent of the Reissner-Nordström metric (2.5) but with a negative root here. As a consequence, the black hole exhibits a single event horizon, of radius $r_+$, in contrast with the Reissner-Nordström geometry.

As for the scalar field, its kinetic term is given by

$$X(r) = A(r)\phi'^2(r) = \frac{f_1^2}{p_1^2 r^4}, \tag{2.18}$$

which is non constant. The scalar field profile can be found explicitly by integrating the equation

$$\phi'(r) = \pm\frac{f_1}{p_1 r\sqrt{(r-r_+)(r+r_-)}}\,, \tag{2.19}$$

yielding[3]

$$\phi(r) = \pm\frac{f_1}{p_1\sqrt{r_+r_-}}\arctan\left[\frac{\mu r + 2r_+r_-}{2\sqrt{r_+r_-}\sqrt{(r-r_+)(r+r_-)}}\right] + \mathrm{cst}\,. \tag{2.20}$$

### 2.2.2. Validity as an EFT

If the Horndeski theory of eq. (2.13) is considered as an EFT, one should check whether the scale of validity of the theory is high enough to contain present GW observations: if this is not the case, the results obtained for the BCL BH will not be comparable to experiments. In order to verify this, we use the EFT formulation of Weakly Broken Galileons given in [137] and reviewed in [138] that reproduces the Horndeski action.

Two scales $\Lambda_2$ and $\Lambda_3$ are introduced: $\Lambda_2$ is the scale associated to $X$ and $\Lambda_3$ is the scale associated to $\phi_{\mu\nu}$. The theory will be valid up to $\Lambda_3$, provided the following relation holds [137]:

$$\Lambda_2 = (\Lambda_3^3 M_\mathrm{P})^{1/4}\,. \tag{2.21}$$

3. The sign of $\phi(r)$ and the constant are physically irrelevant. Notice that the derivative of the scalar field diverges at the horizon. According to [136], this is not a problem as it is a coordinate dependent statement which disappears in the tortoise coordinate $r_*$ such that $\mathrm{d}r/\mathrm{d}r_* = A(r)$ for instance. Furthermore, it was argued in [136] that all physical meaningful quantities are well-defined at the horizon, for e.g. the scalar field itself.

In this dimensionful formulation, the Horndeski action of eq. (1.20) becomes

$$S = \int \mathrm{d}^4x \, \sqrt{-g} \sum_{i=0}^{3} L^{(i)}_{\mathrm{WBG}} \,, \tag{2.22}$$

with

$$\begin{aligned} L^{(0)}_{\mathrm{WBG}} &= \Lambda_2^4 P \,, \quad L^{(1)}_{\mathrm{WBG}} = \frac{\Lambda_2^4}{\Lambda_3^3} Q \,, \\ L^{(2)}_{\mathrm{WBG}} &= \frac{\Lambda_2^8}{\Lambda_3^6} FR + 2\frac{\Lambda_2^4}{\Lambda_3^6} F_X (\phi_{\mu\nu}\phi^{\mu\nu} - (\Box\phi)^2) \,, \\ L^{(3)}_{\mathrm{WBG}} &= \frac{\Lambda_2^8}{\Lambda_3^9} G(X) E^{\mu\nu}\phi_{\mu\nu} + \frac{1}{3}\frac{\Lambda_2^4}{\Lambda_3^9} G_X ((\Box\phi)^3 - 3\Box\phi\phi_{\mu\nu}\phi^{\mu\nu} + 2\phi_{\mu\rho}\phi^{\rho\nu}\phi_\nu{}^\mu) \,. \end{aligned} \tag{2.23}$$

Here, $X$ is also redefined as $\hat{X} = \frac{\phi_\mu\phi^\mu}{\Lambda_2^4}$. All the functions $P$, $Q$, $F$ and $G$ are then functions of $\hat{X}$.

In the case of the BCL solution, one can write

$$-p_1\phi_\mu\phi^\mu = -p_1\Lambda_2^4\hat{X} = \Lambda_2^4 P(\hat{X}) \,, \quad \text{with} \quad P(\hat{X}) = -\hat{p}_1\tilde{X} \,, \quad \hat{p}_1 = p_1 \,. \tag{2.24}$$

One also has

$$\begin{aligned} f_0 + f_1\sqrt{\phi_\mu\phi^\mu} &= \frac{\Lambda_2^8}{\Lambda_3^6}\left(\frac{\Lambda_3^6}{\Lambda_2^8} f_0 + \frac{\Lambda_3^6}{\Lambda_2^6} f_1\sqrt{\hat{X}}\right) = \frac{\Lambda_2^8}{\Lambda_3^6} F \,, \\ F(\hat{X}) &= \hat{f}_0 + \hat{f}_1\sqrt{\hat{X}} \,, \quad \hat{f}_0 = \frac{\Lambda_3^6}{\Lambda_2^8} f_0 \quad \text{and} \quad \hat{f}_1 = \frac{\Lambda_3^6}{\Lambda_2^6} f_1 \,. \end{aligned} \tag{2.25}$$

Finally, one must give the dimension of a length to $\mu$, by writing $\mu = M/M_{\mathrm{P}}^2$, with $M$ a mass related to the one of the BH by a constant factor.

With these definitions in mind, one can compute the value for the parameter $\xi$ that parametrizes deviations from GR, introducing $\hat{\xi} = \hat{f}_1^2 \big/ 2\hat{f}_0\hat{p}_1$:

$$\xi = \hat{\xi}\frac{\Lambda_2^4 M_{\mathrm{P}}^4}{\Lambda_3^6 M^2} \,. \tag{2.26}$$

Using dimensional arguments, one can argue that $\hat{\xi}$ must be of order 1. Then, one can obtain the order of magnitude of $\Lambda_3$ using eq. (2.21):

$$\Lambda_3 \sim \left(\frac{M_{\mathrm{P}}^5}{M^2\xi}\right)^{1/3} = 6.012 \times 10^{-7}\,\mathrm{GeV} \times \xi^{-1/3}\left(\frac{M_\odot}{M}\right)^{2/3} \,. \tag{2.27}$$

One observes that the cutoff of the theory increases when $\xi$ is lower, and becomes infinite in the GR limit $\xi \longrightarrow 0$. This is artificial: of course, in this limit, the cutoff cannot become greater than $M_{\rm P}$ which is the cutoff of GR. As a concrete example, for a BH of 10 solar masses and a deviation $\xi = 1$, one finds that the theory is valid up to frequencies of order $7 \times 10^{11}$ Hz, which is much higher than the current GWs frequencies observed by the LIGO/Virgo collaboration. Henceforth, the physical predictions obtained for the BCL BH will be physically relevant and comparable to experiments.

## 2.3. Stealth solutions

### 2.3.1. Motivation for stealth solutions

Stealth solutions are solutions for which the metric coincides with a vacuum solution of GR, possibly with a cosmological constant. This means that, even if the scalar field profile is non trivial, i.e. $\phi$ non constant, its effective energy-momentum tensor reduces to that of a cosmological constant. These solutions have been actively studied in the context of Horndeski, beyond Horndeski and more generally DHOST theories in the last few years [108, 134, 136, 139, 140, 141, 142, 143, 144] [4].

Stealth solutions are written quite simply since the metric sector is similar to the one of a GR solution, and are therefore also useful as building blocks for new solutions of DHOST theories. Indeed, a new generic method to construct non-stealth solutions in DHOST theories has been introduced recently in [143]. The idea consists in using a known solution $(g_{\mu\nu}, \phi)$ of a given DHOST theory to build, via a disformal transformation (1.78), a new solution $(\tilde{g}_{\mu\nu}, \phi)$ for the disformally related DHOST theory. In general, a stealth solution transforms into a non-stealth one. An interesting result from this method is the construction of the first non-stealth rotating black hole solutions in DHOST theories [148, 149].

### 2.3.2. Conditions to admit a stealth solution

We give in this section the conditions for a DHOST theory to admit stealth solutions, i.e. solutions of modified gravity whose metric coincides with a vacuum solution of GR plus a cosmological constant. The main stealth solutions in shift-symmetric quadratic DHOST theories are described by the Schwarzschild metric

4. Note that stealth solutions were first introduced in the context of three-dimensional gravity [145] and an earlier stealth solution in four-dimensional modified gravity was discovered [146] in the context of ghost condensate [147] (even though it was not named “stealth”).

and a scalar field of the form

$$\phi(t,r) = qt + \psi(r)\,, \tag{2.28}$$

where $q$ is constant. We also assume a constant value for $X \equiv \phi_\mu \phi^\mu$, which we denote $X_0$.

Stealth Schwarzschild solutions can be found in that context with either $X_0 = -q^2$ or $X_0 \neq -q^2$, provided that the functions appearing in eq. (1.25) satisfy the conditions (see eq. (22) of [140])

$$P = P_X = Q_X = A_1 + A_2 = A_{1X} + A_{2X} = 0\,, \tag{2.29}$$
$$(X_0 + q^2)A_1 = (X_0 + q^2)(2A_{1X} + A_3) = 0 \quad (\text{at } X = X_0)\,, \tag{2.30}$$

where all functions are evaluated at $X = X_0$ (one restricts to quadratic theories here so $F_3$ and the $B_i$ do not appear). These conditions were shown to be necessary and sufficient for the equations of motion of the metric to reduce to those of GR for static and spherical symmetric metric [140]. Type Ia DHOST theories verify $A_2(X) = -A_1(X)$, which implies that the last two conditions in eq. (2.29) are automatically satisfied. By contrast, the conditions of eq. (2.30) are more restrictive if $X_0 + q^2 \neq 0$. These two cases were discussed in detail in [140].

One can also look for DHOST theories such that *any* solution of GR (with a cosmological constant $\Lambda$), not only the static spherically symmetric metric solutions, is also solution of the DHOST theory, which imposes much more stringent conditions [150]:

$$\begin{aligned} &P + 2\Lambda F = 0\,, \quad P_X + \Lambda(4F_X - X_0 A_{1X}) = 0\,,\\ &Q_X = 0\,, \quad A_1 = 0 \quad A_3 + 2A_{1X} = 0\,, \end{aligned} \tag{2.31}$$

where all these expressions are evaluated at $X = X_0$. These conditions have been recently generalised to non-shift symmetric theories and to the case where matter is coupled to gravity minimally [150].

### 2.3.3. Stealth Schwarzschild black hole

For shift-symmetric DHOST Ia theories, or more specifically Horndeski theories, one can obtain stealth Schwarzschild solutions with a scalar field satisfying eq. (2.2) if the conditions of eq. (2.29) are verified. Since the equations of motion involve the functions $F$, $P$ and $Q$ up to their second derivatives only evaluated at the background value $X_0 = -q^2$ (see section 2.1.3), if we fix $F(X_0) = 1$ for convenience, then the only theory-dependent parameters that appear in the equations of motion are

$$\alpha \equiv F_X(X_0)\,, \quad \beta \equiv F_{XX}(X_0)\,, \quad \gamma \equiv P_{XX}(X_0)\,, \quad \delta \equiv Q_{XX}(X_0)\,. \tag{2.32}$$

In other words, without loss of generality, we can limit our study to Horndeski theories with

$$F(X) \equiv 1 + \alpha(X + q^2) + \frac{\beta}{2}(X + q^2)^2\,,$$
$$P(X) \equiv \frac{\gamma}{2}(X + q^2)^2\,, \quad Q(X) \equiv \frac{\delta}{2}(X + q^2)^2\,. \tag{2.33}$$

All the other terms in the expansions in powers of $(X + q^2)$ of these functions are irrelevant.

The stealth Schwarzschild solution is then described by the metric (2.1) with

$$A(r) = B(r) = 1 - \frac{\mu}{r}\,, \tag{2.34}$$

where $\mu$ denotes the Schwarzschild radius, and the scalar field (2.2) with [5]

$$\psi'(r) = q\frac{\sqrt{r\,\mu}}{r - \mu}\,, \tag{2.35}$$

which is obtained by solving $X = -q^2$ (see [118]).

## 2.4. EsGB solution

In this section, we present a BH solution of Einstein-scalar-Gauss-Bonnet (EsGB) theories, in which one adds to the usual Einstein-Hilbert term for the metric a non-standard coupling to a scalar field $\phi$ which involves the Gauss-Bonnet term $\mathscr{G}$ introduced in eq. (1.47). Analytical non rotating black hole solutions were found in the case of specific coupling values in [151, 152, 153, 154, 155], and rotating solutions in the same setups in [156, 157, 158]. A solution for any coupling form was obtained in [159]. All these solutions are given as expansions in a small parameter appearing in the coupling function. This small parameter parametrises the deviation from GR.

The EsGB action is given by

$$S[g_{\mu\nu}] = \int \mathrm{d}^4x\,\sqrt{-g}\,\left(R - 2X + f(\phi)\mathscr{G}\right)\,, \tag{2.36}$$

where $f(\phi)$ is an arbitrary function of $\phi$.

5. Note that the equations of motion lead to $\psi'$ up to a global sign. Here we make one choice because it gives a regular expression (in Eddington-Finkelstein coordinates) while the expression with the opposite sign leads to a singular scalar field on the horizon [118]. However, such a singularity has no physical consequences because $X$ itself and the stress-tensor energy are not singular.

Although this action is not manifestly of the form (1.20), its equations of motion can be shown to be second order, which means that the theory can be reformulated as a Horndeski theory [52, 82]. This is explicitly shown in section 1.3.2, working directly at the level of the action. The corresponding Horndeski functions are given by

$$P(\phi,X) = -2X + 2f^{(4)}(\phi)X^2(3-\ln X)\,, \quad Q(\phi,X) = 2f^{(3)}(\phi)X(7-3\ln X)$$
$$F(\phi,X) = 1-2f''(\phi)X(2-\ln X) \quad \text{and} \quad G(\phi,X) = -4f'(\phi)\ln X\,. \tag{2.37}$$

To find a static black hole solution, we start with the ansatz of eq. (2.1), with the additional choice $A(r) = B(r)$. An alternative choice would have been to assume $C = r^2$ and $B \neq A$ (see for example [160]).

When the coupling function $f$ is a constant, the term proportional to $\mathcal{G}$ in the action becomes a total derivative and is thus irrelevant for the equations of motion, which are then the same as in GR with a massless scalar field. One thus immediately obtains as a solution the Schwarzschild metric with a constant and uniform scalar field:

$$A(r) = B(r) = 1-\frac{\mu}{r}\,, \quad C(r) = r^2 \quad \text{and} \quad \psi(r) = \psi_\infty\,, \tag{2.38}$$

where $\psi_\infty$ is an arbitrary constant.

When $f(\phi)$ is not constant, the above configuration is no longer a solution but can nevertheless be considered as the zeroth order expression of the full solution written as a series expansion in terms of the parameter $\varepsilon$, defined by

$$\varepsilon = \frac{f'(\psi_\infty)}{\mu^2}\,, \tag{2.39}$$

and assumed to be small, as it was proposed initially proposed in [151] and recently developed in [159]. Hence, we expand the metric components and scalar field as series in power of $\varepsilon$ (up to some order $N$) as follows:

$$A(r) = 1-\frac{\mu}{r} + \sum_{i=1}^{N} a_i(r)\varepsilon^i + \mathcal{O}(\varepsilon^{N+1})\,, \tag{2.40}$$

$$C(r) = 1 + \sum_{i=1}^{N} c_i(r)\varepsilon^i + \mathcal{O}(\varepsilon^{N+1})\,, \tag{2.41}$$

$$\psi(r) = \psi_\infty + \sum_{i=1}^{N} s_i(r)\varepsilon^i + \mathcal{O}(\varepsilon^{N+1})\,, \tag{2.42}$$

where the functions $a_i$, $c_i$ and $s_i$ can be determined, order by order, by solving the associated differential equations obtained by substituting the above expressions into the equations of motion.

One can see that the metric equations of motion expanded up to order $\varepsilon^N$ involve $a_N(r)$, $c_N(r)$ and $s'_{N-1}(r)$, while the scalar equation of motion at order $\varepsilon^N$ relates $a_{N-1}(r)$, $c_{N-1}(r)$ and $s'_N(r)$. Then, it is possible to use this separation of orders to solve the equations of motion order by order. We need boundary conditions to integrate these equations and we impose that all these functions go to zero at spatial infinity.

At first order in $\varepsilon$, one obtains the equations

$$a_1(r) = -\tau_3 + \frac{1}{r}(\tau_1 + \tau_2) - \frac{\mu\tau_2}{2r^2}, \qquad c_1(r) = \tau_3 - \frac{\tau_2}{r}, \tag{2.43}$$

where the $\tau_i$ are integration constants. The boundary conditions at spatial infinity impose $\tau_3 = 0$. Furthermore, the constant $\tau_1 + \tau_2$, which can be interpreted as a shift of the black hole mass at first order in $\varepsilon$, can be absorbed by redefining $\mu$ as follows:

$$\mu_{\text{new}} = \mu_{\text{old}} - \varepsilon(\tau_1 + \tau_2)\,. \tag{2.44}$$

Finally, the remaining terms proportional to $\tau_2$ can be absorbed by the coordinate change

$$r_{\text{new}} = r_{\text{old}} + \varepsilon\tau_2/2\,. \tag{2.45}$$

As a consequence, at first order in $\varepsilon$, one simply recovers the background solution given in eq. (2.38), up to a change of mass and a change of coordinate, which corresponds to taking

$$a_1(r) = 0 \quad \text{and} \quad c_1(r) = 0\,. \tag{2.46}$$

As for the scalar field, its equation of motion yields, at first order in $\varepsilon$,

$$s_1(r) = \frac{\mu}{r} + \frac{\mu^2}{2r} + \frac{\mu^3}{3r^3} + \nu_1 + \left(1 + \frac{\nu_2}{\mu}\right)\ln\left(1 - \frac{\mu}{r}\right), \tag{2.47}$$

with $\nu_1$ and $\nu_2$ constants. One can obviously absorb the constant $\nu_1$ into a redefinition of $\psi_\infty$ while one chooses $\nu_2$ so that $s_1(r)$ remains regular at the horizon.

At order $\varepsilon^2$, one can repeat the same method to solve for $a_2$, $b_2$ and $s_2$. One can ignore the five integration constants that appear since they can be reabsorbed using the boundary conditions, mass redefinition and coordinate change, as previously. At the end, the metric and scalar functions read

$$a_2(r) = -\left(\frac{\mu^3}{3r^3} - \frac{11\mu^4}{6r^4} + \frac{\mu^5}{30r^5} + \frac{17\mu^7}{15r^7}\right), \tag{2.48}$$

$$c_2(r) = -\left(\frac{\mu^2}{r^2} + \frac{2\mu^3}{3r^3} + \frac{7\mu^4}{6r^4} + \frac{4\mu^5}{5r^5} + \frac{3\mu^6}{5r^6}\right), \tag{2.49}$$

$$s_2(r) = \rho_2\left(\frac{73}{60}\left(\frac{\mu}{r} + \frac{\mu^2}{2r^2} + \frac{\mu^3}{3r^3} + \frac{\mu^4}{4r^4}\right) + \frac{7\mu^5}{75r^5} + \frac{\mu^6}{36r^6}\right), \tag{2.50}$$

where we have introduced the constant $\rho_2$ defined by

$$\rho_2 = \frac{f''(\psi_\infty)}{f'(\psi_\infty)} \,. \tag{2.51}$$

In principle, it is possible to continue this procedure and find all coefficients up to some arbitrary order $\varepsilon^N$ in a finite number of steps, but the complexity of the expressions quickly makes the computations very cumbersome. Here, we stop at order $\varepsilon^2$.

By taking into account the higher order corrections to the metric functions, the black hole horizon is no longer at $r = \mu$ but is slightly shifted to the new value

$$r_h = \mu\left(1 - \frac{\varepsilon^2}{3}\right) + \mathcal{O}(\varepsilon^3) \,. \tag{2.52}$$

Since $r_h$ is known only as a power series of $\varepsilon$, it is more convenient to work with the new radial coordinate dimensionless variable $z$ with

$$z = \frac{r}{r_h} \,, \tag{2.53}$$

in terms of which the horizon is exactly located at $z = 1$ at any order in $\varepsilon$.

## 2.5. 4dEGB solution

In this section, we present another BH solution of a modified theory of gravity that involves the Gauss-Bonnet invariant $\mathcal{G}$ defined in eq. (1.47). The action for this theory was obtained in eq. (1.71); it is given by

$$S[g_{\mu\nu},\phi] = \int \mathrm{d}^4x\,\sqrt{-g}\,\left(R + \alpha(\phi\mathcal{G} + 4E^{\mu\nu}\phi_\mu\phi_\nu - 4X\Box\phi + 2X^2)\right) \,, \tag{2.54}$$

where $\alpha$ is a constant. This action can be obtained as the $4D$ limit, in some specific sense, of the $D$-dimensional EGB action [83], as we presented in section 1.3.2. As for EsGB theories, this theory also belongs to degenerate scalar-tensor theories. It can be recast into a Horndeski theory with the following functions:

$$P(X) = 2\alpha X^2\,, \quad Q(X) = -4\alpha X\,, \quad F(X) = 1 - 2\alpha X \quad \text{and} \quad G(X) = -4\alpha\ln X\,. \tag{2.55}$$

We will also assume that $\alpha > 0$, otherwise $|\alpha|$ is constrained to be extremely small [161].

One can find a simple analytical solution to the equations of motion associated with eq. (2.54), as discussed in [83, 162]. The metric function $A$ is given by

$$A(r) = 1 + \frac{r^2}{2\alpha}\left(1 - \sqrt{1 + \frac{4\alpha\mu}{r^3}}\right) = 1 - \frac{2\mu/r}{1 + \sqrt{1 + \frac{4\alpha\mu}{r^3}}} \,. \tag{2.56}$$

This reduces to the Schwarzschild metric in the limit $\alpha \to 0$, the parameter $\mu$ corresponding to twice the black hole mass in this limit. This analytical solution is called 4-dimensional-Einstein-Gauss-Bonnet (4dEGB).

If $\mu^2 < 4\alpha$, the solution is a naked singularity and is therefore of no interest. If $\mu^2 \geq 4\alpha$, the solution for the metric describes a black hole and its horizons can be found by solving the equation $A(r) = 0$ for $r$. This gives two roots, the largest one corresponding to the outermost horizon,

$$r_h = \frac{1}{2}\left(\mu + \sqrt{\mu^2 - 4\alpha}\right) . \tag{2.57}$$

The equation for the scalar field gives two different branches:

$$\phi'(r) = \frac{\sigma + \sqrt{A(r)}}{r\sqrt{A(r)}} \quad \text{with} \quad \sigma = \pm 1 . \tag{2.58}$$

Integrating this equation in the limit where $r$ is large (i.e. $r \gg r_h$), one obtains

$$\phi(r) \simeq \frac{\mu}{2r} \quad \text{if } \sigma = -1 , \qquad \phi(r) \simeq 2 \ln\left(\frac{r}{\mu}\right) \quad \text{if } \sigma = +1 . \tag{2.59}$$

Hence, the branch $\sigma = +1$ leads to a divergent behaviour of the scalar field at spatial infinity. In this branch, moreover, $\phi$ does not vanish when the black hole mass goes to zero and we will see later in chapter 7 that the perturbations feature also a pathological behaviour. For these reasons, we will mostly restrict our analysis to the branch $\sigma = -1$.

In the rest of the manuscript, when studying this solution, it will be convenient to use the dimensionless quantities

$$z = \frac{r}{r_h} \quad \text{and} \quad \beta = \frac{\alpha}{r_h^2} . \tag{2.60}$$

According to these definitions and (2.57), one can replace $\mu$ by $(1+\beta) r_h$. Note that

$$0 \leq \beta = \frac{\mu - r_h}{r_h} \leq 1 , \tag{2.61}$$

as $0 \leq r_h \leq \mu$. One can notice that both bounds can be reached: the case $\beta = 0$ is the GR limit, while the case $\beta = 1$ is an extremal black hole, as both horizons merge into one located at $r_h = \sqrt{\alpha}$. The parameter $\beta$ is therefore similar to the extremality parameter $Q/M$ for a charged black hole, and it is interesting to use it instead of $\alpha$ when studying the present family of black hole solutions [6].

6. One can note that $\beta$ is also a parameter for the stealth solutions presented in section 2.3.3; however, it will always be possible to differentiate between the two by looking at the solution that is being studied.

Moreover, the outermost horizon is now at $z = 1$ and the new metric function is

$$A(z) = 1 + \frac{z^2}{2\beta}\left(1 - \sqrt{1 + \frac{4\beta(1+\beta)}{z^3}}\right) = 1 - \frac{2(1+\beta)}{z\left(1 + \sqrt{1 + \frac{4\beta(1+\beta)}{z^3}}\right)} . \tag{2.62}$$

Since $\phi'$ depends on $\sqrt{A}$, as shown in eq. (2.58), it is also convenient to introduce the new function

$$f(z) = \sqrt{A(z)} . \tag{2.63}$$

## 2.6. Rotating solutions

Obtaining rotating BH solutions in GR is already a hard task; in modified gravity theories, this becomes very complicated. Very few rotating BH solutions have been obtained in the literature for scalar-tensor theories. In most cases, the solution is obtained using a perturbation expansion in the rotation parameter [133, 156, 157, 158]. A stealth Kerr solution was obtained nonperturbatively in [119], and its perturbations were studied in [163]. Another approach is to obtain solutions numerically; while this is less useful for analytical computations of some properties, it is already enough to prove that the no hair theorem does not hold, and can pave the way for an analytical resolution of the background equations of motion. Such a resolution was done in [164] in the case of a cubic Galileon theory.

CHAPTER 3

# GEOMETRIC FORMULATION OF DHOST THEORIES

CONTENTS

THE family of DHOST theories presented in eq. (1.25) constitutes the most general scalar-tensor theory propagating three degrees of freedom, and containing up to cubic terms in $\phi_{\mu\nu}$ in its action. However, the action of such theories is very involved, and the degeneracy conditions presented in section 1.2.3.3 seem overly complicated when compared to the simple requirement that the theory be degenerate.

In this chapter, we identify a "frame" where the action of quadratic DHOST theories takes a remarkably simple form, with a natural geometric interpretation based on the three-dimensional hypersurface $\Sigma_\phi$ where the scalar field $\phi$ is constant. We also show how the classification of quadratic DHOST theories becomes transparent in this new formulation, as well as the reason behind the instability of several classes of DHOST theories.

Note that three-dimensional quantities based on the uniform scalar field hypersurfaces have already been used in several earlier works, for instance in the context of the EFT of inflation [165], of Horava's gravity and its extensions [166], in the study of the cosmology of Horndeski and Beyond Horndeski theories [167, 168]. The novelty here is that we combine this three-dimensional formalism with the systematic exploitation of disformal transformations, in order to simplify the description of DHOST theories.

The chapter is organized as follows. We start by defining "weakly degenerate" theories which are DHOST degenerate only in the unitary gauge (in which $\phi$ is a function of time only), as well as the parametrisations that have been introduced in previous works to describe them. In section 3.2, we present the new formulation of quadratic DHOST and weakly degenerate actions. In section 3.3, we revisit the classification of DHOST theories in this new perspective and present a very simple argument based on this new formulation which proves that DHOST theories in the classes II and III are plagued by instabilities (or do not propagate gravitational waves). Finally, we compute the equations of motion in the new formulation in section 3.4. This chapter is based on the paper [169].

## 3.1. Higher-order scalar-tensor theories and degeneracy

We work in this chapter with quadratic DHOST theories: these correspond to eq. (1.25) with $F_3 = 0$ and $B_i = 0$. We do not assume that these theories are shift-symmetric: their action takes the form

$$S[g_{\mu\nu},\phi] = \int \mathrm{d}^4x \, \sqrt{-g} \left[ F_2(\phi,X)\,{}^{(4)}\!R + \sum_{i=1}^{5} A_i(\phi,X) L_i^{(2)} \right], \tag{3.1}$$

where the $L_i^{(2)}$ are defined in eq. (1.27). We do not consider the lower order Lagrangians of the form $P(\phi,X)$ and $Q(\phi,X)\Box\phi$ since they do not affect the degeneracy of the action.

Instead of working directly with DHOST theories, it will be convenient at this stage to consider a larger family of theories, which are degenerate in a weaker sense as we now explain. Assuming that $\varepsilon = \mathrm{sgn}(X) = -1$, one can introduce the so-called unitary gauge, where the scalar field is spatially uniform and thus depends only on time. We will name *weakly degenerate* all the theories that are degenerate in the unitary gauge. This includes obviously the DHOST theories, which are degenerate in any gauge and thus *a fortiori* in the unitary gauge, but also theories that are degenerate only in the unitary gauge, dubbed U-degenerate theories in [170]. Since U-degenerate theories are not DHOST theories, they contain an extra degree of freedom, but this scalar mode is not propagating in the unitary gauge, as it satisfies an elliptic partial differential equation.

Weakly degenerate theories have been classified in [170]. For quadratic theories, they satisfy a *single* degeneracy condition (to be contrasted with the *three* conditions of eq. (1.34) obeyed by the quadratic DHOST theories), which reads [12]

$$D_0(X) - XD_1(X) + X^2 D_2(X) = 0\,. \tag{3.2}$$

Hence only five out of the six functions in eq. (3.1) are independent. In [170], the unitary degeneracy condition (3.2) was solved by expressing the five functions $A_i$ in terms of $F_2$ and four independent functions $\kappa_1$, $\kappa_2$, $\alpha$ and $\sigma$ as follows:

$$A_1 = \kappa_1 + \frac{F_2}{X}\,, \quad A_2 = \kappa_2 - \frac{F_2}{X}\,,$$

$$
\begin{aligned}
A_3 &= \frac{2F_2 - 4XF_{2X}}{X^2} + 2\sigma\kappa_1 + 2\left(3\sigma - \frac{1}{X}\right)\kappa_2\,, \\
A_4 &= \alpha + 2\frac{X(F_{2X} - \kappa_1) - F_2}{X^2}\,, \\
A_5 &= \frac{2F_{2X} - X\alpha}{X^2} + \kappa_1\left(\frac{1}{X^2} + 3\sigma^2 - \frac{2\sigma}{X}\right) + \kappa_2\left(3\sigma - \frac{1}{X}\right)^2 .
\end{aligned} \tag{3.3}
$$

Quadratic DHOST theories can be inferred from weakly degenerate theories by imposing two more additional conditions (say $D_0 = 0$ and $D_1 = 0$) and therefore can be parametrised in terms of three independent functions only. Quadratic DHOST theories contain several subclasses and the explicit form of their parametrisation depends on the specific subclass considered [12].

## 3.2. DHOST theories from geometrical quantities

### 3.2.1. Uniform scalar field hypersurfaces $\Sigma_\phi$ and the geometric frame

#### 3.2.1.1. Geometry of uniform scalar field hypersurfaces

The scalar field $\phi$ present in DHOST theories naturally induces a preferred slicing of spacetime, which we are going to exploit. Let us first introduce various geometric tensors associated with this slicing. The intrinsic geometry of any constant $\phi$ hypersurface $\Sigma_\phi$ is characterized by the three-dimensional induced metric

$$
h_{\mu\nu} \equiv g_{\mu\nu} - \varepsilon\, n_\mu n_\nu\,, \quad n_\mu \equiv \frac{\phi_\mu}{\sqrt{|X|}}\,, \tag{3.4}
$$

where $n_\mu$ is the unit vector orthogonal to $\Sigma_\phi$. One can also introduce the Riemann tensor ${}^{(3)}R_{\mu\nu\rho\sigma}$ associated with $h_{\mu\nu}$, the extrinsic curvature tensor and the "acceleration" vector, the components of the latter being respectively given by

$$
K_{\mu\nu} \equiv h^\alpha_\mu h^\beta_\nu \nabla_\alpha n_\beta\,, \quad a_\mu \equiv n^\nu \nabla_\nu n_\mu\,. \tag{3.5}
$$

The explicit expressions of these tensors in terms of $\phi$, its first derivatives $\phi_\mu$ and its second derivatives $\phi_{\mu\nu}$ can be easily computed:

$$
a_\mu = \frac{1}{2|X|} h_{\mu\nu} X^\nu\,, \quad K_{\mu\nu} = \frac{1}{\sqrt{|X|}}\left[\phi_{\mu\nu} + \frac{\phi^\alpha X_\alpha}{2X^2}\phi_\mu\phi_\nu - \frac{1}{2X}\left(\phi_\mu X_\nu + \phi_\nu X_\mu\right)\right], \tag{3.6}
$$

where $X_\mu = \partial_\mu X = 2\phi_{\mu\nu}\phi^\nu$.

The induced three-dimensional Riemann tensor is also of great importance. It is given by

$$
{}^{(3)}R_{\mu\nu\rho\sigma} = h^\alpha_\mu h^\beta_\nu h^\gamma_\rho h^\delta_\sigma\, {}^{(4)}R_{\alpha\beta\gamma\delta} + \varepsilon\left(K_{\mu\rho}K_{\nu\sigma} - K_{\nu\rho}K_{\mu\sigma}\right). \tag{3.7}
$$

This relation enables us to compute the three-dimensional Ricci tensor and the Ricci scalar which is given by the Gauss-Codazzi relation,

$$ {}^{(3)}R = {}^{(4)}R - \varepsilon \left[ K^2 - K_{\mu\nu}K^{\mu\nu} + 2\nabla_\mu (a^\mu - K n^\mu) \right] . \tag{3.8} $$

To obtain a more explicit form, we can use the following equations:

$$ \varepsilon K^2 = \frac{1}{X} L_2^{(2)} - \frac{2}{X^2} L_3^{(2)} + \frac{1}{X^3} L_5^{(2)} , \tag{3.9} $$

$$ \varepsilon K_{\mu\nu} K^{\mu\nu} = \frac{1}{X} L_1^{(2)} - \frac{2}{X^2} L_4^{(2)} + \frac{1}{X^3} L_5^{(2)} , \tag{3.10} $$

$$ a^2 = \frac{1}{X^2} L_4^{(2)} - \frac{1}{X^3} L_5^{(2)} , \tag{3.11} $$

where the $L_i^{(2)}$ are the elementary quadratic Lagrangians defined in eq. (1.27), together with the relation

$$ \varepsilon \int \mathrm{d}^4x \, \sqrt{-g} f \nabla_\mu (a^\mu - K n^\mu) = \int \mathrm{d}^4x \, \sqrt{-g} \Big[ \frac{2f_X}{X} L_3^{(2)} - \frac{2f_X}{X} L_4^{(2)} + \frac{f_\phi}{X} (X \Box \phi - \phi_{\mu\nu} \phi^\mu \phi^\nu) \Big] , \tag{3.12} $$

where $f$ is an arbitrary function of $\phi$ and $X$, $f_X$ its derivative with respect to $X$ and $f_\phi$ its derivative with respect to $\phi$.

Finally, one can express the determinant $g$ of the metric $g_{\mu\nu}$ in terms of the determinant $h$ of $h_{\mu\nu}$ and $X$ as follows

$$ \begin{aligned} g &= \frac{1}{24} \varepsilon^{\mu_1\nu_1\rho_1\sigma_1} \varepsilon^{\mu_2\nu_2\rho_2\sigma_2} g_{\mu_1\mu_2} g_{\nu_1\nu_2} g_{\rho_1\rho_2} g_{\sigma_1\sigma_2} \\ &= \frac{1}{6X} \varepsilon^{\mu_1\nu_1\rho_1\sigma_1} \varepsilon^{\mu_2\nu_2\rho_2\sigma_2} h_{\mu_1\mu_2} h_{\nu_2\nu_2} h_{\rho_1\rho_2} \phi_{\sigma_1} \phi_{\sigma_2} = \frac{h}{X} , \end{aligned} \tag{3.13} $$

which follows from the very definition of the determinant, and where $\varepsilon_{\mu\nu\rho\sigma}$ is the fully antisymmetric four-dimensional tensor.

#### 3.2.1.2. DHOST theories in the geometric frame

One notes that both $K_{\mu\nu}$ and $a^\mu$ are linear in $\phi_{\mu\nu}$ as shown in eq. (3.6). This suggests to rewrite the quadratic Lagrangian of DHOST theories in terms of the square of these quantities. In the following, we will therefore examine theories whose action reads

$$ S[g_{\mu\nu}, \phi] = \int \mathrm{d}^4x \, \sqrt{-g} \left( \frac{M_{\rm P}^2}{2} \, {}^{(4)}R + L_\phi \right) , \tag{3.14} $$

with a Lagrangian term $L_\phi$ of the form [1]

$$L_\phi = \lambda_1\, {}^{(3)}\!R + \lambda_2\, \varepsilon\, K^2 + \lambda_3\, a^2 + \lambda_4\, \varepsilon\, K_{\mu\nu} K^{\mu\nu}\,, \tag{3.15}$$

where $\lambda_A$ are arbitrary functions of $\phi$ and $X$. We have also included a dependence on the three-dimensional scalar curvature of $\Sigma_\phi$. It turns out that using only the first three terms, i.e. choosing $\lambda_4 = 0$, will be sufficient for our purpose as we will explain later in section 3.2.2. The total action is thus [2]

$$S[g_{\mu\nu},\phi] = \int \mathrm{d}^4x\, \sqrt{-g} \left( \frac{1}{2} M_{\rm P}^2\, {}^{(4)}\!R + \lambda_1\, {}^{(3)}\!R + \lambda_2\, \varepsilon\, K^2 + \lambda_3\, a^2 \right) . \tag{3.16}$$

As a consequence to the results of section 3.2.1, it can be reformulated as a sum of a quadratic action in the more usual form (3.1) with the coefficients

$$\begin{aligned} &F_2 = \frac{M_{\rm P}^2}{2} + \lambda_1\,, \quad A_1 = \frac{\lambda_1}{X}\,, \quad A_2 = \frac{\lambda_2 - \lambda_1}{X}\,, \\ &A_3 = \frac{2\lambda_1 - 4X\lambda_{1X} - 2\lambda_2}{X^2}\,, \quad A_4 = \frac{4X\lambda_{1X} - 2\lambda_1 + \lambda_3}{X^2}\,, \quad A_5 = \frac{\lambda_2 - \lambda_3}{X^3}\,, \end{aligned} \tag{3.17}$$

supplemented with a k-essence and a cubic galileon action given by

$$\int \mathrm{d}^4x\, \sqrt{-g}\, \frac{2\lambda_{1\phi}}{X} \left( X\Box\phi - \phi_{\mu\nu}\phi^\mu\phi^\nu \right) = \int \mathrm{d}^4x\, \sqrt{-g}\, \left( (2X\beta_X + \beta)\Box\phi + \beta_\phi X \right) , \tag{3.18}$$

with $X\beta = \lambda_{1\phi}$, where the last result follows from the relation

$$2 \int \mathrm{d}^4x\, \sqrt{-g}\, \beta_X \phi_{\mu\nu}\phi^\mu\phi^\nu = - \int \mathrm{d}^4x\, \sqrt{-g}\, \left( X\beta_\phi + \beta\Box\phi \right) , \tag{3.19}$$

for any function $\beta(\phi, X)$.

We have therefore proven that any theory of the form (3.16) can be cast into a quadratic DHOST theory.

#### 3.2.1.3. Geometric formulation of the cubic Galileon

As we have shown that a quadratic DHOST action can be rewritten in a more geometrical way, it is natural to look for a way to replace the cubic galileon term $Q(\phi, X)\Box\phi$ by a combination of $K_{\mu\nu}$ and $a_\mu$ as well. The only combination of these objects that is linear in $\phi$ is the trace of $K_{\mu\nu}$. Therefore, we now consider the action

$$S[g_{\mu\nu},\phi] = \int \mathrm{d}^4x\, \sqrt{-g}\, (\nu_0 + \nu_1 K) \ , \tag{3.20}$$

1. The sign $\varepsilon$ can be absorbed into a redefinition of $\lambda_2$ but we leave it here for later convenience.
2. Notice that the action (3.16) is a special case of the so-called spatially covariant gravity actions introduced in [171] and studied further in [172].

where $v_0$ and $v_1$ are two arbitrary functions of $X$ and $\phi$. Using (3.6), this action can be written as

$$S[g_{\mu\nu},\phi] = \int d^4x\,\sqrt{-g}\left[v_0 + \alpha(\phi^\mu\phi^\nu\phi_{\mu\nu} - X\Box\phi)\right]\,, \quad \alpha \equiv -\frac{\varepsilon v_1}{|X|^{3/2}}\,. \tag{3.21}$$

Using (3.19), we can show that, for any arbitrary function $\alpha(\phi,X)$,

$$\int d^4x\,\sqrt{-g}\,\alpha(\phi^\mu\phi^\nu\phi_{\mu\nu} - X\Box\phi) = -\frac{1}{2}\int d^4x\,\sqrt{-g}\left[XA_\phi + (A+2\alpha X)\,\Box\phi\right]\,, \tag{3.22}$$

where $A$ is such that $A_X = \alpha$. As a consequence, the action (3.20) can be written as

$$S[g_{\mu\nu},\phi] = \frac{1}{2}\int d^4x\,\sqrt{-g}\left(2\pi - XA_\phi - (A+2\alpha X)\,\Box\phi\right)\,. \tag{3.23}$$

One recovers the well-known k-essence and cubic galileon terms associated with the functions $P$ and $Q$ given by,

$$P(\phi,X) = v_0 - \frac{X}{2}A_\phi\,, \quad Q(\phi,X) = -\frac{1}{2}\,(A+2\alpha X)\,. \tag{3.24}$$

If the quadratic action of eq. (3.16) is also taken into account, it gives a linear contribution given by eq. (3.18). One must therefore replace $\alpha$ in (3.21) by

$$\alpha = -\frac{\varepsilon v_1}{|X|^{3/2}} + \frac{2\lambda_{1\phi}}{X}\,. \tag{3.25}$$

### 3.2.2. Disformal transformations

Let us now study in details the action of eq. (3.16), with the conditions of eq. (3.17). We do not consider the lower-order terms appearing in eq. (3.18) here. It is immediate to check that this quadratic Lagangian is indeed weakly degenerate, as it is of the form (3.3) with

$$F_2 = \frac{M_P^2}{2} + \lambda_1\,, \quad \kappa_1 = -\frac{M_P^2}{2X}\,, \quad \kappa_2 = \frac{M_P^2 + 2\lambda_2}{2X}\,, \quad \sigma = 0\,, \quad \alpha = \frac{\lambda_3 + 2X\lambda_{1X}}{X^2}\,. \tag{3.26}$$

We now perform a disformal transformation (see eq. (1.78)) of the action (3.16), using the formulas derived in [57] and recalled in section 1.4.1. The quadratic part of the new action is still weakly degenerate, i.e. of the form (3.3) with the coefficients

$$\begin{aligned} &\sigma = \frac{A_X}{A}, \quad \kappa_1 = -\frac{M_P^2}{2X}\frac{A^{3/2}}{\sqrt{A+BX}}, \quad \kappa_2 = \frac{M_P^2 + 2\lambda_2}{2X}\frac{A^{3/2}}{\sqrt{A+BX}}, \\ &F_2 = \frac{M_P^2 + 2\lambda_1}{2}\sqrt{A(A+BX)}, \quad \alpha = \mathscr{A}(\lambda_1,\lambda_3,A,B)\,, \end{aligned} \tag{3.27}$$

where we do not write explicity $\mathscr{A}$ as it is rather cumbersome (but it can be deduced from the expression (3.28) given below[3]). Interestingly, the new theory (3.27) is parametrised by five independent functions, as many functions as required to span the whole family of weakly degenerate theories.

Conversely, given a weakly degenerate theory defined by the set of functions $(F_2, \kappa_1, \kappa_2, \sigma, \alpha)$, one can successively invert the relations (3.27) and determine $A$ from $\sigma$, $B$ from $\kappa_1$, $\lambda_2$ from $\kappa_2$ and $\lambda_1$ from $F_2$. The last relation yields $\lambda_3$, which reads

$$\lambda_3 = \frac{2\left(X\alpha - 2f\sigma(2+X\sigma) + 2f_X(1+4X\sigma)\right)X^2\kappa_1^3}{M^2A^2\left(3\kappa_1(X\sigma-1) - 2X\kappa_{1X}\right)^2} \,. \tag{3.28}$$

This proves that any weakly degenerate action can be obtained from a disformal transformation of the action (3.16). As a conclusion, eq. (3.16) provides us with a complete parametrization, up to disformal transformations, of quadratic weakly degenerate theories.

Since the family of weakly degenerate theories is parametrised by five free functions and disformal transformations depend on two free functions, only three functions are needed to parametrise the geometric frame[4].

Finally, let us note that, although our intuitive reasoning was based on the unitary gauge, which implicitly assumes that the hypersurfaces $\Sigma_\phi$ are spacelike, i.e. $\epsilon = -1$, the relation between the actions (3.1) with coefficients (3.3) and (3.16) can be obviously extended to the case $\epsilon = +1$ even if they can no longer be interpreted as weakly degenerate actions.

## 3.3. DHOST classification revisited

All the subclasses of DHOST theories are stable under disformal transformations, as shown in [57] and section 1.4.1. This implies that once the DHOST theories in the geometric frame have been classified, the classification can immediately be extended to the whole family of DHOST theories, as they are all generated by disformal transformations from the geometric frame actions. As the classification

3. Indeed, the expression of $A$ can be obtained directly from (3.28) which gives

$$\alpha = \frac{M^2A^2\left(3\kappa_1(X\sigma-1) - 2X\kappa_{1X}\right)^2}{2X^3\kappa_1^3}\lambda_3 + \frac{2F_2\sigma(2+X\sigma) - 2F_{2X}(1+4X\sigma)}{X} \,.$$

Then, one substitutes $F_2$, $\kappa_1$ and $\sigma$ by their expressions (3.27) in terms of $A$, $B$ and $\lambda_1$ to get $\mathscr{A}$ explicitly.

4. This is the reason why we chose $\lambda_4 = 0$ in eq. (3.15). Note that one could have made different choices to reduce the number of free functions, e.g. one could impose $\lambda_2 = 0$, $\lambda_1 = 0$ or another linear relation between the four functions $\lambda_A$. However, one would lose the remarkable simplicity of the geometric frame, specially for class I which is explicitly written in eq. (3.31).

relies on the several ways one can solve the degeneracy conditions, we start by giving these conditions with the definitions of eq. (3.17).

### 3.3.1. Degeneracy conditions for quadratic DHOST theories

The three degeneracy conditions for the quadratic DHOST theories are given in eq. (1.35). These expressions simplify quite a lot when they are written in term of $\lambda_1$, $\lambda_2$ and $\lambda_3$, substituting eq. (3.17):

$$\begin{aligned} D_0(X) &= \frac{2\lambda_2}{X}\left[\left(M_{\rm P}^2+2\lambda_1\right)\left(M_{\rm P}^2+2\lambda_1-\lambda_3-8X\lambda_{1X}\right)+16X^2\lambda_{1X}^2\right], \\ D_1(X) &= -\frac{2M_{\rm P}^4}{X^2}\lambda_3+\frac{2\lambda_2}{X^2}\left[8\lambda_1^2-5M_{\rm P}^2\lambda_3+4\lambda_1(2M_{\rm P}^2-\lambda_3-8X\lambda_{1X})\right. \\ &\qquad\left.+2(M_{\rm P}^2-4X\lambda_{1X})^2\right], \\ D_2(X) &= -\frac{2M_{\rm P}^4}{X^2}\lambda_3+\frac{2\lambda_2}{X^2}\left[4\lambda_1^2-4M_{\rm P}^2\lambda_3+2\lambda_1(2M_{\rm P}^2-\lambda_3-8X\lambda_{1X})\right. \\ &\qquad\left.+(M_{\rm P}^2-4X\lambda_{1X})^2\right]. \end{aligned} \tag{3.29}$$

Weakly degenerate theories satisfy the single condition [12]

$$D_0(X)-XD_1(X)+X^2D_2(X)=0\,. \tag{3.30}$$

### 3.3.2. DHOST theories in class I

Class I is characterized by the relation $A_1=-A_2$, which is equivalent to $\lambda_2=0$ according to eq. (3.17). Under this assumption, the first degeneracy condition $D_0=0$ is automatically satisfied and the two other conditions, remarkably, both reduce to $\lambda_3=0$, as can be seen from eq. (3.29). DHOST theories in class I, when expressed in the geometric frame, are thus of the very simple form

$$S_{\rm I}=\int d^4x\sqrt{-g}\left(\frac{1}{2}M_{\rm P}^2\,{}^{(4)}\!R+\lambda_1\,{}^{(3)}\!R\right). \tag{3.31}$$

Within the class I, one finds the subclass Ib characterized by $\alpha_1=F_2/X$. It is then clear from the first relation in eq. (3.17) that this subclass corresponds to the action (3.31) with $M_{\rm P}=0$, i.e. a pure three-dimensional curvature term, which does not contain tensor modes [57] .

Let us discuss further the subclass Ia, which is the most interesting from a physical point of view. The simplest theory in the geometric frame is obviously GR, with $\lambda_1=0$. Via disformal transformations, it generates a family of DHOST theories parametrised by two functions, $A$ and $B$. Let us stress that the speed $c_g$ of gravitational waves is modified via a disformal transformation as the causal structure is modified. From an initial value $c_g^2 = 1 + 2\lambda_1/M_{\rm P}^2$ in the geometric frame [5], one gets after disformal transformation

$$c_g^2 = \left(1 + \frac{BX}{A}\right)\left(1 + \frac{2\lambda_1}{M_{\rm P}^2}\right) . \tag{3.33}$$

As expected, only conformal transformations, i.e. with $B = 0$, leave $c_g$ invariant. Consequently, theories in subclass Ia such that $c_g = c$ (for any solution) can be obtained from a theory with arbitrary $\lambda_1$ in the geometric frame (for which $c_g \neq c$) via a disformal transformation that compensates the initial detuning of $c_g$ from $c$ so that the final DHOST theory verifies $c_g = c$ (which is equivalent to the condition $\alpha_1 = 0$ in the original DHOST formulation [111, 173, 174]).

Note that the DHOST theories that can play the role of dark energy while satisfying both $c_g = c$ and the GW decay constraint [101], as suggested by the GW170817 observation [99] (see section 1.5.2.2), correspond to the theories generated via conformal transformations ($B = 0$) from GR in the geometric frame, i.e. with $\lambda_1 = 0$. The cosmology of such theories has been explored in [175]. However, since the LIGO-Virgo measurements probe wavelengths many orders of magnitude smaller than cosmological scales, these constraints do not necessarily apply on cosmological scales (see e.g. [117]), leaving the other DHOST theories still relevant for cosmology [176].

In addition to the geometric frame introduced in the present work, another convenient frame for the theories of subclass Ia is the "Horndeski frame" where the corresponding action is of the Horndeski form (up to lower order terms), i.e.

$$S_H[g_{\mu\nu}, \phi] = \int \mathrm{d}^4x \sqrt{-g} \left( F\, {}^{(4)}\!R + 2F_X (L_1^{(2)} - L_2^{(2)}) \right) . \tag{3.34}$$

It is thus interesting to derive explicity the disformal transformation that relates these two frames. Given any (quadratic) Horndeski action, characterized by a function $F(\phi, X)$, one can define

$$\lambda_1 = -\frac{M_{\rm P}^2}{2} + \frac{2}{M_{\rm P}^2} F (F - 2XF_X) , \tag{3.35}$$

5. The quantity $c_g^2$ is easily obtained from the coefficients of the time derivatives and spatial gradients of the tensor modes in the action, using the same method as in section 1.5.2.2. In practice, using the Gauss-Codazzi identity given in eq. (3.8), we express ${}^{(4)}\!R$ in terms of ${}^{(3)}\!R$ and $K_{\mu\nu}$ so that the action (3.31) becomes

$$S_{\rm I} = \frac{1}{2} M_{\rm P}^2 \int \mathrm{d}^4x \sqrt{-g} \left[ (1 + 2\lambda_1/M_{\rm P}^2)\ {}^{(3)}\!R + \varepsilon \left( K^2 - K_{\mu\nu}K^{\mu\nu} \right) \right] , \tag{3.32}$$

and then we obtain $c_g^2$ as the ratio of the coefficient of ${}^{(3)}\!R$ with the coefficient of $-\varepsilon K_{\mu\nu}K^{\mu\nu}$.

and verify that the disformal transformation of eq. (3.31) with $A = \mathrm{sgn}(F - 2XF_X) \in \{+1, -1\}$ and

$$AB = \frac{M_{\mathrm{P}}^4}{4X(2XF_X - F)^2} - \frac{1}{X}, \tag{3.36}$$

gives exactly eq. (3.34). Note that we have necessarily $2XF_X - F \neq 0$ since we are not in the subclass Ib.

### 3.3.3. DHOST theories in the classes II and III

#### 3.3.3.1. Classification

Let us first discuss DHOST theories in the class III, characterized by $F_2 = 0$, i.e.

$$\lambda_1 = -\frac{M_{\mathrm{P}}^2}{2}. \tag{3.37}$$

If $M_{\mathrm{P}} \neq 0$, the degeneracy conditions $D_1 = 0$ and $D_2 = 0$ then imply

$$\lambda_3 = 0 \quad \text{or} \quad \lambda_2 = -\frac{M_{\mathrm{P}}^2}{3}, \tag{3.38}$$

corresponding to the subclasses IIIa and IIIb respectively. Finally, the case $M_{\mathrm{P}} = 0$ and thus $\lambda_1 = 0$, with $\lambda_3$ and $\lambda_2$ free, corresponds to the subclass IIIc, which does not contain tensor modes.

Let us now turn to the class II, which contains the theories that are neither in class I or in class III. From the degeneracy conditions given in eq. (3.29), one finds that the subclass IIa is characterized by

$$\lambda_1 + \frac{M_{\mathrm{P}}^2}{2} = F_2(X) = \xi\sqrt{|X|}, \quad \lambda_3 = 0, \tag{3.39}$$

where $\xi$ is a constant. The class IIb is characterized by two free functions, $\lambda_1$ and $\lambda_2$, and the conditions

$$M_{\mathrm{P}} = 0, \quad \lambda_3 = 2\frac{(\lambda_1 - 2X\lambda_{1X})^2}{\lambda_1}, \tag{3.40}$$

which solve the degeneracy conditions.

As a consequence, any DHOST theory in the class II can be described by the action

$$S_{\mathrm{II}}[g_{\mu\nu}, \phi] = \xi \int \mathrm{d}^4x\, \sqrt{-g|X|}\, {}^{(3)}\!R - \frac{M_{\mathrm{P}}^2}{2}\varepsilon \int \mathrm{d}^4x\, \sqrt{-g}\big(K_{\mu\nu}K^{\mu\nu} - (1-\mu_2)\,K^2\big) \tag{3.41}$$

$$= \xi \int \mathrm{d}^4x \sqrt{|h|}\, {}^{(3)}\!R - \frac{M_{\mathrm{P}}^2}{2}\varepsilon \int \mathrm{d}^4x \sqrt{-g}\big(K_{\mu\nu}K^{\mu\nu} - (1-\mu_2)K^2\big)\,, \tag{3.42}$$

where we used eq. (3.13) to relate the determinants $g$ and $h$ and we introduced the dimensionless parameter $\mu_2 \equiv 2\lambda_2/M_{\mathrm{P}}^2$. Interestingly the three-dimensional Ricci term in this action reduces exactly to the three-dimensional Einstein-Hilbert Lagrangian for the metric $h_{\mu\nu}$ but integrated over the four-dimensional space-time.

#### 3.3.3.2. Instabilities

The geometric frame reformulation (3.42) of DHOST theories in class II is particularly convenient to see that these theories are plagued by instabilities, as originally shown in [60] (see section 1.5.2.1). Indeed, if one considers solutions with $\varepsilon = -1$ in the unitary gauge, spatial gradients in the equations of motion of the (scalar and tensor) fields can only originate from the ${}^{(3)}\!R$ term in the action.

Let us consider a homogeneous and isotropic background, with scale factor $a(t)$ and scalar field $\phi(t)$. In the unitary gauge, the perturbations about such a background are fully encoded in the scalar perturbation $\zeta$ and the tensor perturbation $\gamma_{ij}$ of the three-dimensional metric $h_{ij}$,

$$h_{ij} = a^2(t)e^{2\zeta}(\delta_{ij} + \gamma_{ij})\,, \tag{3.43}$$

where latin letters $(i,j,\cdots)$ hold for spatial indices. When we substitute this expression into the the ${}^{(3)}\!R$ term which appears in (3.42), we obtain at quadratic order in the perturbations

$$\int \mathrm{d}^4x \sqrt{h}\, {}^{(3)}\!R = \int \mathrm{d}^4x\, a\, \left(-\frac{1}{4}\partial_k\gamma_{ij}\,\partial^k\gamma^{ij} + \partial_i\zeta\,\partial^i\zeta + o(\gamma^2,\zeta^2,\gamma\zeta)\right)\,. \tag{3.44}$$

Hence, we immediately see that the gradient term of the scalar mode $\zeta$ has an opposite sign compared to the tensor modes and therefore there will be necessarily a gradient instability either in the tensor sector or in the scalar sector[6]. Thus, we recover very easily the result that DHOST theories in class II are unstable [60]. We see that this instability is closely related to the form (3.39) of $\lambda_1$ which is necessary to select a DHOST theory among weakly degenerate theories.

6. In the case where $\varepsilon = +1$, one cannot take the unitary gauge anymore, but instead one could fix $\phi$ to be one of the three spatial coordinates. Hence, a similar analysis would lead to a quadratic action of the form (3.44) with $(i,j,k)$ three-dimensional space-time indices while the terms with $K_{\mu\nu}$ in (3.42) would involve now (space-like) gradients of the fields only. Therefore, the only kinetic terms of the scalar field and of the tensor modes would be contained in eq. (3.44) and we clearly see that they have opposite signs, which means that there would be a ghost instability.

## 3.4. Equations of motion in the geometric frame

The expression and geometrical interpretation of DHOST theories are much simpler in this novel geometric frame than in the Horndeski frame. The equations of motion are also simpler and can be formulated in geometrical terms which could help finding solutions.

In this section, we compute the equations of motion of the class I action given in eq. (3.31), which reads

$$S = \int \mathrm{d}^4x \, \sqrt{-g} \left( \frac{1}{2} M_\mathrm{P}^2 \, {}^{(4)}\!R + \lambda_1 \, {}^{(3)}\!R + \nu_1 K + \nu_0 \right) . \tag{3.45}$$

We added for purposes of generality a k-essence term $S_{\nu_0}$ associated with the function $\nu_0(\phi, X)$ and a K-term $S_{\nu_1}$ associated with the function $\nu_1(\phi, X)$ (see eq. (3.20)). As we proved in section 3.2.1, the K-term can be equivalently reformulated as a cubic Galileon up to k-essence terms.

Using the formulae

$$\delta X = 2\phi^\mu \partial_\mu \delta\phi - \phi^\mu \phi^\nu \delta g_{\mu\nu} , \tag{3.46}$$

$$\delta h_{\mu\nu} = \delta g_{\mu\nu} + \frac{\delta X}{X^2} \phi_\mu \phi_\nu - \frac{\phi_\mu \partial_\nu \delta\phi + \phi_\nu \partial_\mu \delta\phi}{X} , \tag{3.47}$$

for the infinitesimal variations of $X$ and $h_{\mu\nu}$ together with eq. (3.13), we show by a direct calculation that the equation of motion for the metric takes the standard form,

$$\frac{M_\mathrm{P}^2}{2} \, {}^{(4)}G_{\mu\nu} = T_{\mu\nu}^{(\lambda_1)} + T_{\mu\nu}^{(\nu_1)} + T_{\mu\nu}^{(\nu_0)} , \tag{3.48}$$

where $T_{\mu\nu}^{(\nu_0)}$ and $T_{\mu\nu}^{(\nu_1)}$ are the usual stress-energy tensors associated to the actions $S_{\nu_0}$ and $S_{\nu_1}$ respectively:

$$T_{\mu\nu}^{(\nu_0)} = \frac{\nu_0}{2} g_{\mu\nu} - \nu_{0X} \phi_\mu \phi_\nu , \quad T_{\mu\nu}^{(\nu_1)} = \frac{1}{2q} \left( \nu_{1\phi} - 2\nu_{1X} \Box\phi \right) \phi_\mu \phi_\nu . \tag{3.49}$$

where $q \equiv |X|^{1/2}$. The tensor $T_{\mu\nu}^{(\lambda_1)}$ is the stress-tensor energy associated to the three-dimensional Ricci term in eq. (3.31):

$$T_{\mu\nu}^{(\lambda_1)} = -q \left[ \mu_1 \, {}^{(3)}G_{\mu\nu} + \mu_{1X} {}^{(3)}\!R \, \phi_\mu \phi_\nu + ( h_{\mu\nu} {}^3\Box - {}^3\nabla_\mu {}^3\nabla_\nu ) \mu_1 \right] , \tag{3.50}$$

where ${}^3\nabla_\mu$ is the 3-dimensional covariant derivative with respect to $h_{\mu\nu}$ and we introduced the notation $\mu_1 \equiv \lambda_1 / q$ for simplicity.

Even though the equation of motion for the scalar field is redundant as it can be deduced from the previous one, it is nonetheless useful to give its expression which takes the form

$$\nabla_\mu J^\mu + \Phi = 0 , \tag{3.51}$$

where the current $J^\mu$ and the source $\Phi$ are given by

$$J^\mu = J^\mu_{(\lambda_1)} + J^\mu_{(\nu_1)} + J^\mu_{(\nu_0)}\,, \quad \Phi = \Phi_{(\lambda_1)} + \Phi_{(\nu_1)} + \Phi_{(\nu_0)}\,, \tag{3.52}$$

and each components of $J^\mu$ and $\Phi$ are given by

$$\begin{aligned} &J^\mu_{(\lambda_1)} = \lambda_{1X}{}^{(3)}\!R\phi^\mu\,, \quad J^\mu_{(\nu_1)} = \frac{1}{2q}(2q\nu_{1X}K - \nu_{1\phi})\phi^\mu + q\nu_{1X}a^\mu\,, \quad J^\mu_{(\nu_0)} = \nu_{0X}\phi^\mu\,, \\ &\Phi_{(\lambda_1)} = -\frac{\lambda_{1\phi}}{2}\,{}^{(3)}\!R\,, \quad \Phi_{(\nu_1)} = \frac{1}{2q}\left[X\nu_{1\phi\phi} + 2X\nu_{1\phi X}(\Box\phi - qK)\right]\,, \\ &\Phi_{(\nu_0)} = -\frac{\nu_{0\phi}}{2}\,. \end{aligned} \tag{3.53}$$

If the theory is shift symmetric, i.e. the functions $\lambda_1$, $\nu_1$ and $\nu_0$ do not depend on $\phi$, we recover the well-known fact that the equation of the scalar field reduces to a conservation equation for the curent $J^\mu$.

Notice that the equation for the metric involves third order derivatives of the scalar field and, similarly, the equation for the scalar field involves third order derivatives of the metric components. This is expected as the action is not formulated in the Horndeski frame. However, in the case where $\varepsilon = -1$, these higher order terms are all spatial derivatives which is consistent with the fact that there is no ghost propagating in the theory. These equations have a simple form compared to Horndeski theories, which might be useful to find new solutions.

Let us see how the conditions for having stealth solutions given in section 2.3.2 are formulated in the geometric frame. We assume that $X$ is a constant $X_0$. Thus, the equations of motion simplify drastically and we have (for the metric only)

$$M_P^2\,{}^{(4)}G_{\mu\nu} - \nu_0 g_{\mu\nu} + 2q\,\mu_1\,{}^{(3)}G_{\mu\nu} + 2\left(q\,\mu_{1X}\,{}^{(3)}\!R + \nu_{0X} + \frac{\nu_{1X}}{q}\Box\phi\right)\phi_\mu\phi_\nu = 0\,. \tag{3.54}$$

Now, we see that, when the theory satisfies the following conditions [7],

$$\mu_1(X_0) = 0\,, \quad \mu_{1X}(X_0) = 0\,, \quad \nu_{0X}(X_0) = 0\,, \quad \nu_{1X}(X_0) = 0\,, \tag{3.55}$$

7. Notice that these conditions are consistent with those obtained in [150] for DHOST theories when their actions are written in the usual form of eq. (3.1) supplemented with a k-essence term and a cubic galileon term:

$$P + 2\Lambda F_2 = 0\,, \quad P_X + \Lambda(4F_{2X} - XA_{1X}) = 0\,, \quad Q_X = 0\,, \quad A_1 = 0\,, \quad A_3 + 2A_{1X} = 0\,.$$

When we replace the coefficients $A_i$ by their expressions given in eq. (3.17) in terms of $\lambda_1$ and $P$ and $Q$ by their expressions in terms of $\nu_0$ and $\nu_1$ obtained in eq. (3.24), we recover immediately the stealth conditions (3.55). In particular, we show that $Q_X = \nu_{1X}/q$, which immediately implies the equivalence between the conditions $Q_X = 0$ and $\nu_{1X} = 0$.

it admits stealth solutions which satisfy the usual Einstein equation for general relativity,

$$^{(4)}G_{\mu\nu} + \Lambda g_{\mu\nu} = 0\,, \quad M_{\rm P}^2 \Lambda + \nu_0(X_0) = 0\,. \tag{3.56}$$

The conditions of eq. (3.55) are sufficient but not necessary for the existence of stealth solutions. The theory could admit stealth Schwarzschild solutions without satisfying these conditions as shown in [150] (and references therein) for instance.

## Conclusion

In this chapter, we have presented a strikingly simple reformulation of quadratic DHOST theories (and of weakly degenerate theories), based on a Lagrangian involving a few geometrical terms associated with the three-dimensional constant $\phi$ hypersurfaces, in addition to the standard Einstein-Hilbert term. This geometric frame action describes only a subset of DHOST theories but the rest of the family can be "generated" from this subset via disformal transformations.

Moreover, since the various subclasses of DHOST theories are stable under disformal transformations, it is sufficient to classify the subset of theories in the geometric frame to automatically generalise this classification to the whole family. A compelling illustration is given by the subclass Ia, the most interesting class from a phenomenological perspective, which includes Horndeski's theories. In the original classification of DHOST theories, this subclass is parametrised by the three functions $F_2, A_1$ and $A_2$, while the other functions are expressed in terms of these, with rather ugly expressions for $A_4$ and $A_5$. By contrast, in this new geometric perspective, the subclass Ia arises from an geometric frame action that depends on a single function $\lambda_1$ multiplying the three-dimensional curvature, all the other theories being obtained via disformal transformations. The subclass Ia is thus parametrised by the three functions $\lambda_1, A$ and $B$. One can proceed similarly for the other DHOST subclasses, as well as for all the quadratic weakly degenerate theories which are also of the form (3.16).

As we have argued in this chapter, the geometric frame perspective is appealing to understand and analyse the underlying dynamical structure of DHOST theories, similarly to the Einstein frame in traditional scalar-tensor theories. However we should stress that, in the presence of matter, it is much more convenient in general, from a practical point of view, to stick to the physical frame where matter is minimally coupled to the metric rather than to move to the geometric frame. Indeed, matter would be disformally coupled to the geometric frame metric, with the functions $A$ and $B$ usually defined only implicitly. This would make the calculations in this frame very cumbersome. For concrete applications, it is thus more appropriate to work directly with the original formulation and to add matter minimally coupled to the metric in eq. (3.1). Scalar-tensor theories that are disformally related thus correspond to physically distinct theories since matter is minimally coupled to both theories.

By contrast, the geometric frame approach should be useful to get a better intuitive understanding of DHOST theories as well as for their classification, or to study their generic properties invariant under disformal transformations, such as the instabilities in classes II and III. In this respect, it would be interesting to extend the geometric frame description to include the cubic DHOST theories, whose classification in the standard formulation is even more involved than for the quadratic case.

CHAPTER 4

# BLACK HOLE PERTURBATIONS

Contents

Oscillations of black holes have been studied theoretically for several decades. Today, with the first observations of gravitational waves emitted by BH mergers, one can now hope to observe directly these oscillations via their GW signatures, especially in the ringdown phase of the signal when the post-merger black hole relaxes to a Kerr black hole, according to GR. One of the major goals of future detections will be to check whether the observed oscillations coincide with the predictions based on GR (see e.g. [177, 178]). This is also an ideal playground to test alternative theories of gravitation. Indeed, even if the background BH solution may coincide with that of GR, the linear perturbations in general obey different equations of motion.

During the ringdown phase, at least in the linear regime, the GW signal is expected to mainly consist of a superposition of the so-called quasi-normal modes which have been excited by the merger and then decay via GW radiation: these modes correspond to the proper oscillation modes of the black hole and are characterised by a complex frequency $\omega$, whose imaginary part quantifies their damping rate.

In the simplest case of nonrotating black holes, i.e. Schwarzschild black holes, the computation of QNMs is based on the classical papers by Regge & Wheeler [179] and later Zerilli [180], who reformulated the linearised Einstein equations in the frequency domain, which are *first-order* with respect to the radial coordinate, as a *second-order* Schrödinger-like equation. This familiar equation, with a specific potential for axial and polar metric perturbations, is the standard starting point for the numerical calculations or semi-analytical treatments of QNMs, using for instance well-known methods in quantum mechanics.

In this chapter, we define QNMs and give the mathematical procedure used to compute them. We then review the classical computations of perturbations around a Schwarzschild BH in GR, from the perturbation of Einstein's equations to their reformulation as Schrödinger-like equations for two scalars built from the perturbation quantities. We conclude with a review of the new difficulties that appear when one studies perturbations around a BH in modified gravity theories. This chapter is based on the paper [181].

## 4.1. Quasi-normal modes

### 4.1.1. Definition and theoretical computation

From a perturbative point of view, physical systems behave very similarly in many different domains of physics. Indeed, the linear response of any system to a small "kick" will be to oscillate around its position of equilibrium, the possible oscillation frequencies being part of a discrete set called the *eigenmodes* of the system. The study of these modes gives a lot of information about the system and is therefore very interesting in order to characterize its behaviour and test the theories that describe it.

Let us consider the example of a guitar string of length $L$. At equilibrium, it has a straight shape, due to a tension that is applied to it on both ends. One can then study the dynamics of a small perturbation $y(t,x)$ along the string. After a few computations, one obtains an evolution equation for this perturbation:

$$\frac{\partial^2 y}{\partial t^2} - c_{\mathrm{s}}^2 \frac{\partial^2 y}{\partial x^2} = 0\,. \tag{4.1}$$

This equation describes the propagation of a wave along the string, at speed $c_{\mathrm{s}}$. Finding the modes then requires one to impose $y(t,0) = y(t,L) = 0$, since the string is fixed at both ends. The only possible frequencies are then the $f_n$ such that

$$f_n = \frac{n c_{\mathrm{s}}}{2L} \quad \text{with} \quad n \in \mathbf{N}^*\,. \tag{4.2}$$

One therefore recovers a discrete set of frequencies: the response to any perturbation of the equilibrium position will be a sum of sinusoids with frequencies $f_n$. One can observe that the computed frequencies depend on the system via the length $L$ as well as the theory describing the dynamics through the speed $c_{\mathrm{s}}$.

The fact that perturbations have discrete frequencies and depend on both the solution and the theory is also true in the case of BH physics. Therefore, the study of these perturbations is very relevant in the context of tests of GR, since it will allow one to test both the BH background *and* the underlying theory of gravity. This way, even stealth solutions such as the ones described in section 2.3 will show deviations from the usual GR solutions.

A BH being described by the metric of spacetime around it, the perturbation will be in the form of a rank-2 symmetric tensor that will play the role of a metric perturbation. The computation will be similar to what is done for a guitar string:

1. computation of the equations of motion for the perturbations;
2. casting of these equations into a wave propagation equation;
3. derivation of physically relevant boundary conditions;
4. numerical computation.

The first two steps, while quite simple in the case of the string, become technical in the case of GR, and are even more complicated in a modified theory of gravity. In this chapter, we describe the procedure for the former case; chapters 6 and 7 will be devoted to the latter. We describe the relevant boundary conditions below, and the numerical treatment will be done in chapter 8.

### 4.1.2. Boundary conditions

Let us assume we have obtained a propagation equation for the perturbation. In the case of a guitar string, the boundary conditions come from a physical argument: the string is fixed to the guitar at both ends, so it cannot be displaced from its equilibrium position $y = 0$; in the case of a BH, we must perform a similar reasoning.

A BH is defined by the presence of an event horizon, separating spacetime into two zones that cannot be both causally linked to the other: any event happening inside the horizon will never have any effect on what is happening outside. Therefore, it is not possible for information to travel out of the BH event horizon towards the rest of spacetime: one must impose *ingoing* boundary conditions at the event horizon. This means that the metric perturbation must behave as $e^{-i(\omega t+kr)}$, with $r$ some radial coordinate.

Similarly, no information can come from infinity: the only possible direction of propagation at infinity is outwards, meaning that one must impose *outgoing* boundary conditions at infinity: the metric perturbation must behave as $e^{-i(\omega t-kr)}$. The presence of two boundary conditions, combined with one propagation equation, is enough to compute the allowed set of frequencies: more details about the numerical methods will be given in chapter 8.

One can note a fundamental difference between a problem of the kind of the guitar string perturbations and the BH perturbations: in the former case, the problem is *self-adjoint*, since both the propagation equation and the boundary conditions are real. This implies that the frequencies obtained are real. In the latter case however, the boundary conditions are not real: the problem cannot be self-adjoint, and the frequencies obtained will be complex. This means that one will not get stationary waves oscillating through spacetime, but damped or exponentially growing sinusoids. This can be understood as a loss of energy of the waves towards infinity and the horizon.

### 4.1.3. Physical realisation and experimental measurement

QNMs appear during the last phase of a BBH merger. Indeed, the BH created after the two parents have merged is initially in an excited state: its surface is deformed, and it needs to lose its excess of energy by radiating GWs through spacetime. While the exact nature of the sourcing of perturbations by the deformation of the BH horizon is extremely complicated, the frequencies at which GWs can be emitted are restricted to the set of QNMs. This means that modeling the ringdown signal by a sum of damped sinusoids will be sufficient to measure the QNM frequencies [182].

A typical signal from a BBH can be observed on fig. 4.1. One can observe that the ringdown signal dampens very quickly: only a few oscillations will be measurable, which will limit the precision. Furthermore, since the beginning of the ringdown is not clearly defined, one must make further assumptions about its starting time in order to extract QNMs, which will lead to more uncertainties.

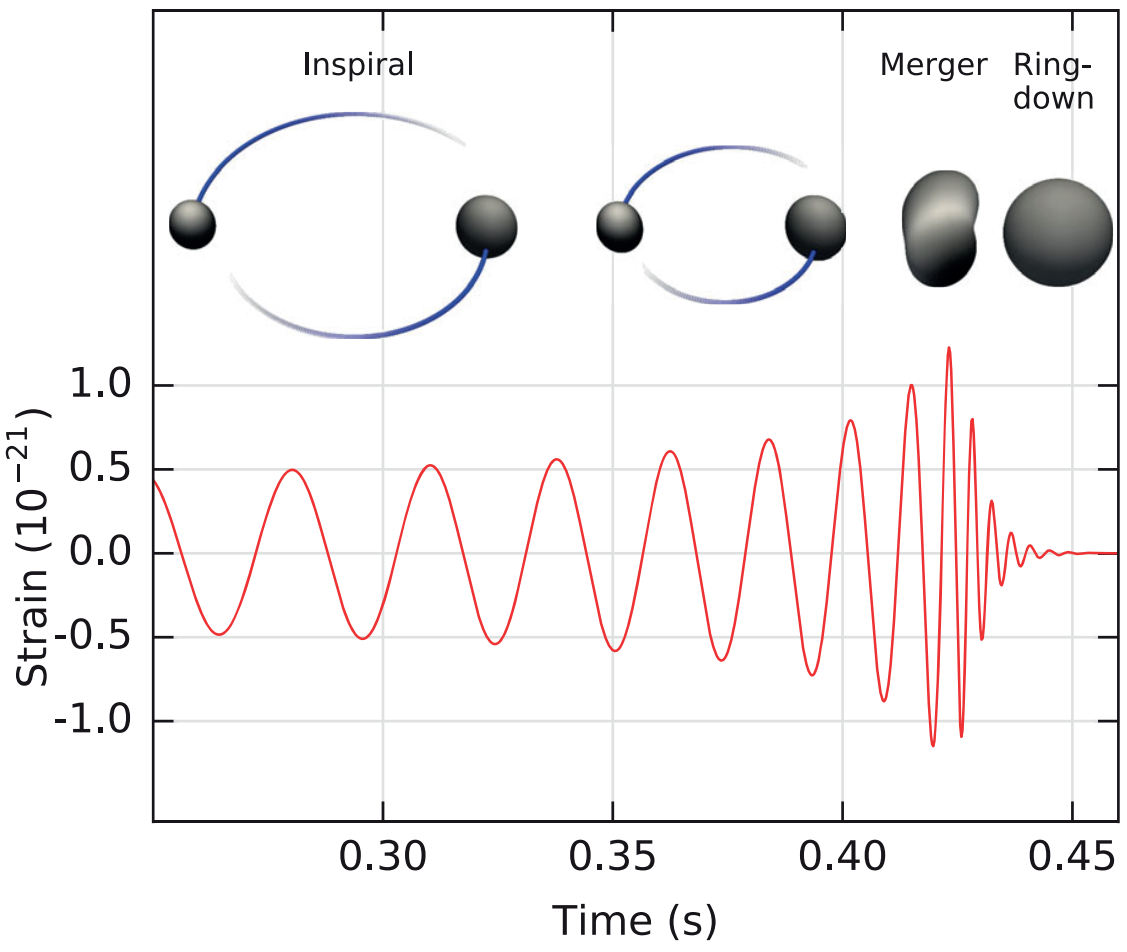

Figure 4.1. – GW signal from a BBH merger. Figure adapted from [14].

Tests of GR using ringdown signal have been performed since the detection of GW150914 [183]. Improvements using the measurement of overtones were proposed in [184, 185], but it was shown in [186] that the presence of these overtones might be artificially induced by noise. Further constraints on the fundamental QNM were obtained in [187], and improved using a second detection in [188]. Presently, GR is in agreement with all observations; in the future, improved GW detectors will provide better precision and constraints will be tightened [189, 190].

It is also possible to test GR using the inspiral phase. However, in order to provide constraints on existing modified gravity theories, this requires a modeling of the gravitational waveform emitted by a BBH system in the context of such theories. Computations of this kind are very hard and existing results are very recent [191, 192]. From a phenomenological point of view, it is also possible to introduce *ad hoc* deviations from GR in the expression of the signal in the inspiral phase in order to measure them [193, 194, 195].

## 4.2. Linearized GR equations

In this section, we give the framework for the computation of the equations of motion of a metric perturbation over a Schwarzschild background, paving the way for the derivation of a wave propagation equation. Let us start with the four-dimensional Einstein-Hilbert action in vacuum (with no cosmological constant) for the metric $g_{\mu\nu}$,

$$S[g_{\mu\nu}] = \frac{M_\mathrm{P}^2}{2}\int \mathrm{d}^4x\,\sqrt{-g}\,R\,, \tag{4.3}$$

where $g \equiv \det\left(g_{\mu\nu}\right)$ is the determinant of the metric, $R$ the four-dimensional Ricci scalar and $M_\mathrm{P}$ denotes the Planck mass, which actually will not show up in the equations of motion since we are not considering any matter field here.

Given any background metric $\bar{g}_{\mu\nu}$ solution to the Einstein equations, one can introduce the perturbed metric

$$g_{\mu\nu} = \bar{g}_{\mu\nu} + h_{\mu\nu} \tag{4.4}$$

where the $h_{\mu\nu}$ denote the linear perturbations of the metric. In order to derive the linear equations of motion that govern the evolution of $h_{\mu\nu}$, one expands the Einstein-Hilbert action (4.3) up to the second order in $h_{\mu\nu}$. The Euler-Lagrange equations associated with the *quadratic* part of this expansion then provide the linearised equations of motion for $h_{\mu\nu}$.

By expanding (4.3), one obtains the following quadratic action for $h_{\mu\nu}$,

$$\begin{aligned} S_\mathrm{quad}[h_{\mu\nu}] = \frac{M_\mathrm{P}^2}{2}\int \mathrm{d}^4x\,\sqrt{-\bar{g}}\Big[ & -\frac{1}{2}h_{\mu\nu}h^{\mu\nu}\bar{R} + \frac{1}{4}h^2\bar{R} + hh_{\mu\nu}\bar{R}^{\mu\nu} \\ & + 4h_\mu{}^\rho h^{\mu\nu}\bar{R}_{\nu\rho} - 2h^{\mu\nu}h^{\rho\sigma}\bar{R}_{\mu\rho\nu\sigma} + \frac{1}{2}(\bar{\nabla}_\mu h)(\bar{\nabla}^\mu h) \\ & - 2(\bar{\nabla}_\mu h^\mu{}_\nu)(\bar{\nabla}_\rho h_\nu{}^\rho) - (\bar{\nabla}_\mu h)(\bar{\nabla}_\nu h^{\mu\nu}) \end{aligned}$$

$$+3(\bar{\nabla}_\nu h_{\mu\rho})(\bar{\nabla}^\rho h^{\mu\nu}) - \frac{1}{2}(\bar{\nabla}_\rho h_{\mu\nu})(\bar{\nabla}^\rho h^{\mu\nu})\Big], \quad (4.5)$$

where $\bar{R}_{\mu\nu\rho\sigma}$, $\bar{R}_{\mu\nu}$, $\bar{R}$ and $\bar{\nabla}_\mu$ are respectively the Riemann tensor, the Ricci tensor, the Ricci scalar and the covariant derivative associated with the *background* metric $\bar{g}_{\mu\nu}$. The indices are lowered or raised with $\bar{g}_{\mu\nu}$ and $h \equiv \bar{g}^{\mu\nu}h_{\mu\nu}$ denotes the trace of the metric perturbation. The linearised Einstein equations are then given by the Euler-Lagrange equations of (4.5) and can be written in the form

$$\mathcal{E}_{\mu\nu} \equiv \bar{\nabla}_\sigma\bar{\nabla}^\sigma h_{\mu\nu} + \bar{\nabla}_\mu\bar{\nabla}_\nu h + (\bar{\nabla}_\alpha\bar{\nabla}_\beta h^{\alpha\beta} - \bar{\nabla}_\sigma\bar{\nabla}^\sigma h)\bar{g}_{\mu\nu} + 2\bar{\nabla}_{(\mu}\bar{\nabla}_\alpha h^\alpha_{\nu)} - 6\bar{\nabla}_\alpha\bar{\nabla}_{(\mu}h^\alpha_{\nu)}$$
$$+\bar{R}_{\mu\nu}h - \bar{R}h_{\mu\nu} + \frac{1}{2}\bar{R}\bar{g}_{\mu\nu}h + \bar{R}^{\alpha\beta}\bar{g}_{\mu\nu}h_{\alpha\beta} + 8\bar{R}_{\alpha(\mu}h^\alpha_{\nu)} = 0\,, \quad (4.6)$$

where use the standard notation $A_{(\mu\nu)} \equiv (A_{\mu\nu} + A_{\nu\mu})/2$ for the symmetrisation of any rank-2 tensor $A_{\mu\nu}$.

Let us now specialise these equations to the case where the background metric is the Schwarzschild metric, expressed as

$$\bar{g}_{\mu\nu}\,\mathrm{d}x^\mu\,\mathrm{d}x^\nu = -\left(1-\frac{\mu}{r}\right)\mathrm{d}t^2 + \left(1-\frac{\mu}{r}\right)^{-1}\mathrm{d}r^2 + r^2\left(\mathrm{d}\theta^2 + \sin^2\theta\,\mathrm{d}\varphi^2\right)\,, \quad (4.7)$$

where $\mu = 2M_s$ is the Schwarzschild radius, $M_s$ being the mass of the black hole.

Given the spherical symmetry of the background solution, it is convenient to decompose the metric perturbations $h_{\mu\nu}$ into scalar, vectorial and tensorial spherical harmonics that are defined from the standard $Y_{\ell m}(\theta,\varphi)$ functions and their derivatives with respect to $\theta$ and $\varphi$. They are labelled by the two multipole integers $\ell$ and $m$ (with $\ell \geq 0$ and $-\ell \leq m \leq \ell$).

Furthermore, one can distinguish axial and polar modes, which behave differently under the parity transformation $\vec{r} \to -\vec{r}$: the polar, or even-parity, modes transform as $(-1)^\ell$, similarly to the scalar spherical harmonics $Y_{\ell m}(\theta,\varphi)$, whereas the axial, or odd-parity, modes transform as $(-1)^{\ell+1}$. These modes can be treated separately as they are decoupled at linear order. Moreover, we consider here only the modes $\ell \geq 2$. The particular cases of the $\ell = 0$ and $\ell = 1$ modes are briefly discussed in section 4.3.3.

Since the background metric is static, it is also convenient to decompose the time dependence of the perturbations into Fourier modes,

$$F(t,r) = \int_{-\infty}^{+\infty}\mathrm{d}\omega\,\tilde{F}(\omega,r)e^{-i\omega t}\,. \quad (4.8)$$

In the rest of this chapter, we will use the same notation for the time-dependent function $F$ and its Fourier transform, as there will be no ambiguity. From a practical point of view, we simply replace every time derivative by a multiplication by $-i\omega$ in the linearised equations, which leads to a system of ordinary differential equations with respect to the radial variable $r$.

In both axial and polar sectors, the equations of motion can be reduced to a system of two first order ordinary differential equations, as we will show below.

## 4.3. Regge-Wheeler-Zerilli gauge

### 4.3.1. Axial perturbations

Axial perturbations are parametrised by three families of functions $h_0^{\ell m}$, $h_1^{\ell m}$ and $h_2^{\ell m}$ of the variables $(r,t)$, according to

$$\begin{aligned}
h_{t\theta} &= \frac{1}{\sin\theta}\sum_{\ell,m} h_0^{\ell m}(t,r)\,\partial_\varphi Y_{\ell m}(\theta,\varphi)\,, \qquad h_{t\varphi} = -\sin\theta\sum_{\ell,m} h_0^{\ell m}(t,r)\,\partial_\theta Y_{\ell m}(\theta,\varphi)\,,\\
h_{r\theta} &= \frac{1}{\sin\theta}\sum_{\ell,m} h_1^{\ell m}(t,r)\,\partial_\varphi Y_{\ell m}(\theta,\varphi)\,, \qquad h_{r\varphi} = -\sin\theta\sum_{\ell,m} h_1^{\ell m}(t,r)\,\partial_\theta Y_{\ell m}(\theta,\varphi)\,,\\
h_{ab} &= \sin\theta\sum_{\ell,m} h_2^{\ell m}(t,r)\,\epsilon_{c(a}D^c\partial_{b)}Y_{\ell m}(\theta,\varphi)\,,
\end{aligned} \tag{4.9}$$

where, in the last equation, the indices $a$ and $b$ belong to the set $\{\theta,\varphi\}$, $\epsilon_{ab}$ is the totally antisymmetric symbol such that $\epsilon_{\theta\varphi} = +1$ and $D_a$ is the 2-dimensional covariant derivative associated with the metric of the 2-sphere $d\theta^2 + \sin^2\theta\, d\varphi^2$. More explicitely, the angular components of the metric can be written

$$\begin{aligned}
h_{\theta\theta} &= \sum_{\ell,m} \frac{1}{\sin\theta} h_2^{\ell m}(t,r)\left(\partial_\theta\partial_\varphi - \text{cotan}\,\theta\,\partial_\varphi\right) Y_{\ell m}(\theta,\varphi)\,,\\
h_{\theta\varphi} = h_{\varphi\theta} &= -\sum_{\ell,m} \sin\theta\, h_2^{\ell m}(t,r)\left(\frac{\ell(\ell+1)}{2} + \partial_\theta^2\right) Y_{\ell m}(\theta,\varphi)\,,\\
h_{\varphi\varphi} &= -\sum_{\ell,m} h_2^{\ell m}(t,r)\sin\theta\left(\partial_\theta\partial_\varphi - \text{cotan}\,\theta\,\partial_\varphi\right) Y_{\ell m}(\theta,\varphi)\,.
\end{aligned} \tag{4.10}$$

All the other components of the axial perturbations vanish.

Due to the invariance of the theory under space-time diffeomorphisms, the parametrization with the functions $h_0^{\ell m}$, $h_1^{\ell m}$ and $h_2^{\ell m}$ is redundant. To prove this, let us consider an infinitesimal change of coordinates $x^\mu \to x^\mu + \xi^\mu$. This induces the transformation

$$h_{\mu\nu} \to h_{\mu\nu} + \nabla_\mu\xi_\nu + \nabla_\nu\xi_\mu \tag{4.11}$$

at the linear level. In the axial sector, the nonzero components of the generator $\xi^\mu$ that preserves the odd parity of the perturbations can be decomposed into spherical harmonics as follows:

$$\xi_\theta = \sum_{\ell,m} \xi^{\ell m}(t,r)\,\partial_\theta Y_{\ell,m}(\theta,\varphi)\,, \qquad \xi_\varphi = \sum_{\ell,m} \xi^{\ell m}(t,r)\,\partial_\varphi Y_{\ell,m}(\theta,\varphi)\,, \tag{4.12}$$

and the induced gauge transformations on the functions $h_0$, $h_1$ and $h_2$ are given, according to eq. (4.11), by

$$h_0 \to h_0 - \dot{\xi}\,, \qquad h_1 \to h_1 - \xi' + \frac{2}{r}\xi\,, \quad h_2 \;\to\; h_2 - 2\xi\,, \tag{4.13}$$

where we have dropped the indices $(\ell m)$ for simplicity. A dot and a prime denote a derivative with respect to $t$ and $r$, respectively.

As a consequence, one can always choose a gauge in which $h_2^{\ell m} = 0$ which is the well-known Regge-Wheeler (RW) gauge for the axial perturbations [179]. Notice that this gauge choice is possible for $\ell \geq 2$ only (the cases $\ell = 0$ and $\ell = 1$ will be discussed later below). We drop the indices $\ell$ and $m$ in the following for clarity, since at the linear level no coupling between modes with different values of these parameters is expected.

This choice allows us to recover the usual RW gauge [179] to describe the axial modes:

$$\begin{aligned} h_{t\theta} &= \frac{1}{\sin\theta}\sum_{\ell,m} h_0^{\ell m}\partial_\varphi Y_{\ell m}(\theta,\varphi)\,, \qquad & h_{t\varphi} &= -\sin\theta\sum_{\ell,m} h_0^{\ell m}\partial_\theta Y_{\ell m}(\theta,\varphi)\,,\\ h_{r\theta} &= \frac{1}{\sin\theta}\sum_{\ell,m} h_1^{\ell m}\partial_\varphi Y_{\ell m}(\theta,\varphi)\,, \qquad & h_{r\varphi} &= -\sin\theta\sum_{\ell,m} h_1^{\ell m}\partial_\theta Y_{\ell m}(\theta,\varphi)\,, \end{aligned} \tag{4.14}$$

while the other components vanish. For these perturbations, the equations of motion (4.6) reduce to the following three equations:

$$\begin{aligned} \mathcal{E}_{t\theta} &= 2\left(\frac{\mu}{r} - 1 - \lambda\right)h_0(t,r) + r(r-\mu)\frac{\partial^2 h_0}{\partial r^2} - 2(r-\mu)\frac{\partial h_1}{\partial t}\\ &\qquad\qquad\qquad\qquad - r(r-\mu)\frac{\partial^2 h_1}{\partial t\partial r} = 0\,,\\ \mathcal{E}_{r\theta} &= -2\lambda h_1(t,r) - \frac{2r^2}{r-\mu}\frac{\partial h_0}{\partial t} + \frac{r^3}{r-\mu}\frac{\partial^2 h_0}{\partial t\partial r} - \frac{r^3}{r-\mu}\frac{\partial^2 h_1}{\partial t^2} = 0\,,\\ \mathcal{E}_{\theta\theta} &= 2\mu h_1(t,r) + 2r(r-\mu)\frac{\partial h_1}{\partial r} - \frac{2r^3}{r-\mu}\frac{\partial h_0}{\partial t} = 0\,, \end{aligned} \tag{4.15}$$

where we have introduced the notation

$$2\lambda \;\equiv\; \ell(\ell+1) - 2\,, \tag{4.16}$$

as the equations $\mathcal{E}_{t\varphi} = 0$, $\mathcal{E}_{r\varphi} = 0$, $\mathcal{E}_{\varphi\varphi} = 0$ and $\mathcal{E}_{\theta\varphi} = 0$ are identical to the above ones.

Since there are only two independent functions, $h_0$ and $h_1$, one expects one of the above equations to be redundant. This is indeed verified by noting the following relation between the equations (4.15) and their derivatives, written now in the frequency domain,

$$\frac{\mathrm{d}\mathcal{E}_{r\theta}}{\mathrm{d}r} + \frac{ir^2\omega}{(r-\mu)^2}\mathcal{E}_{t\theta} + \frac{\mu}{r(r-\mu)}\mathcal{E}_{r\theta} + \frac{\lambda}{r(r-\mu)}\mathcal{E}_{\theta\theta} = 0\,. \tag{4.17}$$

This shows that the two equations $\mathscr{E}_{r\theta} = 0$ and $\mathscr{E}_{\theta\theta} = 0$ are sufficient to fully describe the dynamics of axial perturbations. As a consequence, the initial system (4.15) reduces to

$$\frac{\mathrm{d}Y}{\mathrm{d}r} = M(r)Y, \quad M(r) = \begin{pmatrix} 2/r & 2i\lambda(r-\mu)/r^3 - i\omega^2 \\ -ir^2/(r-\mu)^2 & -\mu/r(r-\mu) \end{pmatrix}, \tag{4.18}$$

where the two components of the column vector $Y \equiv {}^T(Y_1, Y_2)$ are $Y_1(r) \equiv h_0(r)$ and $Y_2(r) \equiv h_1(r)/\omega$. Notice that we divided the variable $h_1(r)$ by $\omega$ in the definition of $Y_2$ in order to get a system which does not involve powers of $\omega$ higher than 2, or equivalently which is at most second order in time if one inverts the Fourier transform (4.8).

### 4.3.2. Polar perturbations

Before gauge fixing, polar perturbations of the metric are parametrised by seven families of functions $H_0^{\ell m}, H_1^{\ell m}, H_2^{\ell m}$, $\alpha^{\ell m}$, $\beta^{\ell m}$, $K^{\ell m}$ and $G^{\ell m}$ of the variables $(r,t)$ which appear in the components of the metric perturbations as follows:

$$\begin{aligned}
h_{tt} &= A(r)\sum_{\ell,m} H_0^{\ell m}(t,r)Y_{\ell m}(\theta,\varphi)\,, \quad h_{tr} = \sum_{\ell,m} H_1^{\ell m}(t,r)Y_{\ell m}(\theta,\varphi)\,,\\
h_{rr} &= \frac{1}{A(r)}\sum_{\ell,m} H_2^{\ell m}(t,r)Y_{\ell m}(\theta,\varphi)\,,\\
h_{ta} &= \sum_{\ell,m} \beta^{\ell m}(t,r)\,\partial_a Y_{\ell m}(\theta,\varphi)\,, \quad h_{ra} = \sum_{\ell,m} \alpha^{\ell m}(t,r)\,\partial_a Y_{\ell m}(\theta,\varphi)\,,\\
h_{ab} &= \sum_{\ell,m} K^{\ell m}(t,r) g_{ab} Y_{\ell m}(\theta,\varphi) + \sum_{\ell,m} G^{\ell m}(t,r) D_a D_b Y_{\ell m}(\theta,\varphi)\,.
\end{aligned} \tag{4.19}$$

More precisely, the angular part of the metric can be written as

$$\begin{aligned}
h_{\theta\theta} &= \sum_{\ell,m} K^{\ell m}(t,r)Y_{\ell m}(\theta,\varphi) + \sum_{\ell,m} G^{\ell m}(t,r)\,\partial_\theta^2 Y_{\ell m}(\theta,\varphi)\,,\\
h_{\theta\varphi} &= h_{\varphi\theta} = -\sum_{\ell,m} G^{\ell m}(t,r)\,\mathrm{cotan}\,\theta\,\partial_\varphi Y_{\ell m}(\theta,\varphi)\,,\\
h_{\varphi\varphi} &= \sum_{\ell,m} \sin^2\theta K^{\ell m}(t,r)Y_{\ell m}(\theta,\varphi)\\
&\quad + \sum_{\ell,m} G^{\ell m}(t,r)\left(\partial_\varphi^2 + \sin\theta\cos\theta\,\partial_\theta\right)Y_{\ell m}(\theta,\varphi)\,.
\end{aligned} \tag{4.20}$$

Similarly to the axial sector, this parametrisation is redundant and can be simplified by gauge fixing. Now, linear diffeomorphisms which preserve even-parity of the metric components are generated by vector fields $\xi$ whose components decompose into spherical harmonics as follows,

$$\xi_t = \sum_{\ell,m} T^{\ell m}(t,r) Y_{\ell m}(\theta,\varphi)\,, \quad \xi_r = \sum_{\ell,m} R^{\ell m}(t,r) Y_{\ell m}(\theta,\varphi)\,,$$
$$\xi_\theta = \sum_{\ell,m} \Theta^{\ell m}(t,r)\,\partial_\theta Y_{\ell m}(\theta,\varphi)\,, \quad \xi_\varphi = \sum_{\ell,m} \Theta^{\ell m}(t,r)\,\partial_\varphi Y_{\ell m}(\theta,\varphi)\,. \tag{4.21}$$

Here $T^{\ell m}$, $R^{\ell m}$ and $\Theta^{\ell m}$ are arbitrary functions of $(t,r)$. These linear diffeomorphisms induce gauge transformations on the functions that parametrise metric perturbations according to

$$\begin{aligned}
H_0^{\ell m}(t,r) &\longrightarrow H_0^{\ell m}(t,r) + \frac{2}{A(r)}\dot{T}^{\ell m}(t,r) + A'(r)R^{\ell m}(t,r),\\
H_1^{\ell m}(t,r) &\longrightarrow H_1^{\ell m}(t,r) + \dot{R}^{\ell m}(t,r) + T'^{\ell m}(t,r) + \frac{A'(r)}{A(r)}T^{\ell m}(t,r),\\
H_2^{\ell m}(t,r) &\longrightarrow H_2^{\ell m}(t,r) + 2A(r)R'^{\ell m}(t,r) - A'(r)R_{\ell m}(t,r),\\
\beta^{\ell m}(t,r) &\longrightarrow \beta^{\ell m}(t,r) + T^{\ell m}(t,r) + \dot{\Theta}^{\ell m}(t,r),\\
\alpha^{\ell m}(t,r) &\longrightarrow \alpha^{\ell m}(t,r) + R^{\ell m}(t,r) + \Theta'^{\ell m}(t,r) - \frac{2}{r}\Theta^{\ell m}(t,r),\\
K^{\ell m}(t,r) &\longrightarrow K^{\ell m}(t,r) + \frac{2A(r)}{r}R^{\ell m}(t,r),\\
G^{\ell m}(t,r) &\longrightarrow G^{\ell m}(t,r) + 2\Theta^{\ell m}(t,r)\,.
\end{aligned} \tag{4.22}$$

An immediate consequence of the gauge transformations is that one can choose the gauge parameter $\xi$ such that $G^{\ell m} = 0$ by fixing $\Theta^{\ell m}$, then $\alpha^{\ell m} = 0$ and $\beta^{\ell m} = 0$ by fixing $R^{\ell m}$ and $T^{\ell m}$ respectively, in the case where $\ell \geq 2$. This gauge is known as the Zerilli gauge [180] (see [196] for a recent presentation in the context of modified gravity); the nonvanishing metric perturbations then read

$$h_{tt} = A(r) \sum_{\ell,m} H_0^{\ell m}(t,r) Y_{\ell m}(\theta,\varphi), \quad h_{tr} = \sum_{\ell,m} H_1^{\ell m}(t,r) Y_{\ell m}(\theta,\varphi), \tag{4.23}$$
$$h_{rr} = \frac{1}{A(r)} \sum_{\ell,m} H_2^{\ell m}(t,r) Y_{\ell m}(\theta,\varphi), \quad h_{ab} = \sum_{\ell,m} K^{\ell m}(t,r) g_{ab} Y_{\ell m}(\theta,\varphi), \tag{4.24}$$

where $A(r) \equiv 1 - \mu/r$ is included in the definitions for later convenience, and the indices $a$ or $b$ in the last equation are the angles $\theta$ or $\varphi$. Similarly to the axial case, we drop the indices $\ell$ and $m$ in the following.

The linearised Einstein's equations yield seven distinct equations:

$$\begin{aligned}
\mathscr{E}_{tt} = &-2(\lambda+2)\left(1-\frac{\mu}{r}\right)H_2(t,r) - 2\lambda\left(1-\frac{\mu}{r}\right)K(t,r) - \frac{2}{r}(r-\mu)^2\frac{\partial H_2}{\partial r}\\
&+\left(6r - 11\mu + \frac{5\mu^2}{r}\right)\frac{\partial K}{\partial r} + 2(r-\mu)^2\frac{\partial^2 K}{\partial r^2} = 0\,,
\end{aligned}$$

$$\mathscr{E}_{tr} = -2(\lambda+1)H_1(t,r) - 2r\frac{\partial H_2}{\partial t} + r\frac{2r-3\mu}{r-\mu}\frac{\partial K}{\partial t} + 2r^2\frac{\partial^2 K}{\partial t\partial r} = 0\,,$$

$$
\begin{aligned}
\mathscr{E}_{rr} = & -2\frac{\lambda+1}{1-\mu/r}H_0(t,r) + \frac{2}{1-\mu/r}H_2(t,r) + \frac{2\lambda}{1-\mu/r}K(t,r) + 2r\frac{\partial H_0}{\partial r} \\
& - r\frac{2r-\mu}{2(r-\mu)}\frac{\partial K}{\partial r} - \frac{4r^2}{r-\mu}\frac{\partial H_1}{\partial t} + \frac{2r^4}{(r-\mu)^2}\frac{\partial^2 K}{\partial t^2} = 0\,, \\
\mathscr{E}_{t\theta} = & -\frac{\mu}{r}H_1(t,r) - (r-\mu)\frac{\partial H_1}{\partial r} + r\frac{\partial H_2}{\partial t} + r\frac{\partial K}{\partial t} = 0\,, \\
\mathscr{E}_{r\theta} = & \frac{2r-3\mu}{2(r-\mu)}H_0(t,r) - \frac{2r-\mu}{2(r-\mu)}H_2(t,r) - r\frac{\partial H_0}{\partial r} + r\frac{\partial K}{\partial r} + \frac{r^2}{r-\mu}\frac{\partial H_1}{\partial t} = 0\,, \\
\mathscr{E}_{\theta\theta} = & \frac{2r+\mu}{2}\frac{\partial H_0}{\partial r} + \frac{2r-\mu}{2}\frac{\partial H_2}{\partial r} - (2r-\mu)\frac{\partial K}{\partial r} + r(r-\mu)\frac{\partial^2 H_0}{\partial r^2} \\
& - r(r-\mu)\frac{\partial^2 K}{\partial r^2} - r\frac{2r-\mu}{r-\mu}\frac{\partial H_1}{\partial t} - 2r^2\frac{\partial^2 H_1}{\partial t\partial r} \\
& + \frac{r^3}{r-\mu}\frac{\partial^2 H_2}{\partial t^2} + \frac{r^3}{r-\mu}\frac{\partial^2 K}{\partial t^2} = 0\,, \\
\mathscr{E}_{\theta\varphi} = & H_0(t,r) - H_2(t,r) = 0\,.
\end{aligned}
\tag{4.25}
$$

The equations of motion $\mathscr{E}_{t\varphi} = 0$, $\mathscr{E}_{r\varphi} = 0$ and $\mathscr{E}_{\varphi\varphi} = 0$ are identical to $\mathscr{E}_{t\theta} = 0$, $\mathscr{E}_{r\theta} = 0$ and $\mathscr{E}_{\theta\theta} = 0$, respectively.

We can immediately solve the last equation of the system (4.25) and replace $H_2$ by $H_0$ in all the other equations. We thus get six equations for only three independent functions $K$, $H_0$ and $H_1$, and we want to extract three "simple" independent equations out of them. One can then note that the combination

$$
\mathscr{E} \equiv \frac{i\mu}{4\omega r(r-\mu)}\mathscr{E}_{tr} + \frac{1}{2}\mathscr{E}_{rr} + \mathscr{E}_{r\theta} \tag{4.26}
$$

is purely algebraic, i.e. it does not involve any derivatives of the functions. Moreover, we find that the system $\mathscr{E}_{tr}$, $\mathscr{E}_{t\theta}$, $\mathscr{E}_{r\theta}$, $\mathscr{E}$ enables us to recover $\mathscr{E}_{tt}$ and $\mathscr{E}_{\theta\theta}$ so that we can restrict immediately to the system formed by these four equations which, after some simple calculations, are given by the system of differential equations

$$
\begin{aligned}
& K'(r) - \frac{1}{r}H_0(r) - \frac{i(\lambda+1)}{\omega r^2}H_1(r) + \frac{1}{r}\frac{2r-3\mu}{2(r-\mu)}K(r) = 0\,, \\
& H_1'(r) + \frac{i\omega r}{r-\mu}H_0(r) + \frac{\mu}{r(r-\mu)}H_1(r) + \frac{i\omega r}{r-\mu}K(r) = 0\,, \\
& H_0'(r) - K'(r) + \frac{\mu}{r(r-\mu)}H_0(r) + \frac{i\omega r}{r-\mu}H_1(r) = 0\,,
\end{aligned}
\tag{4.27}
$$

$$
\tag{4.28}
$$

together with the algebraic equation

$$
\left(\frac{3\mu}{r} + 2\lambda\right)H_0(r) + \left(\frac{i\mu(\lambda+1)}{\omega r^2} - 2i\omega r\right)H_1(r)
$$

$$+\frac{3\mu^2+2\mu(2\lambda-1)r-4\lambda r^2+4\omega^2r^4}{2r(r-\mu)}K(r)=0\,. \quad (4.29)$$

One equation is still redundant. However, we can solve the algebraic equation for $H_0$ and substitute its expression into the first three equations. This shows that the third is not independent from the first two. Finally, we obtain a system of the form (4.18):

$$\frac{\mathrm{d}Y}{\mathrm{d}r}=M(r)Y\,,$$

$$M(r)=\frac{1}{3\mu+2\lambda r}\begin{pmatrix}\frac{\mu(3\mu+(\lambda-2)r)-2r^4\omega^2}{r(r-\mu)} & \frac{2i(\lambda+1)(\mu+\lambda r)+2ir^3\omega^2}{r^2}\\ \frac{ir(9\mu^2-8\lambda r^2+8(\lambda-1)\mu r)+4ir^5\omega^2}{2(r-\mu)^2} & \frac{2r^4\omega^2-\mu(3\mu+3\lambda r+r)}{r(r-\mu)}\end{pmatrix}, \quad (4.30)$$

where now the two components of $Y$ are defined by $Y_1(r)\equiv K(r)$ and $Y_2(r)\equiv H_1(r)/\omega$. Similarly to the axial sector, the definition of $Y_2$ is motivated by the fact that the resulting system involves at most $\omega^2$ terms.

### 4.3.3. Monopole and dipole perturbations

We consider here the special cases $\ell=0$ and $\ell=1$.

#### 4.3.3.1. Axial modes

For the axial modes, the components $h_{ab}$ vanish identically for $\ell=1$ (axial perturbations do not have $\ell=0$ components) which means that $h_2$ does not show up in the components of the metric. Hence, when $\ell=1$, it is necessary to make a different gauge choice. In general, one chooses $h_1=0$ which fixes the gauge parameter $\xi$ up to a function of the form $C(t)r^2$. Therefore, $h_0$ inherits a residual gauge invariance given by $h_0\to h_0+F(t)r^2$ where $F(t)$ is an arbitrary function. Then $h_0$ can be shown to satisfy the equation of motion

$$2h_0(r)-rh_0'(r)=0\,. \quad (4.31)$$

Therefore, the mode $h_0$ is not propagating.

#### 4.3.3.2. Polar modes

Let us now turn to polar perturbations. In the case $\ell=0$, $H_0$, $H_1$, $H_2$ and $K$ are the only non-vanishing components of the metric perturbations whereas $T$ and $R$ are the only non-vanishing components of the gauge parameter (so that the gauge transformation preserves the monopole). As in the general case, one can choose $R$ to fix $K=0$. Then, one can in principle make use of $T$ to get rid of $H_1$ (we could have also set $H_0 = 0$). Finally, we are left with only two non-vanishing functions which are either $H_2$ or $H_0$. The equations of motion simplify drastically and, after some calculations, yield

$$H_0(r) - H_2(r) = 0\,, \quad H_2(r) + (r-\mu)H_2'(r) = 0\,. \tag{4.32}$$

The solution reads $H_2(r) = C/(r-\mu)$ and the mode is not propagating.

Concerning the gauge fixing, the main difference between the general case and the case $\ell = 1$ lies in the fact that, in the latter, $h_{ab}$ can be shown to depend on the difference $G - K$ only, so that one can fix $K = 0$ without loss of generality. Furthermore, one can make the gauge fixing $G = 0$ by an appropriate choice of $\Theta$. Then, one makes use of $T$ to fix $\beta = 0$. Finally, one uses the remaining free gauge function $R$ to fix $\alpha = 0$. At the end, we are left with the three non-vanishing functions $H_0$, $H_1$ and $H_2$. These functions satisfy the three independent equations

$$\begin{aligned} &2H_2(r) + (r-\mu)H_2'(r) = 0\,, \quad H_1(r) + i\omega H_2(r) = 0\,, \\ &H_0(r) + (\mu - r)H_0'(r) - 2ir\omega H_1(r) + H_2(r) = 0\,. \end{aligned} \tag{4.33}$$

Indeed, the full set of the original Einstein equations is equivalent to this one which can easily be solved explicitly but its solution is not relevant for our purpose. Nonetheless, we see immediately from the equations that, like the monopole, the polar dipole does not propagate. This is why we do not consider it in this manuscript.

## 4.4. Schrödinger equation formulation

In both axial and polar sectors, the equations of motion have been recast in the form of a system consisting simply of two first-order differential equations (with respect to the radial variable), namely eq. (4.18) for axial perturbations and eq. (4.30) for polar perturbations. In both cases, we now recall how this system can be rewritten as a Schrödinger-like equation.

### 4.4.1. From the first order system to the Schrödinger-like equation

As shown in [179] and [180], one can rewrite these systems as a single second order (in radial derivatives) Schrödinger-like equation for a unique dynamical variable. Reformulating a first order system of this kind as a Schrödinger equation is, in general, not an easy task because one has to ensure that the Schrödinger equation is second order in time and in space. It requires, in particular, a decoupling of the dynamical variables involved in the original first order system and a “clever” choice for the dynamical variable that should satisfy the second order Schrödinger equation. For now, we obtain the Schrödinger-like form for the GR systems of eqs. (4.18) and (4.30) using a clever change of variables; we shall describe in chapter 6 the general procedure to obtain a Schrödinger-like equation from the first-order system describing axial perturbations.

The systems we consider take the general form

$$\frac{\mathrm{d}Y}{\mathrm{d}r} = M(r)Y, \tag{4.34}$$

where the coefficients of the matrix $M$ are polynomials (of degree at most 2) in $\omega$ and rational functions in $r$. First, we consider the general (linear) change of vector

$$Y(r) = P(r)\hat{Y}(r), \tag{4.35}$$

where $\hat{Y}$ is a new column vector and the two dimensional invertible matrix $P$ has not been fixed at this stage. We also define a new radial coordinate $r_*$ and introduce the "Jacobian" of the transformation $n(r) \equiv \mathrm{d}r/\mathrm{d}r_*$. Now, the idea is to show that it is possible to find a matrix $P$ such that the new system satisfied by $\hat{Y}$ takes the canonical form

$$\frac{\mathrm{d}\hat{Y}}{\mathrm{d}r_*} = \begin{pmatrix} 0 & 1 \\ V(r) - \omega^2 & 0 \end{pmatrix} \hat{Y}, \tag{4.36}$$

where the potential $V(r)$ depends on $r$, but not on $\omega$. Somehow, the first component $\hat{Y}_1$ plays the role of the "momentum" conjugate to the second component $\hat{Y}_2$ which immediately implies that $\hat{Y}_1$ is the "canonical" variable satisfying a Schrödinger-like equation

$$\frac{\mathrm{d}^2\hat{Y}_1}{\mathrm{d}r_*^2} + \left(\omega^2 - V(r)\right)\hat{Y}_1 = 0. \tag{4.37}$$

The frequency $\omega$ appears quadratically in eq. (4.37), which thus corresponds to a propagation equation if one replaces $\omega$ with $-i\,\partial/\partial t$. Applying this procedure to both axial and polar modes will lead to two propagation equations: these will correspond to the two degrees of freedom present in GR, and will prove very useful for the computation of QNMs.

### 4.4.2. Axial modes

Applying this procedure to the system (4.18) for the axial perturbations is rather simple. Indeed, when one changes variables according to (4.35), the new variable $\hat{Y}$ satisfies the differential equation

$$\frac{\mathrm{d}\hat{Y}}{\mathrm{d}r_*} = \hat{M}\hat{Y}, \quad \hat{M} \equiv n(r)\left(P^{-1}MP - P^{-1}P'\right), \tag{4.38}$$

where $P'$ is the derivative of $P$ with respect to $r$, $M$ is the matrix introduced in (4.18) while $\hat{M}$ is the matrix entering in the system (4.36). They take a similar form: $M = M_{[0]} + \omega^2 M_{[2]}$ and $\hat{M} = \hat{M}_{[0]} + \omega^2 \hat{M}_{[2]}$ where the expressions of $M_{[0]}, M_{[2]}, \hat{M}_{[0]}$ and $\hat{M}_{[2]}$ are trivially obtained. As $P$ does not depend on $\omega$, the relation between $M$ and $\hat{M}$ translates into the two matricial relations $\hat{M}_{[2]} = n(r)P^{-1}M_{[2]}P$ and $\hat{M}_{[0]} = n(r)(P^{-1}M_{[0]}P - P^{-1}P')$, which can be viewed as 8 equations for the 6 unknowns $n(r)$, $V(r)$ together with the four components of $P$. Interestingly, the system is not overdetermined and admits a solution for $P$ (4.39), for the potential $V(r)$ (4.41) and for the function $n(r)$ which can be shown to be associated with the tortoise coordinate (4.40). We show only the result here; details can be found in section 6.2 which describes the procedure to obtain the Schrödinger-like equation for any cubic DHOST theory.

This procedure yields the following transition matrix:

$$P(r) = \begin{pmatrix} 1 - \mu/r & r \\ -ir^2/(r-\mu) & 0 \end{pmatrix}, \tag{4.39}$$

while $n(r) = 1 - \mu/r$, which means that $r_*$ is the "tortoise" coordinate,

$$r_* \equiv \int \frac{\mathrm{d}r}{1-\mu/r} = r + \mu \ln(r/\mu - 1). \tag{4.40}$$

Finally the effective potential $V_{\text{odd}}(r)$ for the axial perturbations takes the form

$$V_{\text{odd}}(r) = \left(1 - \frac{\mu}{r}\right) \frac{2(\lambda+1)r - 3\mu}{r^3}. \tag{4.41}$$

Note that this potential vanishes both at spatial infinity ($r \to +\infty$) and at the horizon ($r \to \mu$).

### 4.4.3. Polar modes

The case of polar perturbations is slightly more involved. Starting from the system (4.30), we find using the method described previously that the transition matrix leading to a canonical form (4.36) is given by

$$P = \begin{pmatrix} \frac{3\mu^2+3\lambda\mu r+2r^2\lambda(\lambda+1)}{2r^2(3\mu+2\lambda r)} & 1 \\ -i + \frac{i\mu}{2(r-\mu)} + \frac{3i\mu}{3\mu+2\lambda r} & -\frac{ir^2}{r-\mu} \end{pmatrix}, \tag{4.42}$$

with, in addition, $n(r) = 1 - \mu/r$, which means that $r_*$ is still the tortoise coordinate (4.40). Finally, the corresponding potential $V_{\text{even}}(r)$ reads

$$V_{\text{even}}(r) = \left(1 - \frac{\mu}{r}\right) \frac{9\mu^3 + 18\mu^2 r\lambda + 12\mu r^2\lambda^2 + 8r^3\lambda^2(1+\lambda)}{r^3(3\mu+2r\lambda)^2}. \tag{4.43}$$

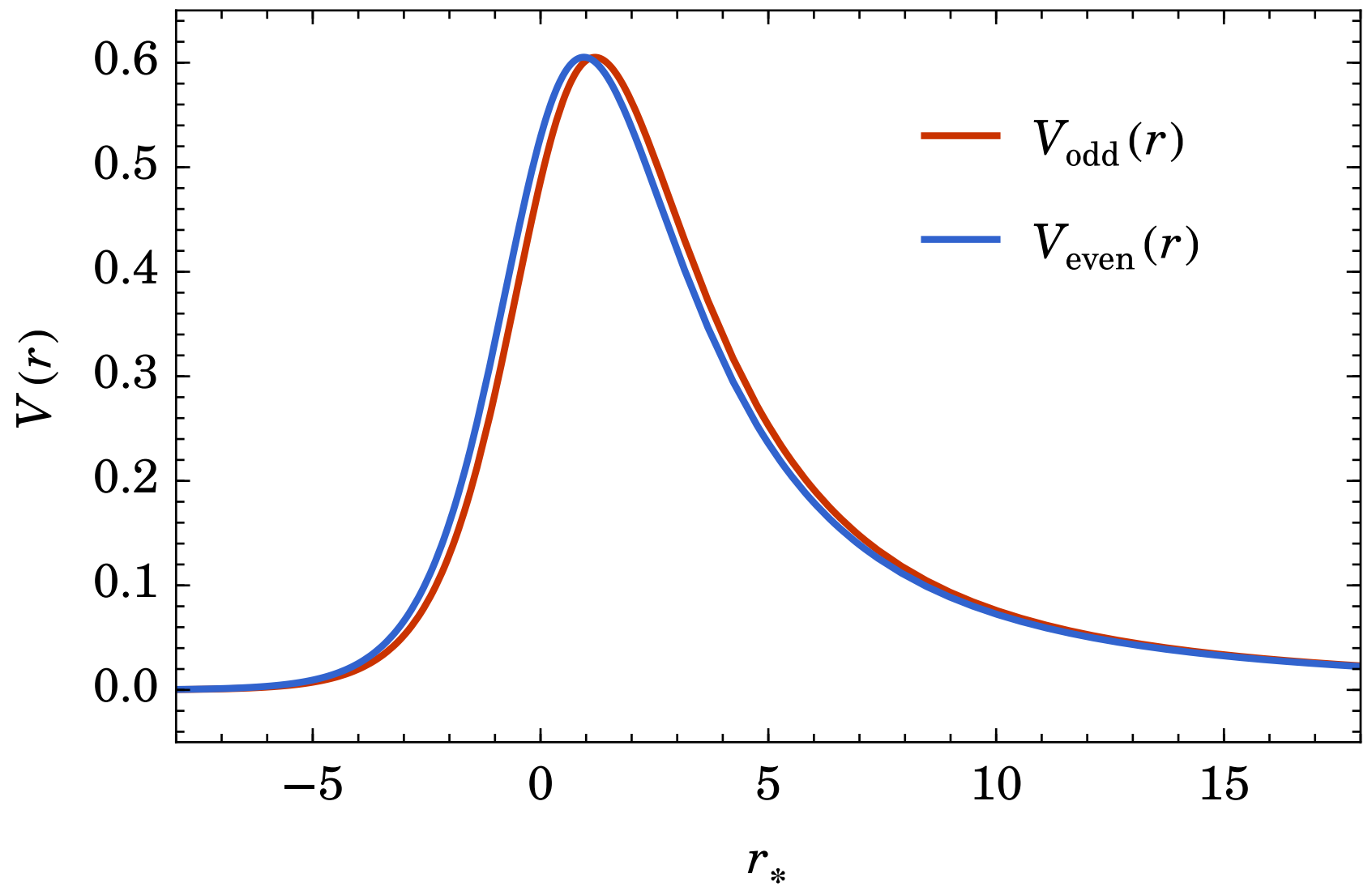

Figure 4.2. – Illustration of the effective potentials (for axial and polar modes) for a Schwarzschild black hole. The parameters are such that $\mu = 1$ (i.e. the mass of the black is 1/2 in natural units) and $\ell = 2$ here.

Despite their different analytic forms, we notice in fig. 4.2 that the potentials $V_{\text{odd}}(r)$ and $V_{\text{even}}(r)$ are quite similar, although distinct. In fact, there exists an underlying symmetry between these two potentials, further explained in [197], leading to the isospectrality theorem which states that the spectra of axial and polar perturbations are exactly the same.

### 4.4.4. Boundary conditions from the Schrödinger equation

As explained in section 4.1.2, finding quasi-normal modes requires to impose the appropriate boundary conditions: the modes must be outgoing at infinity and ingoing at the horizon. We present here how the Schrödinger form obtained in section 4.4.1 for the perturbations' equations of motion is useful for the obtention of such boundary conditions.

Since both $V_{\text{odd}}$ and $V_{\text{even}}$ go to zero at infinity and at the horizon, equation (4.37) becomes asymptotically

$$\frac{\mathrm{d}^2\hat{Y}_1}{\mathrm{d}r_*^2} + \omega^2\hat{Y}_1 \approx 0 \qquad (r_* \to \pm\infty)\,, \tag{4.44}$$

where $\approx$ is an equality up to sub-leading corrections [1]. Therefore, at both bound-

1. Near the horizon, $V = \mathcal{O}(r-\mu)$ for both potentials, hence we assume $\mu^2\omega^2 \gg r/\mu - 1$. At

aries, the function $\hat{Y}_1$ behaves like

$$\hat{Y}_1(r) \approx \mathscr{A}\, e^{i\omega r_*} + \mathscr{B}\, e^{-i\omega r_*}\,, \tag{4.45}$$

where $\mathscr{A}$ and $\mathscr{B}$ are integration constants which take different values at the horizon and at infinity.

The physical interpretation of these modes is more transparent if we include their time dependence explicitly, which gives

$$\begin{aligned} \hat{Y}_1(t,r) &\approx \mathscr{A}_{\rm hor}\, e^{-i\omega(t-r_*)} + \mathscr{B}_{\rm hor}\, e^{-i\omega(t+r_*)} \quad \text{when} \quad r \longrightarrow \mu\,, \\ \hat{Y}_1(t,r) &\approx \mathscr{A}_\infty\, e^{-i\omega(t-r_*)} + \mathscr{B}_\infty\, e^{-i\omega(t+r_*)} \quad \text{when} \quad r \longrightarrow \infty\,. \end{aligned} \tag{4.46}$$

We can interpret each term as a radially propagating wave: the terms proportional to $\mathscr{A}_{\rm hor}$ and $\mathscr{A}_\infty$ are outgoing while the terms proportional to $\mathscr{B}_{\rm hor}$ and $\mathscr{B}_\infty$ are ingoing. Imposing a purely outgoing behaviour at infinity and a purely ingoing behaviour at the horizon, i.e. such that $\mathscr{A}_{\rm hor} = 0$ and $\mathscr{B}_\infty = 0$ severely restricts the possible values of $\omega$. These values can be found numerically by integrating the Schrödinger-like equation (see [198] and the reviews [182, 199, 200, 201], along with the detailed discussion given in chapter 8).

Finally, one can easily deduce the asymptotic expansion of the original gravitational perturbations using the transformations (4.35). For the axial modes, the leading order terms at infinity are thus given by

$$\begin{aligned} h_0(r) &\approx i\omega r \left(\mathscr{A}_\infty\, e^{i\omega r_*} - \mathscr{B}_\infty\, e^{-i\omega r_*}\right)\,, \\ h_1(r) &\approx -i\omega r \left(\mathscr{A}_\infty\, e^{i\omega r_*} + \mathscr{B}_\infty\, e^{-i\omega r_*}\right)\,, \end{aligned} \tag{4.47}$$

while the leading order terms at the horizon read

$$\begin{aligned} h_0(r) &\approx i\omega\mu \left(\mathscr{A}_{\rm hor}\, e^{i\omega r_*} - \mathscr{B}_{\rm hor}\, e^{-i\omega r_*}\right)\,, \\ h_1(r) &\approx -\frac{i\omega\mu^2}{\varepsilon} \left(\mathscr{A}_{\rm hor}\, e^{i\omega r_*} + \mathscr{B}_{\rm hor}\, e^{-i\omega r_*}\right)\,, \end{aligned} \tag{4.48}$$

where we have introduced the variable $\varepsilon \equiv r - \mu$ which satisfies $\varepsilon \ll \mu$ near the horizon.

For the polar modes, the leading order terms at infinity are

$$\begin{aligned} K(r) &\approx i\omega \left(\mathscr{A}_\infty\, e^{i\omega r_*} - \mathscr{B}_\infty\, e^{-i\omega r_*}\right)\,, \\ H_1(r) &\approx r\omega^2 \left(\mathscr{A}_\infty\, e^{i\omega r_*} - \mathscr{B}_\infty\, e^{-i\omega r_*}\right)\,, \end{aligned} \tag{4.49}$$

while the leading terms at the horizon are a bit more involved and read

$$K(r) \approx \frac{\lambda + 1 + 2i\omega\mu}{\mu} \mathscr{A}_{\rm hor}\, e^{i\omega r_*} + \frac{\lambda + 1 - 2i\omega\mu}{\mu} \mathscr{B}_{\rm hor}\, e^{-i\omega r_*}\,, \tag{4.50}$$

$$H_1(r) \approx \frac{i\mu\omega(1 - 2i\omega\mu)}{2\varepsilon} \mathscr{A}_{\rm hor}\, e^{i\omega r_*} + \frac{i\mu\omega(1 + 2i\omega\mu)}{2\varepsilon} \mathscr{B}_{\rm hor}\, e^{-i\omega r_*}\,. \tag{4.51}$$

infinity, $V = \mathscr{O}(1/r^2)$ for both potentials as well, hence we assume $\omega^2 r^2 \gg 1$ in this limit.

## 4.5. Challenges in modified gravity

When the equations of motion of the perturbations are written as a second-order Schrödinger equation, obtaining their asymptotic behaviour is immediate, as it simply depends on the asymptotic behaviour of the effective potential. For example, in the case of a Schwarzschild BH, imposing outgoing waves at infinity and ingoing waves at the horizon would mean setting $\mathscr{A}_{\rm hor} = 0$ and $\mathscr{B}_\infty = 0$ in the equations of section 4.4.4. It is then possible to recover the asymptotic behaviours of the metric perturbation functions from eqs. (4.47) to (4.51). Numerical methods can then be used to find the allowed values for $\omega$.

In the context of modified gravity however, and more particularly scalar-tensor theories, the problem can become more involved for several reasons. First, the background metric can differ from the standard GR solutions, i.e. be different from Schwarzschild in the nonrotating case. Moreover, scalar-tensor theories involve an additional scalar field: the number of degrees of freedom increases and therefore the linear equations of motion are more complex. Indeed, one must also consider the pertubations of the scalar field, and one expects to obtain a third wave equation for the scalar degree of freedom.

In scalar-tensor theories such as DHOST, the procedure to obtain the perturbations equations of motion is similar to the procedure used in GR: one starts with the action, computes the quadratic action for perturbations and uses the Euler-Lagrange equations for this action in order to obtain the equations of motion. In order to get the quadratic action, one must however introduce not only a metric perturbation $h_{\mu\nu}$ but also a scalar field perturbation $\delta\phi$:

$$g_{\mu\nu} = \bar{g}_{\mu\nu} + h_{\mu\nu} \quad \text{and} \quad \phi = \bar{\phi} + \delta\phi \,. \tag{4.52}$$

By expanding eq. (1.25), one then gets a quadratic action $S_{\rm quad}[h_{\mu\nu}, \delta\phi]$ that generalizes eq. (4.5). This action depends on two fields: the equations of motion are

$$\mathscr{E}_{\mu\nu} \equiv \frac{\delta S_{\rm quad}}{\delta h_{\mu\nu}} = 0 \quad \text{and} \quad \mathscr{E}_\phi \equiv \frac{\delta S_{\rm quad}}{\delta\, \delta\phi} = 0\,. \tag{4.53}$$

The equation $\mathscr{E}_\phi = 0$ turns out to be redundant as a consequence of Bianchi's identities, so we just need to account the 10 metric equations $\mathscr{E}_{\mu\nu} = 0$, similarly to GR. This is a consequence of the relationship between the background equations of motion given in section 2.1.3.

In the case of a spherically symmetric background, the scalar field perturbation is parametrised by one family of functions according to

$$\delta\phi = \sum_{\ell,m} \delta\phi^{\ell m}(t,r) Y_{\ell m}(\theta,\varphi)\,. \tag{4.54}$$

One drops the indices $\ell$ and $m$ for clarity, similarly to the GR case. The metric perturbations are unchanged: they are still given by eqs. (4.14) and (4.23). One can note that the scalar perturbations must be of even parity: this means that no coupling between the scalar degree of freedom and the gravitational axial degree of freedom is expected.

As a result, the procedure presented in section 4.4.1 will no longer be applicable in general. Indeed, separation by parity will still lead to two systems of the form (4.34), but while the axial system will still be 2-dimensional, the polar system will be 4-dimensional [2].

In some cases, the equations of motion for the perturbations can be rewritten as a generalised $N$-dimensional matrix Schrödinger-like system for $N$ fields $\Psi_i$, of the typical form (see e.g. [202])

$$f(r)\frac{\mathrm{d}}{\mathrm{d}r}\left(f(r)\frac{\mathrm{d}\Psi_i}{\mathrm{d}r}\right) + \left(\omega^2 - f(r)\,V_{ij}\right)\Psi_j = 0\,, \tag{4.55}$$

where $f(r)$ depends on the background and the $N \times N$ matrix $V_{ij}$ of radial potentials usually vanishes or becomes a constant diagonal matrix asymptotically. Such a system describes the propagation of $N$ coupled degrees of freedom. The boundary conditions are still easy to infer from such a differential system. However, one could also encounter more general situations where such a simple reformulation of the equations of motion is not available or would require an involved and lengthy procedure. Explicit examples will be given in chapter 7. To tackle more general situations, it would be very useful to be able to analyse directly the first-order system of equations in its original form and to extract directly from it the asymptotic behaviour of perturbations. This could then be used for numerical computations. In chapter 5, we present an algorithm that allows us to do precisely this; we apply it in chapters 6 and 7 and use its results to compute QNMs numerically in chapter 8.

2. 2 first order equations correspond to one degree of freedom.

CHAPTER 5

# BLACK HOLE PERTURBATIONS FROM THE FIRST-ORDER SYSTEM

CONTENTS

UNDERSTANDING the asymptotic behaviour of BH perturbations at the horizon and at spatial infinity is crucial for the computation of QNMs, which are defined by very specific boundary conditions. Indeed, they correspond to purely outgoing radiation at spatial infinity and ingoing radiation at the horizon. Imposing these specific boundary conditions leads to a discrete set of allowed frequencies.

The equations of motion for the perturbations can be written as several decoupled first-order systems, which can in turn be cast into Schrödinger-like equations in the case of GR (see chapter 4). The derivation of the asymptotic behaviour of perturbations is easy when the equations of motion are written in such a form (see section 4.4.4). However, in the context of modified gravity, the problem can become more involved, for reasons exposed in section 4.5. It would therefore be useful to be able to recover the asymptotic behaviour of the perturbations from the first-order system directly. The purpose of this chapter is to present an algorithm that allows one to obtain asymptotical behaviour from any first-order differential system.

In order to reach this goal, we use recent developments that appeared in the mathematical literature. These results enable us to determine, via a systematic algorithm, the asymptotic structure of the solutions of a generic first-order differential system. For pedagogical reasons, we use here this algorithm to recover the asymptotic solutions for the axial — or odd-parity — modes and for the polar — or even-parity — modes of the standard Schwarzschild solution, that were already presented in chapter 4. The method presented in this chapter will then be used in the context of DHOST theories in chapters 6 and 7.

The outline of the chapter is the following. In the next section, we present the algorithm in the specific case of the first-order systems obtained for a Schwarzschild BH in chapter 4 and show explicitly how this new method enables us to recover the usual asymptotic solution, working directly with the first order system. We then present, in section 5.2, the general algorithm, carefully listing the various steps of the algorithm depending on the structure of the system. This chapter is based on [181].

## 5.1. First order approach to Schwarzschild perturbations

As we have seen in section 4.4.1, finding a (second-order) Schrödinger-like equation for the metric perturbations starting from the Einstein equations requires some manipulations of the equations of motion and an appropriate choice of the function that verifies the Schrödinger-like equation. Such manipulations are actually not needed if one only wants to obtain the asymptotic behaviour of perturbations. The general method will be described in a systematic way in the next section; as it is somewhat tedious, we present it first here in a pedestrian way for the perturbations of a Schwarzschild BH in GR. A more mathematically-minded reader might prefer to jump directly to the next section and later come back to this section to find a particular application of the general method.

### 5.1.1. Method

Ignoring the traditional Schrödinger reformulation presented in chapter 4, we now go back to the original first-order systems for axial and polar perturbations given in eqs. (4.18) and (4.30). Schematically, we thus have a first-order system of the form

$$\frac{\mathrm{d}Y}{\mathrm{d}r} = M(r)Y\,, \tag{5.1}$$

where $Y(r)$ is a column vector and $M(r)$ a square matrix. In order to study the system at spatial infinity, say, i.e. when $r \to \infty$, one can expand the matrix $M(r)$ in powers of $r$:

$$M(r) = M_p\, r^p + \cdots + M_0 + M_{-1}\,\frac{1}{r} + \mathcal{O}\left(\frac{1}{r^2}\right), \tag{5.2}$$

where all the matrix coefficients $M_i$ are $r$-independent. We stop here the expansion at order $1/r$, which is sufficient for the simplest cases, but higher orders might be needed in general.

If *all* matrices $M_i$ are *diagonal*, it is immediate to integrate the truncated system, which then consists of $n$ ordinary differential equations of the form

$$y'(r) = \left(\lambda_p r^p + \cdots + \lambda_0 + \frac{\mu}{r}\right) y(r)\,, \tag{5.3}$$

whose solution is

$$y(r) = y_0 e^{q(r)} r^{\mu} \quad \text{with} \quad q(r) = \frac{\lambda_p}{p+1} r^{p+1} + \cdots + \lambda_0 r \,. \tag{5.4}$$

Putting together these $n$ solutions, we thus get the solution to the system (5.1), assuming all matrices $M_i$ in eq. (5.2) are diagonal, in the form

$$Y(r) = e^{\mathbf{\Upsilon}(r)} r^{\mathbf{\Delta}} \mathbf{F}(r) Y_0 \tag{5.5}$$

where $Y_0$ is a constant vector, corresponding to the $n$ integration constants, $\mathbf{\Upsilon}$ is a diagonal matrix whose coefficients are polynomials of degree at most $p+1$, $\mathbf{\Delta}$ is a constant diagonal matrix and $\mathbf{F}(r)$ is a matrix which is regular at infinity (i.e. whose limit is finite).

Of course, in general, the matrices $M_i$ are not diagonal but, remarkably, it is always possible to transform the truncated system into a fully diagonal system, in a *finite* number of steps following an algorithm introduced in [203, 204, 205, 206], which we will present in full details in the next section.

At each step in the algorithm, one introduces a new vector $\tilde{Y}$, related to the vector $Y$ of the previous step by

$$Y = P\tilde{Y} \,, \tag{5.6}$$

where $P$ is an invertible matrix so that the previous system (5.1) is transformed into a new, but equivalent, system of the form

$$\frac{\mathrm{d}\tilde{Y}}{\mathrm{d}r} = \tilde{M}(r)\tilde{Y}, \quad \tilde{M}(r) \equiv P^{-1}MP - P^{-1}\frac{\mathrm{d}P}{\mathrm{d}r} \,. \tag{5.7}$$

The idea is then to choose an appropriate transfer matrix $P$ at each step in order to diagonalise, order by order, the matrices that appear in the expansion of $M$. Once all the matrices are diagonalised, one can integrate directly the diagonal system, as we have seen earlier, and obtain the general asymptotic solution of the system.

For the asymptotic behaviour near the horizon, one proceeds in the same way by noting that the variable $z = 1/(r-\mu)$ goes to infinity when $r \to \mu$. In the rest of this section, we will illustrate the algorithm by considering in turn the asymptotic behaviours of the axial and polar modes.

### 5.1.2. Axial modes

The analysis of the asymptotic behaviour of the first order system (4.18) is relatively simple and instructive. We recall that the system is of the form

$$\frac{\mathrm{d}Y}{\mathrm{d}r} = M(r)Y \,, \tag{5.8}$$

$$Y(r) \equiv \begin{pmatrix} h_0(r) \\ h_1(r)/\omega \end{pmatrix}, \quad M(r) \equiv \begin{pmatrix} 2/r & 2i\lambda(r-\mu)/r^3 - i\omega^2 \\ -ir^2/(r-\mu)^2 & -\mu/r(r-\mu) \end{pmatrix}. \quad (5.9)$$

#### 5.1.2.1. Asymptotic analysis at spatial infinity

We first study the asymptotic behaviour at spatial infinity, i.e. when $r \to \infty$.The asymptotic expansion of the matrix $M(r)$ at large $r$ reads

$$M(r) = M_0 + \frac{1}{r}M_{-1} + \mathcal{O}\left(\frac{1}{r^2}\right), \quad M_0 \equiv -i\begin{pmatrix} 0 & \omega^2 \\ 1 & 0 \end{pmatrix}, \quad M_{-1} \equiv 2\begin{pmatrix} 1 & 0 \\ -i\mu & 0 \end{pmatrix}. \quad (5.10)$$

The leading term $M_0$ is diagonalisable and one can go to a basis where it is diagonal, by introducing the new vector $Y^{(1)}$ defined by

$$Y \equiv P_{(1)}Y^{(1)}, \quad P_{(1)} = \begin{pmatrix} \omega & -\omega \\ 1 & 1 \end{pmatrix}. \quad (5.11)$$

According to eq. (5.7), this gives the new system

$$\frac{\mathrm{d}Y^{(1)}}{\mathrm{d}r} = M^{(1)}Y^{(1)}, \quad M^{(1)}(r) = M_0^{(1)} + \frac{1}{r}M_{-1}^{(1)} + \mathcal{O}\left(\frac{1}{r^2}\right), \quad (5.12)$$

with

$$M_0^{(1)} \equiv \begin{pmatrix} -i\omega & 0 \\ 0 & i\omega \end{pmatrix}, \quad M_{-1}^{(1)} \equiv \begin{pmatrix} -i\omega\mu + 1 & i\omega\mu - 1 \\ -i\omega\mu - 1 & i\omega\mu + 1 \end{pmatrix}. \quad (5.13)$$

We need some extra work to diagonalise the next-to-leading order matrix $M_{-1}^{(1)}$ while keeping the leading order matrix diagonal.

This can be achieved by introducing a new vector $Y^{(2)}$ defined by

$$Y^{(1)} \equiv P_{(2)}Y^{(2)}, \quad P_{(2)} = I + \frac{1}{r}\Xi, \quad (5.14)$$

where $I$ is the identity matrix and $\Xi$ a constant matrix. Indeed, it is immediate to see that such a change of variable leads to the equivalent differential system

$$\frac{\mathrm{d}Y^{(2)}}{\mathrm{d}r} = M^{(2)}Y^{(2)}, \quad M^{(2)}(r) = M_0^{(2)} + \frac{1}{r}M_{-1}^{(2)} + \mathcal{O}\left(\frac{1}{r^2}\right), \quad (5.15)$$

with

$$M_0^{(2)} = M_0^{(1)}, \quad M_{-1}^{(2)} = M_{-1}^{(1)} + [M_0^{(1)}, \Xi]. \quad (5.16)$$

The leading matrix remains unchanged while one can easily find a matrix $\Xi$ so that $M_{-1}^{(2)}$ is diagonal. Notice that $\Xi$ appears in eq. (5.16) only in a commutator with the diagonal matrix $M_0^{(1)}$, hence the diagonal part of $\Xi$ is irrelevant and we can already fix the diagonal terms of $\Xi$ to 0. In this case, the solution to eq. (5.16) with $M_{-1}^{(2)}$ diagonal is unique and given by

$$\Xi = \frac{1}{2i\omega}\begin{pmatrix} 0 & i\omega\mu - 1 \\ i\omega\mu + 1 & 0 \end{pmatrix} . \tag{5.17}$$

We have thus managed to obtain a fully diagonalised system, up to order $1/r$, with the matrix

$$M^{(2)}(r) = \begin{pmatrix} -i\omega & 0 \\ 0 & i\omega \end{pmatrix} + \frac{1}{r}\begin{pmatrix} 1 - i\omega\mu & 0 \\ 0 & 1 + i\omega\mu \end{pmatrix} + \mathcal{O}\left(\frac{1}{r^2}\right) . \tag{5.18}$$

This system can be immediately integrated in the form of eq. (5.5), and the asymptotic solution reads

$$Y^{(2)}(r) = (1 + \mathcal{O}(1/r))\begin{pmatrix} c_- \, \mathfrak{a}_+^\infty(r) \\ c_+ \, \mathfrak{a}_-^\infty(r) \end{pmatrix} , \tag{5.19}$$

where $c_\pm$ are integration constants and the components $\mathfrak{a}_\pm^\infty$ are such that

$$\mathfrak{a}_\pm^\infty(r) = e^{\pm i\omega r} r^{1 \pm i\omega\mu} . \tag{5.20}$$

Taking into account the time dependency $e^{-i\omega t}$ of the modes, the two components of $Y^{(2)}$ are of the form

$$e^{-i\omega t}\, \mathfrak{a}_\mp^\infty(r) = (1 + \mathcal{O}(1/r))\, c_\pm r e^{-i\omega(t \mp (r + \mu \ln r))} = c_\pm \left(r + \mathcal{O}(1)\right) e^{-i\omega(t \mp r_*)} , \tag{5.21}$$

where it is convenient to use the "tortoise" coordinate $r_*$, introduced in eq. (4.40), noting that

$$r_* = r + \mu \ln(r/\mu - 1) = r + \mu \ln r + \mathcal{O}(1) . \tag{5.22}$$

As a consequence, one can identify $\mathfrak{a}_-^\infty(r)$ as an ingoing mode and $\mathfrak{a}_+^\infty(r)$ as an outgoing mode at spatial infinity.

Finally, we can return to the original vector $Y$ thanks to the transformation

$$Y = P_{(1)} P_{(2)} Y^{(2)} = \begin{pmatrix} \omega & -\omega \\ 1 & 1 \end{pmatrix} \left(1 + \frac{\Xi}{r}\right) Y^{(2)} , \tag{5.23}$$

in order to obtain the asymptotic expansion of the two original gravitational perturbations $h_0$ and $h_1$ at spatial infinity,

$$\begin{aligned} h_0(r) &= \omega \left(c_- e^{-i\omega r_*} - c_+ e^{+i\omega r_*}\right) \left(r + \mathcal{O}(1)\right) , \\ h_1(r) &= \omega \left(c_- e^{-i\omega r_*} + c_+ e^{+i\omega r_*}\right) \left(r + \mathcal{O}(1)\right) . \end{aligned} \tag{5.24}$$

One can immediately check that these expressions agree with the asymptotic expansion (4.47) obtained from the Schrödinger-like equation (with $c_- = -i\mathscr{B}_\infty$ and $c_+ = -i\mathscr{A}_\infty$).

#### 5.1.2.2. Asymptotic analysis near the black hole horizon

Let us now study the behaviour of the axial modes near the horizon. In this case, it is convenient to introduce the new radial variable $\varepsilon \equiv r - \mu$ and expand the matrix $M$ for the system (5.9) in powers of $\varepsilon$. One finds [1]

$$M(\varepsilon) = \frac{1}{\varepsilon^2} M_2 + \frac{1}{\varepsilon} M_1 + M_0 + \mathcal{O}(\varepsilon) , \tag{5.25}$$

with the matrix coefficients

$$M_2 \equiv \begin{pmatrix} 0 & 0 \\ -i\mu^2 & 0 \end{pmatrix} , \quad M_1 \equiv \begin{pmatrix} 0 & 0 \\ -2i\mu & -1 \end{pmatrix} , \quad M_0 \equiv \begin{pmatrix} 2/\mu & -i\omega^2 \\ -i & 1/\mu \end{pmatrix} . \tag{5.26}$$

An important difference with the previous situation is that the leading term $M_2$ is no longer diagonalisable but nilpotent instead. We thus need to first perform a transformation that yields a diagonalisable leading matrix [2], taking advantage of the derivative term in eq. (5.7). This can be done with the transformation

$$Y \equiv P_{(1)} Y^{(1)} , \quad P_{(1)}(\varepsilon) \equiv \begin{pmatrix} 1 & 0 \\ 0 & 1/\varepsilon \end{pmatrix} , \tag{5.27}$$

leading to the new system

$$\frac{\mathrm{d}Y^{(1)}}{\mathrm{d}\varepsilon} = M^{(1)} Y^{(1)} , \quad M^{(1)}(\varepsilon) = -\frac{1}{\varepsilon} \begin{pmatrix} 0 & i\omega^2 \\ i\mu^2 & 0 \end{pmatrix} + \mathcal{O}(1) . \tag{5.28}$$

The transformation (5.27) has eliminated the term in $1/\varepsilon^2$ in the expansion and the leading term $M_1^{(1)}$ is now diagonalisable, so that only the expansion of $M^{(1)}$ up to order $1/\varepsilon$ is required (see discussion in the footnote). It is worth noticing that $M_1^{(1)}$ receives contributions from $M_2$, $M_1$ and $M_0$. In particular, some of its coefficients involve the frequency $\omega$ which is originally present only in $M_0$.

The final step of the analysis consists in diagonalising the system (5.28), via the transformation

$$Y^{(1)} = P_{(2)} Y^{(2)} , \quad P_{(2)} \equiv \begin{pmatrix} \omega & -\omega \\ \mu & \mu \end{pmatrix} , \tag{5.29}$$

leading to

$$\frac{\mathrm{d}Y^{(2)}}{\mathrm{d}\varepsilon} = M^{(2)} Y^{(2)} , \quad M^{(2)}(\varepsilon) \equiv \frac{1}{\varepsilon} \begin{pmatrix} -i\omega\mu & 0 \\ 0 & i\omega\mu \end{pmatrix} + \mathcal{O}(1) . \tag{5.30}$$

---

1. Note that $\varepsilon$ goes to zero here, in contrast to the previous case where the variable $r$ was going to infinity. One could work in a fully analogous system by using the variable $z = 1/\varepsilon$, with the system

$$\frac{\mathrm{d}Y}{\mathrm{d}z} = \tilde{M}(z) Y , \quad \tilde{M} = -\frac{1}{z^2} M(z^{-1}) = -M_2 - M_1 \frac{1}{z} - M_0 \frac{1}{z^2} .$$

In the present case, one must push the expansion up to order $1/z^2$ because the leading matrix $M_2$ is nilpotent.

2. Indeed, integrating naively the system $\mathrm{d}Y/\mathrm{d}\varepsilon = M_2 Y/\varepsilon^2$ yields $Y \approx c \begin{pmatrix} 1 & i\mu^2/\varepsilon \end{pmatrix}^\top$ with $c$ a constant, and one does not recover at all the results of section 4.4.4.

Integrating this equation yields

$$Y^{(2)}(\varepsilon) = (1+\mathcal{O}(\varepsilon))\begin{pmatrix} c_-\varepsilon^{-i\omega\mu} \\ c_+\varepsilon^{+i\omega\mu}\end{pmatrix} = (1+\mathcal{O}(\varepsilon))\begin{pmatrix} c_-\mathfrak{a}^{\mathrm{h}}_-(r_*) \\ c_+\mathfrak{a}^{\mathrm{h}}_+(r_*)\end{pmatrix}, \tag{5.31}$$

with

$$\mathfrak{a}^{\mathrm{h}}_{\pm}(r_*) = e^{\pm i\omega r_*}, \tag{5.32}$$

where we have again expressed the result in terms of the tortoise coordinate $r_*$, which behaves as $r_* = \mu\ln\varepsilon + \mathcal{O}(1)$ near the horizon. One can immediately recognize the ingoing and outgoing modes at the horizon.

Finally, one can return to the original functions, via $Y = P_{(1)}P_{(2)}Y^{(2)}$, and derive the expressions

$$\begin{aligned} h_0(r) &= \omega\left(c_-e^{-i\omega r_*} - c_+e^{+i\omega r_*}\right)(1+\mathcal{O}(\varepsilon)), \\ h_1(r) &= \frac{\omega\mu}{\varepsilon}\left(c_-e^{-i\omega r_*} + c_+e^{+i\omega r_*}\right)(1+\mathcal{O}(\varepsilon)), \end{aligned} \tag{5.33}$$

which coincide with the asymptotic expansions (4.48) obtained from the Schrödinger-like equation (with $c_- = -i\mu\mathscr{B}_{\mathrm{hor}}$, $c_+ = -i\mu\mathscr{A}_{\mathrm{hor}}$).

### 5.1.3. Polar modes

The dynamics of the polar perturbations is described by the first-order system (4.30), of the form

$$\frac{\mathrm{d}Y}{\mathrm{d}r} = M(r)Y, \quad \text{with} \quad Y(r) \equiv \begin{pmatrix} K(r) \\ H_1(r)/\omega \end{pmatrix}, \tag{5.34}$$

and the matrix

$$M(r) = \frac{1}{3\mu+2\lambda r}\begin{pmatrix} \frac{\mu(3\mu+(\lambda-2)r)-2r^4\omega^2}{r(r-\mu)} & \frac{2i(\lambda+1)(\mu+\lambda r)+2ir^3\omega^2}{r^2} \\ \frac{ir(9\mu^2-8\lambda r^2+8(\lambda-1)\mu r)+4ir^5\omega^2}{2(r-\mu)^2} & \frac{2r^4\omega^2-\mu(3\mu+3\lambda r+r)}{r(r-\mu)} \end{pmatrix}. \tag{5.35}$$

#### 5.1.3.1. Asymptotic analysis at spatial infinity

Expanding eq. (5.35) in powers of $r$, one gets

$$\begin{aligned} M(r) = &\begin{pmatrix} 0 & 0 \\ \frac{i\omega^2}{\lambda} & 0 \end{pmatrix} r^2 + \begin{pmatrix} -\frac{\omega^2}{\lambda} & 0 \\ \frac{i\mu\omega^2(4\lambda-3)}{2\lambda^2} & \frac{\omega^2}{\lambda} \end{pmatrix} r \\ &+ \begin{pmatrix} -\frac{(2\lambda-3)\mu\omega^2}{2\lambda^2} & \frac{i\omega^2}{\lambda} \\ -2i + \frac{3i(4\lambda^2-4\lambda+3)\mu^2\omega^2}{4\lambda^3} & \frac{(2\lambda-3)\mu\omega^2}{2\lambda^2} \end{pmatrix} \end{aligned}$$

$$+\frac{1}{r}\begin{pmatrix} -\frac{(4\lambda^2-6\lambda+9)\mu^2\omega^2}{4\lambda^3} & -\frac{3i\mu\omega^2}{2\lambda^2} \\ \frac{i(8(1-2\lambda)\lambda^3\mu-(27-4\lambda(\lambda(8\lambda-9)+9))\mu^3\omega^2)}{8\lambda^4} & \frac{(4\lambda^2-6\lambda+9)\mu^2\omega^2}{4\lambda^3} \end{pmatrix} + \mathcal{O}\left(\frac{1}{r^2}\right) . \tag{5.36}$$

In contrast with the axial modes at spatial infinity, the leading matrix is of order $r^2$ and is nilpotent. So, in principle, one needs to apply a procedure similar to the near-horizon analysis of axial modes, which will be presented in full generality in the next section, and then diagonalise in turn all subsequent orders. All this involves many steps which are straightforward but rather tedious to describe.

To shorten our discussion, we provide directly the transformation that combines all these intermediate steps, given by

$$Y = P\tilde{Y}\,, \quad P = \begin{pmatrix} \mathcal{S}+\mathcal{T} & \mathcal{S}-\mathcal{T} \\ \mathcal{U}-\mathcal{V} & u+\mathcal{V} \end{pmatrix}, \tag{5.37}$$

with the functions

$$\mathcal{S}(r) \equiv \frac{i(r-\mu)((2\lambda-3)\mu+4\lambda r)}{4\lambda r} + \frac{i\lambda}{2r\omega^2}\,, \quad \mathcal{T}(r) \equiv \frac{(1-2\lambda)\mu+2(1+2\lambda)r}{4r\omega}\,,$$
$$\mathcal{U}(r) \equiv r^2 + \frac{2\lambda-3}{4\lambda}\mu r\,, \quad \mathcal{V}(r) \equiv \frac{ir}{2\omega}\,. \tag{5.38}$$

This leads to the new system

$$\frac{\mathrm{d}\tilde{Y}}{\mathrm{d}r} = \tilde{M}(r)\tilde{Y}\,, \quad \tilde{M}(r) = \begin{pmatrix} i\omega & 0 \\ 0 & -i\omega \end{pmatrix} + \begin{pmatrix} -1+i\omega\mu & 0 \\ 0 & -1-i\omega\mu \end{pmatrix}\frac{1}{r} + \mathcal{O}\left(\frac{1}{r^2}\right), \tag{5.39}$$

which is diagonal and whose solution is

$$\tilde{Y}(r) = \begin{pmatrix} c_-\,\mathfrak{g}_-^\infty(r) \\ c_+\,\mathfrak{g}_+^\infty(r) \end{pmatrix}(1+\mathcal{O}(1/r))\,, \tag{5.40}$$

with

$$\mathfrak{g}_\pm^\infty(r) = e^{\pm i\omega r} r^{-1\pm i\omega\mu} = \frac{1}{r}e^{\pm i\omega r_*}\,. \tag{5.41}$$

This result is very similar to that obtained for axial perturbations in eq. (5.19), even though the asymptotic expansion of the matrix $M$ is rather different. In terms of the original functions, we find

$$K(r) = \frac{i}{\omega}H_1(r) = i(c_-\,e^{-i\omega r_*} + c_+\,e^{+i\omega r_*})\,(1+\mathcal{O}(1/r))\,, \tag{5.42}$$

which agree with eq. (4.49) (with $c_- = -\omega\mathcal{B}_\infty$ and $c_+ = \omega\mathcal{A}_\infty$).

#### 5.1.3.2. Asymptotic analysis at the black hole horizon

We finally turn to the near-horizon behaviour of polar modes. The expansion of the matrix (5.35) in terms of the small parameter $\varepsilon \equiv r - \mu$ yields

$$\begin{aligned} M(\varepsilon) &= \frac{1}{\varepsilon^2} M_2 + \frac{1}{\varepsilon} M_1 + M_0 + \mathcal{O}(\varepsilon)\,, \\ M_2 &= \begin{pmatrix} 0 & 0 \\ \gamma_2 & 0 \end{pmatrix}, \; M_1 = \begin{pmatrix} \alpha_1 & 0 \\ \gamma_1 & \delta_1 \end{pmatrix}, \; M_0 = \begin{pmatrix} \alpha_0 & \beta_0 \\ \gamma_0 & \delta_0 \end{pmatrix}, \end{aligned} \tag{5.43}$$

where only a few of the coefficients $\alpha_I$, $\beta_I$ and $\gamma_I$ will be needed explicitly.

Once more, the dominant $M_2$ is a nilpotent matrix and, as in the axial case, we use the transformation

$$Y = P_{(1)} Y^{(1)} \quad \text{with} \quad P_{(1)}(\varepsilon) \equiv \begin{pmatrix} 1 & 0 \\ 0 & 1/\varepsilon \end{pmatrix}, \tag{5.44}$$

which gives the new system

$$\frac{\mathrm{d}Y^{(1)}}{\mathrm{d}\varepsilon} = M^{(1)} Y^{(1)}\,, \quad M^{(1)}(\varepsilon) = \frac{1}{\varepsilon} \begin{pmatrix} \alpha_1 & \beta_0 \\ \gamma_2 & 1+\delta_1 \end{pmatrix} + \mathcal{O}(1)\,, \tag{5.45}$$

with the coefficients

$$\begin{aligned} \alpha_1 &= -(1+\delta_1) = \frac{1+\lambda-2\mu^2\omega^2}{3+2\lambda}\,, \\ \beta_0 &= \frac{2i}{\mu^2} \frac{(\lambda+1)^2+\mu^2\omega^2}{3+2\lambda}\,, \\ \gamma_2 &= \frac{i\mu^2}{2} \frac{1+4\mu^2\omega^2}{3+2\lambda}\,. \end{aligned} \tag{5.46}$$

The leading matrix can now be diagonalised via the transformation

$$Y^{(1)} = P_{(2)} Y^{(2)}\,, \quad \text{with} \quad P_{(2)} = \begin{pmatrix} \alpha-\beta & \alpha+\beta \\ 1 & 1 \end{pmatrix} \quad \text{and} \quad \alpha = \frac{\alpha_1}{\gamma_2}\,, \quad \beta = \frac{i\omega\mu}{\gamma_2}\,, \tag{5.47}$$

leading to the system

$$\frac{\mathrm{d}Y^{(2)}}{\mathrm{d}\varepsilon} = M^{(2)} Y^{(2)}\,, \quad M^{(2)} = \frac{1}{\varepsilon} \begin{pmatrix} -i\omega\mu & 0 \\ 0 & i\omega\mu \end{pmatrix} + \mathcal{O}(1)\,. \tag{5.48}$$

Note that this expression is extremely simple and does not involve $\lambda$, as expected, even though it appears explicitly in $M^{(1)}$. We obtain immediately the asymptotic behaviour of $X^{(2)}$ near the horizon:

$$Y^{(2)}(\varepsilon) = (1+\mathcal{O}(\varepsilon)) \begin{pmatrix} c_- \, \mathfrak{g}^{\mathrm{h}}_-(r_*) \\ c_+ \, \mathfrak{g}^{\mathrm{h}}_+(r_*) \end{pmatrix}, \tag{5.49}$$

with
$$\mathfrak{g}^{\mathrm{h}}_{\pm}(r_*) = e^{\pm i\omega r_*}\,, \tag{5.50}$$
which reproduces the same result as for the axial mode (5.31). In terms of the original gravitational functions $H_1(r)$ and $K(r)$, using the transformation $Y = P_{(1)}P_{(2)}Y^{(2)}$, we recover the result of eq. (4.51), with
$$c_+ = \frac{i}{2}\mu(1-2i\omega\mu)\mathscr{A}_{\mathrm{hor}}\,, \quad c_- = -\frac{i}{2}\mu(1+2i\omega\mu)\mathscr{B}_{\mathrm{hor}}\,. \tag{5.51}$$
This completes our study of all asymptotic behaviours of Schwarzschild perturbations, demonstrating that one can recover the standard results directly from the linearised Einstein's equations, without resorting to the Schrödinger-like reformulation of the system.

## 5.2. General analysis

As we have seen in the previous section, it is possible to compute the asymptotic behaviour of BH perturbations in GR without reformulating the linearised Einstein equations in terms of a Schrödinger-like equation. The advantage of this method is that it can be straightforwardly generalised to the study of BHs in theories of modified gravity where it might be difficult or impossible to reduce the linearised equations to a Schrödinger-like form.

In this section, we present a systematic algorithm for a generic first-order system of the form (5.8), which has been developed in the mathematics literature, first in [203] and more recently in [204, 205, 206, 207, 208]. The various steps of the algorithm presented in this section are summarised in the flowchart diagram depicted in section 5.3.

### 5.2.1. Asymptotic solution: overview

We consider a general system of first-order ordinary differential equations of the form
$$\frac{\mathrm{d}Y}{\mathrm{d}z} = M(z)Y\,, \tag{5.52}$$
where $Y$ is a $n$-dimensional column vector, $M$ an $n\times n$-dimensional matrix and $z$ a real variable defined in some interval. In the following, we will consider only the asymptotic behaviour when $z \to +\infty$, but it is straightforward to extend the algorithm near a finite value $z_0$ where the system is singular, by a suitable change of the variable $z$.

We then assume that one can expand $M$ in powers of $z$, up to some order (depending on the required precision of the asymptotic expansion) as follows:
$$M(z) = M_r z^r + \cdots + M_0 + \ldots M_{r-f} z^{r-f} + \mathscr{O}(z^{r-f-1})\,,$$

$$= z^r \sum_{k=0}^{f} M_{r-k} z^{-k} + \mathcal{O}(z^{r-f-1}) \,, \tag{5.53}$$

where the integer $r$ is called the Poincaré rank of the system, and the $M_i$ are $z$-independent matrices. In most cases[3], the general solution to the system (5.52) admits an asymptotic expansion of the form [203]

$$Y(z) = e^{\boldsymbol{\Upsilon}(z)} r^{\boldsymbol{\Delta}} \mathbf{F}(z) Y_0 \,, \tag{5.54}$$

where $Y_0$ is a constant vector, corresponding to $n$ integration constants, $\boldsymbol{\Upsilon}$ is a diagonal matrix whose coefficients are polynomials of degree at most $r+1$, $\boldsymbol{\Delta}$ is a constant diagonal matrix and $\mathbf{F}(z)$ is a matrix which is regular at infinity.

The goal of the algorithm presented below is to determine explicitly the expression (5.54) up to some irrelevant sub-leading terms. As we have already seen in the previous section, the guiding principle in order to obtain this expression is to fully diagonalise the differential system, up to the appropriate order, by using iteratively transformations of the vector $Y$ into a new vector $\tilde{Y}$, of the form

$$Y(z) = P(z)\tilde{Y}(z) \,,$$

where $P$ is an invertible matrix. The system (5.52) is then transformed into a new but equivalent differential system, given by

$$\frac{\mathrm{d}\tilde{Y}}{\mathrm{d}z} = \tilde{M}(z)\tilde{Y}, \quad \tilde{M}(z) \equiv P^{-1}MP - P^{-1}\frac{\mathrm{d}P}{\mathrm{d}z} \,. \tag{5.55}$$

The end point of this procedure is a system where the matrix coefficients in the expansion of the form (5.53) are diagonal at each order. It is then immediate to integrate the system and to find the solution in the form (5.54), as discussed in section 5.1.1.

In the following subsections, we describe the algorithm step by step. We have also inserted two subsections that contain examples chosen to illustrate some of the finer points of the algorithm. The algorithm contains several branches, depending on whether the leading term $M_r$ in the expansion of $M(z)$ is diagonalisable or not.

### 5.2.2. Case 1: the leading term is diagonalisable

The simplest situation is when the leading matrix $M_r$ is diagonalisable, with each eigenvalue of multiplicity 1. In this case, one first uses the transformation

3. Note that, in some cases, the variable $z$ in the expression (5.54) differs from the variable $z$ in the original system (5.52), because a change of variable is necessary, as will be discussed around eq. (5.71). Morever, the special case where $M(z) = M_{-1}/z + \mathcal{O}(z^{-2})$ with $M_{-1}$ nilpotent leads to a $\ln(z)$ behaviour at large $z$, as discussed at the end of section 5.2.3.

$Y = P_{(1)} Y^{(1)}$ where $P_{(1)}$ is a constant matrix that diagonalises $M_r$, which gives the new system

$$\frac{\mathrm{d}Y^{(1)}}{\mathrm{d}z} = M^{(1)} Y^{(1)} ,$$
$$M^{(1)}(z) = D_r z^r + M^{(1)}_{r-1} z^{r-1} + \cdots + M^{(1)}_0 + M^{(1)}_{-1} \frac{1}{z} + \mathcal{O}\left(\frac{1}{z^2}\right) , \tag{5.56}$$

where the matrix $D_r$ is diagonal.

One then seeks to transform the next-to-leading matrix $M^{(1)}_{r-1}$ into a diagonal matrix (if it is not already) without affecting the diagonal form of the leading order. This can be accomplished with a new transformation

$$Y^{(1)} = P_{(2)} Y^{(2)} , \quad P_{(2)}(z) = I + \frac{1}{z} \Xi_{(2)} , \tag{5.57}$$

where $\Xi_{(2)}$ is a constant matrix. Indeed, this yields the new system

$$\frac{\mathrm{d}Y^{(2)}}{\mathrm{d}z} = M^{(2)} Y^{(1)} ,$$
$$M^{(2)}(z) = D_r z^r + D_{r-1} z^{r-1} + M^{(2)}_{r-2} z^{r-2} + \cdots + M^{(2)}_{-1} \frac{1}{z} + \mathcal{O}\left(\frac{1}{z^2}\right) , \tag{5.58}$$

with

$$D_{r-1} = M^{(1)}_{r-1} + [D_r, \Xi^{(2)}] , \tag{5.59}$$

which is imposed to be diagonal via an appropriate choice [4] for $\Xi_{(2)}$. Furthermore, $D_{r-1}$ is the diagonal part of $M^{(1)}_{r-1}$.

One can proceed similarly to "diagonalise" all the other terms, order by order, until one gets a system of the form [5]

$$\frac{\mathrm{d}Y^{(r+2)}}{\mathrm{d}z} = M^{(r+2)} Y^{(r+2)} ,$$
$$M^{(r+2)}(z) = D_r z^r + \cdots + D_0 + D_{-1} \frac{1}{z} + \mathcal{O}\left(\frac{1}{z^2}\right) , \tag{5.60}$$

where all matrices are diagonal up to order $1/z$. The system can then be immediately integrated, to yield

$$Y^{(r+2)}(z) = e^{\boldsymbol{\Upsilon}(z)} z^{\boldsymbol{\Delta}} \, \mathbf{F}(z) \, Y_0 , \quad \boldsymbol{\Delta} \equiv D_{-1} , \quad \boldsymbol{\Upsilon}(z) \equiv D_r \frac{z^{r+1}}{r+1} + \cdots + D_0 z , \tag{5.61}$$

4. To find $\Xi$ such that the matrix $\tilde{D} = M + [D, \Xi]$ is diagonal, $M$ being arbitrary and $D$ diagonal, one notices that $[D, \Xi]_{ij} = (d_i - d_j) \Xi_{ij}$ where $d_i$ are the eigenvalues of $D$. Consequently, $\tilde{D}$ is given by the diagonal component of $M$ and the coefficients of $\Xi$ satisfy $(d_i - d_j)\Xi_{ij} + M_{ij} = 0$, which always admit at least one solution for each $\Xi_{ij}$ as long as all $d_i$ are different.

5. Note that we could have proceeded in a single step by introducing the new variable $\tilde{Y}$ defined by $Y = P(z)\tilde{Y}$ with $P(z) = P_0 + \frac{1}{z} P_1 + \cdots + \frac{1}{z^{r+1}} P_{r+1}$ and determining the constant matrices $P_j$ so that $\tilde{M}(z)$ is equal to (5.60). The calculation we have just done proves this is possible with $\tilde{Y} = Y^{(r+2)}$.

where $Y_0$ is a constant vector.

The asymptotic expansion of the original vector $Y$ can be simply deduced from the combined transformations, i.e.

$$Y = P_{(1)}P_{(2)} \cdots P_{(r+2)}Y^{(r+2)} . \tag{5.62}$$

Since the $P_{(j)}$ are polynomials of $1/z$, $Y$ has exactly the same exponential behaviour (in its asymptotic expansion) as $Y^{(r+2)}$.

The above procedure is not directly applicable if the leading matrix $M_r$ has eigenvalues of multiplicity higher than one. In such a case, writing $M_r$ in a block diagonal form, with eigenvalues $\lambda_i$ of multiplicity $m_i$, one applies a transformation

$$Y^{(1)} = P_{(2)}Y^{(2)} , \tag{5.63}$$

where $P_{(2)}$ has the same block structure as $M_r$, with the blocks $B_i$ of size $m_i \times m_i$ defined as $B_i = \exp\left(\frac{\lambda_i}{r+1}z^{r+1}\right)$ if $m_i \geq 2$ and $B_i = 1$ if $m_i = 1$. For example, if the leading matrix is $M_r = \mathrm{Diag}(\lambda_1, \lambda_1, \lambda_2)$, with $r = 1$, then the transformation is $P_{(2)} = \mathrm{Diag}(\exp\left(\lambda_1\frac{z^2}{2}\right), \exp\left(\lambda_1\frac{z^2}{2}\right), 1)$.

Such a transformation puts the multi-dimensional blocks to zero, allowing one to pursue the algorithm with the subleading terms. One must however be careful when coming back to the original variable $Y^{(1)}$, since the transformation $P_{(2)}$ will greatly affect the computed asymptotic behaviour.

### 5.2.3. Case 2: the leading term is non-diagonalisable, similar to a single-block Jordan matrix

Solving asymptotically a system where the dominant term $M_r$ is not diagonalisable is more challenging. The basic idea consists in finding a transformation where the leading term of the new matrix becomes diagonalisable. This can be done by reducing progressively the Poincaré rank of the system until the leading term is diagonalisable, in which case the procedure of the previous subsection becomes applicable. If the leading term never gets diagonalisable down to the rank $r = -1$, then the general formula (5.54) for the asymptotic expansion is not valid but the system can nevertheless be integrated explicitly.

The reduction of the Poincaré rank together with the diagonalisation of the leading term is done in different steps, which we now describe, first when the leading term is similar to a Jordan matrix with a single block. The case of a Jordan matrix with several blocks will be discussed later, in section 5.2.5.

#### 5.2.3.1. Step 1. Transformation to a Jordan block

Starting from the asymptotic expansion (5.53) of the matrix $M$, we use the transformation $X = P_{(1)}X^{(1)}$ to write $M_r^{(1)} = P_{(1)}^{-1}M_rP_{(1)}$ in a Jordan canonical form (although with a lower triangular matrix). We assume here that $M_r^{(1)}$ contains a single (lower triangular) Jordan block with eigenvalue $\lambda$, i.e. of the form

$$M_r^{(1)} = \begin{pmatrix} \lambda & 0 & \cdots & & \\ 1 & \lambda & 0 & \cdots & \\ 0 & 1 & \lambda & 0 & \cdots \\ \vdots & & & & \end{pmatrix} \equiv \lambda I + J(n)\,, \tag{5.64}$$

where $J(n)$ has the property to be nilpotent (we recall that $n$ is the dimension of the matrix).

#### 5.2.3.2. Step 2. Transformation to a nilpotent matrix

We then apply the transformation

$$Y^{(1)} = P_{(2)}Y^{(2)}\,, \quad P_{(2)}(z) \equiv \exp\left(\frac{\lambda}{r+1}z^{r+1}\right)I\,, \tag{5.65}$$

which renders the leading term nilpotent [6]:

$$M^{(2)}(z) = J(n)z^r + M_{r-1}^{(2)}z^{r-1} + \cdots + M_0^{(2)} + M_{-1}^{(2)}\frac{1}{z} + \mathcal{O}\left(\frac{1}{z^2}\right)\,. \tag{5.66}$$

#### 5.2.3.3. Step 3. Normalisation and reduction of the Poincaré rank

The next step consists in reducing the Poincaré rank of the system by using the transition matrix

$$P(z) = D(n,z) \equiv \begin{pmatrix} 1 & 0 & 0 & \cdots & \cdots & 0 \\ 0 & z & 0 & \cdots & \cdots & 0 \\ 0 & 0 & z^2 & 0 & \cdots & 0 \\ \vdots & & & \ddots & & \vdots \\ 0 & \cdots & & & \cdots & z^{n-1} \end{pmatrix}\,, \tag{5.67}$$

which satisfies the useful property

$$P^{-1}J(n)P = \frac{1}{z}J(n)\,. \tag{5.68}$$

6. This follows from the relation

$$P_{(2)}^{-1}\left(z^r(\lambda I + J(n))\right)P_{(2)} - P_{(2)}^{-1}\frac{\mathrm{d}P_{(2)}}{\mathrm{d}z} = z^r(M_r^{(1)} - \lambda I) = z^rJ(n)\,.$$

A transformation with the above $P$ will thus reduce the order of the leading term $J(n)z^r$, but will also affect the sub-dominant terms in the expansion (5.66) of $M^{(2)}$, in particular $M^{(2)}_{r-1}$ which could generate terms whose order is higher than $r-1$ in the new matrix.

To avoid this situation, we need first to "normalise" the system, with the transformation

$$P_{(3)}(z) = I + \frac{1}{z}\Lambda_{(3)}\,, \tag{5.69}$$

where $\Lambda_{(3)}$ is a constant matrix, chosen such that such that the next-to-leading order matrix $M^{(3)}_{r-1}$ in the new matrix expansion contains only zeros except possibly in the first row. Let us stress that this transformation leaves the leading term of the expansion unchanged. The new system associated with $M^{(3)}$ is said to be *normalised*.

One can then perform the transformation generated by the transition matrix

$$P_{(4)}(z) = D(n,z)\,, \tag{5.70}$$

which, in *most* cases, gives a reduced Poincaré rank. There are however a few exceptions where the reduction does not work. These special cases require a more general transformation, with a transition matrix of the form

$$P_{(4)}(z) = D(n,z^{p/q}) \quad \text{with} \quad 1 \le p \le q \le n\,, \tag{5.71}$$

where $p$ and $q$ are co-prime integers. For example, when $n = 4$, the possible choices are $\{1/4, 1/3, 1/2, 2/3, 3/4, 1\}$, where the last value corresponds to the generic case (5.70). To identify the appropriate value of $p/q$, one must test successively the possible values, in decreasing order, until the transformation (5.71) effectively leads to a system with a lower Poincaré rank. The corresponding value of $p/q$ is said to be "admissible". In practice, this can be understood as a change of variable [7], $z$ being replaced by $u = z^{p/q}$.

#### 5.2.3.4. Step 4. Diagonalisable or not diagonalisable?

The next step depends on the nature of the system $(Y^{(4)}, M^{(4)})$, which possesses a lower Poincaré rank than the initial system. If the leading term of $M^{(4)}$ is diagonalisable, one proceeds as in section 5.2.2.

If $M^{(4)}$ is not diagonalisable, one needs to reduce again the Poincaré rank of the system, unless one has already reached $r = -1$, in which case one can jump directly to the next paragraph. Otherwise, one must distinguish the following different cases.

7. In this case, the asymptotic expansion of the solution may have an exponential behaviour where the argument $\Upsilon(z)$ is not a polynomial of $z$ but rather a polynomial of $z^{1/q}$.

– If the leading term is similar to a single-block Jordan matrix and we took $p/q = 1$ in the previous step, we repeat the procedure of this subsection.
– If the leading term is similar to a single-block Jordan matrix but we took $p/q < 1$ in the previous step, we discard the last step, and start again with the normalised system $M^{(3)}$. However, this time, we normalise the system up to second order: after having normalised $M_{-1}$, we repeat the procedure with $z^2$ instead of $z$ in $P_{(3)}$ (5.69) and require that $M_{-2}$ has a specific form. Details can be found in [204]. If necessary, one can pursue the normalisation to higher orders.
– If the Jordan canonical form of the leading term contains several blocks, we go to section 5.2.5.

Eventually we obtain either a system with a diagonalisable leading term, which can be solved following section 5.2.2, or a system of Poincaré rank $r = -1$ with a nilpotent leading term. In the latter case, the solution is equivalent to a polynomial of $\ln(z)$ at large $z$. Indeed, a system of the form

$$\frac{\mathrm{d}Y}{\mathrm{d}z} = \frac{\mu_0}{z}\begin{pmatrix} 0 & 0 & \cdots & & \\ 1 & 0 & 0 & \cdots & \\ 0 & 1 & 0 & 0 & \cdots \\ \vdots & & & & \end{pmatrix} Y, \tag{5.72}$$

where $\mu_0$ is an arbitrary constant, is easily integrated. The components $Y_i$ (for $1 \le i \le n$) are obtained iteratively and are given by $Y_1(z) = \xi_1$, $Y_2(z) = \xi_1 \ln z + \xi_2$ and more generally,

$$Y_i(z) = \sum_{j=1}^{i} \frac{\xi_j}{(i-j)!}(\mu_0 \ln(z))^{i-j}, \tag{5.73}$$

where the $\xi_i$ are $n$ constants of integration. All the components of $Y$ are thus polynomials of $\ln(z)$ at large $z$.

### 5.2.4. An example with a nilpotent leading term

Let us give a concrete example of the procedure used for systems with a nilpotent leading term. We consider the two-dimensional system defined by

$$\frac{\mathrm{d}Y}{\mathrm{d}z} = M(z)Y, \quad M(z) = \begin{pmatrix} 0 & 1 \\ 0 & 0 \end{pmatrix} z^2 + \begin{pmatrix} 1 & 0 \\ 0 & -1 \end{pmatrix}. \tag{5.74}$$

Let us determine its asymptotic solution at large $z$, following the algorithm described above.

We first put the leading term in its lower triangular Jordan form:

$$P_{(1)} = \begin{pmatrix} 0 & 1 \\ 1 & 0 \end{pmatrix} \quad \Longrightarrow \quad M^{(1)}(z) = \begin{pmatrix} 0 & 0 \\ 1 & 0 \end{pmatrix} z^2 + \begin{pmatrix} -1 & 0 \\ 0 & 1 \end{pmatrix}. \tag{5.75}$$

Since the leading term is already nilpotent, step 2 is irrelevant. Moreover, the system is already normalised since the next-to-leading order term vanishes.

We can thus move directly to the reduction of the order of the system and consider the transformation of the form (5.67):

$$P_{(2)}(z) = \begin{pmatrix} 1 & 0 \\ 0 & z \end{pmatrix} \quad \Longrightarrow \quad M^{(2)}(z) = \begin{pmatrix} 0 & 0 \\ 1 & 0 \end{pmatrix} z + \begin{pmatrix} -1 & 0 \\ 0 & 1 \end{pmatrix} . \tag{5.76}$$

The order has been reduced but the leading term is still nilpotent. Since the reduction was obtained via a transformation with $p/q = 1$, we continue the process by doing a new iteration of the algorithm. We first normalise the system with a transformation of the form (5.69):

$$P_{(3)}(z) = I + \frac{1}{z}\begin{pmatrix} 0 & -1 \\ 0 & 0 \end{pmatrix} \quad \Longrightarrow \quad M^{(3)}(z) = \begin{pmatrix} 0 & 0 \\ 1 & 0 \end{pmatrix} z + \begin{pmatrix} 0 & 1 \\ 0 & -1 \end{pmatrix} \frac{1}{z} + \begin{pmatrix} 0 & -2 \\ 0 & 0 \end{pmatrix} \frac{1}{z^2} . \tag{5.77}$$

We can then reduce the order of the system again with the transformation

$$P_{(4)}(z) = \begin{pmatrix} 1 & 0 \\ 0 & z \end{pmatrix} \quad \Longrightarrow \quad M^{(4)}(z) = \begin{pmatrix} 0 & 1 \\ 1 & 0 \end{pmatrix} + \begin{pmatrix} 0 & -2 \\ 0 & -2 \end{pmatrix} \frac{1}{z} . \tag{5.78}$$

The leading term is now diagonalisable. We diagonalise it explicitly, via

$$P_{(5)} = \begin{pmatrix} -1 & 1 \\ 1 & 1 \end{pmatrix} \quad \Longrightarrow \quad M^{(5)}(z) = \begin{pmatrix} -1 & 0 \\ 0 & 1 \end{pmatrix} + \begin{pmatrix} 0 & 0 \\ -2 & -2 \end{pmatrix} \frac{1}{z} , \tag{5.79}$$

then we diagonalise the next-to-leading term with a transformation of the form (5.57):

$$P_{(6)}(z) = \begin{pmatrix} 1 & 0 \\ 1/z & 1 \end{pmatrix} \quad \Longrightarrow \quad M^{(6)}(z) = \begin{pmatrix} -1 & 0 \\ 0 & 1 \end{pmatrix} + \begin{pmatrix} 0 & 0 \\ 0 & -2 \end{pmatrix} \frac{1}{z} + \mathcal{O}\left(\frac{1}{z^2}\right) . \tag{5.80}$$

We have thus managed to fully diagonalise the system, which immediately gives us the asymptotic solution

$$Y^{(6)}(z) = (1 + \mathcal{O}(1/z)) \begin{pmatrix} \exp(-z) & 0 \\ 0 & \frac{1}{z^2}\exp(z) \end{pmatrix} Y_0, \quad Y_0 \equiv \begin{pmatrix} \xi_1 \\ \xi_2 \end{pmatrix} , \tag{5.81}$$

where $Y_0$ is a constant column vector. As a consequence, to obtain the behaviour of $Y$ in the original system, we use the combined transformations

$$Y = \left( \prod_{j=1}^{6} P_{(j)} \right) Y^{(6)}, \tag{5.82}$$

which implies

$$Y(z) = \left(1 + \mathcal{O}\left(1/z\right)\right) \begin{pmatrix} \xi_1 e^{-z} z^2 + \xi_2 e^z \\ -2\xi_1 e^{-z} \end{pmatrix}. \tag{5.83}$$

For this particular example, it turns out that the original system (5.74) can be solved exactly, with the solution

$$Y(z) = \begin{pmatrix} \frac{1}{2}\xi_1 e^{-z}\left(1 + 2z + 2z^2\right) + \xi_2 e^z \\ -2\xi_1 e^{-z} \end{pmatrix}. \tag{5.84}$$

One can thus check that the asymptotic solution (5.83) agrees with the asymptotic behaviour of the exact solution.

### 5.2.5. Case 3: $M_r$ is similar to a Jordan matrix with several blocks

We now briefly discuss (without entering into too many details, which can be found in [204]) the more general case where $M_r$ is block diagonalisable and its canonical Jordan form admits several Jordan blocks. The first two steps of section 5.2.3 still apply to this case and one can find a transformation (with a constant matrix $P$) such that the new system associated with $M^{(2)}$ (we use the same notation as in section 5.2.3) has a block diagonal leading term $M_r^{(2)}$ with Jordan lower triangular blocks, each block being either nilpotent or 1-dimensional:

$$M_r^{(2)} = \begin{pmatrix} J(n_1) & 0 & \cdots & & & \\ 0 & J(n_2) & 0 & \cdots & & \\ \vdots & 0 & \ddots & 0 & \cdots & \\ & \vdots & 0 & \lambda_1 & 0 & \cdots \\ & & \vdots & 0 & \lambda_2 & 0 \\ & & & \vdots & 0 & \ddots \end{pmatrix}, \quad J(n) \equiv \begin{pmatrix} 0 & 0 & \cdots & & & \\ 1 & 0 & 0 & \cdots & & \\ 0 & 1 & 0 & 0 & \cdots & \\ \vdots & & & & & \end{pmatrix}. \tag{5.85}$$

The Jordan form is chosen so that the blocks $J(n)$ are ordered by decreasing size ($n_1 \geq n_2 \geq \cdots$). We will use this block structure as a layout for the block structure of the other matrices that appear in the expansion of $M^{(2)}$. Each block will be denoted by two indices, $(KL)$, corresponding to a submatrix of dimensions $n_K \times n_L$.

The principle of the diagonalisation procedure is similar to what was done in sections 5.2.2 and 5.2.3. However, it is now possible to have both diagonalisable blocks and nilpotent blocks. Those must be dealt with separately to get the full asymptotic behaviour of the system. In order to do this, one can generalise the order-by-order procedure of section 5.2.2: this is called the “Splitting Lemma” in [204]. It is not detailed here, but can be understood by considering blocks instead of scalars in the computations of section 5.2.2 [8].

One can use this lemma to block diagonalise $M^{(2)}$, order by order : the two global blocks considered will be the nilpotent part of $M_r^{(2)}$ and its diagonalisable part. The latter can be dealt with using the procedure given in section 5.2.2, while the former must be addressed using a generalised version of the procedure given in section 5.2.3. We give here more details about the last part and, in the rest of this section, assume without loss of generality that $M_r^{(2)}$ contains only nilpotent blocks, such that

$$M_r^{(2)} = \begin{pmatrix} J(n_1) & 0 & \cdots & & \\ 0 & J(n_2) & 0 & \cdots & \\ \vdots & 0 & J(n_3) & 0 & \cdots \\ & \vdots & & & \ddots \end{pmatrix} \quad (\text{with } n_1 \geq n_2 \geq \cdots \geq n_{\text{last}})\,. \tag{5.86}$$

The procedure in such a case requires to put the system in a specific normalized form. For a matrix $M$, obtained at a generic step in the algorithm, one says that the matrix is "normalized up to order $s$" if all its leading terms $M_r, \cdots M_{r-s}$ have their $(KL)$ blocks verifying the following properties:

- either all rows are equal to zero except possibly the first one if $K \leq L$,
- or all columns are equal to zero except possibly the last one if $K > L$.

In order to reach this normalized form, one must use a succession of transformations [9] $P_{\text{norm}}(k)$ of the form

$$P_{\text{norm}}(k) = I + \frac{1}{z^k}\Lambda\,, \tag{5.87}$$

where $k$ varies from 1 to $s$. The matrix $\Lambda$ is a constant matrix, whose coefficients must be chosen, similarly to $\Xi$ in (5.57), such that the new matrix $M$ is normalised, in the sense defined above ($\Lambda$ is uniquely defined if one requires that all its blocks $\Lambda^{KL}$ have zero last row if $K \leq L$ and zero first column if $K > L$). The procedure is iterative: if the system is normalized up to order $k$, it is possible to normalize it up to order $k+1$ by applying a transformation $P_{\text{norm}}(k+1)$. Indeed, this transformation will not modify any term of order higher than $r-k-1$.

8. In the case where $M_r^{(2)}$ consists of a 2-block Jordan matrix, one would use a transformation of the form

$$P = \begin{pmatrix} I & \sum_{j=1}^{p} \Xi_j z^{-j} \\ \sum_{j=1}^{p} \Lambda_j z^{-j} & I \end{pmatrix}\,,$$

where the $\Xi_j$ and $\Lambda_j$ are constant matrices. Such a transformation, which generalises eq. (5.57), enables us to transform each $M_{r-j}^{(2)}$ in the same block diagonal form as $M_r^{(2)}$ with a convenient choice of $\Xi_i$ and $\Lambda_i$. Therefore, the initial system gives two decoupled sub-systems and, for each one, we proceed along the same lines as in the previous section.

9. Let us emphasize on the fact that the hierarchy $n_1 \geq n_2 \geq \cdots$ is crucial for this step to succeed.

The complete procedure to reduce the Poincaré rank of the matrix is then the following:

1. one starts with $s = 1$ ;
2. one normalizes the system up to order $s$ using $P_{\text{norm}}(k)$ transformations ;
3. if $M_{r-s}$ is not block-diagonal, one uses a transformation

$$P_u(n) = \text{Diag}(I_{n_1}, I_{n_2}, \cdots, z^s I_{n_{\text{last}}}) \tag{5.88}$$

   and one goes back to step 1;[10]
4. if it is block-diagonal, one uses a $P_{p/q}$ transformation, which is a block form of (5.67) or (5.71):

$$P_{p/q} = \begin{pmatrix} D(n_1, z^{p/q}) & 0 & \cdots & & \\ 0 & D(n_2, z^{p/q}) & 0 & \cdots & \\ \vdots & 0 & D(n_3, z^{p/q}) & 0 & \cdots \\ & \vdots & & & \ddots \end{pmatrix}, \tag{5.89}$$

   where the matrices $D(n,z)$ have been defined in eq. (5.67) and $p$ and $q$ are either co-prime integers (with $1 \leq p \leq q \leq n_1$) or equal in the case $p/q = 1$ ;
5. if no $P_{p/q}$ transformation is admissible (see the definition after eq. (5.71)), one goes back to step 1 with $s$ increased by one. Otherwise, one stops here.

Thanks to the above procedure, one obtains either a system depending on $z$ with a reduced Poincaré rank, or a new system depending on $z^{p/q}$ with a non-nilpotent leading term. In the former case, one can simply pursue with the algorithm. In the latter case, one can change variables by writing $w = z^{p/q}$ and start the algorithm again.

### 5.2.6. A higher dimensional example with $p/q \neq 1$

We now present a higher dimensional ($n = 5$) example, adapted from [206], where the dominant term in the asymptotic expansion of the matrix $M$ has a non trivial canonical Jordan form with two Jordan blocks. The matrix $M(z)$ is given by

$$M(z) = \begin{pmatrix} 0 & z^3 & -z & 1 & 2z \\ -z^2 & z & 0 & -z & 0 \\ z & 1 & 0 & z^3 & 1 \\ 1 & -z & 1 & z & z^3 \\ z & 0 & -3z & 0 & -1 \end{pmatrix} \equiv M_3 z^3 + M_2 z^2 + M_1 z + M_0 , \tag{5.90}$$

10. It is proved in [204] that after a finite number of steps, one always gets a block-diagonal subleading term, which means that this procedure stops at some point and that one can go on with step 4.

where the leading term $M_3$ is nilpotent and has a 2-block Jordan structure.

We perform a first transformation $Y = P_{(1)}Y^{(1)}$ so that the leading term has now the following Jordan (lower triangular) canonical form (the matrix $P_{(1)}$ can easily been deduced):

$$M^{(1)}(z) = \begin{pmatrix} -1 & 0 & -3z & 0 & z \\ z^3 & z & 1 & -z & 1 \\ 1 & z^3 & 0 & 1 & z \\ 0 & -z & 0 & z & -z^2 \\ 2z & 1 & -z & z^3 & 0 \end{pmatrix} \implies M_3^{(1)} = \begin{pmatrix} 0 & 0 & 0 & 0 & 0 \\ 1 & 0 & 0 & 0 & 0 \\ 0 & 1 & 0 & 0 & 0 \\ 0 & 0 & 0 & 0 & 0 \\ 0 & 0 & 0 & 1 & 0 \end{pmatrix}. \tag{5.91}$$

The block structure of $M_3^{(1)}$ defines the layout that we will be using to compute the asymptotic expansion of the solution.

We notice that the next-to-leading term $M_2^{(1)}$ in the expansion of $M^{(1)}$ is already normalised. Therefore, we can immediately try to reduce the order of the system thanks to a new transformation $Y^{(1)} = P_{(2)}Y^{(2)}$,

$$P_{(2)} = \begin{pmatrix} 1 & 0 & 0 & 0 & 0 \\ 0 & z & 0 & 0 & 0 \\ 0 & 0 & z^2 & 0 & 0 \\ 0 & 0 & 0 & 1 & 0 \\ 0 & 0 & 0 & 0 & z \end{pmatrix} \implies M^{(2)} = \begin{pmatrix} -1 & 0 & -3z^3 & 0 & z^2 \\ z^2 & z - \frac{1}{z} & z & -1 & 1 \\ \frac{1}{z^2} & z^2 & -\frac{2}{z} & \frac{1}{z^2} & 1 \\ 0 & -z^2 & 0 & z & -z^3 \\ 2 & 1 & -z^2 & z^2 & -\frac{1}{z} \end{pmatrix}. \tag{5.92}$$

However, we immediately see that the order of the system has not diminished. This example falls in the cases where we need to change the variable $z$ or, equivalently, to make a transformation of the form (5.71) for each Jordan block. We must therefore cancel the previous transformation (5.92) and instead consider $Y^{(1)} = \tilde{P}_{(2)}\tilde{Y}^{(2)}$, with

$$\tilde{P}_{(2)}(z) = \begin{pmatrix} 1 & 0 & 0 & 0 & 0 \\ 0 & z^{p/q} & 0 & 0 & 0 \\ 0 & 0 & z^{2p/q} & 0 & 0 \\ 0 & 0 & 0 & 1 & 0 \\ 0 & 0 & 0 & 0 & z^{p/q} \end{pmatrix}. \tag{5.93}$$

Following the method described below eq. (5.71), we note that the largest Jordan block is of dimension 3, therefore we should take 2 co-prime integers between 1 and 3 for $p$ and $q$ with $p \leq q$. The possible choices for the ratio $p/q$ belong to the set $\{1/3, 1/2, 2/3\}$, since $p/q = 1$ does not work. The largest value is $p/q = 2/3$, which gives for the matrix $M^{(3)}$ the expression

$$\begin{pmatrix} -1 & 0 & -3z^{7/3} & 0 & z^{5/3} \\ z^{7/3} & z-\frac{2}{3z} & z^{2/3} & -z^{1/3} & 1 \\ \frac{1}{z^{4/3}} & z^{7/3} & -\frac{4}{3z} & \frac{1}{z^{4/3}} & z^{1/3} \\ 0 & -z^{5/3} & 0 & z & -z^{8/3} \\ 2z^{1/3} & 1 & -z^{5/3} & z^{7/3} & -\frac{2}{3z} \end{pmatrix}. \tag{5.94}$$

We observe that the subdiagonal terms have order $7/3$. To keep this value of $p/q$, we must make sure that no other term behaves like $z^{\alpha}$ with $\alpha > 7/3$. However in this case there is a $z^{8/3}$ term. Therefore, the value $2/3$ is not admissible and we have to consider the next possible choice which is $p/q = 1/2$. Such a change of variable leads to the matrix

$$\tilde{M}^{(2)} = \begin{pmatrix} -1 & 0 & -3z^{2} & 0 & z^{3/2} \\ z^{5/2} & z-\frac{1}{2z} & \sqrt{z} & -\sqrt{z} & 1 \\ \frac{1}{z} & z^{5/2} & -\frac{1}{z} & \frac{1}{z} & \sqrt{z} \\ 0 & -z^{3/2} & 0 & z & -z^{5/2} \\ 2\sqrt{z} & 1 & -z^{3/2} & z^{5/2} & -\frac{1}{2z} \end{pmatrix}. \tag{5.95}$$

Now, it verifies the requirements and we thus keep the value $p/q = 1/2$ and continue the process.

The previous change of variable leads to a differential system where the coefficients of $M^{(3)}$ are non-integer powers functions of $z$. To apply the algorithm, we have to make a change of coordinate so that the system involves only integer powers of $z$. This can easily be done by introducing the new coordinate $u$ defined by $z = u^2$. As a consequence, the new differential system is now given by

$$\frac{\mathrm{d}Y^{(3)}}{\mathrm{d}u} = M^{(3)}(u)Y^{(3)}, \quad M^{(3)}(u) = \begin{pmatrix} -2u & 0 & -6u^{5} & 0 & 2u^{4} \\ 2u^{6} & \frac{2u^{4}-1}{u} & 2u^{2} & -2u^{2} & 2u \\ \frac{2}{u} & 2u^{6} & -\frac{2}{u} & \frac{2}{u} & 2u^{2} \\ 0 & -2u^{4} & 0 & 2u^{3} & -2u^{6} \\ 4u^{2} & 2u & -2u^{4} & 2u^{6} & -\frac{1}{u} \end{pmatrix}, \tag{5.96}$$

where $Y^{(3)}(u) \equiv \tilde{Y}^{(2)}(z)$ and $M^{(3)}(u) \equiv 2u\tilde{M}^{(2)}(z)$ with $z = u^2$. As the leading term is not nilpotent, we keep the value of $p/q$. If it had been nilpotent, we would have had to go back one step and normalise up to the next order.

We can continue the algorithm with this new system: we will to do a new change of variables, reduce the order, and decouple the system... We will not present more steps as the rest of the computations is similar to what was done here and in previous sections. Nonetheless, for the sake of completeness, we give the final result. We show that, after enough steps of the algorithm, the initial system can be equivalently reformulated as

$$\frac{\mathrm{d}Y^{(4)}}{\mathrm{d}w} = M^{(4)}(w)Y^{(4)}\,, \tag{5.97}$$

where $w = z^{1/6}$ and $M^{(4)}(w)$ is the following diagonal matrix

$$\begin{aligned} M^{(4)}(w) = \mathrm{Diag}[&3^{4/3}(1-i\sqrt{3})w^{19} + 2w^{11}, -2\times 3^{4/3}w^{19} + 2w^{11},\\ &3^{4/3}(1+i\sqrt{3})w^{19} + 2w^{11}, 6iw^{20} + 3w^{11},\\ &-6iw^{20} + 3w^{11}] + \mathcal{O}(w^9)\,, \end{aligned} \tag{5.98}$$

up to order $\mathcal{O}(w^9)$. Integrating such a system is immediate and yields the leading orders of the asymptotic expansion of $Y^{(4)}$ from which we can extract the asymptotic expansion of the original variable $Y$.

## 5.3. Flowchart for the algorithm

In this section, we draw a flowchart to illustrate the algorithm that we are using to compute the asymptotic behaviour of a solution of a first order system. This flowchart is presented in fig. 5.1.

It should be noted that, in principle, one can skip the first question *"Is the leading term diagonalisable?"* and put directly the leading order term in its Jordan form. Indeed, when the leading term is diagonalisable, putting it into its Jordan form is equivalent to diagonalising it and the resulting Jordan matrix is made of $d$ one-dimensional blocks where $d$ is the dimension of the system, thus of the matrix. Therefore, the procedure for splitting the system into several subsystems described in section 5.2.5 is in this case equivalent to the procedure described in section 5.2.2 where we are treating several blocks.

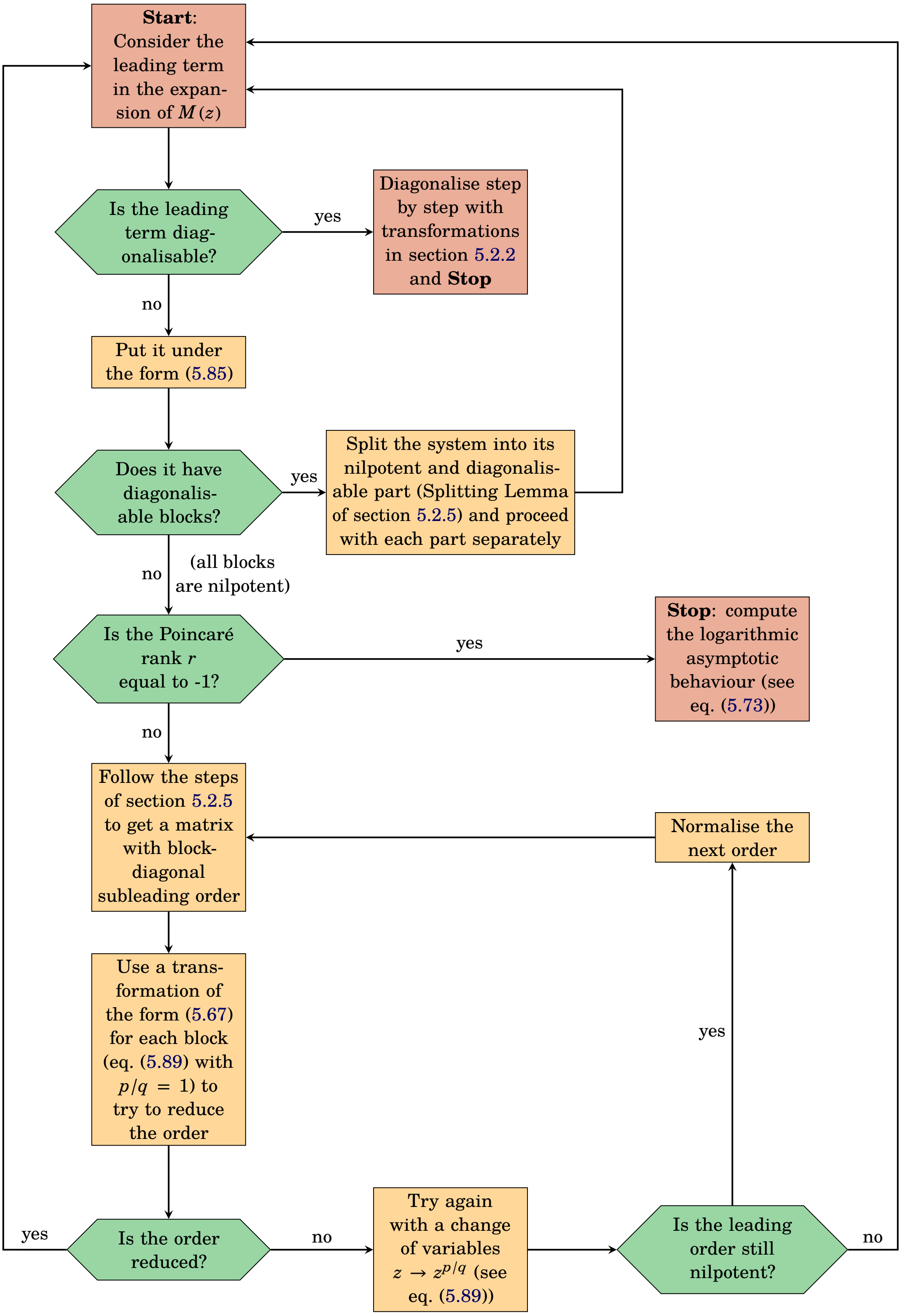

Figure 5.1. – Flowchart for the algorithm described in chapter 5.

## Conclusion

In this chapter, we have studied the asymptotic behaviours, both at spatial infinity and near the horizon, of the linear perturbations about Schwarzschild black holes. Instead of following the traditional approach described in chapter 4 that consists in rewriting the equations of motion in the form of a stationary Schrödinger-like equation, which is second-order with respect to the radial coordinate, we have worked directly with the first-order equations of motion (in the frequency domain). For this direct approach to the asymptotic behaviour, we have used an algorithm that has been developed in several recent articles published in mathematical journals.

The principle of this algorithm is to transform the differential system, via successive changes of functions, until it can be written in an explicitly diagonal form, up to the required order (in the small parameter characterising the asymptotic regime). This procedure automatically provides the combination of the metric perturbations that encapsulates the physical degree of freedom in this asymptotic region and enables one to separate the ingoing and outgoing physical modes. Although we have worked in the standard RW gauge, the same approach would work similarly for any other gauge choice.

This novel approach will be especially useful in the context of BH solutions of DHOST theories, since such systems are expected to contain couplings between gravitational and scalar degrees of freedom that will not be easy to disentangle (see section 4.5). In the following chapters, we will apply this algorithm to the solutions given in chapter 2. The same method could be applied to the study of other types of BHs, or even completely different physical systems: this is why we have presented a few illustrative examples in sections 5.2.4 and 5.2.6.

CHAPTER 6

# AXIAL BLACK HOLE PERTURBATIONS

Contents

Axial perturbations of BHs being of odd parity, they cannot couple to the new degree of freedom present in scalar-tensor theories, since this degree of freedom is a scalar and as such can only be of even parity (see section 4.5). Therefore, their treatment in the case of modified gravity theories such as DHOST should not differ so much from their treatment in the case of GR. This reasoning motivates us to look for a way to generalise the results of chapter 4 for axial perturbations to the much broader case of cubic DHOST theories. This is what we present in this chapter.

We start by describing the dynamics of the perturbations in terms of a system of two linear first order equations (in the radial coordinate) for two unknowns, generalising the results obtained in the case of GR in chapter 4. Then, we recover the well-known Schrödinger-like equation for axial perturbations with the explicit form of the speed and potential in any cubic DHOST theory. This result can be seen to be consistent with what has been obtained in several other works [141, 209, 210] from the calculation of the quadratic Lagrangian. We then use this result to study the axial perturbations of the BH solutions presented in chapter 2. Finally, we are able to compute the effective metric in which such perturbations propagate and to use it to make statements about the stability of the studied solutions. This chapter is based on [211, 212, 213].

## 6.1. Canonical system for axial BH perturbations

In this section, we concentrate on axial linear perturbations about a spherically symmetric background solution of the form (2.1) for the metric sector and (2.2) for the scalar sector, and we compute their equations of motion.

### 6.1.1. Perturbation setup

We first define a metric perturbation $h_{\mu\nu}$ and a scalar perturbation $\delta\phi$ by

$$g_{\mu\nu} = \bar{g}_{\mu\nu} + h_{\mu\nu} \quad \text{and} \quad \phi = \bar{\phi} + \delta\phi\,, \tag{6.1}$$

where a bar denotes a background solution. We proceed as in the case of GR presented in chapter 4: we start with the expansion of the action (1.25) at the quadratic order in the perturbations $h_{\mu\nu}$ and $\delta\phi$. We then compute the associated equations of motion. We restrict ourselves to the study of axial perturbations by setting the axial Regge-Wheeler gauge (see section 4.3.1 and references [179, 209]) where the only non-vanishing components of the perturbations are

$$\begin{aligned}
h_{t\theta} &= \frac{1}{\sin\theta}\sum_{\ell,m} h_0^{\ell m}(t,r)\,\partial_\varphi Y_{\ell m}(\theta,\varphi)\,, \qquad & h_{t\varphi} &= -\sin\theta\sum_{\ell,m} h_0^{\ell m}(t,r)\,\partial_\theta Y_{\ell m}(\theta,\varphi)\,,\\
h_{r\theta} &= \frac{1}{\sin\theta}\sum_{\ell,m} h_1^{\ell m}(t,r)\,\partial_\varphi Y_{\ell m}(\theta,\varphi)\,, \qquad & h_{r\varphi} &= -\sin\theta\sum_{\ell,m} h_1^{\ell m}(t,r)\,\partial_\theta Y_{\ell m}(\theta,\varphi)\,,
\end{aligned} \tag{6.2}$$

which were expanded in spherical harmonics $Y_{\ell m}(\theta,\varphi)$ ($\ell$ is an integer greater than 1 for axial perturbations and $-\ell \leq m \leq \ell$) because of the spherical symmetry of the background. Notice that $\delta\phi$ vanishes identically when one considers axial perturbations, since scalar perturbations are even-parity.

In the following, since perturbations with different values of $\ell$ and $m$ will not couple at the linear level, we drop the indices $\ell$ and $m$ for clarity. We will also be using the notation $\lambda$ defined by

$$2\lambda = \ell(\ell+1) - 2\,, \tag{6.3}$$

instead of $\ell$ itself. Finally, since the background metric is static, it is convenient to make use of the Fourier transforms of the perturbation functions $h_0$ and $h_1$ to simplify notations. More precisely, for a given function $f$, we define

$$f(t,r) = \int_{-\infty}^{+\infty} \mathrm{d}\omega\, f(\omega,r)e^{-i\omega t}\,. \tag{6.4}$$

In practice, this implies that all partial derivatives with respect to time become, in Fourier space, multiplications by $-i\omega$ . The equations of motion for the Fourier modes, which we will denote $\mathscr{E}_{\mu\nu} = 0$, therefore consist of a system of ordinary differential equations, with only derivatives with respect to the variable $r$. In the following, we do not explicitly write the $\omega$ dependency of the Fourier transforms.

### 6.1.2. Perturbed Einstein's equations

The perturbation of Einsteins' equations yields 10 equations. In the case of axial perturbations, components $\mathscr{E}_{tt}$, $\mathscr{E}_{tr}$ and $\mathscr{E}_{rr}$ are identically zero and the associated equations are therefore trivial. Moreover, due to spherical symmetry of the background, equations for $\mathscr{E}_{t\varphi}$, $\mathscr{E}_{r\varphi}$ and $\mathscr{E}_{\varphi\varphi}$ are equivalent respectively to the equations for $\mathscr{E}_{t\theta}$, $\mathscr{E}_{r\theta}$ and $\mathscr{E}_{\theta\theta}$. This leaves four non trivial independent equations for two independent functions $h_0$ and $h_1$. One can thus expect that two of these equations are redundant, which is indeed the case.

First, one has $\mathscr{E}_{\theta\theta} + 2\mathscr{E}_{\theta\varphi} = 0$. Then, one can notice that, out of these four equations, $\mathscr{E}_{t\theta}$ contains second-order derivatives of $h_0$ and $h_1$ while the others contain at most first order derivatives. This is an indication that $\mathscr{E}_{t\theta}$ is redundant and, as expected, one can show that a combination of $\mathscr{E}_{t\theta}$, $\mathscr{E}_{\theta\theta}$, $\mathscr{E}_{r\theta}$ and their derivatives vanishes. As a consequence, the dynamics of the axial perturbations is fully determined by the system consisting of the two equations

$$\mathscr{E}_{r\theta} = 0, \quad \mathscr{E}_{\theta\theta} = 0\,, \tag{6.5}$$

for the two variables $h_0$ and $h_1$, with both equations being first-order in $r$. More precisely, the system of perturbation equations can be written in the form

$$\begin{aligned} 0 &= \omega a_1(r)h_0'(r) + \big(\lambda a_2(r) + \omega^2 a_3(r)\big)h_1(r) + (q\lambda a_4(r) + \omega a_5(r))h_0(r)\,, \\ 0 &= q a_6(r)h_0'(r) + a_7(r)h_1'(r) + (q a_8(r) + \omega a_9(r))h_0(r) \\ &\qquad + (a_{10}(r) + q\omega a_{11}(r))h_1(r)\,, \end{aligned} \tag{6.6}$$

where the coefficients $a_i$, whose expressions are too cumbersome to be written here, are functions of $r$ (but not of $\lambda$, $\omega$ or $q$) and depend on the Lagrangian of the theory and on the background solution. They satisfy the following properties:

$$\begin{aligned} &a_3 = ia_1\,, \quad Ca_5 = -C'a_1\,, \quad a_{11} = -ia_6\,, \\ &a_2a_6 = a_4a_7\,, \quad a_6(a_{10} - a_7') = a_7(a_8 - a_6')\,, \end{aligned} \tag{6.7}$$

for any choice of functions $F_2, A_1, F_3, B_2$ and $B_6$, even if they do not satisfy the degeneracy conditions given in eq. (1.40).

### 6.1.3. Canonical first-order system

We now wish to reformulate the system (6.6) in the canonical form

$$\frac{\mathrm{d}Y}{\mathrm{d}r} = (M_0 + \omega M_1 + \omega^2 M_2)Y, \tag{6.8}$$

where the components of the vector $Y$ are independent linear combinations of $h_0$ and $h_1$, and the three matrices $M_0$, $M_1$ and $M_2$ do not depend on $\omega$. To do so, let us try the following ansatz:

$$Y = \begin{pmatrix} Y_1 \\ Y_2 \end{pmatrix} \quad \text{with} \quad Y_1 = h_0 \,, \quad \omega Y_2 = h_1 - qfh_0 \,, \tag{6.9}$$

where $f$ is an undetermined function at this stage. Equation (6.6) implies that the differential system satisfied by $Y$ is given by

$$\begin{pmatrix} -\omega a_1 & 0 \\ -q(a_6 + fa_7) & -\omega a_7 \end{pmatrix} \frac{\mathrm{d}Y}{\mathrm{d}r} = \begin{pmatrix} q\lambda(a_4 + fa_2) + \omega(a_5 + q\omega f a_3) & \omega(\lambda a_2 + \omega^2 a_3) \\ q(a_8 + fa_{10} + f'a_7) + \omega(a_9 + q^2 f a_{11}) & \omega(a_{10} + q\omega a_{11}) \end{pmatrix} Y. \tag{6.10}$$

To remove the off-diagonal term in the left-hand side matrix and simplify the system, one chooses

$$f = -\frac{a_6}{a_7}\,, \tag{6.11}$$

which also implies

$$a_4 + fa_2 = 0\,, \quad a_8 + fa_{10} + f'a_7 = 0\,, \tag{6.12}$$

due to the last two relations in eq. (6.7). Using the remaining three relations in eq. (6.7), the system further reduces to

$$\frac{\mathrm{d}Y}{\mathrm{d}r} = MY\,, \quad \text{with} \quad M = \begin{pmatrix} C'/C + i\omega\Psi & -i\omega^2 + 2i\lambda\Phi/C \\ -i\Gamma & \Delta + i\omega\Psi \end{pmatrix}, \tag{6.13}$$

where the functions $\Psi$, $\Phi$, $\Gamma$ and $\Delta$ are given by

$$\begin{aligned}
\mathscr{F} &= AF_2 - (q^2 + AX)A_1 - \frac{1}{2}AB\psi'X'F_{3X} \\
&\quad - \frac{1}{2}B\psi'(AX)'B_2 - \frac{A}{2B}(B\psi')^3X'B_6\,, \\
\frac{\mathscr{F}}{\Phi} &= F_2 - XA_1 - \frac{1}{2}B\psi'X'F_{3X} - \frac{1}{2}B\psi'\frac{(CX)'}{C}B_2 - \frac{1}{2}B\psi'XX'B_6\,, \\
\mathscr{F}\Psi &= q\psi'A_1 + \frac{q}{2}\left(B\psi'^2\right)'F_{3X} + \frac{q}{2}\frac{(AX)'}{A}B_2 + \frac{q}{4}\left(B^2\psi'^4\right)'B_6\,, \\
\Gamma &= \Psi^2 + \frac{1}{2AB\mathscr{F}}\big(2q^2A_1 + 2AF_2 + AB\psi'X'F_{3X} \\
&\qquad + q^2\frac{(AX)'}{A\psi'}B_2 + q^2B\psi'X'B_6\big)\,, \\
\Delta &= -\frac{\mathscr{F}'}{\mathscr{F}} - \frac{B'}{2B} + \frac{A'}{2A} = -\frac{\mathrm{d}}{\mathrm{d}r}\left(\ln\left(\sqrt{\frac{B}{A}}\mathscr{F}\right)\right).
\end{aligned} \tag{6.14}$$

Here, $f'$ denotes the derivative of any function $f$ with respect to $r$ and we imposed the degeneracy condition $3B_3 + 2B_2 = 0$ from eq. (1.40) to simplify the expressions. One can note that only the functions $F_2$, $A_1$, $B_2$ and $B_6$ appear in the perturbations ($B_3$ also appears but was removed using the degeneracy condition): this could be expected from the ADM decomposition of the DHOST action given in eq. (1.88), since only the terms containing contractions of the extrinsic curvature tensor of the form $K^{ij}K_{ij}$ or $K^{ij}K_{jk}K_k{}^i$ contain couplings with the axial modes (this can be understood by looking at the quadratic action for tensor modes given in [60, 168]).

The first-order system of eq. (6.13) is the generalisation of eq. (4.18) for any cubic DHOST theory. In the case of GR, one finds

$$\Psi = 0\,, \quad \Phi = A\,, \quad \Gamma = \frac{1}{AB} \quad \text{and} \quad \Delta = -\frac{A'}{2A} - \frac{B'}{2B}\,, \tag{6.15}$$

which means that the first-order system can be written as

$$\frac{\mathrm{d}Y}{\mathrm{d}r} = \begin{pmatrix} C'/C & -i\omega^2 + 2i\lambda A/C \\ -i/AB & -A'/2A - B'/2B \end{pmatrix} Y\,. \tag{6.16}$$

To summarize, axial perturbations about a general static and spherically symmetric background in cubic DHOST theories are fully described in terms of the first order system (6.13). Let us note that this system describes the dynamics of a single degree of freedom whereas it has been obtained without imposing any degeneracy condition. Hence, the Ostrogradsky ghost does not show up in the axial sector of the perturbations and should appear when one considers polar perturbations. This result is fully consistent with the analysis of [210] based on the computation of the quadratic Lagrangian. Furthermore, it is similar to what happens in the context of cosmological perturbations where the Ostrogradsky ghost can be seen to appear in the scalar sector and not in the tensorial sector [60].

## 6.2. Schrödinger formulation

In this section, we show how to recover for axial[1] perturbations a Schrödinger-like equation very similar to the one obtained in GR for axial and polar perturbations (see eq. (4.37)). The results we will present are consistent with those obtained recently from the quadratic action directly in [210, 214].

1. The case of polar perturbations, much more involved due to coupling between the gravitational and scalar degrees of freedom, will be presented in chapter 7.

We start from eq. (6.13), which we can slightly simplify by making a change of coordinates. In order to see this, we go back to the time variable $t$, yielding

$$\frac{\partial Y}{\partial r} + \Psi \frac{\partial Y}{\partial t} = M_{[0]}\, Y + M_{[2]}\, \frac{\partial^2 Y}{\partial t^2}\,, \tag{6.17}$$

where $M_{[0]}$ and $M_{[2]}$ are matrices that one can directly obtain from eq. (6.13). We then absorb the first time derivative of $Y$ using a change of time coordinate of the form [2]

$$t_* = t - \int \mathrm{d}r\, \Psi(r)\,, \tag{6.18}$$

which transforms eq. (6.17) into

$$\frac{\partial Y}{\partial r} = M_{[0]}\, Y + M_{[2]}\, \frac{\partial^2 Y}{\partial t_*^2}\,. \tag{6.19}$$

Then, we develop in Fourier modes using the variable $t_*$ (and not the variable $t$ used in eq. (6.4)), and obtain the first order differential system in the radial variable:

$$\frac{\mathrm{d}Y}{\mathrm{d}r} = MY\,, \quad \text{with} \quad M = \begin{pmatrix} C'/C & -i\omega^2 + 2i\lambda\Phi/C \\ -i\Gamma & \Delta \end{pmatrix}\,. \tag{6.20}$$

This means that one can eliminate the function $\Psi$ from the differential system (6.13) by a simple redefinition of the time coordinate. We consider this simplified system in the following.

### 6.2.1. Using a specific transfer matrix

In this section, we recover the Schrödinger-like equation obtained for axial perturbation of cubic DHOST theories in eq. (6.23) using a specific choice of transfer matrix $P$, similarly to what was done in the case of GR in section 4.4.1. We consider the first-order differential system of eq. (6.20). This system leads to a Schrödinger-like equation of the form (6.23) if one can find a new vector $\hat{Y}$ related to $Y$ by the transformation $Y = \hat{P}\hat{Y}$, where the transfer matrix $\hat{P}$ depends on $r$ but not on $\omega$, leading to a system of the form

$$\frac{\mathrm{d}\hat{Y}}{\mathrm{d}r} = \hat{M}\hat{Y}\,, \quad \text{with} \quad \hat{M}(r) = \frac{1}{n(r)} \begin{pmatrix} 0 & 1 \\ V(r) - \omega^2/c_*^2(r) & 0 \end{pmatrix}\,, \tag{6.21}$$

where $V$ and $c_*$ are functions of $r$ and $n$ is related to $r$ by

$$\frac{\mathrm{d}r}{\mathrm{d}r_*} = n(r)\,, \tag{6.22}$$

with $n(r)$ is an arbitrary (monotonic) function at this stage and $r_*$ a new radial coordinate called the *tortoise coordinate*. This system implies that the variable

2. This change of coordinate is well-defined only in the domain where $\Psi(r)$ is integrable.

$\hat{Y}_1$ verifies the following Schrödinger-like equation in the coordinate system $(t_*, r_*)$:

$$\frac{\mathrm{d}^2\hat{Y}_1}{\mathrm{d}r_*^2} + \left[\frac{\omega^2}{c_*^2(r)} - V(r)\right]\hat{Y}_1 = 0\,. \tag{6.23}$$

Let us find such a transfer matrix. Using similar notations as in eq. (6.19), we can decompose $\hat{M}$ as

$$\hat{M}(r) = \hat{M}_{[0]}(r) + \omega^2\hat{M}_{[2]}(r)\,, \tag{6.24}$$

where the invidual matrices can be read off from eq. (6.21) and are related to the matrices in eq. (6.19) by

$$\hat{M}_{[2]} = \hat{P}^{-1}M_{[2]}\hat{P}\,,\quad \hat{M}_{[0]} = \hat{P}^{-1}M_{[0]}\hat{P} - \hat{P}^{-1}\hat{P}'\,, \tag{6.25}$$

where $\hat{P}'$ denotes the derivative of $\hat{P}$ with respect to $r$. Let us consider the first equation in eq. (6.25): one can easily check that the most general transfer matrix $\hat{P}$ that satisfies this equation, given eq. (6.20), is

$$\hat{P} = \begin{pmatrix} y & z \\ x & 0 \end{pmatrix}, \tag{6.26}$$

where $x$, $y$ and $z$ are arbitrary functions such that neither $x$ nor $z$ are zero.

These can be determined by requesting that the initial matrix $M_{[0]}$ is transformed into the requested form $\hat{M}_{[0]}$. Using the second transformation relation in (6.25), one obtains

$$\begin{aligned} &x' - x\Delta + iy\Gamma = 0\,,\\ &x + inz\Gamma = 0\,,\\ &\frac{z'}{z} - \frac{C'}{C} - i\Gamma\frac{y}{x} = 0\,,\\ &x^2\frac{2i\lambda\Phi}{C} + xy\left(\frac{C'}{C} - \Delta\right) + y^2\left(i\Gamma + \left(\frac{x}{y}\right)'\right) = \frac{xz}{n}V\,. \end{aligned} \tag{6.27}$$

The first and second equations can be solved by

$$x = -inz\Gamma \quad \text{and} \quad y = i\frac{x' - x\Delta}{\Gamma}\,. \tag{6.28}$$

The third equation then becomes

$$2\frac{z'}{z} = -\frac{n'}{n} + \Delta + \frac{C'}{C} - \frac{\Gamma'}{\Gamma}\,, \tag{6.29}$$

and can be solved using the expression of $\Delta$ given in eq. (6.14):

$$z(r) = z_0\sqrt{\frac{C}{n\mathscr{F}\Gamma}\sqrt{\frac{A}{B}}}\,, \tag{6.30}$$

where $z_0$ is a constant.

The resolution of eq. (6.27) then implies that the potential $V$ apearing in eq. (6.23) is of the form

$$V = 2n^2\lambda\frac{\Gamma\Phi}{C} + n^2 V_0\,, \tag{6.31}$$

with

$$V_0 = \frac{1}{4}\Big[\Delta^2 + 2\Delta' - 2\Delta\left(\frac{\Gamma'}{\Gamma} + \frac{C'}{C}\right) + 2\frac{\Gamma' C'}{\Gamma C} + 3\left(\frac{\Gamma'}{\Gamma}\right)^2 + \left(\frac{n'}{n}\right)^2 + 3\left(\frac{C'}{C}\right)^2 - 2\left(\frac{\Gamma''}{\Gamma} + \frac{n''}{n} + \frac{C''}{C}\right)\Big]\,. \tag{6.32}$$

One can also obtain the expression of the tortoise coordinate $r_*$ and the radial velocity $c_*$ associated to the coordinate system $(t_*, r_*)$ by looking at the expression of $\widehat{M}_{[2]}$:

$$\frac{\mathrm{d}r_*}{\mathrm{d}r} = \frac{1}{n(r)} \quad \text{and} \quad n^2\Gamma c_*^2 = 1\,. \tag{6.33}$$

It is interesting to consider a coordinate system where the mode propagates at speed $c_* = 1$. In that case, the free function $n$ is fixed by the relation $n^2\Gamma = 1$ and the expression of $V$ simplifies slightly.

By using the expressions of $\Gamma$ and $\Delta$ given for the Schwarzschild solution in eq. (6.15), one finds that the transfer matrix $\hat{P}$ in this case is given by

$$\hat{P} = z_0 \begin{pmatrix} 1 - \mu/r & 1 \\ -ir^2/(r-\mu)^2 & 0 \end{pmatrix}, \tag{6.34}$$

with $z_0$ a constant. One recovers the transfer matrix given in eq. (4.39) with the choice $z_0 = 1$.

As a conclusion, we see that it is possible to recover the Schrödinger equation for axial perturbations using a specific transfer matrix $\hat{P}$ that generalises the choice of transfer matrix used in the case of GR in chapter 4.

### 6.2.2. From the resolution of constraints

We can also notice that the two equations in (6.20) have different interpretations. The equation for $Y_1$ is clearly a dynamical equation of motion as it involves second time derivatives of the variables:

$$\frac{\mathrm{d}Y_1}{\mathrm{d}r} = \frac{C'}{C}Y_1 + i\Big(\frac{2\lambda\Phi}{C} - \omega^2\Big)Y_2\,, \tag{6.35}$$

while the second equation, which does not involve any time derivative, is in fact a constraint in the sense of Hamiltonian dynamics:

$$\frac{\mathrm{d}Y_2}{\mathrm{d}r} = -i\Gamma Y_1 + \Delta Y_2\,. \tag{6.36}$$

This second equation allows us to eliminate the variable $Y_1$ by expressing it in terms of $Y_2$ and its radial derivative. Hence, one obtains a second order (in both variables $t_*$ and $r$) partial differential equation satisfied by the variable $Y_2$ which is given, after a direct calculation, by

$$\frac{\mathrm{d}^2 Y_2}{\mathrm{d}r^2} - \left(\Delta + \frac{\Gamma'}{\Gamma} + \frac{C'}{C}\right)\frac{\mathrm{d}Y_2}{\mathrm{d}r} + \left[\Gamma\left(\omega^2 - 2\lambda\frac{\Phi}{C}\right) + \Delta\left(\frac{\Gamma'}{\Gamma} + \frac{C'}{C}\right)\right] Y_2 = 0\,. \tag{6.37}$$

To extract a Schrödinger-like equation from this, we proceed in two more steps. First, we make a change of radial coordinate by introducing the tortoise coordinate defined in eq. (6.22). Second, we normalise the variable $Y_2$ by introducing the variable $\mathscr{Y}$ defined by

$$Y_2(r) = N(r)\,\mathscr{Y}(r)\,. \tag{6.38}$$

We choose the normalisation function $N(r)$ such that the term proportional to the first radial derivative $\mathrm{d}\mathscr{Y}/\mathrm{d}r_*$ in the equation of motion disappears. An immediate calculation shows that this condition leads to the relation

$$2\frac{N'}{N} = \frac{n'}{n} + \Delta + \frac{\Gamma'}{\Gamma} + \frac{C'}{C}\,. \tag{6.39}$$

As announced, $\mathscr{Y}$ then satisfies a Schrödinger-like equation of the form given in eq. (6.23).

### 6.2.3. Effect of a change of coordinates

Up to now, we have described the background metric in the static coordinates $(t,r)$ and showed that axial perturbations satisfy a Schrödinger-like equation in the coordinates $(t_*,r_*)$ defined in eqs. (6.18) and (6.22). We now want to check that this result is independent of the original choice of coordinates.

To this end, consider a general change of coordinates $(t,r) \longrightarrow (t_g,r_g)$ defined by

$$\mathrm{d}t_g = b_4(t,r)\,\mathrm{d}t + b_1(t,r)\,\mathrm{d}r\,, \quad \mathrm{d}r_g = b_3(t,r)\,\mathrm{d}t + b_2(t,r)\,\mathrm{d}r\,, \tag{6.40}$$

where $b_1$, $b_2$, $b_3$ and $b_4$ are arbitrary functions at this stage. We restrict ourselves to “static” changes of coordinates where all these functions do not depend on $t$. In that case, by virtue of the Schwarz theorem on cross derivatives, $b_4$ and $b_3$ must be constant and, without loss of generality, we can fix $b_4 = 1$. One then obtains

$$\mathrm{d}t_g = \mathrm{d}t + b_1(r)\,\mathrm{d}r\,, \qquad \mathrm{d}r_g = b_3\,\mathrm{d}t + b_2(r)\,\mathrm{d}r\,. \tag{6.41}$$

After a direct calculation, we show that the Schrödinger operator in the coordinate system $(t_*,r_*)$ transforms as

$$\frac{1}{n^2}\left(\frac{\partial^2}{\partial r_*^2} - \frac{1}{c_*^2}\frac{\partial^2}{\partial t_*^2}\right) = \left(\tilde{b}_2^2 - \frac{b_3^2}{n^2 c_*^2}\right)\frac{\partial^2}{\partial r_g^2} + \left(\tilde{b}_1^2 - \frac{1}{n^2 c_*^2}\right)\frac{\partial^2}{\partial t_g^2}$$

$$+ 2\left(\tilde{b}_1\tilde{b}_2 - \frac{b_3}{n^2 c_*^2}\right)\frac{\partial^2}{\partial t_g \partial r_g}$$
$$+ \left(\tilde{b}_1' + \frac{n'}{n}\tilde{b}_1\right)\frac{\partial}{\partial t_g} + \left(\tilde{b}_2' + \frac{n'}{n}\tilde{b}_2\right)\frac{\partial}{\partial r_g}, \tag{6.42}$$

where we introduced

$$\tilde{b}_2 = b_2 + b_3\Psi \quad \text{and} \quad \tilde{b}_1 = b_1 + \Psi\,. \tag{6.43}$$

As expected, in general, the equation in the new coordinate system is no longer a Schrödinger-like equation, which implies that the wave propagating towards infinity and the wave propagating from infinity are not travelling at the same speed.

The necessary and sufficient conditions on $b_1$, $b_2$ and $b_3$ in order to obtain a Schrödinger-like equation for the perturbation in the $(t_g, r_g)$ coordinate system are

$$\tilde{b}_1 n = C_1\,, \qquad \tilde{b}_2 n = C_2\,, \qquad n^2\tilde{b}_1\tilde{b}_2 = \frac{b_3}{c_*^2}\,, \tag{6.44}$$

where $C_1$ and $C_2$ are constants. Let us solve these conditions.

First, we consider the generic case where $c_*$ is not a constant and depends on $r$. It is then immediate to see that the last condition in eq. (6.44) implies $b_3 = 0$ which in turn implies $C_1 = 0$ or $C_2 = 0$.

- The case $C_2 = 0$ leads to $b_2 = 0$ and then the change of coordinates (6.41) is singular.
- The case $C_1 = 0$ leads to $\mathrm{d}t_g = \mathrm{d}t_*$ and $\mathrm{d}r_g = \mathrm{d}r_*$ which means that we recover the $(t_*, r_*)$ coordinate system defined from the original $(t, r)$ coordinate system by

  $$\mathrm{d}t_* = \mathrm{d}t - \Psi(r)\,\mathrm{d}r \quad \text{and} \quad \mathrm{d}r_* = \frac{\mathrm{d}r}{n(r)}\,, \tag{6.45}$$

  where $n(r)$ is an arbitrary function here.

As a consequence, when $c_*$ is not a constant, the perturbation satisfies a Schrödinger-like equation in the coordinate system $(t_*, r_*)$ only.

Second, we assume that $c_*$ is a constant and without loss of generality we fix $c_* = 1$. In that case, it is straightforward to show that the new coordinate system $(t_g, r_g)$ satisfies

$$\mathrm{d}t_g = \mathrm{d}t_* + C_1\,\mathrm{d}r_*\,, \qquad \mathrm{d}r_g = b_3\left(\mathrm{d}t_* + \frac{\mathrm{d}r_*}{C_1}\right), \tag{6.46}$$

which means that $t_g$ and $r_g$ are both constant linear combinations of $t_*$ and $r_*$.

On can go further and ask the question whether it is possible or not to recover a Schrödinger-like equation in the coordinate system $(t_g, r_g)$ not for the perturbation itself $\psi$ but for a renormalized perturbation, say $\tilde{\psi}$, such that $\psi = N\tilde{\psi}$. The renormalisation factor $N$ should depend on $r_g$ only in order to avoid a $t_g$ dependency in the potential.

After a direct calculation, we show that $\tilde{\psi}$ satisfies a Schrödinger-like equation if the first and the third conditions in eq. (6.44) are satisfied and then, to get rid of the term proportional to $\frac{\partial}{\partial r_g}$, one obtains an equation for $N$:

$$\frac{1}{N}\frac{\partial N}{\partial r_g} = Q\,, \quad Q(r) = -\frac{n^2}{2}\frac{\tilde{b}_2' + \tilde{b}_2 n'/n}{n^2\tilde{b}_2^2 - b_3^2/c_*^2}\,. \tag{6.47}$$

As $N$ is supposed to depend on $r_g$ only, the function $Q$ must depend on $r_g$ only as well. However, $Q$ is a function of $r$ and the fact that

$$b_3\frac{\partial Q}{\partial r} = (b_3 b_1 - b_2)\frac{\partial Q}{\partial t_g} \tag{6.48}$$

implies that $Q$ must be vanishing, which leads to the remaining second condition in (6.44). Hence, we return to the previous analysis.

As a conclusion, any coordinates $t_g$ and $r_g$ such that the perturbation $\psi$ (or a renormalized perturbation) satisfies a Schrödinger-like equation are constant linear combinations of $(t_*, r_*)$. The consequence is that we can rigorously define a speed of propagation and potential for the perturbation only in coordinates which are constant linear combinations of $(t_*, r_*)$.

However, it is possible to show that one can always define a speed in the large frequency limit. The necessary condition for this to be the case is that the second order cross derivative in eq. (6.42) vanishes, i.e. we only get the last condition in eq. (6.44):

$$n^2\tilde{b}_1\tilde{b}_2 = \frac{b_3}{c_*^2}\,. \tag{6.49}$$

In that regime, the new coordinate system is given by

$$\mathrm{d}t_g = \mathrm{d}t_* + n\tilde{b}_1\,\mathrm{d}r_*\,, \quad \mathrm{d}r_g = b_3\left(\mathrm{d}t_* + \frac{1}{n\tilde{b}_1 c_*^2}\mathrm{d}r_*\right)\,, \tag{6.50}$$

where $\tilde{b}_1$ is now a free function. Thus, it is possible to define a speed of propagation $c_g$ for the perturbation in the new coordinate system and one can easily compute it:

$$c_g^2 = \frac{b_3^2}{n^2 c_*^2 \tilde{b}_1^2} = n^2\tilde{b}_2^2 c_*^2\,. \tag{6.51}$$

Obviously, $c_g$ and $c_*$ are different (as they are associated with different coordinate systems), but the stability condition $c_g^2 > 0$ is equivalent to $c_*^2 > 0$. Hence, we obtain the same stability condition in any coordinate system in the large frequency limit as one would have expected.

## 6.3. Application to several BH solutions

The simple form of eq. (6.23) and the expressions of the potential and the speed given in eqs. (6.31) and (6.33) for any cubic DHOST theory allow us to study the axial perturbations of all the BH solutions we presented in chapter 2. In this section, we compute the speed and potential for each of these solutions.

### 6.3.1. BCL solution

Let us first consider the BCL solution described in section 2.2. In this case, the new coordinate $r_*$ is given by

$$r_* = \int \mathrm{d}r \, \frac{r^2}{(r-r_+)(r+r_-)} = r + \frac{r_+^2 \ln(r/r_+ - 1) - r_-^2 \ln(r/r_- + 1)}{r_+ + r_-} \,. \tag{6.52}$$

We find that the functions $\Psi$, $\Phi$, $\Gamma$ and $\Delta$ defined in eq. (6.14) entering in the coefficients of the differential system eq. (6.13) read

$$\begin{aligned} &\Psi = 0\,, \qquad \Phi = A\,, \qquad \Gamma = \frac{F}{f_0 A^2} = \frac{r^2(r^2+2r_+r_-)}{(r-r_+)^2(r+r_-)^2}\,, \\ &\Delta = -\frac{A'}{A} = -\frac{r_+}{r(r-r_+)} + \frac{r_-}{r(r+r_-)}\,, \end{aligned} \tag{6.53}$$

and that we have $\mathscr{F} = f_0 A$.

Furthermore, the potential (6.31) takes the form

$$V(r) = A(r) \frac{V_0 + V_1(\mu/r) + V_2(\mu/r)^2 + V_4(\mu/r)^4 + V_6(\mu/r)^6}{2r^2(1+\xi(\mu/r)^2)^2}\,, \tag{6.54}$$

with the coefficients

$$V_0 = 4(\lambda{+}1)\,, \quad V_1 = -6\,, \quad V_2 = 6(2\lambda{-}1)\xi\,, \quad V_4 = (12\lambda{-}1)\xi^2\,, \quad V_6 = 4\lambda\xi^3\,. \tag{6.55}$$

The constant $\xi$ defined in eq. (2.17) parametrises the deviation from GR (corresponding to the limit $r_- = 0$, i.e. $f_1 = 0$).

One notes that one must have $\xi \geq 0$ to prevent a singularity in the potential. When $\xi = 0$, one recovers the standard RW potential for the Schwarzschild geometry given in eq. (4.41).

Potentials for several values of $\xi$ are shown in fig. 6.1, where one can see that the potential is a deformation, parametrised by $\xi$, of the RW potential. At infinity, the behaviour of the potential is very similar to that of the RW potential, with corrections appearing only at second order in $\mu/r$:

$$V(r) = \frac{1}{\mu^2}\left[2(\lambda+1)\frac{\mu^2}{r^2} - (2\lambda+5)\frac{\mu^3}{r^3} + \mathcal{O}\left(\frac{\mu^4}{r^4}\right)\right]. \tag{6.56}$$

By contrast, the leading order behaviour is modified near the horizon:

$$V(r) = \frac{32\mu_\xi\left(\lambda(3\mu_\xi-1)^2 - \mu_\xi(1+\mu_\xi)\right)}{(1+\mu_\xi)^5(3\mu_\xi-1)\,\mu^3}(r-r_+) + \mathcal{O}((r-r_+)^2), \quad \mu_\xi \equiv \sqrt{1+2\xi}, \tag{6.57}$$

where we have used $r_\pm = \mu(1\pm\mu_\xi)/2$. Notice that the height of the potential also depends on the value of $\xi$.

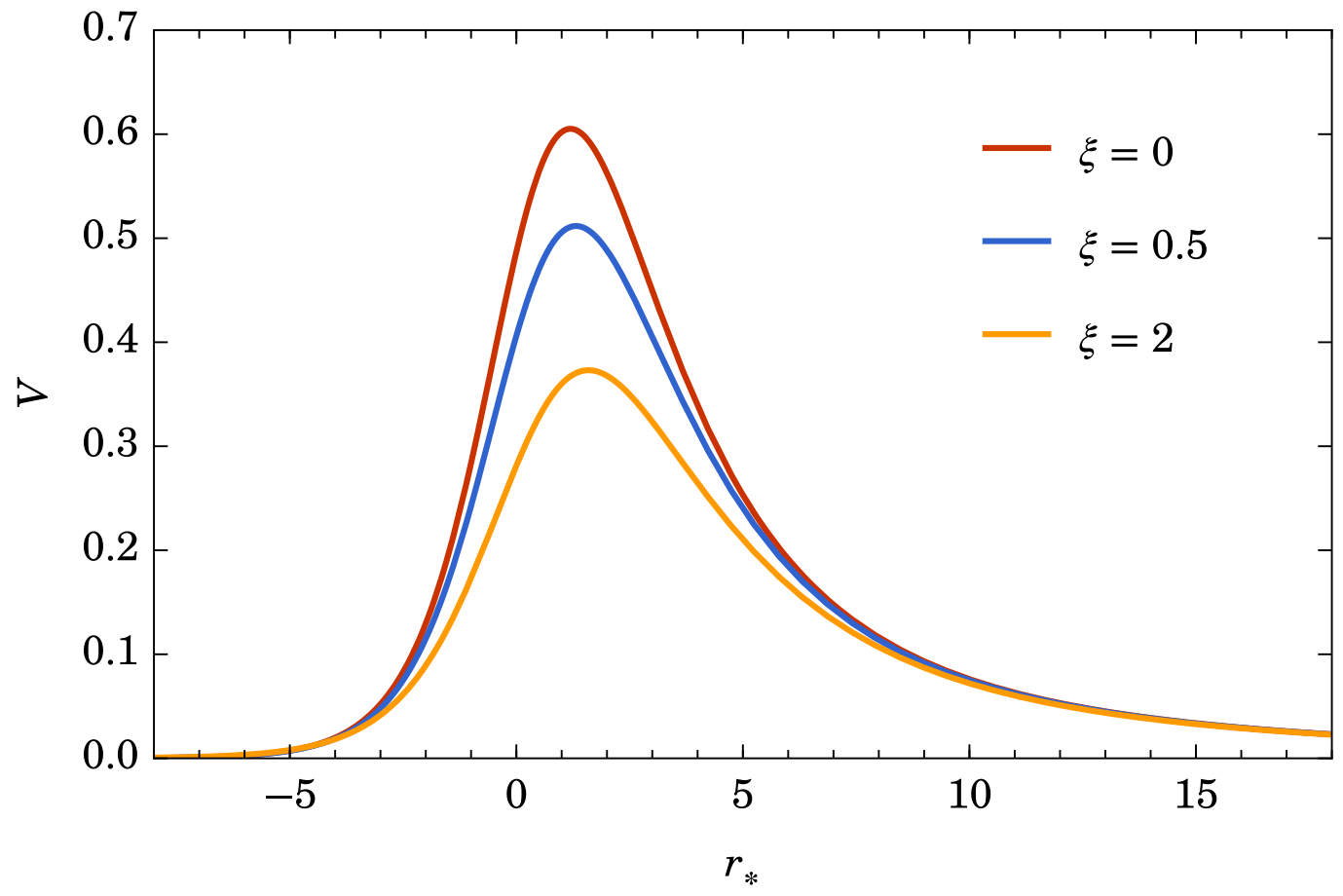

Figure 6.1. – Potential $V(r)$ for the BCL BH for different values of $\xi$ but fixed values of $\mu = 1$ and $\ell = 2$ ($\lambda = 2$).

The propagation speed is given by

$$c(r) = \frac{r}{\sqrt{r^2+\xi\mu^2}}. \tag{6.58}$$

We thus recover the usual value $c = 1$ at spatial infinity (when $r \to +\infty$), but at the horizon we find

$$c(r_+) = \sqrt{\frac{r_+}{r_+ + 2r_-}} = \sqrt{\frac{\mu_\xi+1}{3\mu_\xi-1}} \leq 1. \tag{6.59}$$

### 6.3.2. Stealth solutions

#### 6.3.2.1. Speed and potential

Let us apply the above results to the stealth Schwarzschild solution [3] described in section 2.3. Substituting eqs. (2.33) and (2.34) into eq. (6.14), one finds

$$\Psi = \frac{\zeta\,\mu^{1/2} r^{3/2}}{(r-\mu)\,(r-r_g)}\,, \qquad \Phi = \frac{r-r_g}{(1+\zeta) r}\,, \tag{6.60}$$

$$\Gamma = \frac{(1+\zeta) r^2}{(r-r_g)^2}\,, \qquad \Delta = \frac{1}{r} - \frac{1}{r-r_g}\,, \tag{6.61}$$

where we have introduced the constant parameters

$$\zeta \equiv 2q^2\alpha \geq 0\,, \quad r_g \equiv (1+\zeta)\,\mu\,. \tag{6.62}$$

This dimensionless constant $\zeta$ parametrises the deviation from GR, since one recovers the GR functions given in eq. (6.15) when $\zeta = 0$. The radius $r_g$, which differs from $\mu$ when $\zeta \neq 0$, appears as an extra pole in the above functions, in addition to $\mu$ and 0.

From eqs. (6.31) and (6.33), one can compute the potential $V(r)$ and the propagation speed $c(r)$ that appear in the Schrödinger-like equation. These quantities depend on the choice of the radial coordinate; if one adopts the usual Schwarzschild tortoise coordinate, defined by

$$r_* = \int \mathrm{d}r\,\frac{r}{r-\mu} = r + \mu \ln(r/\mu - 1)\,, \tag{6.63}$$

corresponding to the choice $n = A(r) = 1 - \mu/r$, the potential takes the form

$$V(r) = \frac{V_0 + V_1\,(\mu/r) + V_2\,(\mu/r)^2 + V_3\,(\mu/r)^3 + V_4\,(\mu/r)^4}{(r-r_g)^2}\,, \tag{6.64}$$

with

$$\begin{aligned} &V_0 = 2(\lambda+1)\,, \quad V_1 = -2(\lambda+3)\zeta - 6\lambda - 9\,,\\ &V_2 = (15\zeta + 16\lambda + 70)\zeta/4 + 6\lambda + 15\,,\\ &V_3 = -(1+\zeta)\,(13\zeta/2 + 2\lambda + 11)\,, \quad V_4 = 3(1+\zeta)^2\,, \end{aligned} \tag{6.65}$$

and the propagation speed is given by the expression

$$c(r) = \frac{r-r_g}{\sqrt{1+\zeta}\,(r-\mu)}\,, \tag{6.66}$$

3. We present here the stealth Schwarzschild as a solution of a Horndeski theory. However, since only the DHOST function $A_1$ appears in axial perturbations of quadratic DHOST theories, one could see the stealth solution considered here as a solution of any quadratic DHOST theory and define $\alpha$ via $2\alpha = A_{1X}$. One should note that this is only true for axial perturbations and that this generalisation breaks down in the case of polar perturbations treated in chapter 7.

where one must take $\zeta > -1$ in order to have $c^2 > 0$.

Another possibility is to choose the radial coordinate such that the propagation speed is $c = 1$, i.e.

$$r_* = \int \mathrm{d}r\,\sqrt{\Gamma} = \sqrt{1+\zeta}\left[\, r + r_g \ln\!\left(r/r_g - 1\right)\right] , \tag{6.67}$$

which is very similar to the usual tortoise coordinate, with $r_g$ instead of $\mu$ and a global rescaling. In this case, the potential becomes

$$V_{c=1}(r) = \left(1 - \frac{r_g}{r}\right)\frac{2(\lambda+1)r - 3r_g}{(1+\zeta)\,r^3} , \tag{6.68}$$

which is, quite remarkably, identical to the standard RW potential of eq. (4.41), with $r_g$ instead of $\mu$, up to a global rescaling. One can note that $\mu$ has completely disappeared from the equation of motion and $r_g$ seems to play the role of the horizon that is effectively "seen" by the axial metric perturbations. The same result was obtained recently in [210] by analysing the effective metric that appears in the equation of motion for the axial perturbations. This is recovered in section 6.5 using a different formalism.

#### 6.3.2.2. Case of a change of coordinates

In [214], the authors studied the perturbations of stealth BHs using a different coordinate system than the one used in this manuscript. They introduced the Lemaître coordinates $(\tau, \rho)$ which are related to the original ones by a transformation of the form of eq. (6.40) with $t_g = \tau$, $r_g = \rho$ and

$$b_1 = \frac{\sqrt{r\mu}}{r-\mu}, \quad b_2 = \frac{r^2}{\sqrt{r\mu}(r-\mu)}, \quad b_3 = 1 . \tag{6.69}$$

The authors computed the quadratic Lagrangian for axial perturbations. The associated equations of motion do not reproduce a Schrödinger-like equation in their study; this is consistent with the results of section 6.2.3 since the first two conditions in eq. (6.44) are not satisfied. Indeed, a very quick calculation shows that

$$\tilde{b}_1 = b_1 + \Psi = (1+\zeta)\frac{\sqrt{r_s r}}{r - r_g} \quad \text{and} \quad \tilde{b}_2 = b_2 + b_3\Psi = \frac{r}{r_s}\frac{\sqrt{r_s r}}{r - r_g} , \tag{6.70}$$

which means that $\tilde{b}_1/\tilde{b}_2$ is not a constant.

The perturbation $\psi$ instead satisfies an equation of the form

$$-\frac{\partial}{\partial\tau}\left(s_1\frac{\partial\psi}{\partial\tau}\right) + \frac{\partial}{\partial\rho}\left(s_2\frac{\partial\psi}{\partial\rho}\right) + W\psi = 0 , \tag{6.71}$$

where

$$s_1 = \frac{(1+\zeta)^2 r^6}{\sqrt{\mu/r}}\,, \quad s_2 = \frac{(1+\zeta) r^6}{(\mu/r)^{3/2}}\,, \tag{6.72}$$

and the expression of $W$ is not needed but can be found in [214]. Notice that $r$ depends on $\tau$ and $\rho$, which means that the equation is time-dependent.

One can deduce the speed of propagation associated to this equation in the large frequency limit:

$$c_g^2 = \frac{s_2}{s_1} = \frac{r}{\mu(1+\zeta)}\,, \tag{6.73}$$

which is consistent with eq. (6.51),

$$c_g^2 = n^2 c_*^2 \tilde{b}_2^2 = \frac{\tilde{b}_2^2}{b_3}\,. \tag{6.74}$$

We see that the stability conditions $c_g^2 > 0$ and $c_*^2 > 0$ lead to the same inequality $1+\zeta \geq 0$, which could be expected from the discussion under eq. (6.51).

Finally, one should note that the propagation speed given in [214] is actually defined by the relation

$$c_\rho^2 = -\frac{g_{\rho\rho}}{g_{\tau\tau}} c_g^2\,, \tag{6.75}$$

in order to express the propagation speed in normalised units (i.e. in the normalised basis spanned by the vectors $(-g_{\tau\tau})^{1/2}\partial_\tau$ and $(g_{\rho\rho})^{-1/2}\partial_\rho$).

### 6.3.3. EsGB solution

Let us now turn to the study of axial perturbations about the EsGB BH solution. In terms of the dimensionless radial coordinate $z$ (see section 2.4), the functions defined in eq. (6.14) functions read, up to order $\varepsilon^2$,

$$\begin{aligned}
\Gamma &= \frac{1}{(z-1)^2}\left[z^2 + \frac{10z^5+10z^4-100z^3-95z^2-94z+206}{15z^4}\varepsilon^2\right] + \mathcal{O}(\varepsilon^3)\,,\\
\Phi &= (z-1)\left[\frac{1}{z} - \frac{10z^5+10z^4+140z^3-95z^2-94z-214}{30z^7}\varepsilon^2\right] + \mathcal{O}(\varepsilon^3)\,,\\
\Psi &= 0\,,\\
\Delta &= \frac{1}{z-z^2} + \frac{-5z^5-10z^4-30z^3+190z^2+235z+282}{15z^7}\varepsilon^2 + \mathcal{O}(\varepsilon^3)\,.
\end{aligned} \tag{6.76}$$

When $\varepsilon$ goes to zero, one recovers the standard Schwarzschild expressions (see eq. (6.15)).

By substituting the above expressions into eqs. (6.31) and (6.33) and choosing $n(z) = A(z)$, one can then obtain (up to order $\varepsilon^2$) the propagation speed from

$$c^2 = 1 + 4\varepsilon^2\left(-\frac{4}{z^6} + \frac{1}{z^5} + \frac{1}{z^4} + \frac{2}{z^3}\right) + \mathcal{O}(\varepsilon^3)\,, \tag{6.77}$$

and the potential

$$V = \left(1 - \frac{1}{z}\right)\Bigg[\frac{-3 + 2z(1+\lambda)}{z^3} + \varepsilon^2\Big(\frac{2542}{5}\frac{1}{z^9} + \frac{1}{15}(-8009 + 712\lambda)\frac{1}{z^8} - \frac{2}{15}(-29+\lambda)\frac{1}{z^7} + \frac{2}{3}(-47+\lambda)\frac{1}{z^6} + (70 - 24\lambda)\frac{1}{z^5} + \frac{4}{3}(4+\lambda)\frac{1}{z^4} - \frac{1}{3}(5+2\lambda)\frac{1}{z^3}\Big)\Bigg] + \mathcal{O}(\varepsilon^3)\,. \quad (6.78)$$

These quantities have been illustrated in fig. 6.2 for some values of $\varepsilon$. Note that the potential is plotted as a function of the "tortoise" coordinate $z_*$, defined similarly to $r_*$ in eq. (6.22) with $n = A$:

$$\frac{\mathrm{d}z_*}{\mathrm{d}z} = \frac{1}{n(z)} = \frac{1}{A(z)}\,. \quad (6.79)$$

Substituting the expression of $A(z)$, obtained from eqs. (2.40), (2.48) and (2.53),

$$A(z) = \left(1 - \frac{1}{z}\right) - \varepsilon^2\left(\frac{17}{15z^7} + \frac{1}{30z^5} - \frac{11}{6z^4} + \frac{1}{3z^3} + \frac{1}{3z}\right) + \mathcal{O}(\varepsilon^3)\,, \quad (6.80)$$

one gets

$$z_* = z - \varepsilon^2\left(\frac{17}{60z^4} + \frac{34}{45z^3} + \frac{103}{60z^2} + \frac{83}{30z} - \frac{73}{30}\ln(z)\right) + \ln(z-1)\left(1 - \frac{21}{10}\varepsilon^2\right) + \mathcal{O}(\varepsilon^3)\,. \quad (6.81)$$

This implies, in particular, the asymptotic behaviours at spatial infinity

$$z_* \approx z + \ln(z)\left(1 + \frac{\varepsilon^2}{3}\right) + \mathcal{O}(\varepsilon^3)\,, \quad (6.82)$$

and at the horizon

$$z_* \approx \ln(z-1)\left(1 - \frac{21}{10}\varepsilon^2\right) + \mathcal{O}(\varepsilon^3)\,, \quad (6.83)$$

where the symbol $\approx$ means equality up to sub-dominant terms in the $z$ variable [4].

Noting that since $c$ tends to 1 and $V$ vanishes both at the horizon and at spatial infinity, the asymptotic behaviour of (6.23) is simply given by

$$\frac{\mathrm{d}^2\hat{Y}_1}{\mathrm{d}z_*^2} + \Omega^2\hat{Y}_1 \approx 0\,, \quad (6.85)$$

4. More precisely, given two functions $f(z)$ and $g(z)$, we say that $f(z) \approx g(z)$ at $z_0$ (which can be here $z_0 = \infty$ or $z_0 = 1$) when

$$f(z) \approx g(z) \quad \text{at}\, z \to z_0\, \text{means} \quad \lim_{z\to z_0}\frac{f(z) - g(z)}{f(z)} = 0\,. \quad (6.84)$$

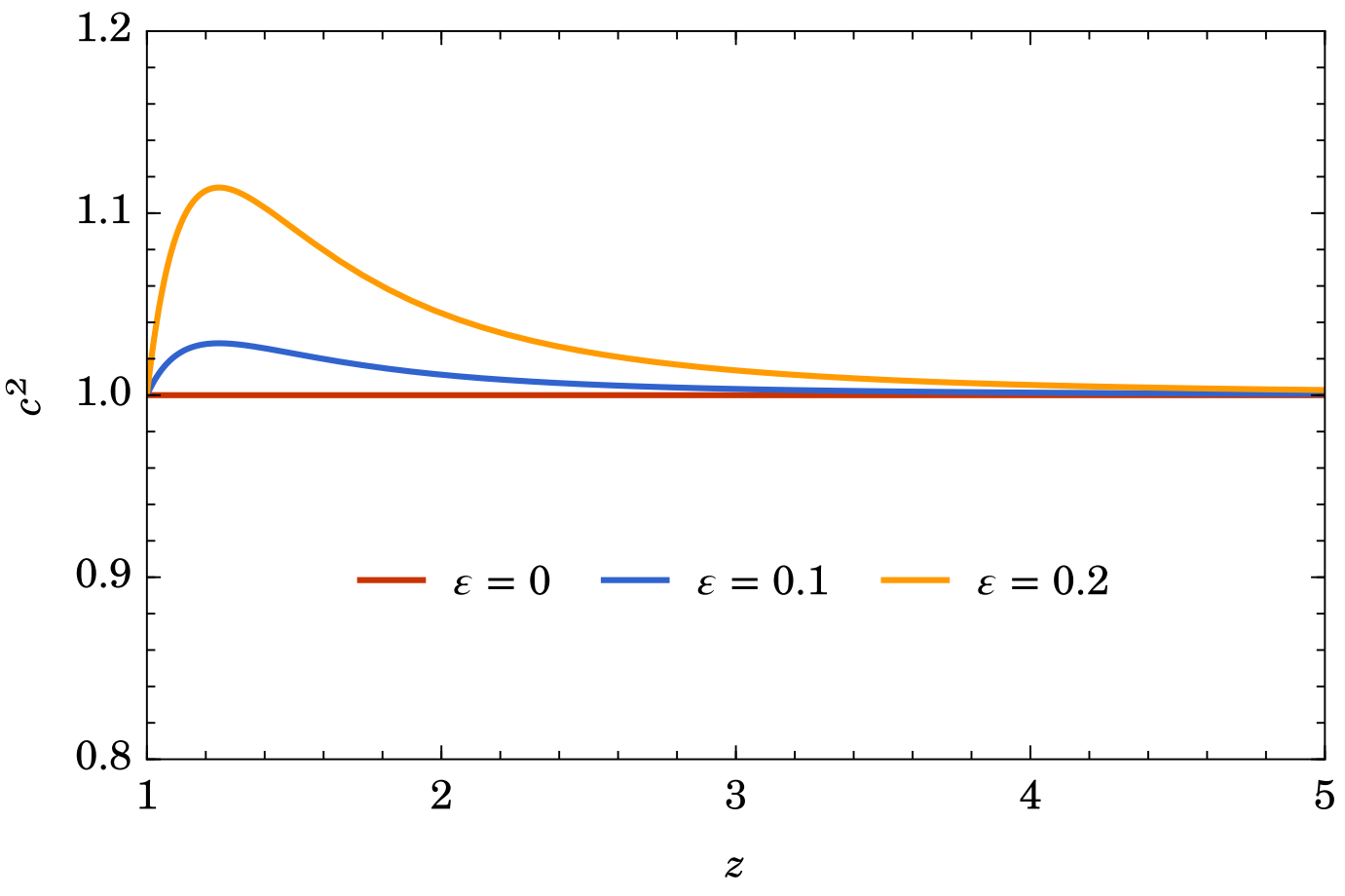

(a) Squared speed

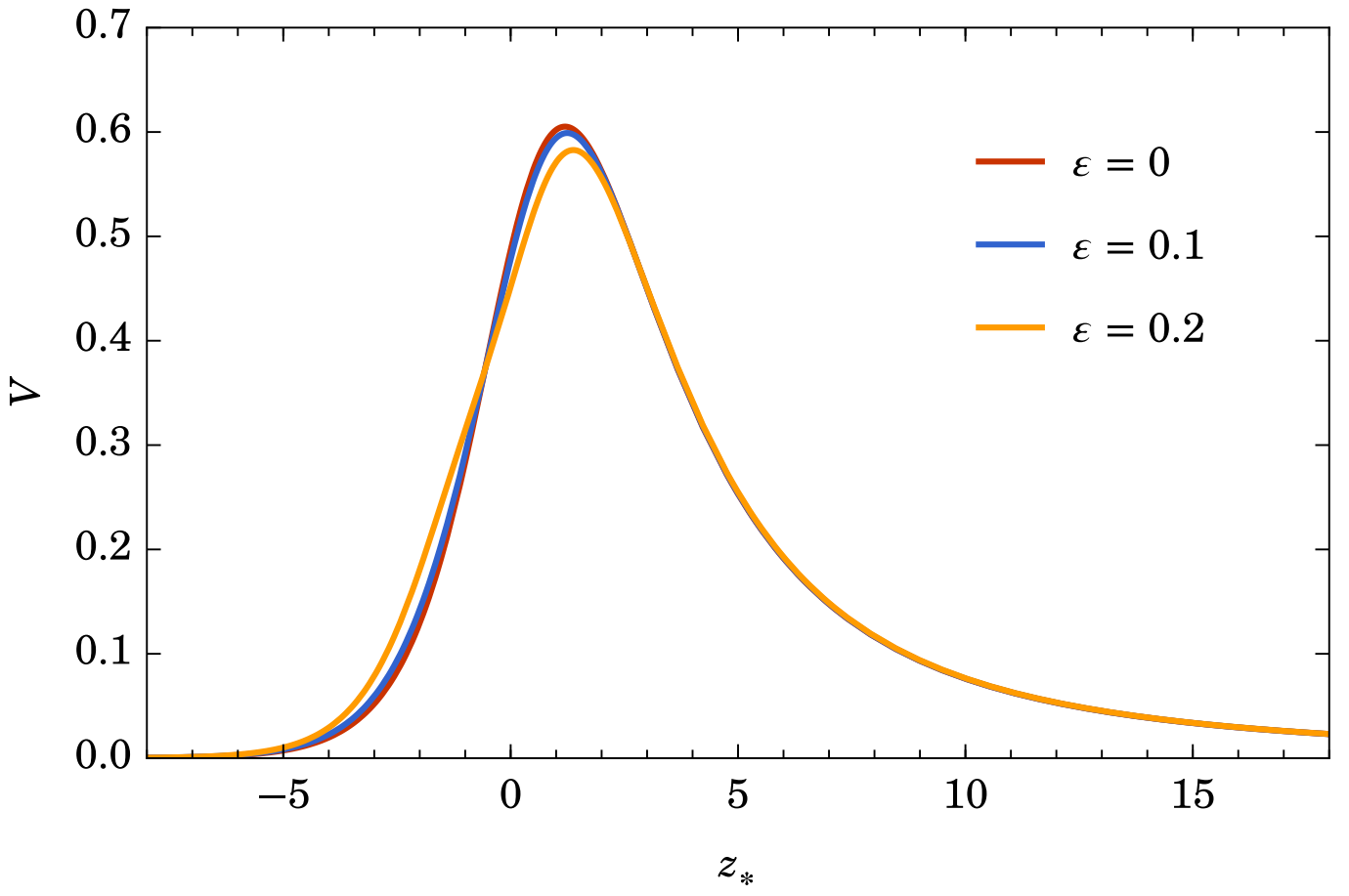

(b) Potential

Figure 6.2. – Plot of the squared propagation speed $c^2$ as a function of $z$ and the potential $V$ as a function of $z_*$ for $\lambda = 2$. Note that the coordinate $z_*$ is defined up to a constant which, in this plot, differs from the choice in (6.81). In the figure, the constant is chosen such that $z_* = 0$ when $z = 1 + W(e^{-1})$, where $W$ is the Lambert function. This corresponds to the definition $z_* = z + \ln(z-1)$ in the GR case ($\varepsilon = 0$).

where we have rescaled the frequency according to

$$\Omega = \omega r_h \,. \tag{6.86}$$

As a consequence, at spatial infinity, using eq. (6.82), the asymptotic solution is

$$\hat{Y}_1 \approx \mathscr{A}_\infty e^{+i\Omega z} z^{i\Omega(1+\varepsilon^2/3)} + \mathscr{B}_\infty e^{-i\Omega z} z^{-i\Omega(1+\varepsilon^2/3)} \,, \tag{6.87}$$

while the solution near the horizon takes the form,

$$\hat{Y}_1 \approx \mathscr{A}_{\text{hor}}(z-1)^{+i\Omega(1-21\varepsilon^2/10)} + \mathscr{B}_{\text{hor}}(z-1)^{-i\Omega(1-21\varepsilon^2/10)}\,, \tag{6.88}$$

where we have used (6.83) when we replace $z_*$ by its expression in terms of $z$. Finally, the constants $\mathscr{A}_\infty$, $\mathscr{B}_\infty$, $\mathscr{A}_{\text{hor}}$ and $\mathscr{B}_{\text{hor}}$ can be fixed or partially fixed by appropriate boundary conditions.

### 6.3.4. 4dEGB solution

Let us now turn to the study of the 4dEGB solution. Substituting eqs. (2.55), (2.58), (2.62) and (2.63) into eq. (6.14) and rescaling all dimensionful quantities by the appropriate powers of $r_h$ to make them dimensionless (or, equivalently, working in units where $r_h = 1$), one gets the following expressions for $\mathscr{F}$, $\Gamma$, $\Phi$ and $\Delta$ (since $q = 0$, $\Psi$ is zero here) :

$$\mathscr{F} = \frac{f^2}{z^2}\left[z^2 + 2\beta(\sigma+f)(\sigma+f-2zf')\right], \tag{6.89}$$

$$\Gamma = \frac{1}{\mathscr{F}f^2z^2}\left[z^2 - 2\beta(1-f^2) - 4z\beta ff'\right] = \frac{z^4 - 2\beta(1+\beta)z}{\mathscr{F}f^2z^2[z^2+2\beta(1-f^2)]}\,, \tag{6.90}$$

$$\Phi = \frac{\mathscr{F}z^2}{z^2+2\beta(1-f^2)}\,, \qquad \Delta = -\frac{\mathscr{F}'}{\mathscr{F}}\,, \tag{6.91}$$

where we have used the explicit definition of $f(z)$ and the expression of its derivative

$$f' = \frac{f^2-1}{zf} + \frac{3(1+\beta)}{2f[z^2-2\beta(f^2-1)]} \tag{6.92}$$

to obtain a simplified expression for $\Gamma$. Here, we have kept the parameter $\sigma$ variable: as we can see, it appears in the expression of $\mathscr{F}$ which means that it becomes relevant for the perturbations of the BH solution.

In the sequel, it will be convenient to express the quantities defined in eqs. (6.89) and (6.91) in terms of the following three functions of $z$:

$$\gamma_1 = f\left[z^2 + 2\beta(\sigma+f)(\sigma+f-2zf')\right], \tag{6.93}$$

$$\gamma_2 = z^4 - 2\beta(1+\beta)z\,, \tag{6.94}$$

$$\gamma_3 = z^2 + 2\beta(1-f^2)\,. \tag{6.95}$$

A short calculation then leads to

$$\mathscr{F} = \frac{f\gamma_1}{z^2}\,, \quad \Gamma = \frac{\gamma_2}{f^3\gamma_1\gamma_3} \quad \text{and} \quad \Phi = \frac{f\gamma_1}{\gamma_3}\,. \tag{6.96}$$

When we study the perturbations and their asymptotics, it is important to look at the zeros and the singularities of the expressions (6.96). For this reason, we quickly discuss the zeros of the functions $\gamma_i$. We note that, for $z > 0$, the function $\gamma_3$, explicitly given by

$$\gamma_3 = z^2\sqrt{1 + \frac{4\beta(1+\beta)}{z^3}}\,, \tag{6.97}$$

is strictly positive and the function $\gamma_2$ vanishes at

$$z_2 = [2\beta(1+\beta)]^{1/3}\,. \tag{6.98}$$

This root is only relevant in our analysis if it lies outside the horizon, i.e. when $z_2 > 1$, which is the case if $\beta \geq \beta_c$, with

$$\beta_c \equiv \frac{\sqrt{3}-1}{2} \simeq 0.366\,. \tag{6.99}$$

Hence, when $\beta < \beta_c$, $\gamma_2$ remains strictly positive outside the horizon. Let us note that at the special value $\beta = \beta_c$, the zeros of $f$ and $\gamma_2$ coincide. Finally, the position of the zeros of $\gamma_1$ depends on the sign of $\sigma$. If $\sigma = -1$, then $\sigma + f \leq 0$ and, since $f' \geq 0$, the product $(\sigma+f)(\sigma+f-2zf')$ is always positive, and therefore $\gamma_1 > 0$ outside the horizon. By contrast, if $\sigma = +1$, one finds numerically that $\gamma_1$ has a zero $z_1 > 1$. This is another reason (in addition to the behaviour of the scalar field at infinity discussed below (2.59)) to restrict our analysis to the case $\sigma = -1$.

Let us summarise. When $\beta < \beta_c$ and $\sigma = -1$, the functions $\gamma_i$ do not vanish outside the horizon and then neither of the three functions $\mathscr{F}$, $\Gamma$ and $\Phi$ vanishes or has a pole for $z > 1$. Near the horizon, these functions behave as follows:

$$\mathscr{F} \approx \frac{6\beta(1+\beta)}{1+2\beta}f, \quad \Gamma \approx \frac{(1+2\beta)(1-2\beta-\beta^2)}{6\beta(1+\beta)}\frac{1}{f^3}, \quad \Phi \approx \frac{6\beta(1+\beta)}{1+2\beta}f, \tag{6.100}$$

with

$$f(z) = \sqrt{\frac{1-\beta}{1+2\beta}}\sqrt{z-1} + \mathscr{O}\left((z-1)^{3/2}\right). \tag{6.101}$$

At infinity, the behaviour is much simpler as the three functions in eq. (6.96) are constant and tend to 1.

A natural choice for $n$ is $n(z) = A(z) = f^2(z)$, in which case $z_*$ is the analog of the Schwarzschild tortoise coordinate. With this choice, one finds, according to eq. (6.96),

$$c^2 = \frac{\gamma_1\gamma_3}{f\gamma_2}\,. \tag{6.102}$$

The condition $\beta < \beta_c$ together with the choice $\sigma = -1$ therefore ensures that $c^2 > 0$ everywhere outside the horizon. The potential is given by

$$V(z) = \frac{z^2 A(\kappa_1 + A\kappa_2)}{\gamma_2^2\gamma_3^4}\,, \quad \text{with} \tag{6.103}$$

$$\kappa_1 = 2(\lambda+1)z^{12} - 3(\beta+1)z^{11} - 2\beta(\beta+1)(2\lambda-7)z^9 - 18\beta(\beta+1)^2 z^8 - 24\beta^2(\beta+1)^2(\lambda+1)z^6 + 54\beta^2(\beta+1)^3 z^5 + 4\beta^3(\beta+1)^3(20\lambda-7)z^3 - 12\beta^3(\beta+1)^4 z^2 - 8\beta^4(\beta+1)^4(8\lambda-1)\,,$$

$$\kappa_2 = 30\beta(\beta+1)z^9 + 126\beta^2(\beta+1)^2 z^6 + 108\beta^3(\beta+1)^3 z^3 + 12\beta^4(\beta+1)^4\,.$$

The propagation speed and the potential for $\lambda = 2$ are represented in fig. 6.3 for three different values of $\beta$, satisfying the condition $\beta < \beta_c$. We observe that the propagation speed diverges at the horizon $z = 1$, while the potential vanishes at this point. The potential can be negative in some region for sufficiently large values of $\beta$. It is difficult to study analytically the sign of the potential but one can compute its derivative when $z_* \to -\infty$ and one finds that it remains positive up to some value $\beta_*(\lambda)$. We find that $\beta_*(\lambda = 2) \simeq 0.162917$ numerically and that $\beta_*(\lambda \to \infty) = \beta_c$.

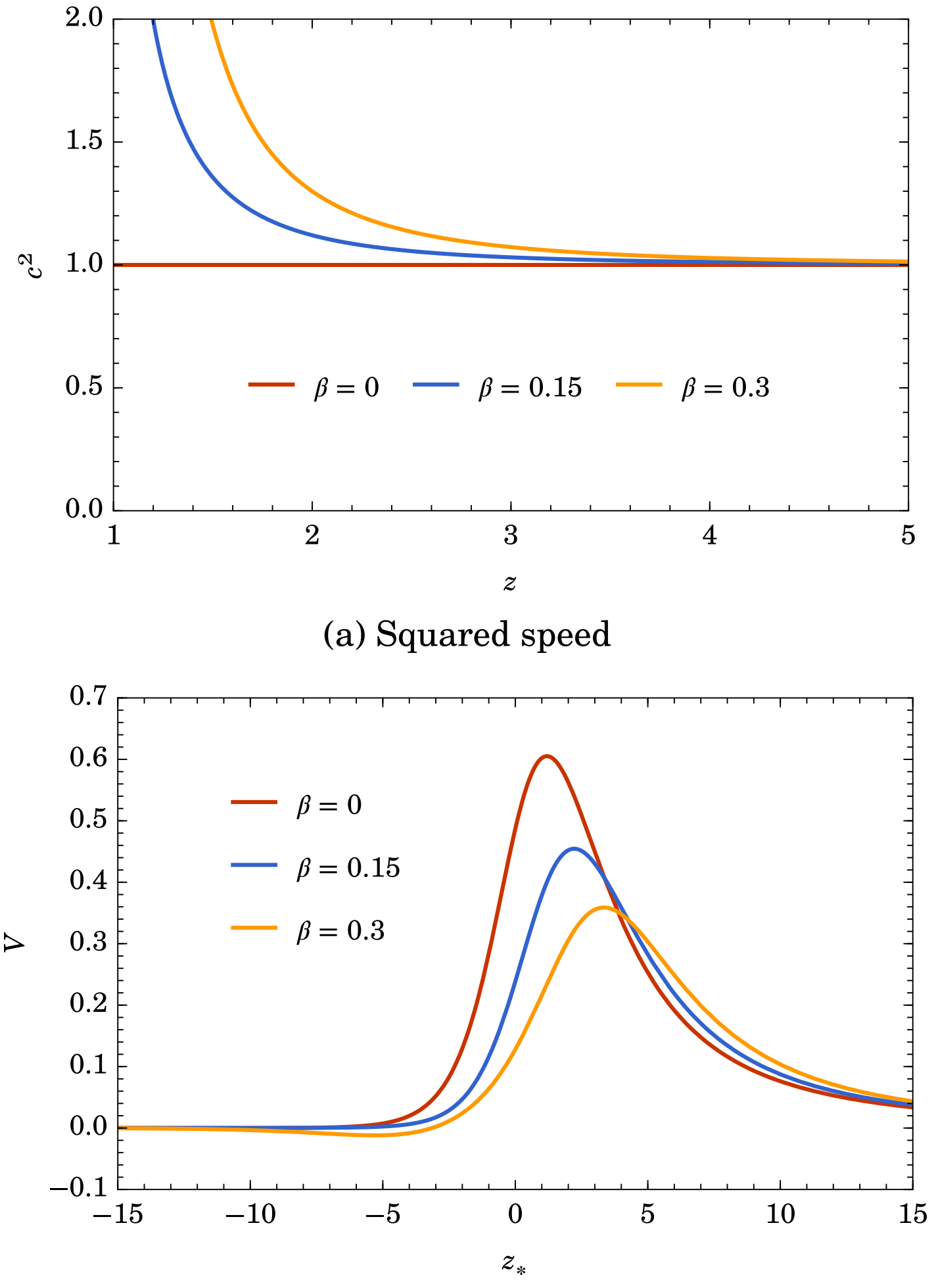

(a) Squared speed

(b) Potential

Figure 6.3. – Plot of the squared speed $c^2$ and the potential $V$ for $\lambda = 2$. We choose the integration constant in the computation of $z_*$ through the same procedure as explained in fig. 6.2.

In terms of the new coordinate $z_*$, the Schrödinger-like equation (6.23) is of the form

$$-\frac{\mathrm{d}^2\hat{Y}_1}{\mathrm{d}z_*^2} + V(z)\chi = w(z)\,\Omega^2\hat{Y}_1\,, \qquad w = f^4\Gamma\,, \tag{6.104}$$

where $\Omega$ is related to $\omega$ similarly to what was done for the EsGB solution in eq. (6.86): $\Omega = \omega r_h$. The left-hand side of this equation can be seen as an operator acting on the space of functions that are square-integrable with respect to the measure $w\,\mathrm{d}z_*$. It is instructive to study the asymptotic behaviour of the solutions of eq. (6.104), near the horizon and at spatial infinity.

Near the horizon ($z \to 1$ or $z_* \to -\infty$), using $\mathrm{d}z_* = \mathrm{d}z/f^2$ and eq. (6.101), one finds

$$z_* \approx \frac{1+2\beta}{1-\beta}\ln(z-1) \iff z-1 \approx e^{\eta z_*}\,, \quad \eta \equiv \frac{1-\beta}{1+2\beta}\,, \tag{6.105}$$

and the asymptotic behaviours for the potential and for $w$ are

$$V(z) \approx C_1(z-1)\,, \qquad w(z) \approx \frac{1-2\beta-2\beta^2}{2\beta\sqrt{(1-\beta)(1+2\beta)}}\sqrt{z-1}\,, \tag{6.106}$$

where $C_1$ is a constant. It is immediate to rewrite these asymptotic expressions in terms of $z_*$, using eq. (6.105).

Near the horizon, for $z_* \to -\infty$, the potential decays faster than the right-hand side of eq. (6.104) so that the differential equation takes the form

$$-\frac{\mathrm{d}^2\hat{Y}_1}{\mathrm{d}z_*^2} + C_1 e^{\eta z_*/2}\hat{Y}_1 \approx 0\,, \tag{6.107}$$

whose solutions are

$$\hat{Y}_1 \approx a_1 I_0\left(\frac{2}{\eta}C_1^{1/2}e^{\eta z_*/4}\right) + a_2 K_0\left(\frac{2}{\eta}C_1^{1/2}e^{\eta z_*/4}\right)\,, \tag{6.108}$$

where $I_0$ and $K_0$ are the modified Bessel functions of order 0 while $a_1$ and $a_2$ are integration constants.

Since $I_0(u) \approx 1$ and $K_0(u) \approx -\ln u$ when $u \to 0$, the general solution behaves as an affine function of $z_*$ when $z_* \to -\infty$ and is therefore square integrable with respect to the measure $w\,\mathrm{d}z_* \approx e^{\eta z_*/2}\,\mathrm{d}z_*$. This means that the endpoint $z_* \to -\infty$ is of limit circle type (according to the standard terminology, see e.g. [215]). Interestingly, the analysis of the axial modes near the horizon in our case is very similar to that near a naked singularity as discussed in [216]. In contrast with the GR case, none of the two axial modes is ingoing or outgoing, which means that the stability analysis of these perturbations differs from the GR one.

For the other endpoint (at spatial infinity), $z_* \approx z \to +\infty$, the asymptotic behaviours of the potential $V$ and the functions $w$, according to eqs. (6.102) and (6.103), are given by

$$V(z) \approx \frac{2(\lambda+1)}{z^2}, \qquad w(z) \approx 1, \tag{6.109}$$

which coincides with the GR behaviour at spatial infinity. In particular, $V$ goes to zero and $w$ goes to one, so that one recovers the usual combination of ingoing and outgoing modes

$$\hat{Y}_1 \approx b_1 e^{i\Omega z_*} + b_2 e^{-i\Omega z_*}, \tag{6.110}$$

where $b_1$ and $b_2$ are constants. If $\Omega$ contains a nonzero imaginary part, then one of the modes is normalisable and then this endpoint is now of limit-point type.

As we have already said previously, the analysis of axial perturbations in this theory is very different from the analysis in GR. The main reason is that we no longer have a distinction between ingoing and outgoing modes at the horizon. The choice of the right behaviour to consider might be guided by regularity properties of the mode. Indeed, if we require the perturbation[5] $\hat{Y}_1$ to be regular when $z_* \to -\infty$, then we have to impose $a_2 = 0$. The problem turns into a Sturm-Liouville problem, which implies that $\Omega^2$ is real. A very similar problem has been studied in another context in [216] where the authors showed that $\Omega^2 > 0$ when $V > 0$, which implies that the perturbations are stable. Here we can make the same analysis as in [216], and we expect the stability result to be true at least in the case where $V > 0$, i.e. when $\beta$ is sufficiently small, as explained in the discussion below eq. (6.103).

Let us close this subsection with a final remark. It is always possible to use, instead of the tortoise coordinate, a different coordinate $z_*$, for example by choosing $n(z)$ such that $c = 1$ everywhere. This corresponds to the choice

$$n(z) = \frac{1}{\sqrt{\Gamma}}. \tag{6.111}$$

In this new frame, the potential is changed and can be written in the form

$$V_{c=1} = \frac{Q(f)}{16z^2 f\gamma_1\gamma_2^3\gamma_3^5}, \tag{6.112}$$

where $Q$ is a polynomial of order 28 of nonzero constant term whose coefficients depend on $z$. This potential is represented on fig. 6.4 for different values of $\beta$.

5. The regularity concerns the metric components themselves and not directly the function $\hat{Y}_1$. The asymptotic behaviour of the metric components will be given in (6.156).

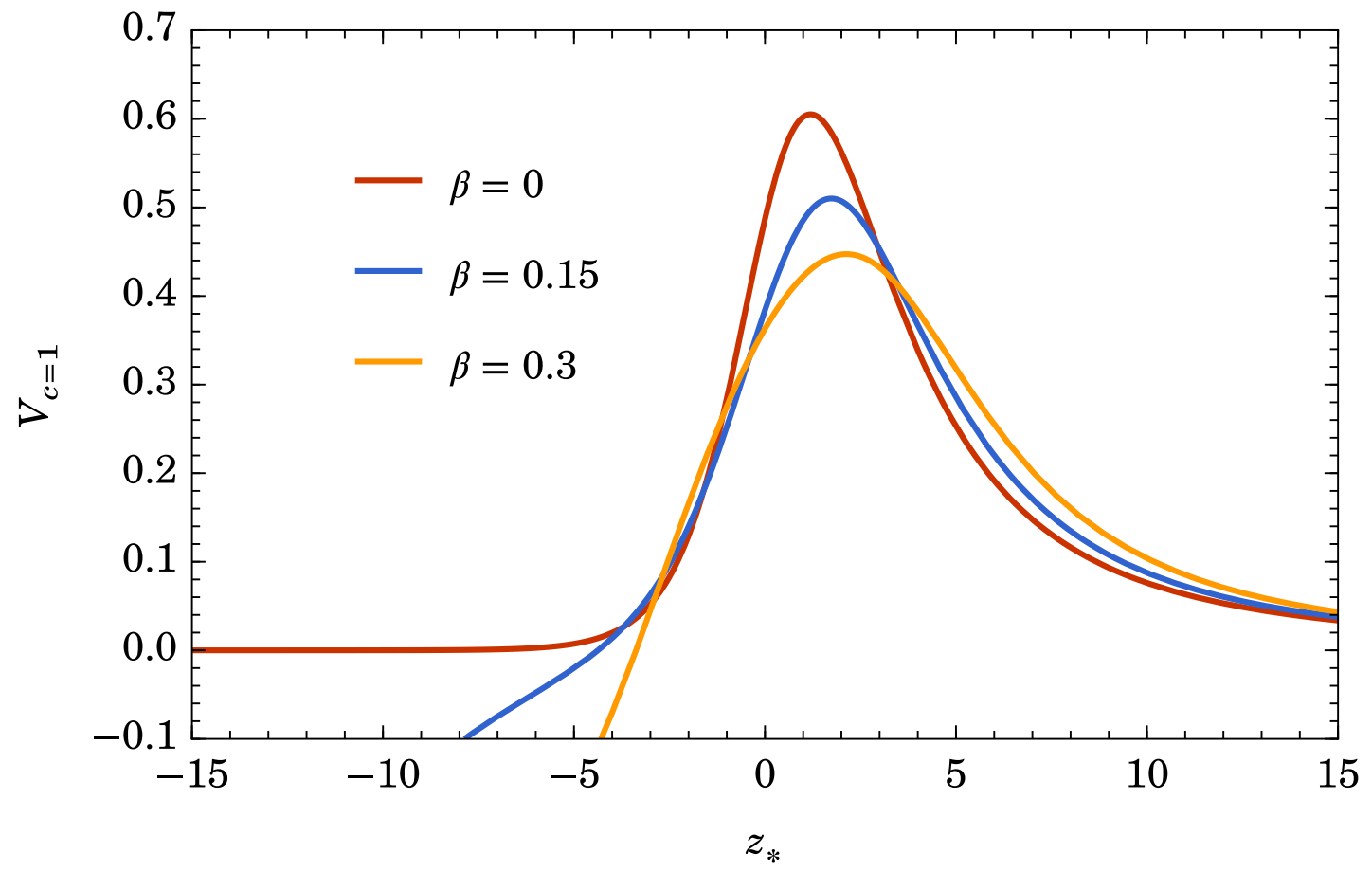

Figure 6.4. – Plot of the potential $V_{c=1}$ for $\lambda = 2$.

## 6.4. Axial perturbations: first-order approach

In this section, we compare the results obtained in section 6.3 to the predictions of the algorithm described in chapter 5 when it is directly applied to the first-order system (6.13). This allows us to check that the algorithm indeed yields the correct results.

### 6.4.1. BCL solution

As found in section 6.3.1, the axial perturbations of the BCL black hole satisfy the system

$$\frac{\mathrm{d}Y}{\mathrm{d}r} = MY, \qquad M(r) = \begin{pmatrix} 2/r & -i\omega^2 + i2\lambda A/r^2 \\ -i\Gamma & \Delta \end{pmatrix}, \tag{6.113}$$

with, according to eq. (6.53),

$$\begin{aligned} A &= \left(1 - \frac{r_+}{r}\right)\left(1 + \frac{r_-}{r}\right), \quad \Gamma = \frac{r^2(r^2 + 2r_+r_-)}{(r - r_+)^2(r + r_-)^2}, \\ \Delta &= -\frac{r_+}{r(r - r_+)} + \frac{r_-}{r(r + r_-)}. \end{aligned} \tag{6.114}$$

#### 6.4.1.1. At spatial infinity

When $r \to +\infty$, the asympotic expansion of the matrix $M(r)$ in eq. (6.113) reads

$$M(r) = M_0 + \frac{1}{r}M_{-1} + \mathcal{O}\left(\frac{1}{r^2}\right),$$

$$M_0 \equiv -i \begin{pmatrix} 0 & \omega^2 \\ 1 & 0 \end{pmatrix}, \quad M_{-1} \equiv 2 \begin{pmatrix} 1 & 0 \\ -i\mu & 0 \end{pmatrix}, \tag{6.115}$$

where we have stopped at order $1/r$, which will be sufficient for our purpose. Note that the two terms in the above expansion do not depend on $\xi$, which means they coincide with the analogous terms in GR. This is consistent with the observation that the asymptotic behaviour of the potential (6.54) at infinity coincides with that of the RW potential (4.41) up to first order in $1/r$.

Since we have already analysed the same asymptotic system in chapter 5 for the axial modes in Schwarzschild, we recall briefly the main result. Using the transformation

$$Y = \tilde{P}\tilde{Y}, \qquad \tilde{P} = \begin{pmatrix} -1+\varpi_+ & 1+\varpi_- \\ 1+\varpi_+ & 1-\varpi_- \end{pmatrix}, \qquad \varpi_\pm \equiv \frac{\pm\,\omega\mu + i}{2\omega r}, \tag{6.116}$$

we obtain the equivalent, and fully diagonalised, system

$$\frac{\mathrm{d}\tilde{Y}}{\mathrm{d}r} = \tilde{M}\tilde{Y}, \quad \tilde{M}(r) = \begin{pmatrix} -i\omega & 0 \\ 0 & i\omega \end{pmatrix} + \frac{1}{r}\begin{pmatrix} 1-i\omega\mu & 0 \\ 0 & 1+i\omega\mu \end{pmatrix} + \mathcal{O}\left(\frac{1}{r^2}\right). \tag{6.117}$$

Direct integration yields the asymptotic solution

$$\tilde{Y}(r) = (1+\mathcal{O}(1/r))\begin{pmatrix} c_-\mathfrak{a}_-^\infty(r) \\ c_+\mathfrak{a}_+^\infty(r) \end{pmatrix}, \quad \mathfrak{a}_\pm^\infty(r) = e^{\pm i\omega r} r^{1\pm i\omega\mu} = r e^{\pm i\omega r_*}, \tag{6.118}$$

where $c_\pm$ are arbitrary constants and we have reintroduced, in the last expression, the variable $r_*$ associated with the BCL solution, defined in eq. (6.52) [6].

Taking into account the time dependence $e^{-i\omega t}$ of the modes, the two physical modes $\mathfrak{a}_\pm^\infty(r)$ take the form

$$e^{-i\omega t}\mathfrak{a}_\pm^\infty(r) = c_\pm r e^{-i\omega(t\mp r_*)}, \tag{6.120}$$

where one recognises the usual ingoing mode (associated with $c_-$) and outgoing mode (associated with $c_+$) at spatial infinity. The values of $c_\pm$ can be restricted by the boundary conditions imposed on the system. For example, requiring that the mode is purely outgoing, as is the case for QNMs, imposes $c_- = 0$.

6. The tortoise coordinate associated with the BCL solution has been computed in eq. (6.52) and its large $r$ expansion reads

$$r_* = r + \mu\ln r - \frac{r_+^2\ln r_+ - r_-^2\ln r_-}{r_+ + r_-} - \frac{r_+^2 + r_-^2 - r_+r_-}{r} + \mathcal{O}\left(\frac{1}{r^2}\right). \tag{6.119}$$

When $\mu = r_+ = 2m$, it coincides with the Schwarzschild tortoise coordinate (6.63) $r_* = r + \mu\ln r$ up to the order $\mathcal{O}(1)$. Hence, one can equivalently use any of the two coordinates in the asymptotic (6.118) which has been given up to $\mathcal{O}(1)$ as well.

#### 6.4.1.2. At the horizon

We now turn to the asymptotic behaviour at the black hole horizon. Introducing the variable

$$\varepsilon \equiv r - r_+ \,, \tag{6.121}$$

the near-horizon asymptotic expansions of the functions $A$, $\Gamma$ and $\Delta$ in eq. (6.114) are given by

$$A = \mathcal{O}(\varepsilon)\,, \qquad \Gamma = i\left(\frac{\Gamma_2}{\varepsilon^2} + \frac{\Gamma_1}{\varepsilon} + \Gamma_0\right) + \mathcal{O}(\varepsilon)\,, \qquad \Delta = \frac{\Delta_1}{\varepsilon} + \Delta_0 + \mathcal{O}(\varepsilon)\,. \tag{6.122}$$

Substituting into eq. (6.113), we obtain the asymptotic expansion of the matrix $M$,

$$M(\varepsilon) = \frac{1}{\varepsilon^2}\begin{pmatrix} 0 & 0 \\ \Gamma_2 & 0 \end{pmatrix} + \frac{1}{\varepsilon}\begin{pmatrix} 0 & 0 \\ \Gamma_1 & \Delta_1 \end{pmatrix} + \begin{pmatrix} 2/r_+ & -i\omega^2 \\ \Gamma_0 & \Delta_0 \end{pmatrix} + \mathcal{O}(\varepsilon)\,, \tag{6.123}$$

where we will need only the explicit expression of the coefficients $\Delta_1$ and $\Gamma_2$,

$$\Delta_1 = -1\,, \qquad \Gamma_2 = -ir_0^2 \quad \text{with} \quad r_0 \equiv r_+ \frac{\sqrt{r_+(r_+ + 2r_-)}}{r_+ + r_-}\,. \tag{6.124}$$

Our system now differs from the GR analog studied in chapter 5. However, the leading order term is still nilpotent, as in GR, and the resolution of the system is very similar to the analysis of chapter 5. According to the algorithm, one first needs to perform the transformation

$$Y \equiv P_{(1)} Y^{(1)}\,, \quad \text{with} \quad P_{(1)}(\varepsilon) \equiv \begin{pmatrix} 1 & 0 \\ 0 & 1/\varepsilon \end{pmatrix}\,, \tag{6.125}$$

which leads to the new system

$$\frac{\mathrm{d}Y^{(1)}}{\mathrm{d}\varepsilon} = M^{(1)} Y^{(1)}\,, \quad M^{(1)} = \frac{1}{\varepsilon}\begin{pmatrix} 0 & -i\omega^2 \\ \Gamma_2 & 1+\Delta_1 \end{pmatrix} + \mathcal{O}(1) = -\frac{i}{\varepsilon}\begin{pmatrix} 0 & \omega^2 \\ r_0^2 & 0 \end{pmatrix} + \mathcal{O}(1)\,. \tag{6.126}$$

The leading term of the new matrix $M^{(1)}$ is now diagonalisable and the system can be explicitly diagonalised via the transformation

$$Y^{(1)} \equiv P_{(2)} Y^{(2)}\,, \quad \text{with} \quad P_{(2)} = \begin{pmatrix} \omega & -\omega \\ r_0 & r_0 \end{pmatrix}\,, \tag{6.127}$$

leading to the new system

$$\frac{\mathrm{d}Y^{(2)}}{\mathrm{d}\varepsilon} = M^{(2)} Y^{(2)}\,, \qquad M^{(2)}(\varepsilon) = \frac{i\omega r_0}{\varepsilon}\begin{pmatrix} -1 & 0 \\ 0 & 1 \end{pmatrix} + \mathcal{O}(1)\,. \tag{6.128}$$

Finally, integrating this system yields

$$Y^{(2)}(\varepsilon) = (1+\mathcal{O}(\varepsilon))\begin{pmatrix} c_- \mathfrak{a}^{\rm h}_-(\varepsilon) \\ c_+ \mathfrak{a}^{\rm h}_+(\varepsilon) \end{pmatrix}, \quad \mathfrak{a}^{\rm h}_\pm(\varepsilon) = \varepsilon^{\pm i\omega r_0} = e^{\pm i\eta\omega r_*}, \tag{6.129}$$

where $c_\mp$ are constants and we have used the asymptotic expansion of the tortoise coordinate (6.52) near the horizon,

$$r_* = \frac{r_+^2}{r_+ + r_-}\ln\varepsilon + \mathcal{O}(1) = \frac{r_0}{\eta}\ln\varepsilon + \mathcal{O}(1), \qquad \eta \equiv \frac{\sqrt{r_+ + 2r_-}}{r_+^{1/2}}. \tag{6.130}$$

Taking into account the time dependence $e^{-i\omega t}$, one thus gets for the two components of $Y^{(2)}$

$$e^{-i\omega t}\mathfrak{a}^{\rm h}_\pm = c_\pm e^{-i\omega(t\mp\eta r_*)}\left(1+\mathcal{O}(\varepsilon)\right), \tag{6.131}$$

where one recognizes the ingoing and outgoing modes, propagating with the velocity $c = \eta^{-1}$, in agreement with the expression (6.59) via the Schrödinger-like equation in section 6.3.1.

### 6.4.2. Stealth solutions

Let us now study the asymptotic behaviour of axial perturbations for the stealth Schwarzschild solution. We use the expressions of the functions $\Psi$, $\Phi$, $\Gamma$ and $\Delta$ given in eq. (6.60). Let us recall that the constant $\zeta$ defined in eq. (6.62) parametrizes the deviation to GR which is recovered in the limit $\zeta \to 0$.

Following our remark at the end of section 6.3.2 that the Schrödinger-like equation for axial modes is equivalent to a standard RW equation, we now show that this property can be seen directly with the first order system, via appropriate rescalings of the time and radial variables. We first perform the time shift given in eq. (6.18) to remove $\Psi$ from the matrix $M(r)$. Then, introducing the new variables

$$\tilde{r} \equiv (1+\zeta)r, \qquad \tilde{r}_g \equiv (1+\zeta)r_g, \qquad \tilde{t} \equiv \sqrt{1+\zeta}\,t \implies \tilde{\omega} = \omega/\sqrt{1+\zeta}, \tag{6.132}$$

one can see that the first order differential system takes exactly the same form as in GR, namely

$$\frac{{\rm d}Y}{{\rm d}\tilde{r}} = \tilde{M}Y, \qquad \tilde{M}(\tilde{r}) = \begin{pmatrix} 2/\tilde{r} & -i\tilde{\omega}^2 + 2i\lambda\frac{\tilde{r}-\tilde{r}_g}{\tilde{r}^3} \\ -i\frac{\tilde{r}^2}{(\tilde{r}-\tilde{r}_g)^2} & -\frac{\tilde{r}_g}{\tilde{r}(\tilde{r}-\tilde{r}_g)} \end{pmatrix}, \tag{6.133}$$

with $\tilde{r}_g$ as Schwarzschild radius.

As a consequence, the asymptotic behaviour of $Y$ is immediately deduced from the GR results given in chapter 4. Both at infinity and near the horizon, the asymptotic behaviours of the two components of $Y$ are linear combinations (with coefficients that can depend on real powers of $r$ or $\varepsilon$) of the following outgoing and ingoing modes:

$$e^{\pm i\tilde{\omega}\tilde{r}_*} = e^{\pm i\omega r_*}, \qquad \tilde{r}_* \equiv \tilde{r} + \tilde{r}_g \ln\left(\tilde{r}/\tilde{r}_g - 1\right), \tag{6.134}$$

where $\tilde{r}_*$ corresponds to the standard tortoise coordinate in Schwarzschild (with radial coordinate $\tilde{r}$ and horizon $\tilde{r}_g$) and $r_*$ is the radial coordinate introduced in eq. (6.67) in order to get $c(r) = 1$.

One can finally reintroduce the time dependence to obtain the asymptotic behaviour in $(t, r)$ coordinates. At spatial infinity, using $\int \mathrm{d}r\, \Psi(r) \approx 2\zeta\sqrt{\mu r}$, one finds that the metric perturbations are combinations of the axial modes given by

$$\mathfrak{a}^{\infty}_{\pm}(r) = e^{2i\omega\zeta\sqrt{\mu r}} e^{\pm i\omega\sqrt{1+\zeta}r} r^{\pm i\omega(1+\zeta)^{3/2}\mu}. \tag{6.135}$$

At the horizon $r = r_g$, using $\int \mathrm{d}r\, \Psi(r) \approx (1+\zeta)^{3/2}\mu \ln\left(r/r_g - 1\right) \approx r_*$, one obtains a linear combination of

$$\mathfrak{a}^{\mathrm{h}}_1(r) = (r - r_g)^{2i\omega\mu(1+\zeta)^{3/2}} \quad \text{and} \quad \mathfrak{a}^{\mathrm{h}}_2(r) = 1. \tag{6.136}$$

In the original coordinate system, only one mode seems to be propagating at the horizon. It is necessary to use a more appropriate time coordinate to identify one outgoing and one ingoing mode.

### 6.4.3. EsGB solution

#### 6.4.3.1. Spatial infinity

At spatial infinity, the coordinate variable is $z$ and the first terms of the initial matrix $M$ in an expansion in power of $z$ read

$$M = \begin{pmatrix} 0 & -i\Omega^2 \\ -i & 0 \end{pmatrix} + \begin{pmatrix} 2 & 0 \\ -2i\left(1+\varepsilon^2/3\right) & 0 \end{pmatrix} \frac{1}{z} + \mathcal{O}\left(\frac{1}{z^2}\right). \tag{6.137}$$

Applying the change of functions (5.6) with

$$\tilde{P} = \begin{pmatrix} \Omega & -\Omega \\ 1 & 1 \end{pmatrix} + \frac{1}{6\Omega} \begin{pmatrix} 3i\Omega - (3+\varepsilon^2)\Omega^2 & 3i\Omega + (3+\varepsilon^2)\Omega^2 \\ -3i + (3+\varepsilon^2)\Omega & 3i + (3+\varepsilon^2)\Omega \end{pmatrix} \frac{1}{z}, \tag{6.138}$$

provided by the algorithm of chapter 5, one obtains the new matrix

$$\tilde{M} = \mathrm{Diag}(-i\Omega, i\Omega) + \frac{1}{z}\mathrm{Diag}\left[1 - i\Omega\left(1+\frac{\varepsilon^2}{3}\right), 1 - +i\Omega\left(1+\frac{\varepsilon^2}{3}\right)\right] + \mathcal{O}\left(\frac{1}{z^2}\right), \tag{6.139}$$

which is diagonal up to order $1/z^2$. Hence, the corresponding system can immediately be integrated and we find that the metric perturbations (6.2) at infinity are a linear combination of the following modes:

$$\mathfrak{a}_{\pm}^{\infty}(z) \approx e^{\pm i\Omega z} z^{\pm i\Omega(1+\varepsilon^2/3)} . \tag{6.140}$$

More precisely, the functions $Y_1$ and $Y_2$ are given by

$$\begin{aligned} Y_1(z) &\approx \Omega \left[ -c_+ e^{i\Omega z} z^{+i\Omega(1+\varepsilon^2/3)} + c_ e^{-i\Omega z} z^{-i\Omega(1+\varepsilon^2/3)} \right] , \\ Y_2(z) &\approx c_+ e^{i\Omega z} z^{+i\Omega(1+\varepsilon^2/3)} + c_ e^{-i\Omega z} z^{-i\Omega(1+\varepsilon^2/3)} , \end{aligned} \tag{6.141}$$

where $c_{\pm}$ are the integration constants. As expected, one recovers the same combination of modes as in (6.87).

#### 6.4.3.2. Near the horizon

To study the asymptotic behaviour near the horizon, it is convenient to use the coordinate $x$ defined by $x = 1/(z-1)$. Then, we study the behaviour, when $x$ goes to infinity, of the system (6.13), reformulated as

$$\frac{\mathrm{d}Y}{\mathrm{d}x} = M_x(x)\, Y, \quad \text{with} \quad M_x(x) = -\frac{1}{x^2} M(1+1/x) . \tag{6.142}$$

The expansion of the matrix $M_x$ in powers of $x^{-1}$ yields

$$\begin{aligned} M_x = &\begin{pmatrix} 0 & 0 \\ i(1-21/5\varepsilon^2) & 0 \end{pmatrix} + \begin{pmatrix} 0 & 0 \\ 2i(1-121/15\varepsilon^2) & 1 \end{pmatrix} \frac{1}{x} \\ &+ \frac{1}{15} \begin{pmatrix} -30-244\varepsilon^2 & 15i\Omega^2 \\ 15i(1+1111\varepsilon^2) & -15-662\varepsilon^2 \end{pmatrix} \frac{1}{x^2} + \mathcal{O}\left(\frac{1}{x^3}\right) . \end{aligned} \tag{6.143}$$

The algorithm provides us with the transfer matrix

$$\begin{aligned} \tilde{P} = &\begin{pmatrix} 0 & 0 \\ 1 & 1 \end{pmatrix} x + \Omega \begin{pmatrix} -1-21\varepsilon^2/10 & 1+21\varepsilon^2/10 \\ 0 & 0 \end{pmatrix} \\ &+ \left[ \begin{pmatrix} 2(i+\Omega) & 2(i-\Omega) \\ 0 & 0 \end{pmatrix} + \frac{\varepsilon^2}{15} \begin{pmatrix} 10i-53\Omega & 10i+53\Omega \\ 0 & 0 \end{pmatrix} \right] \frac{1}{x} + \mathcal{O}\left(\frac{1}{x^2}\right) , \end{aligned} \tag{6.144}$$

and one obtains a new differential system with a diagonal matrix $\tilde{M}_x$,

$$\tilde{M}_x = \frac{1}{x} \mathrm{Diag}\left[ -i\Omega\left(1-\frac{21}{10}\varepsilon^2\right), i\Omega\left(1-\frac{21}{10}\varepsilon^2\right) \right] + \mathcal{O}\left(\frac{1}{x^2}\right) . \tag{6.145}$$

Integrating the system immediately yields that the metric components are a linear combination of the following modes:

$$\mathfrak{a}_{\pm}^{\mathrm{h}}(z) \approx (z-1)^{\pm i\Omega(1-21\varepsilon^2/10)} . \tag{6.146}$$

Expressed in terms of the variable $z$, the metric perturbations are written as

$$\begin{aligned}Y_1(z) &\approx \Omega\left(1+21\varepsilon^2/10\right)\Big[c_+(z-1)^{-1-i\Omega(1-21\varepsilon^2/10)} \\ &\qquad\qquad - c_-(z-1)^{-1+i\Omega(1-21\varepsilon^2/10)}\Big]\,, \\ Y_2(z) &\approx c_+(z-1)^{-1-i\Omega(1-21\varepsilon^2/10)} + c_-(z-1)^{-1+i\Omega(1-21\varepsilon^2/10)}\,, \end{aligned} \tag{6.147}$$

where $c_\pm$ are integration constants (different from those introduced in (6.141)). As expected, we recover the combination of modes found in eq. (6.88) from the Schrödinger-like formulation.

### 6.4.4. 4dEGB solution

Let us turn to the study of asymptotic behaviours of axial perturbations of the metric[7] in the case of the 4dEGB solution. At spatial infinity, the matrix $M$ can be expanded as

$$M(z) = \begin{pmatrix} 0 & -i\Omega^2 \\ -i & 0 \end{pmatrix} + \mathcal{O}\left(\frac{1}{z}\right)\,. \tag{6.148}$$

Therefore, the two components of $Y$ at infinity are immediately found to be a linear combination of the following two modes:

$$\mathfrak{a}^\infty_\pm(z) \approx e^{\pm i\Omega z} = e^{\pm i\omega r}\,. \tag{6.149}$$

Hence, the asymptotic behaviour of the original metric variables $h_0$ and $h_c$ are given by

$$\begin{aligned} Y_1(z) &\approx z\Omega\left[-c_+ e^{i\Omega z} z^{i\Omega(1+\beta)} + c_- e^{-i\Omega z} z^{-i\Omega(1+\beta)}\right]\,, \\ Y_2(z) &\approx z\left[c_+ e^{i\Omega z} z^{i\Omega(1+\beta)} + c_- e^{-i\Omega z} z^{-i\Omega(1+\beta)}\right]\,, \end{aligned} \tag{6.150}$$

where $c_\pm$ are constants.

Near the horizon, we change variables by setting $x = 1/\sqrt{z-1}$, and study the behaviour, when $x$ goes to infinity, of the system (6.13), rewritten as

$$\frac{\mathrm{d}Y}{\mathrm{d}x} = M_x(x)\,Y\,, \quad \text{with} \quad M_x(x) = -\frac{2}{x^3}M(1+1/x^2)\,. \tag{6.151}$$

The algorithm then enables us to simplify the original system, here up to order $x^{-1}$, using the transfer matrix $P$ such that

$$P = \frac{1}{xp_3}\begin{pmatrix} p_1 + xp_2 & p_2 \\ 0 & x^2 p_3 \end{pmatrix}\,, \tag{6.152}$$

7. The change of variables leading from $\omega$ to $\Omega$ requires to rescale $Y_2$ by a factor $r_h$.

with the functions $p_i$ defined by

$$p_1 = (1-\beta)^2\left(1+2\beta+6\beta^2\right), \quad p_2 = 2(1-\beta)^2\beta\sqrt{1+\beta-2\beta^2}\,,$$
$$p_3 = 2i(1+2\beta)^2(1-2\beta(1+\beta))\,. \tag{6.153}$$

The new system is then

$$\frac{\mathrm{d}\tilde{Y}}{\mathrm{d}x} = \tilde{M}_x\,\tilde{Y}\,, \qquad \text{with} \quad \tilde{M}_x = \begin{pmatrix} 0 & 0 \\ 1/x & 0 \end{pmatrix} + \mathcal{O}\left(\frac{1}{x^2}\right). \tag{6.154}$$

Therefore, the solution near the horizon (written in terms of the original variable $z$) is a linear combination of two "modes",

$$\mathfrak{a}^{\mathrm{h}}_1(z) \approx 1 \quad \text{and} \quad \mathfrak{a}^{\mathrm{h}}_2(z) \approx -\frac{1}{2}\ln(z-1)\,. \tag{6.155}$$

Going back to the original variables $Y_1$ and $Y_2$, we get

$$Y_1(z) \approx \frac{p_2}{p_3}c_1 + \sqrt{z-1}\left(\frac{p_1}{p_3}c_1 + \frac{p_2}{p_3}c_2 - \frac{p_2}{2p_3}c_1\ln(z-1)\right),$$
$$Y_2(z) \approx \frac{1}{\sqrt{z-1}}\left(c_2 - \frac{1}{2}c_1\ln(z-1)\right), \tag{6.156}$$

with $c_1$ and $c_2$ two constants [8]. This result is consistent with the asymptotic solution we found from the Schrödinger-like equation in eq. (6.108) when we expand the Bessel functions in power series.

## 6.5. Effective metric

In this section, we show that the dynamics of axial perturbations described as a first-order system in eq. (6.13) is equivalent to the dynamics of the axial component of a massless spin 2 field propagating in an effective metric $\tilde{g}_{\mu\nu}$ that we compute explicitly.

We recover the recent results of [210] where the effective metric has been obtained using the quadratic Lagrangian of axial perturbations. Furthermore, we also recover and discuss the conditions on the theory for the perturbations to be free of gradient instabilities. Finally, in the case of quadratic DHOST theories, we show that the effective metric $\tilde{g}_{\mu\nu}$ is in fact linked to the original background metric $g_{\mu\nu}$ by a disformal transformation, and we interpret the origin of this disformal transformation.

8. We do not call them $c_+$ and $c_-$ as usual here since it is not possible to identify ingoing and outgoing modes.

### 6.5.1. The conformal class of the effective metric

A simple way to determine the effective metric, at least up to a global factor, is to interpret the Schrödinger-like equation as an effective Klein-Gordon equation of the form

$$\tilde{g}^{\mu\nu}\tilde{\nabla}_\mu\tilde{\nabla}_\nu\psi - m_{\text{eff}}^2\psi = 0\,. \tag{6.157}$$

where $\tilde{\nabla}$ denotes the covariant derivative associated with the effective metric $\tilde{g}_{\mu\nu}$. To this end, it is convenient to reformulate the Schrödinger-like equation (6.23) in a more covariant way and then to write the equation satisfied by the full wave function $\psi$ defined by

$$\psi(t_*, r, \theta, \varphi) \;=\; e^{-i\omega t_*}\hat{Y}_1(r)\,Y_{\ell,m}(\theta,\varphi)\,. \tag{6.158}$$

An immediate calculation shows that the full covariant function $\psi$ satisfies the equation

$$-\Gamma\frac{\partial^2\psi}{\partial t_*^2} + \frac{1}{n^2}\frac{\partial^2\psi}{\partial r_*^2} + \frac{\Gamma\Phi}{C}\Delta_S\psi - \left(V_0 - 2\frac{\Gamma\Phi}{C}\right)\psi = 0\,, \tag{6.159}$$

where $\Delta_S$ is the Laplace operator on the two-sphere of radius 1, i.e.

$$\Delta_S \equiv \frac{1}{\sin\theta}\frac{\partial}{\partial\theta}\left(\sin\theta\frac{\partial}{\partial\theta}\right) + \frac{1}{\sin^2\theta}\frac{\partial^2}{\partial\varphi^2}\,. \tag{6.160}$$

This equation shows that the axial gravitational mode corresponds to a perturbation propagating in an effective static and spherically symmetric metric $\tilde{g}_{\mu\nu}$ whose expression in the coordinate system $(t_*, r_*, \theta, \varphi)$ is given, up to a conformal factor $\gamma$ which depends on $r$, by

$$d\tilde{s}^2 = \tilde{g}_{\mu\nu}\,dx^\mu\,dx^\nu = \gamma\left[-\Phi\,dt_*^2 + \Gamma\Phi\,n^2\,dr_*^2 + C\,d\Omega^2\right]\,. \tag{6.161}$$

This result is consistent with [210] but, at this stage, it has not been proven that the perturbation is indeed a spin 2 massless field. This is what we will do in section 6.5.2.

In the original coordinate system, the effective metric is given by

$$d\tilde{s}^2 = \gamma\left[-\Phi(dt - \Psi\,dr)^2 + \Gamma\Phi\,dr^2 + C\,d\Omega^2\right]\,, \tag{6.162}$$

and we immediately see that it does not belong, in general, to the conformal class of the background metric. In particular, it is not clear whether or not the effective metric still describes a BH geometry even though the background solution is itself a black hole. We will discuss more on this aspect later on.

## 6.5.2. Complete effective metric

In this section, we show explicitly that the dynamics of the perturbation describes the axial mode of a massless spin 2 field propagating in an effective metric which obviously belongs to the conformal class of eq. (6.162). Furthermore, we will compute the conformal factor $\gamma$ explicitly as well. We can proceed in two equivalent ways, by starting from the first order system of eq. (6.20) or from the Schrödinger-like equation (6.23).

### 6.5.2.1. From the first-order system

We want to show that eq. (6.20) is equivalent to a first order system of the form (6.16) where the functions $A$, $B$ and $C$ defining the background metric (see eq. (2.1)) are replaced by some functions $\tilde{A}$, $\tilde{B}$ and $\tilde{C}$ which define the effective metric $\tilde{g}_{\mu\nu}$ according to

$$d\tilde{s}^2 = \tilde{g}_{\mu\nu}\, dx^\mu\, dx^\nu = -\tilde{A}(r)\, dt_*^2 + \frac{1}{\tilde{B}(r)}\, dr^2 + \tilde{C}(r)\, d\Omega^2 \,. \tag{6.163}$$

For that purpose, we make a change of variable and introduce $\tilde{Y} = \gamma Y$ where $\gamma$ is an arbitrary function of $r$ at this stage [9]. This new variable satisfies the differential system

$$\frac{d\tilde{Y}}{dr} = \tilde{M}\tilde{Y} \quad \text{with} \quad \tilde{M} = \begin{pmatrix} C'/C + \gamma'/\gamma & -i\omega^2 + 2i\lambda\Phi/C \\ -i\Gamma & \Delta + \gamma'/\gamma \end{pmatrix}, \tag{6.164}$$

and we look for $\tilde{A}$, $\tilde{B}$, $\tilde{C}$ together with $\gamma$ such that $\tilde{M}$ is exactly of the form given in eq. (6.16) , i.e.

$$\tilde{M} = \begin{pmatrix} \tilde{C}'/\tilde{C} & -i\omega^2 + 2i\lambda\tilde{A}/\tilde{C} \\ -i/(\tilde{A}\tilde{B}) & -(\tilde{A}'/\tilde{A} + \tilde{B}'/\tilde{B})/2 \end{pmatrix}. \tag{6.165}$$

We identify each coefficient of the two matrices and we obtain the following three relations between the background metric coefficients and the effective metric coefficients:

$$\tilde{A}\tilde{B} = \frac{1}{\Gamma}, \qquad \tilde{A} = \Phi\frac{\tilde{C}}{C}, \qquad \frac{\tilde{A}'}{\tilde{A}} + \frac{\tilde{B}'}{\tilde{B}} + 2\frac{\tilde{C}'}{\tilde{C}} = 2\frac{C'}{C} - 2\Delta\,, \tag{6.166}$$

while $\gamma$ is obtained by integrating the last equation $\gamma'/\gamma = \tilde{C}'/\tilde{C} - C'/C$ which leads to

$$\gamma = \kappa\frac{\tilde{C}}{C}, \tag{6.167}$$

9. As we are going to see $\gamma$ will be given exactly by the conformal factor in eq. (6.162) and this is the reason why we are using the same notation.

where the integration constant $\kappa$ can be fixed to 1 without loss of generality. From this relation, we see that $\gamma$ is precisely the conformal factor we have introduced in eq. (6.162) (as we said in footnote 9).

Using the expression of $\Delta$ given in eq. (6.14), we can easily integrate the last equation in eq. (6.166), which leads to the new relation

$$\sqrt{\tilde{A}\tilde{B}}\,\tilde{C} = |\mathscr{F}|\sqrt{\frac{B}{A}}\,C\,, \tag{6.168}$$

where we imposed $\tilde{C}$ to be positive in order to ensure a physically meaningful signature for the effective metric. As a consequence, this last equation together with the first two equations in eq. (6.166) enable us to solve the three effective metric coefficients in terms of the background metric coefficients and we obtain

$$\tilde{A} = \gamma\,\Phi\,, \qquad \frac{1}{\tilde{B}} = \gamma\,\Phi\,\Gamma \quad \text{with} \quad \frac{\tilde{C}}{C} = \gamma = |\mathscr{F}|\sqrt{\frac{\Gamma B}{A}}\,. \tag{6.169}$$

As a consequence, the first order differential system satisfied by the axial mode describes indeed the axial mode of a massless spin 2 field which propagates in the effective metric

$$\mathrm{d}\tilde{s}^2 = \tilde{g}_{\mu\nu}\,\mathrm{d}x^\mu\,\mathrm{d}x^\nu = |\mathscr{F}|\sqrt{\frac{\Gamma B}{A}}\left(-\Phi\,\mathrm{d}t_*^2 + \Gamma\Phi\,\mathrm{d}r^2 + C\,\mathrm{d}\Omega^2\right)\,. \tag{6.170}$$

As expected, this metric is consistent with (6.162) and is the same as the one found in [210] from the analysis of the Schrödinger-like equation.

#### 6.5.2.2. From the Schrödinger equation

Let us now recover the result presented in eq. (6.170) starting from the Schrödinger-like equation (6.23). It is proven in [217, 218] that the dynamics of a massless spin-2 field propagating in a background given by eq. (2.1) can be described by a Schrödinger-like equation,

$$\frac{\mathrm{d}^2\psi}{\mathrm{d}r_*^2} + \omega^2\,\psi - V(r)\,\psi = 0\,, \tag{6.171}$$

where the potential $V$ is such that

$$V = 2\lambda\frac{A}{C} + \frac{1}{2}\frac{D^2C'^2}{C} - \frac{1}{2}D\,(C'D)' \quad \text{with} \quad D = \sqrt{AB/C}\,, \tag{6.172}$$

and the coordinate $r_*$ is defined by

$$\mathrm{d}r^2 = \tilde{A}(r)\tilde{B}(r)\;\mathrm{d}r_*^2\,, \tag{6.173}$$

One can notice that this result can be recovered from eq. (6.31) by setting $F_2 = 1$ and all other DHOST functions to zero.

Then, eqs. (6.23) and (6.171) are equivalent if the following conditions hold:

$$n^2 = \frac{1}{\Gamma} = \tilde{A}\tilde{B}\,, \tag{6.174}$$

$$\tilde{A} = \Phi\frac{\tilde{C}}{C}\,, \tag{6.175}$$

$$2\frac{V_0}{\Gamma} = \frac{\tilde{D}^2\tilde{C}'^2}{\tilde{C}} - \tilde{D}\left(\tilde{C}'\tilde{D}\right)'\,. \tag{6.176}$$

The two identities in the first condition are a consequence of the fact that the two tortoise coordinates defined in eqs. (6.33) and (6.173) must coincide along with the relation $c_* = 1$. The last two conditions are obtained by identifying the two potentials.

Furthermore, the first two conditions are exactly the same as to the first two conditions in eq. (6.166) which were obtained from the the first order system. They enable us to express $\tilde{A}$ and $\tilde{B}$ in terms of $\tilde{C}$. Then the last condition can be reformulated as a differential equation for $\tilde{C}$ which reads

$$\frac{3}{2}\left(\frac{\tilde{C}'}{\tilde{C}}\right)^2 - \frac{\tilde{C}''}{\tilde{C}} + \frac{1}{2}\frac{\Gamma'}{\Gamma}\frac{\tilde{C}'}{\tilde{C}} = 2\,V_0\,. \tag{6.177}$$

This equation should be equivalent to the third condition in eq. (6.166) which is much simpler. To verify this equivalence, we consider the solution (6.169) we found from the first order system and we check by a long but direct calculation that it satisfies indeed the previous equation for $\tilde{C}$.

We conclude with the remark that finding the effective metric from the first order system seems technically simpler than from the Schrödinger-like equation.

#### 6.5.2.3. Discussion of the result

The computation of the effective metric used the implicit assumption that $\Gamma$ is positive. This condition is equivalent to the no-gradient instability requirement as the expression of the speed of propagation in the $(t_*, r, \theta, \varphi)$ coordinate system is given by

$$c^2 = -\frac{\tilde{g}_{t_*t_*}}{\tilde{g}_{rr}} = \frac{1}{\Gamma}\,. \tag{6.178}$$

In the simpler case of quadratic DHOST theories where $F_3$ and the $B_i$ are zero, the expression of $\Gamma$ simplifies and the previous condition reduces to

$$F_2(F_2 - XA_1) > 0\,. \tag{6.179}$$

This stability condition coincides with the one obtained in [210].

### 6.5.3. Origin of the effective metric and disformal transformations

If we restrict our analysis to quadratic DHOST theories, the coefficients in eq. (6.14) simplify a lot, and are given by

$$\mathcal{F} = AF_2 - (q^2 + AX)A_1\,, \quad \Phi = \frac{\mathcal{F}}{F_2 - XA_1}\,, \quad \Psi = \frac{q\psi' A_1}{\mathcal{F}}\,,$$
$$\Gamma = \Psi^2 + \frac{q^2 A_1 + AF_2}{AB\mathcal{F}}\,, \quad \Delta = -\frac{\mathrm{d}}{\mathrm{d}r}\left(\ln\left(\sqrt{\frac{B}{A}}\mathcal{F}\right)\right). \tag{6.180}$$

As a consequence, in the special case $q = 0$, the effective metric reduces to

$$\mathrm{d}\tilde{s}^2 = \tilde{g}_{\mu\nu}\,\mathrm{d}x^\mu\,\mathrm{d}x^\nu = \sqrt{F_2(F_2 - XA_1)}\left(-A\,\mathrm{d}t^2 + \frac{F_2}{F_2 - XA_1}\frac{\mathrm{d}r^2}{B} + C\,\mathrm{d}\Omega^2\right). \tag{6.181}$$

This is a simple transformation of the initial background metric, which depends only on $X$ and the corresponding values of $F_2$ and $A_1$. In fact, this transformation can be interpreted as a disformal transformation, as we now explain, and this remains valid in the case $q \neq 0$.

By using the correspondence between DHOST theories due to field redefinitions, it is possible to put the coefficient $A_1$ to zero via a disformal transformation of the metric:

$$\hat{g}_{\mu\nu} = \varkappa g_{\mu\nu} + \varpi\,\phi_\mu\phi_\nu\,. \tag{6.182}$$

Indeed, any quadratic DHOST action $\hat{S}$ written as a functional of $\hat{g}_{\mu\nu}$ and $\phi$ is related to another quadratic DHOST action $S$ for $g_{\mu\nu}$ and $\phi$, defined by

$$S[g_{\mu\nu},\phi] \equiv \hat{S}[\hat{g}_{\mu\nu} = \varkappa g_{\mu\nu} + \varpi\,\phi_\mu\phi_\nu,\phi]\,. \tag{6.183}$$

The relations between the quadratic-order coefficients in the respective actions are given by the expressions (see section 1.4.1)

$$\hat{F}_2 = [\varkappa^2(1 + X\varpi/\varkappa)]^{-1/2}F_2\,, \tag{6.184}$$
$$\hat{A}_1 = (1 + X\varpi/\varkappa)^{3/2}\,(A_1 - \frac{\varpi}{\varkappa + \varpi X}F_2)\,, \tag{6.185}$$

and we do not need here the analogous expressions for the other coefficients.

It is easy to check that one obtains $\hat{F}_2 = \mathrm{sgn}(F_2)$ and $\hat{A}_1 = 0$ by choosing

$$\varkappa = \sqrt{F_2(F_2 - XA_1)}\,, \quad \varpi = \frac{F_2 A_1}{\sqrt{F_2(F_2 - XA_1)}}\,, \tag{6.186}$$

corresponding to the metric

$$\hat{g}_{\mu\nu} = \sqrt{F_2(F_2 - XA_1)}\left(g_{\mu\nu} + \frac{A_1}{F_2 - XA_1}\phi_\mu\phi_\nu\right). \tag{6.187}$$

In other words, the part of the original DHOST Lagrangian for $g_{\mu\nu}$ that determines the dynamics of axial perturbations is equivalent to another Lagrangian for $\hat{g}_{\mu\nu}$ where the relevant part is the same as in GR (the other coefficients of the Lagrangian are also modified in the disformal transformation and can remain nonzero, but they are irrelevant for axial perturbations).

By comparing this statement with the result of the previous subsection, it is clear that the effective metric obtained previously should coincide with the disformally related metric for which the dynamics of the axial perturbations is the same as in GR. Let us check this explicitly. The disformal transformation (6.182) applied to the static spherically metric (2.1) and scalar field (2.2) yields

$$\mathrm{d}\hat{s}^2 = -(\varkappa A - \varpi q^2)\left(\mathrm{d}t - \varpi\frac{q\psi'}{\varkappa A - \varpi q^2}\,\mathrm{d}r\right)^2 + \varkappa\left(\frac{1}{B} + \varpi\frac{A\psi'^2}{\varkappa A - \varpi q^2}\right)\mathrm{d}r^2 + \varkappa C\,\mathrm{d}\Omega^2\,. \quad (6.188)$$

Substituting (6.186) and noting

$$\sqrt{F_2(F_2 - XA_1)} = |\mathscr{F}|\sqrt{\frac{\Gamma B}{A}}\,, \qquad \frac{F_2 A_1}{\sqrt{F_2(F_2 - XA_1)}} = \frac{\varkappa A_1}{F_2 - XA_1}\,, \qquad (6.189)$$

one recovers the effective metric of eq. (6.170).

Note that the conformal factor in eq. (6.187) is well defined if

$$F_2(F_2 - XA_1) > 0\,. \qquad (6.190)$$

One can check that the same condition guarantees that $\Gamma > 0$. Moreover, after disformal transformation, one recovers the GR action provided $\hat{F}_2 > 0$, otherwise the gravitons are ghost-like. According to eq. (6.184), one must therefore impose the condition

$$F_2 > 0 \qquad (6.191)$$

to avoid ghost instabilities. This agrees with the no-ghost condition given in [210] and [214].

### 6.5.4. Effective metric for different BH solutions

In this section, we discuss the effective metric of the solutions presented in chapter 2, after some general considerations on the properties of the effective metric. For simplicity, we restrict our analysis to the case where the background metric is a BH with $A(r) = B(r)$.

#### 6.5.4.1. Comparison of causal structures

As mentioned earlier, non-gravitational fields (e.g. photons, or any type of matter) are minimally coupled to the metric $g_{\mu\nu}$ and therefore propagate in the background geometry. By contrast, axial gravitons behave as if they propagate (in the GR sense) in the effective metric $\tilde{g}_{\mu\nu}$, as we have seen previously.

The fact that gravitational perturbations and other fields effectively "live" in different geometries might lead to interesting new physical effects or inconsistencies. A simple and straightforward analysis consists in checking that the causal structures associated with the two metrics are compatible, following an analogous analysis [10] in [219].

According to eq. (6.170), the lightcone and the time-like region delimited by it are defined, for the effective metric, by

$$-\Phi(\mathrm{d}t - A\Psi\,\mathrm{d}r_*)^2 + \Phi\Gamma A^2\,\mathrm{d}{r_*}^2 \leq 0\,, \tag{6.192}$$

which is equivalent to

$$\Phi\,(\mathrm{d}t - a_+\,\mathrm{d}r_*)\,(\mathrm{d}t - a_-\,\mathrm{d}r_*) \geq 0\,, \tag{6.193}$$

where we have introduced the coefficients $a_-$ and $a_+$ defined by

$$a_\pm(r) = A(\Psi \pm \sqrt{\Gamma})\,. \tag{6.194}$$

#### 6.5.4.2. Stealth solutions

For the stealth solution, the coefficients in eq. (6.14) are given in eq. (6.60). Substituting these coefficients into eq. (6.170), one finds that the effective metric can be written as another Schwarzschild metric:

$$\mathrm{d}\tilde{s}^2 = -\left(1 - \frac{R_g}{R}\right)\mathrm{d}T^2 + \left(1 - \frac{R_g}{R}\right)^{-1}\mathrm{d}R^2 + R^2\mathrm{d}\Omega^2\,, \tag{6.195}$$

in the new coordinates $(T, R)$ defined as

$$R = (1+\zeta)^{1/4} r\,, \quad T = (1+\zeta)^{-1/4} t_*\,, \quad R_g = (1+\zeta)^{1/4} r_g\,, \tag{6.196}$$

This metric describes a Schwarzschild solution whose horizon is located at $r = r_g$ and therefore is shifted with respect to the horizon of the background metric $\mu$. Note that a similar double-horizon structure was previously studied in [220].

The coefficients $a_\pm$ associated with the lightcone of the effective metric are given here by

$$a_\pm(r) = A(\Psi \pm \sqrt{\Gamma}) = \frac{\zeta\sqrt{r\mu}}{r - r_g} \pm \sqrt{1 + \zeta}\frac{r-\mu}{|r - r_g|}\,. \tag{6.197}$$

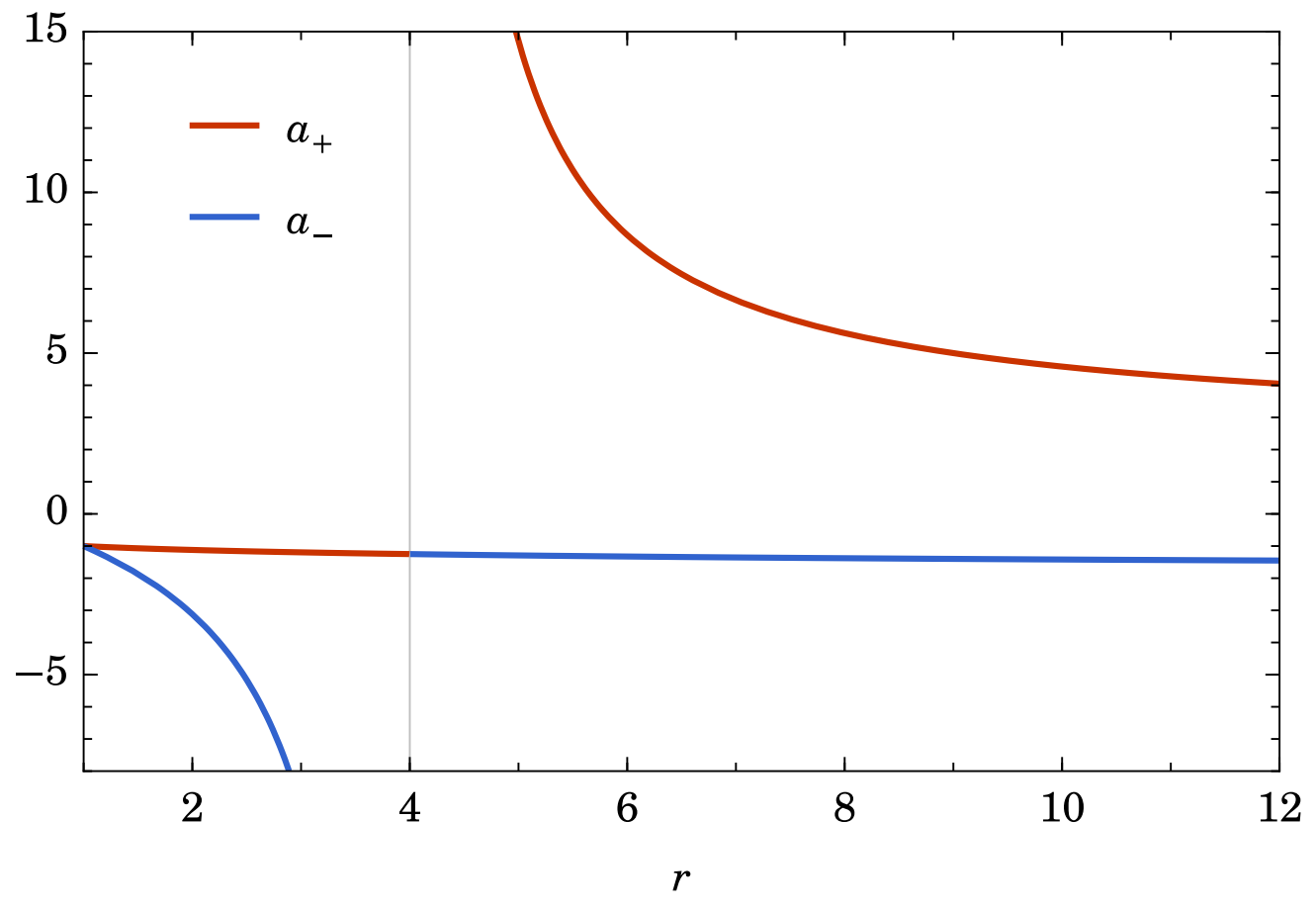

Figure 6.5. – Coefficients $a_\pm(r)$ as functions of $r$, with the choice of parameters $r_s = 1$ and $\zeta = 3$. The vertical line is placed at $r = r_g$.

Their radial dependence is shown in fig. 6.5.

We have also plotted the corresponding lightcones, inside and outside the effective horizon $r_g$, in fig. 6.6. In the same figure, the lightcones associated with the background stealth BH coincides with the standard Minkowski lightcone, since the metric is conformally related to Minkowski in the coordinates $(t, r_*)$ which we are using. We find that the relative position of the lightcones is the same inside and outside $r_g$. This means that the causal structures are compatible since it is possible to define a common spatial hypersurface.

#### 6.5.4.3. BCL solution

We consider now the BCL solution. The effective metric reads

$$\mathrm{d}\tilde{s}^2 = -f_0\sqrt{1+\xi\frac{\mu^2}{r^2}}\left[-A(r)\,\mathrm{d}t^2 + \frac{1}{A(r)}\left(1+\xi\frac{\mu^2}{r^2}\right)\mathrm{d}r^2 + r^2\,\mathrm{d}\Omega^2\right], \quad (6.198)$$

Even though the effective metric is different from the background metric, it still describes a BH geometry whose horizon is the same as the background BH horizon located at $r = r_+$, which is a consequence of the fact that $q$ is 0 for the scalar field [11]. Hence, the effective metric is regular in the domain $]r_+, +\infty[$ and the Schrödinger equation satisfied by the axial graviton can be solved following the usual strategy of GR. One does not expect any instability for the axial gravitational perturbations.

10. In [219], the authors compared the effective metric of the radial scalar perturbation with the physical metric where non-gravitational fields propagate.

11. A more thorough study of the behaviour of axial perturbations at the horizon and its link with stability was proposed in [221].

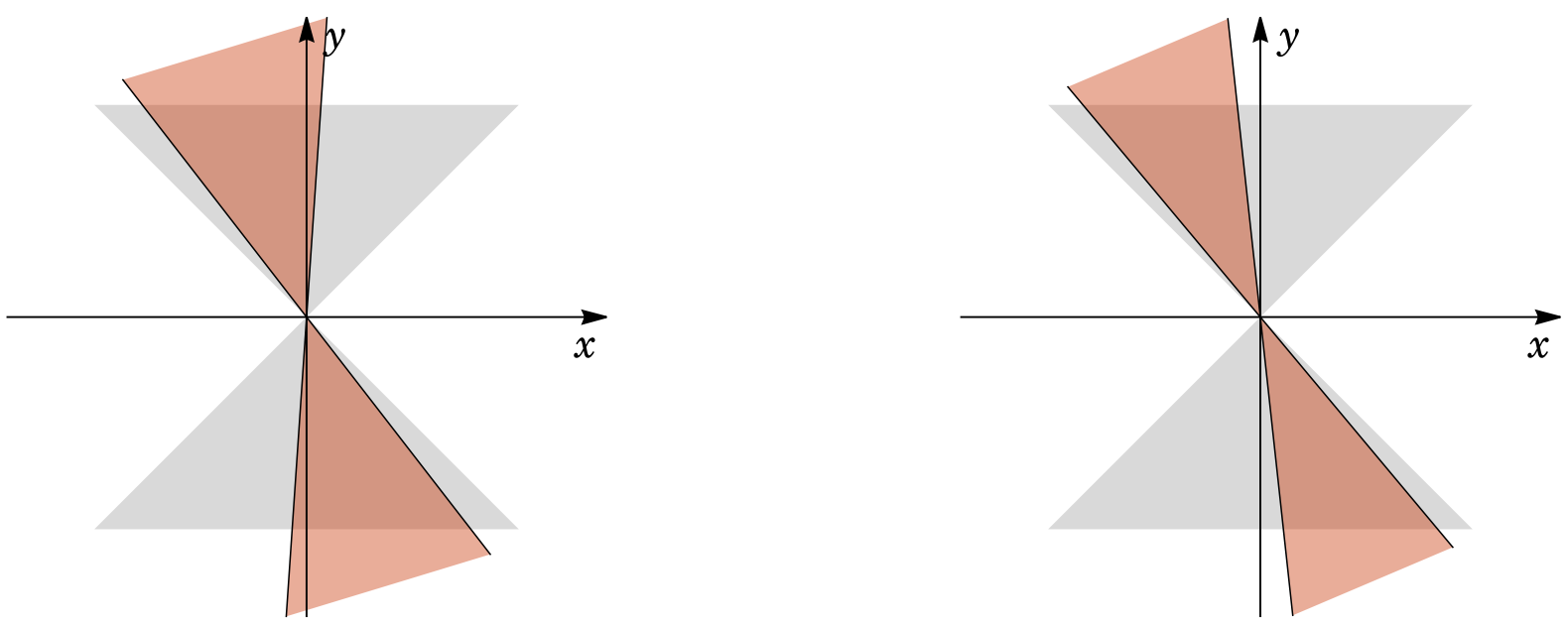

(a) Lightcones for $r > r_g$. (b) Lightcones for $\mu < r < r_g$.

Figure 6.6. – At some point $(t, r_*)$ in spacetime, we plot the lightcones in the $(x = \mathrm{d}r_*, y = \mathrm{d}t)$ plane for a stealth black hole with $\mu = 1$ and $\zeta = 3$. The background lightcones for the background metric are shown in gray. The two cases shown here correspond respectively to $r = 5$ and $r = 3$.

#### 6.5.4.4. 4dEGB solution

Finally, we compute the effective metric of the 4dEGB solution. An immediate calculation leads to

$$\mathrm{d}\tilde{s}^2 = -\frac{1}{z^2}\sqrt{\frac{f\gamma_1^3\gamma_2}{\gamma_3^3}}\,\mathrm{d}t^2 + \frac{1}{z^2}\sqrt{\frac{\gamma_1\gamma_2^3}{f^5\gamma_3^5}}\,\mathrm{d}z^2 + \sqrt{\frac{\gamma_1\gamma_2}{f\gamma_3}}\,\mathrm{d}\Omega^2\,. \qquad (6.199)$$

The effective metric is obviously very different from the background metric and its causal structure can be determined from the analysis of the functions $\gamma_i$, which was done in section 6.3.4. First, as $\gamma_1$ and $\gamma_3$ are positive, one can notice that eq. (6.199) only makes sense when $z \leq z_2$, which imposes $\beta < \beta_c$.

In that case, the behaviour of the effective metric near the outer horizon can be shown to take the form

$$\mathrm{d}\tilde{s}^2 \approx -c_1(z-1)^{1/4}\,\mathrm{d}t^2 + c_2(z-1)^{-5/4}\,\mathrm{d}z^2 + c_3(z-1)^{-1/4}\,\mathrm{d}\Omega^2\,, \qquad (6.200)$$

where the $c_i$ are constants since the functions $\gamma_i$ tends to a constant value at the horizon while $f(z) \approx f_0(z-1)^{1/4}$ with $f_0$ constant.

The Ricci scalar associated with the effective metric (6.200) behaves as $R \approx (z-1)^{-3/4}$ close to the horizon. Henceforth, the effective metric describes a naked singularity. The consequence is that the dynamics of axial modes is very different from the usual case of GR and, in particular, one cannot define ingoing and outgoing modes at the horizon: we recover the results of section 6.4.4. This might lead to stability issues associated with spatial divergences of the perturbations of the metric components.

## Conclusion

In this chapter, we have studied axial perturbations around static and spherically symmetric BH solutions of cubic DHOST theories. We have generalised the Schrödinger equation found in the case of GR in chapter 4 and shown how it can be interpreted as a propagation equation over an effective metric that differs in general from the background metric. Furthermore, we proved that this effective metric can be disformally linked to the background metric in the case of quadratic DHOST theories. This allowed us to recover various stability conditions described in the literature. For concreteness, we applied the results of the chapter to the study of the four BH solutions described in chapter 2, and checked that the results were consistent with the outputs of the algorithm described in chapter 5.

For axial perturbations of the BCL solution, we showed that the behaviours at infinity and at the horizon are perfectly healthy. The effective metric in which GWs propagate is still a BH with the same horizon as the background metric, and no instability is expected in this sector. The conclusions were identical for the EsGB solution.

In the case of the stealth solution, we showed that the axial sector of perturbations behaves similarly to the GR case with a shift of the BH horizon. While this could be expected to lead to instabilities at the Hamiltonian level, due to different orientations of the geodesics of the background and effective metrics, we showed that this was not the case. This does not prove that no other types of instabilities are present in this kind of setup; indeed, some could still appear when one considers the polar sector of perturbations which will be described in chapter 7.

The 4dEGB metric was shown to exhibit pathological perturbations behaviour, for two reasons: first, it is not possible to discriminate between ingoing and outgoing modes at the BH horizon; second, the effective metric seen by axial perturbations possesses a naked singularity at the horizon. The lack of ingoing and outgoing modes at the horizon means that it is not possible to define QNMs. However, it is not necessarily a proof of instability: we were able to link the stability analysis to the one of a different solution described in [216], which also exhibits a naked singularity, and the conclusion was that the perturbations can still be stable.

CHAPTER 7

# POLAR BLACK HOLE PERTURBATIONS

CONTENTS

LET us now turn to the case of polar perturbations. We will study successively the different solutions presented in chapter 2. The main difference with the case of axial perturbations studied in chapter 6 is that the first order system is now four-dimensional since it contains a scalar mode and a gravitational mode, which are coupled. By contrast with the axial case, we have not been able to reduce the system to a 2-dimensional Schrödinger-like equation, so the only option available to us in this case is the asymptotic analysis of the first-order system. We thus use the algorithm given in chapter 5 to obtain the behaviour of the solutions of the system near the horizon and at spatial infinity. This chapter is based on [211, 212].

## 7.1. Perturbation formalism

We choose the same (Zerilli) gauge fixing as usually adopted in GR (see chapter 4), thus the metric perturbations are parametrised by four families of functions $H_0^{\ell m}, H_1^{\ell m}, H_2^{\ell m}$ and $K^{\ell m}$ ($\ell$ and $m$ are integers with $\ell \geq 0$ and $-\ell \leq m \leq \ell$) such that the non-vanishing components of the metric are

$$h_{tt} = A(r)\sum_{\ell,m} H_0^{\ell m}(t,r)Y_{\ell m}(\theta,\varphi), \quad h_{rr} = B(r)^{-1}\sum_{\ell,m} H_2^{\ell m}(t,r)Y_{\ell m}(\theta,\varphi),$$
$$h_{tr} = \sum_{\ell,m} H_1^{\ell m}(t,r)Y_{\ell m}(\theta,\varphi), \quad h_{ab} = \sum_{\ell,m} K^{\ell m}(t,r)g_{ab}Y_{\ell m}(\theta,\varphi)\,, \tag{7.1}$$

where the indices $a, b$ belong to $\{\theta, \varphi\}$. The scalar field perturbation is parametrised by one family of functions according to

$$\delta\phi = \sum_{\ell,m} \delta\phi^{\ell m}(t,r) Y_{\ell m}(\theta, \varphi) \,. \tag{7.2}$$

In the following we will consider only the modes $\ell \geq 2$. The monopole $\ell = 0$ and the dipole $\ell = 1$ require different gauge fixing conditions, which have been explicitly given in the case of GR in section 4.3.3. One can note that while the monopole and dipole did not yield propagating solutions in the case of GR, they would still lead to the propagation of one degree of freedom out of the two contained in the polar sector in the case of DHOST theories, since the scalar mode perturbations would propagate.

In the frequency domain, the linear equations of motion can be written as $\mathscr{E}_{\mu\nu} = 0$ (see section 4.5). Due to spherical symmetry, the equations $\mathscr{E}_{t\varphi}$, $\mathscr{E}_{r\varphi}$ and $\mathscr{E}_{\varphi\varphi}$ are obviously equivalent to $\mathscr{E}_{t\theta}$, $\mathscr{E}_{r\theta}$ and $\mathscr{E}_{\theta\theta}$ respectively. The remaining seven equations depend on five functions: $H_0$, $H_1$, $H_2$, $K$ and $\delta\phi$. Moreover, the equation $\mathscr{E}_{\theta\varphi}$ is algebraic, as in GR, and yields $H_2$.

Among the remaining six equations for four independent functions, it turns out that the four equations $\mathscr{E}_{tr}$, $\mathscr{E}_{rr}$, $\mathscr{E}_{t\theta}$ and $\mathscr{E}_{r\theta}$ are independent, first-order with respect to the radial coordinate and that they imply the last two ones, $\mathscr{E}_{tt}$ and $\mathscr{E}_{\theta\theta}$.

Contrary to GR, the remaining four equations cannot be reduced further because the system now contains two coupled degrees of freedom, the usual polar gravitational mode and the scalar mode. Hence, we obtain a system of four first order equations for the four functions $H_0$, $H_1$, $K$ and $\delta\phi$, of the usual form

$$\frac{\mathrm{d}Y}{\mathrm{d}r} = MY, \tag{7.3}$$

with the column vector

$$Y = \begin{pmatrix} K, & \delta\phi, & H_1, & H_0 \end{pmatrix}^{\top} \,. \tag{7.4}$$

One can note than in some cases (see sections 7.2 and 7.4), it will be useful to renormalise the scalar perturbation $\delta\phi$ to simplify the perturbation equations. In such cases, this renormalised perturbation is called $\chi$ and it will be given explicitly in each case.

For each BH solution presented in chapter 2, we will write the asymptotic expansion of the four-dimensional column vector $Y(r)$ using the algorithm given in chapter 5 as a linear combination of four modes, which we will denote $\mathfrak{g}(r)$ for the modes analogous to the axial gravitational modes and $\mathfrak{s}(r)$ for the additional modes coming from the scalar degree of freedom [1]. There will be two families of such modes, one at spatial infinity and the other one near the horizon, which will be distinguished by the subscript ∞ or h, respectively. When possible, we will give to each mode a subscript + or − depending on the direction of propagation of the associated wave. When no such dichotomy is possible, we will simply use subscripts 1 and 2. We will focus on the *leading order* behaviour of the modes.

## 7.2. BCL solution

### 7.2.1. First-order system

In the case of the BCL solution, the algebraic equation for $H_2$ yields

$$H_2 = \frac{\mu(r+\xi\mu)}{r^3}\delta\phi + \frac{r^2+\xi\mu^2}{r^2}H_0\,. \tag{7.5}$$

One then writes the first-order system (7.3) using the renormalized scalar perturbation $\chi$ defined as

$$\chi(r) \equiv \frac{f_1}{f_0\sqrt{A(r)}}\delta\phi(r)\,. \tag{7.6}$$

The explicit form of the square matrix $M$ can then be read off from the equations of motion:

$$M = \begin{pmatrix} -\frac{1}{r}+\frac{\mathcal{U}}{2r^3A} & \frac{\mathcal{U}}{r^4} & \frac{i(1+\lambda)}{\omega r^2} & \frac{\mathcal{V}}{r^3} \\ \frac{\omega^2r^2}{A^2}-\frac{\lambda}{A}-\frac{\mu}{2rA}+\frac{\mu^2\mathcal{S}}{4r^4A^2} & -\frac{2}{r}-\frac{\mathcal{U}\mathcal{V}}{2r^5A} & -\frac{i\omega r}{A}+\frac{i(1+\lambda)\mathcal{U}}{2r^3\omega A} & -\frac{\lambda}{A}-\frac{3\mathcal{U}}{2r^3A}-\frac{\xi^2\mu^4}{2r^4A} \\ -\frac{i\omega\mathcal{V}}{r^2A} & \frac{2i\omega}{r}-\frac{i\omega\mathcal{U}}{r^3A} & -\frac{\mathcal{U}}{r^3A} & -\frac{i\omega\mathcal{V}}{r^2A} \\ -\frac{1}{r}+\frac{\mathcal{U}}{2r^3A} & \frac{2}{r^2}-\frac{\mathcal{U}^2}{2r^6A} & -\frac{i\omega}{A}+\frac{i(1+\lambda)}{\omega r^2} & \frac{1}{r}-\frac{\mathcal{U}}{2r^3A}-\frac{\mathcal{U}\mathcal{V}}{2r^5A} \end{pmatrix}, \tag{7.7}$$

where we have introduced the functions

$$\begin{aligned} \mathcal{U}(r) &\equiv \mu(r+\xi\mu)\,, \\ \mathcal{V}(r) &\equiv r^2+\xi\mu^2\,, \\ \mathcal{S}(r) &\equiv r^2+2\xi r(2\mu-r)+2\xi^2\mu^2\,. \end{aligned} \tag{7.8}$$

We analyse below the asymptotic behaviours of the above system, first at spatial infinity and then near the horizon.

1. If it is not possible to differentiate between these two families, we will write the modes using the general notation $\mathfrak{p}(r)$.

### 7.2.2. At spatial infinity

The expansion of the matrix $M$ in eq. (7.7) at spatial infinity is of the form

$$M(r) = r^2 M_2 + r M_1 + M_0 + \frac{1}{r} M_{-1} + \mathcal{O}\left(\frac{1}{r^2}\right), \tag{7.9}$$

where the matrices $M_i$ can easily be inferred from eq. (7.7).

The leading matrix $M_2$ contains a single non-zero entry, $(M_2)_{21} = \omega^2$, and is thus nilpotent. To diagonalise the system, one can follow step by step the algorithm presented in chapter 5. Here, however, in order to shorten the procedure, we first adopt a "customised" strategy by considering a transformation of the form

$$Y = P_{(1)} Y^{(1)}, \qquad P_{(1)} = \mathrm{Diag}(r^{p_1}, r^{p_2}, r^{p_3}, r^{p_4}) \tag{7.10}$$

and choosing the powers $p_i$ that simplify the system the most. With the choice

$$p_1 = 0, \quad p_2 = 2, \quad p_3 = p_4 = 1, \tag{7.11}$$

one finds that the system becomes

$$\frac{\mathrm{d}Y^{(1)}}{\mathrm{d}r} = M^{(1)} Y^{(1)}, \quad M^{(1)} = M_0^{(1)} + \frac{1}{r} M_{-1}^{(1)} + \mathcal{O}\left(\frac{1}{r^2}\right), \tag{7.12}$$

where the two matrices $M_0^{(1)}$ and $M_{-1}^{(1)}$ have the simple expressions

$$M_0^{(1)} = \begin{pmatrix} 0 & 0 & 0 & 1 \\ \omega^2 & 0 & -i\omega & 0 \\ 0 & 2i\omega & 0 & -i\omega \\ 0 & 0 & -i\omega & 0 \end{pmatrix},$$

$$M_{-1}^{(1)} = \begin{pmatrix} -1 & \mu & i(1+\lambda)/\omega & 0 \\ 2\omega^2\mu & -4 & -i\omega\mu & \lambda \\ -i\omega & -i\omega\mu & -1 & -i\omega\mu \\ 0 & 2 & -i\omega\mu & 0 \end{pmatrix}. \tag{7.13}$$

Following now the algorithm of chapter 5, two additional steps are needed to obtain a fully diagonalised system (up to order $r^0$), given by

$$\frac{\mathrm{d}\tilde{Y}}{\mathrm{d}r} = \tilde{M}\tilde{Y} \quad \text{with} \quad Y = \tilde{P}\tilde{Y}, \tag{7.14}$$

where the (combined) transfer matrix $\tilde{P}$ and the expansion of $\tilde{M}$ are given by

$$\tilde{P} = \begin{pmatrix} p_1+q_1 & p_1-q_1 & r_1+s_1 & r_1-s_1 \\ 0 & 0 & r_2+s_2 & r_2-s_2 \\ p_3+q_3 & p_3-q_3 & r_3+s_3 & r_3-s_3 \\ p_3+q_3 & -p_3+q_3 & r_4+s_4 & r_4-s_4 \end{pmatrix}, \tag{7.15}$$

with the coefficients defined by

$$
\begin{aligned}
p_1 &= -\frac{2\lambda}{3r\omega^2}, \quad q_1 = \frac{i(3r-2\mu)}{3r\omega}, \quad r_1 = \frac{27-10\lambda}{12r\omega^2}, \quad s_1 = -\frac{\sqrt{2}(12r+7\mu)}{24r\omega}, \\
r_2 &= \frac{\sqrt{2}(3-2\lambda)r}{8\omega}, \quad s_2 = \frac{(12r+7\mu)r}{8}, \quad p_3 = \frac{3r+\mu}{3}, \quad q_3 = -\frac{i\lambda}{3\omega}, \\
r_3 &= \frac{i(2\lambda-9)}{6\omega}, \quad s_3 = \frac{i\sqrt{2}(11\mu-12r)}{12}, \\
r_4 &= \frac{12r-5\mu}{12}, \quad s_4 = -\frac{\sqrt{2}(27+2\lambda)}{12\omega}.
\end{aligned}
\tag{7.16}
$$

Integrating this asymptotic system yields

$$
\tilde{Y}(r) \approx \left(c_-\,\mathfrak{g}_-^\infty(r),\ \ c_+\,\mathfrak{g}_+^\infty(r),\ \ d_-\,\mathfrak{s}_-^\infty(r),\ \ d_+\,\mathfrak{s}_+^\infty(r)\right)^{\mathsf{T}}, \tag{7.17}
$$

where $c_\pm$ and $d_\pm$ are constants and

$$
\mathfrak{g}_\pm^\infty(r) = e^{\pm i\omega r} r^{\pm i\omega\mu}, \quad \mathfrak{s}_\pm^\infty(r) = e^{\pm\sqrt{2}\omega r} r^{-3+\omega\mu/\sqrt{2}}. \tag{7.18}
$$

The first two components are very similar to the components of the asymptotic solution obtained in the axial sector (see section 6.4.1) and it is therefore natural to identify these modes with the usual outgoing and ingoing gravitational modes. By contrast, the last two components have an unusual form. If we return to the original variables, via eqs. (7.10) and (7.15), we find that the asymptotic behavior of the (renormalized) scalar perturbation $\chi$ (7.6) reads

$$
\chi(r) = \frac{3}{2r}\left[d_-\, r^{-\frac{\omega\mu}{\sqrt{2}}}\, e^{-\sqrt{2}\omega r} - d_+\, r^{\frac{\omega\mu}{\sqrt{2}}}\, e^{\sqrt{2}\omega r}\right]\left(1+\mathcal{O}(1/r)\right). \tag{7.19}
$$

The behaviour exhibited by this perturbation appears problematic, as it is associated with an effective metric which does not possess the appropriate causal structure. Indeed, the asymptotic solution (7.19) can be related to an equation of motion for $\tilde{\chi} \equiv r\chi$ of the form

$$
\frac{\partial^2\tilde{\chi}}{\partial t^2} + \frac{\partial^2\tilde{\chi}}{\partial\tilde{r}^2} \approx 0, \quad \text{with} \quad \tilde{r} = \sqrt{2}\left(r + \frac{\mu}{2}\ln r\right), \tag{7.20}
$$

which does not correspond to a wave equation. This non-hyperbolicity is usually associated with a ghost or gradient instability.

For a more direct — although less rigorous — approach to this problem, it is instructive to examine the perturbations of the scalar field on the fixed background geometry, in other words to ignore the backreaction of the scalar field perturbations on the metric. In this case, the equation of motion for the scalar field perturbation $\chi$ is of the form

$$
\frac{\partial^2\chi}{\partial t^2} + \frac{1}{2}A(r)\frac{\partial^2\chi}{\partial r^2} + \frac{1}{r}\left(1+\frac{\xi\mu^2}{2r^2}\right)\frac{\partial\chi}{\partial r} - W(r)\chi = 0, \tag{7.21}
$$

where $W(r)$ is some potential, given explicitly in section 7.6. Since $A > 0$, this equation has the structure of an elliptic equation, similar to eq. (7.20). In fact, it is even possible to show that the asymptotic behaviour (7.19) can be directly recovered from eq. (7.21), as shown in section 7.6.

Finally, one can recover the asymptotic behaviour or $Y$ at infinity using eq. (7.17) and the $r \longrightarrow +\infty$ limit of $\tilde{P}$. However, expanding the matrix $M^{(1)}$ up to order $1/r$, as was done in eq. (7.12), while sufficient to obtain the asymptotic behaviour, is not enough to obtain the asymptotic expression of all components of $Y$. Indeed, two entries in $\tilde{P}$ are zero, which means that the contributions are of subleading order. In order to obtain the asymptotic behaviour of $Y$, one executes two more steps of the algorithm and obtains

$$Y \approx \begin{pmatrix} \frac{i}{\omega} & -\frac{i}{\omega} & -\frac{1}{\sqrt{2}\omega} & \frac{1}{\sqrt{2}\omega} \\ \frac{i\mu^2\xi}{2\omega r} & -\frac{i\mu^2\xi}{2\omega r} & \frac{3r^2}{2} & \frac{3r^2}{2} \\ r & -r & -i\sqrt{2}r & i\sqrt{2}r \\ r & r & r & r \end{pmatrix} \begin{pmatrix} c_- \, \mathfrak{g}_-^\infty(r) \\ c_+ \, \mathfrak{g}_+^\infty(r) \\ d_- \, \mathfrak{s}_-^\infty(r) \\ d_+ \, \mathfrak{s}_+^\infty(r) \end{pmatrix} . \tag{7.22}$$

### 7.2.3. Near the horizon

To obtain the asymptotic behaviour near the horizon, we use the small parameter $\varepsilon$ defined by $\varepsilon \equiv r - r_+$. It is convenient to make the following initial change of vector to simplify the analysis:

$$Y = P_{(1)} Y^{(1)} , \qquad P_{(1)} = \begin{pmatrix} 1 & 0 & 0 & 0 \\ 0 & 1/\varepsilon & 0 & 0 \\ 0 & 0 & 1/\varepsilon & 0 \\ 0 & 0 & 0 & 1/\varepsilon \end{pmatrix} . \tag{7.23}$$

The matrix $M^{(1)}$ associated to the system for $Y^{(1)}$ admits a very simple asymptotic expansion, of the form

$$M^{(1)} = \frac{1}{\varepsilon} M_0^{(1)} + \mathcal{O}(1) , \tag{7.24}$$

where the matrix $M_0^{(1)}$ is given by

$$M_0^{(1)} = \begin{pmatrix} \frac{1}{2} & \frac{\eta}{r_0 r_+} & i\frac{1+\lambda}{\omega r_+^2} & \frac{\eta^2}{r_+} \\ \frac{r_+^2}{4} + \frac{\omega^2 r_0^2 r_+^2}{\eta^2} & \frac{\eta^2}{2} & i\frac{1+\lambda}{2\omega} - \frac{i\omega r_+ r_0}{\eta} & \frac{5-\eta^2+2\eta^4+4\lambda}{4\eta} r_0 \\ 0 & -i\omega & 0 & -i\eta\omega r_0 \\ 0 & -\frac{\eta}{2r_0} & -\frac{i\omega r_0}{\eta} & \frac{1-\eta^2}{2} . \end{pmatrix} . \tag{7.25}$$

Even though the expression of $M^{(1)}$ is relatively complex, it can be transformed into a simple Jordan block form with two Jordan blocks. Indeed, we make a new change of variable $Y^{(1)} = P_{(2)} Y^{(2)}$ where $P_{(2)}$ transforms $M_0^{(1)}$ according to

$$M_0^{(1)} = P_{(2)} \begin{pmatrix} -i\omega r_0 & 0 & 0 & 0 \\ 0 & +i\omega r_0 & 0 & 0 \\ 0 & 0 & 1/2 & 1 \\ 0 & 0 & 0 & 1/2 \end{pmatrix} P_{(2)}^{-1} , \tag{7.26}$$

The solution for $Y^{(2)}$ is obtained immediately and reads

$$Y^{(2)}(r) \approx \left( c_- \mathfrak{g}_-^{\mathrm{h}}(\varepsilon), \quad c_+ \mathfrak{g}_+^{\mathrm{h}}(\varepsilon), \quad d_1 \mathfrak{s}_1^{\mathrm{h}}(\varepsilon) + d_2 \mathfrak{s}_2^{\mathrm{h}}(\varepsilon), \quad d_2 \mathfrak{s}_1^{\mathrm{h}}(\varepsilon) \right)^{\top} , \tag{7.27}$$

where again $c_\pm$, $d_1$ and $d_2$ are constants and

$$\mathfrak{g}_\pm^{\mathrm{h}}(\varepsilon) = \varepsilon^{\pm i\omega r_0} , \quad \mathfrak{s}_1^{\mathrm{h}}(\varepsilon) = \sqrt{\varepsilon} \quad \text{and} \quad \mathfrak{s}_2^{\mathrm{h}}(\varepsilon) = \sqrt{\varepsilon} \ln \varepsilon . \tag{7.28}$$

The asymptotic expansion at the horizon of the original variable $Y$ whose components are the metric and scalar perturbations (7.7) is obtained directly from the matrix of change of variables $P$ such that $Y = PY^{(2)}$. It is given by the product $P = P_{(1)} P_{(2)}$ which reads after a direct calculation

$$P = \frac{1}{\varepsilon} \begin{pmatrix} -\frac{2\rho\varepsilon(i\eta r_+\omega+1+\lambda)}{\omega r_+^{3/2}\Delta_1} & \frac{2\rho\varepsilon(i\eta r_+\omega-1-\lambda)}{\omega r_+^{3/2}\Delta_2} & -\frac{2\rho\varepsilon((3+2\lambda)r_++r_-)}{r_+\Delta_3} & i\varepsilon\frac{\Delta_4}{r_+^2\Delta_3} \\ -\frac{2i\eta r_- r_+^{3/2}}{\Delta_1} & \frac{2i\eta r_- r_+^{3/2}}{\Delta_2} & -\frac{r_+(r_++2r_-)}{\rho} & \frac{i}{2\omega} \\ -\frac{ir_+^{1/2}(\rho+2i\eta r_+^2\omega)}{\Delta_1} & -\frac{ir_+^{1/2}(\rho+2i\eta r_+^2\omega)}{\Delta_2} & 0 & 1 \\ 1 & 1 & 1 & 0 \end{pmatrix} , \tag{7.29}$$

where we introduced the notations $\rho \equiv r_+ + r_-$ and

$$\begin{aligned} &\Delta_1 \equiv \sqrt{r_+}(2\omega r_+^2 + i\eta\rho) , \quad \Delta_2 \equiv \sqrt{r_+}(2\omega r_+^2 - i\eta\rho) , \quad \Delta_3 \equiv \rho^2 + 4\omega^2 r_+^4 , \\ &\Delta_4 \equiv 4(2r_+ + 3r_-)r_+^3\omega^2 - (1+\lambda)\rho^2 . \end{aligned} \tag{7.30}$$

The behaviour of the first two components in eq. (7.27) is the same as in the axial case, and one can thus identify them with the ingoing and outgoing gravitational modes. By contrast, the behaviour of the last two components is very peculiar and is related to the presence of the scalar field degree of freedom. As in the spatial infinity limit, these modes do not seem to correspond to a second-order equation respecting the usual four-dimensional causal structure, which indicates that the effective metric near the horizon, in which the perturbations propagate, is pathological.

## 7.3. Stealth solution

### 7.3.1. First-order system

The asymptotic behaviour of polar perturbations for stealth Schwarzschild can be computed with the same procedure as in the BCL case, even if it turns out to be technically more involved, with rather tedious calculations. Since the details are not very illuminating, we simply give the final results in this section (more details can be found in appendix A.1). Furthermore, to simplify the analysis, we will consider theories where only one of the parameters $\alpha$, $\beta$ or $\gamma$ defined in section 2.3.3 is nonzero.

In the case of stealth solutions, the equation $\mathscr{E}_{\theta\varphi}$ gives

$$\frac{r(1+2q^2\alpha)-\mu}{r-\mu}H_0 - 4q^2\alpha\frac{\sqrt{r\mu}}{r-\mu}H_1 - \frac{r-(1+2q^2\alpha)\mu}{r-\mu}H_2 - 2q\alpha\sqrt{\frac{\mu}{r^3}}\delta\varphi = 0\,. \tag{7.31}$$

The four equations $\mathscr{E}_{tr}$, $\mathscr{E}_{rr}$, $\mathscr{E}_{t\theta}$ and $\mathscr{E}_{r\theta}$ therefore form a complete dynamical system for $H_0$, $H_1$, $K$ and $\delta\phi$, as explained in section 7.1. It can be written in the form

$$M_A\frac{\mathrm{d}X}{\mathrm{d}r} = M_B X\,, \qquad X \equiv {}^T\left(K,\ \ \delta\phi,\ \ H_1,\ \ H_0\right)\,, \tag{7.32}$$

where the expressions of the matrices $M_A$ and $M_B$ are quite cumbersome. They are given explicitly in appendix A.1.

### 7.3.2. Asymptotic behaviour

For the theories with $\beta \neq 0$ or $\gamma \neq 0$, we find the following common behaviours:

- at spatial infinity:

$$\mathfrak{g}^{\infty}_{\pm}(r) \approx e^{\pm i\omega r} r^{\pm i\omega\mu}\,, \tag{7.33}$$

- near the horizon:

$$\mathfrak{g}^{\mathrm{h}}_{\pm}(\varepsilon) \approx \varepsilon^{\pm i\omega\mu}\,, \quad \mathfrak{s}^{\mathrm{h}}_{1}(\varepsilon) \approx \varepsilon^{-i\omega\mu}\,, \quad \mathfrak{s}^{\mathrm{h}}_{2}(\varepsilon) \approx \varepsilon^{-i\omega\mu} \quad . \tag{7.34}$$

By contrast, the behaviours of the "scalar" modes at spatial infinity are different in the two cases:

$$\begin{aligned} \beta \neq 0: &\quad \mathfrak{s}^{\infty}_{1}(r) \approx e^{-2i\omega\mu z} z^{-7+2i\sqrt{\lambda}}\,, \quad \mathfrak{s}^{\infty}_{2}(r) \approx e^{-2i\omega\mu z} z^{-7-2i\sqrt{\lambda}} \\ \gamma \neq 0: &\quad \mathfrak{s}^{\infty}_{1}(r) \approx e^{-2i\omega\mu z(z^2/3+1)} z^{-5}\,, \quad \mathfrak{s}^{\infty}_{2}(r) \approx e^{-2i\omega\mu z(z^2/3+1)} z^{-5}\ln z\,, \end{aligned} \tag{7.35}$$

where we defined $z$ via $z \equiv \sqrt{r/\mu}$.

One observes that, in some cases, the + and − modes share exactly the same leading behaviour at spatial infinity or near the horizon. As a consequence, the usual distinction between ingoing and outgoing modes becomes impossible, at least at leading order, and might require to consider the next orders in the asymptotic expansion. It is also worth noting that, in the cases $\gamma \neq 0$ and $\beta \neq 0$, the equations for the perturbations drastically simplify, as shown in section 7.3.3 for $\gamma \neq 0$, and the asymptotic behaviour of the scalar field can be obtained from the perturbed conservation equation

$$\nabla_\mu(\delta X \phi^\mu) = \frac{1}{\sqrt{-g}}\partial_\mu\left(\sqrt{-g}\,\delta X \phi^\mu\right) = 0\,, \tag{7.36}$$

where $g_{\mu\nu}$ is the Schwarzschild metric and $\delta X$ is the perturbation of $X = \phi_\mu\phi^\mu$. Remarkably this equation can be solved explicitly (at least in the case $\gamma \neq 0$) and its solution reproduces exactly the asymptotic behaviour of the scalar field derived from the analysis of the first order system.

Finally, in the case $\alpha \neq 0$, we find the following asymptotic behaviours at spatial infinity:

$$\begin{aligned} &\mathfrak{g}^\infty_\pm(r) \approx e^{\pm i\omega r_* + 2i\omega\zeta\sqrt{\mu r}}\,, \\ &\mathfrak{s}^\infty_1(z) \approx e^{-2i\omega\mu z} z^{-7+2i\sqrt{\lambda}}\,, \quad \mathfrak{s}^\infty_2(z) \approx e^{-2i\omega\mu z} z^{-7-2i\sqrt{\lambda}}\,, \end{aligned} \tag{7.37}$$

where $r_*$ is the coordinate introduced in eq. (6.67). For the "gravitational" modes, one can clearly identify the ingoing and outgoing modes, and the term proportional to $\sqrt{\mu r}$ in the exponential of $\mathfrak{g}^\infty_\pm(r)$ could be absorbed by a time redefinition of the form (6.18). At the horizon, the study of the asymptotic behaviour is more subtle because the axial modes see a shifted horizon at $r = r_g$, as proven in section 6.5.

We will therefore restrict our discussion here to this horizon, where the axial modes behave as in GR. Near $r = r_g$, we find

$$\mathfrak{g}^{\mathrm{h}}_+(\varepsilon) \approx \varepsilon^{2i\omega(1+\zeta)^{3/2}\mu} \approx e^{2i\omega r_*} \quad \text{and} \quad \mathfrak{g}^{\mathrm{h}}_-(\varepsilon) \approx 1\,, \quad \text{where} \quad \varepsilon \equiv r - r_g\,. \tag{7.38}$$

We thus recover exactly the same behaviour as for the axial modes obtained in section 6.4.2. Performing the time shift used in this section and detailed in eq. (6.18), the above modes in eq. (7.38) would become

$$\mathfrak{g}^{\mathrm{h}}_\pm(\varepsilon) \approx e^{\pm i\omega r_*}\,, \tag{7.39}$$

which can be interpreted as ingoing and outgoing modes. In summary, the polar and axial "gravitational" modes have similar asymptotic properties, which are more easily interpreted in the effective metric with horizon at $r = r_g$. This strongly indicates that the gravitational modes, whether they are polar or axial, propagate in the same effective metric in this case. However, we find for scalar perturbations that the modes contain no singularity at $r = r_g$: one has

$$\mathfrak{s}^{\mathrm{h}}_1(\varepsilon) = \mathfrak{s}^{\mathrm{h}}_2(\varepsilon) = 1\,, \tag{7.40}$$

which seems to indicate that scalar modes do not see the shifted horizon at $r = r_g$.

We conclude by mentioning that a detuning of the DHOST degeneracy conditions (see section 3.3.1), called "scordatura", was proposed in [222] as a solution to the strong coupling problem of the stealth solutions. In order to include this type of model, the method developed here would need to be extended. Indeed, if the degeneracy conditions are not satisfied, the perturbation system contains higher order equations. They can nevertheless be recast into a higher-dimensional first-order system, to which we can apply our method.

### 7.3.3. K-essence case

We study here in more details the case $\gamma \neq 0$. In this case, the action (1.25) reduces to the sum of the Einstein-Hilbert term supplemented with a so-called K-essence term and simply reads

$$S[g_{\mu\nu}, \phi] = \int \mathrm{d}^4x \sqrt{-g} \left( R + \frac{\gamma}{2}(X + q^2)^2 \right) . \tag{7.41}$$

Following the notations and the procedure we described previously, we can compute the corresponding polar perturbations equations about the stealth solution. As expected, they can be cast into a form very similar to those of GR, with three first order equations

$$\begin{aligned}
&K' - \frac{1}{r}H_0 - \frac{i(\lambda+1)}{r^2\omega}H_1 + \frac{2r-3\mu}{2r(r-\mu)}K = \frac{iq^2\gamma\sqrt{r\mu}}{\omega(r-\mu)}\delta X,\\
&H_1' + \frac{ir\omega}{r-\mu}H_0 + \frac{\mu}{r(r-\mu)}H_1 + \frac{ir\omega}{r-\mu}K = 0,\\
&H_0' - K' + \frac{\mu}{r(r-\mu)}H_0 + \frac{ir\omega}{r-\mu}H_1 = 0.
\end{aligned} \tag{7.42}$$

along with one algebraic relation,

$$\begin{aligned}
0 = \left(2\lambda + \frac{3\mu}{r}\right)H_0 + \left(\frac{i(\lambda+1)\mu}{r^2\omega} - 2ir\omega\right)H_1 \\
+ \left(-\frac{4r\lambda(r-\mu) + 2r\mu - 3\mu^2}{2r(r-\mu)} + \frac{2r^3\omega^2}{r-\mu}\right)K \\
+ \left(\frac{2q^2\gamma r^2\mu}{r-\mu} + \frac{iq^2\gamma\sqrt{r\mu}\mu}{\omega(r-\mu)}\right)\delta X,
\end{aligned} \tag{7.43}$$

where we have chosen to keep explicitly $\delta X$, the linear perturbation of $X$. The perturbation $\delta X$ can also be expressed in terms of $\delta\phi$, $H_0$, $H_1$ and $K$:

$$\delta X = -\frac{q^2(\mu+r)}{r-\mu}H_0 + \frac{2q^2\sqrt{\mu r}}{r-\mu}H_1 + 2q\sqrt{\frac{\mu}{r}}\delta\phi' + \frac{2iqr\omega}{r-\mu}\delta\phi . \tag{7.44}$$

At this stage, it is possible to treat eqs. (7.42) and (7.43) in the same way we have treated the system for polar perturbations in GR (see chapter 4). We first solve the algebraic equation (7.43) for $H_0$ and then substitute the solution into the first two differential equations (7.42). Hence, we obtain a system of the form

$$\frac{\mathrm{d}Y}{\mathrm{d}r} - M(r)Y = \frac{q^2\gamma\,\delta X}{(r-\mu)(2r\lambda+3\mu)}\begin{pmatrix} 2r^2\mu - 2i\sqrt{r\mu}(r\lambda+\mu)/\omega \\ \mu r^2\sqrt{r\mu}/(r-\mu) - 2ir^4\mu\omega/(r-\mu)\end{pmatrix}, \tag{7.45}$$

where $Y \equiv \begin{pmatrix} K, & H_1\end{pmatrix}^\top$ and $M(r)$ is the matrix entering in the dynamical system of polar perturbations in GR whose expression has been computed in chapter 4:

$$M(r) = \frac{1}{3\mu+2\lambda r}\begin{pmatrix} \frac{\mu(3\mu+(\lambda-2)r)-2r^4\omega^2}{r(r-\mu)} & \frac{2i(\lambda+1)(\mu+\lambda r)+2ir^3\omega^2}{r^2} \\ \frac{ir(9\mu^2-8\lambda r^2+8(\lambda-1)\mu r)+4ir^5\omega^2}{2(r-\mu)^2} & \frac{2r^4\omega^2-\mu(3\mu+3\lambda r+r)}{r(r-\mu)}\end{pmatrix}. \tag{7.46}$$

We can therefore interpret the system (7.45) as describing the dynamics of unmodified polar perturbations in GR on which the scalar field acts like a source.

Finally, it is possible to obtain a fully decoupled equation for the perturbation $\delta X$. For that one replaces the expressions of $H_0'$, $H_1'$ and $K'$ (computed from eq. (7.45) or the algebraic equation) into (7.42). After a direct calculation, one obtains

$$\begin{aligned} &ir^2\left(\sqrt{r\mu}-2ir^2\omega\right)\delta X'(r) \\ &\quad+\left(\frac{3}{2}ir\sqrt{r\mu}+r^3\left(3-\frac{r}{r-\mu}\right)\omega+\frac{2ir^5}{r-\mu}\sqrt{\frac{r}{\mu}}\omega^2\right)\delta X(r)=0\,, \end{aligned} \tag{7.47}$$

which, after some simplifications, becomes [2]

$$2\sqrt{\mu}(r-r_s)r\,\delta X'(r) + \sqrt{r_s}\left(3(r-r_s)+2ir^2\sqrt{r/r_s}\right)\delta X(r) = 0\,. \tag{7.49}$$

The equation for $\delta X(r)$ can be solved explicitly and one finds

$$\delta X(r) = \frac{C}{r^{3/2}}\left(\frac{\sqrt{r}+\sqrt{\mu}}{\sqrt{r}-\sqrt{\mu}}\right)^{i\omega\mu}\exp\left(-\frac{2}{3}i\omega(r+3\mu)\sqrt{r/\mu}\right), \tag{7.50}$$

where $C$ is an integration constant. Hence, the asymptotics of $\delta X$ are deduced immediately and one obtains

$$\delta X(r) \approx \frac{C}{z^3}\exp\bigl(-2i\omega z\mu(z^2/3+1)\bigr)(1+\mathcal{O}(1/z))\,, \tag{7.51}$$

2. Notice that such a decoupled equation for $\delta X$ was expected. Indeed, we can directly check that it is exactly the same as the well-known conservation equation (for linear perturbations) in shift-symmetric theories,

$$\nabla_\mu\left(\sqrt{-g}\,\delta X e^{-i\omega t}\phi^\mu\right) = \frac{1}{\sqrt{-g}}\partial_\mu\left(\sqrt{-g}\,\delta X e^{-i\omega t}\phi^\mu\right) = 0\,. \tag{7.48}$$

at infinity, and

$$\delta X(r) \approx D(r-\mu)^{-i\omega\mu}(1+\mathcal{O}(r-\mu)) \,, \tag{7.52}$$

near the horizon, where $D$ is a constant that can be computed trivially.

In order to compute the asymptotic behavior of $\delta\phi$, we need to solve eq. (7.44). At this stage, it is already remarkable to observe that the asymptotic behaviour of $\delta X$ agrees with the asymptotic behaviour of $\delta\phi$ computed in eqs. (7.34) and (7.35) from the first order system.

But, for completeness, let us consider eq. (7.44) which can be viewed as a first order equation for $\delta\phi$ with three sources proportional to $H_0$, $H_1$ and $\delta X$. The first two can be computed from (7.45) and the algebraic equation while the third one has just been computed above. By superposition, the general solution is a combination of three particular solutions (solutions where only one of the three sources is turned on) and one homogeneous solution.

The homogeneous equation is

$$\delta\phi'(r) + \sqrt{\frac{r}{\mu}}\frac{i\omega r}{r-\mu}\delta\phi(r) = 0 \,. \tag{7.53}$$

It can be fully integrated, and the solution is

$$\delta\phi = C\left(\frac{z+1}{z-1}\right)^{i\omega\mu} \exp\left(-2i\omega z\mu(z^2/3+1)\right) \,, \quad z \equiv \sqrt{r/\mu} \,, \tag{7.54}$$

where $C$ is also a constant. We observe that the homogeneous solution for $\delta\phi$ is almost the same as $\delta X$ obtained in eq. (7.51). These two functions only differ by an overall factor $r^{3/2}$. Hence, their behaviours at infinity and at the horizon are exactly the same (up to some integers powers of $z$ that play no role). This means that the homogeneous solution and the particular solution associated with $\delta X$ have the same asymptotics. Moreover, the functions $H_0$ and $H_1$ have their asymptotic behaviour fixed by the modified GR system (7.45): they both behave like GR metric modes at infinity and at the horizon.

In conclusion, $\delta\phi$ can have two different behaviors at infinity and at the horizon (or any linear combination of these two): it can either behave exactly like a metric mode, similarly to $H_0$ and $H_1$; or it can have the behaviour of $\delta X$ computed previously.

These behaviours are exactly the ones found for the decoupled modes (7.35). We understand now why the branches $\mathfrak{s}_1^\infty$ and $\mathfrak{s}_2^\infty$ were the same, as well as the branches $\mathfrak{s}_1^{\mathrm{h}}$ and $\mathfrak{s}_2^{\mathrm{h}}$ : the asymptotic scalar behaviour is set by $\delta X$, and $\delta X$ does not verify a second-order equation but a first-order one. A similar behaviour was found for the theory where $\alpha \neq 0$, which means that such a simplification of the equations may also exist in that case.

## 7.4. EsGB solution

For the EsGB solution, the first-order system describing polar perturbations can be written in the form

$$\frac{\mathrm{d}Y}{\mathrm{d}r} = S\,Y, \quad \text{with} \quad Y = \left(K, \ \ \delta\phi, \ \ H_1, \ \ H_0\right)^\top, \tag{7.55}$$

but the matrix $S$ is singular in the GR limit when $\varepsilon \to 0$. This problem can be avoided by using the functions

$$\chi = \varepsilon\,\delta\phi \quad \text{and} \quad Y = \left(K, \ \ \chi, \ \ H_1, \ \ H_0\right)^\top, \tag{7.56}$$

leading to a well-defined system in the GR limit which takes the form

$$\frac{\mathrm{d}Y}{\mathrm{d}z} = M(z)\,Y, \tag{7.57}$$

where $z$ is the dimensionless coordinate introduced in eq. (2.53). The explicit expression of the equations in given in appendix A.2. Let us now consider in turn the two asymptotic limits.

### 7.4.1. Spatial infinity

We start by computing the expansion of $M$ in powers of $z$; we obtain

$$M = M_{-2}z^2 + M_{-1}z + M_0 + \frac{M_1}{z} + \mathcal{O}\left(\frac{1}{z^2}\right). \tag{7.58}$$

The matrices $M_{-2}$, $M_{-1}$ and $M_0$ take the simple expressions

$$M_{-2} = \begin{pmatrix} 0 & 0 & 0 & 0 \\ a & 0 & 0 & 0 \\ 0 & 0 & 0 & 0 \\ 0 & 0 & 0 & 0 \end{pmatrix}, \qquad M_{-1} = \begin{pmatrix} 0 & 0 & 0 & 0 \\ a + \varepsilon^2\Omega^2/6 & 0 & a/i\Omega & 0 \\ 0 & 0 & 0 & 0 \\ 0 & 0 & 0 & 0 \end{pmatrix}, \tag{7.59}$$

$$M_0 = \begin{pmatrix} 0 & 0 & 0 & 0 \\ a\left(1 - \lambda/\Omega^2\right) - \varepsilon^2\Omega^2/3 & 0 & 0 & a\lambda/\Omega^2 \\ -i\Omega & 0 & 0 & -i\Omega \\ 0 & 0 & -i\Omega & 0 \end{pmatrix}, \tag{7.60}$$

which depend on the coefficient $a$ defined by

$$a = -\frac{\Omega^2}{2} + \frac{73}{120}\rho_2\Omega^2\varepsilon + \frac{\Omega^2\left(13201\rho_2^2 + 62555\rho_3 + 209160\right)}{151200}\varepsilon^2$$
$$\text{and} \quad \rho_3 = \frac{f'''(\psi_\infty)}{f'(\psi_\infty)}, \tag{7.61}$$

while $\rho_2$ has been defined in eq. (2.51). The matrices $M_i$ with $i \geq 1$ are more involved than the three above matrices and we do not give their expressions here. Nonetheless, some of them enter in the algorithm described in chapter 5.

The asymptotical diagonal form at infinity cannot immediately be obtained from eq. (7.57), as the leading order matrix $M_{-2}$ is nilpotent. As discussed in chapter 5, for this special subcase of the algorithm, one must first obtain a *diagonalisable* leading order term, by applying a change of functions parametrised by the matrix

$$P^{(1)} = \mathrm{Diag}(z^{-2}, 1, z^{-2}, z^{-2}) \,, \tag{7.62}$$

which gives a new matrix $M^{(1)}$ whose leading order term is now diagonalisable. The diagonalisation of the leading term can be performed using the transformation

$$P^{(2)} = \begin{pmatrix} 0 & -1 & 0 & -1 \\ 0 & -ia/\Omega & 0 & ia/\Omega \\ 1 & 0 & -1 & 0 \\ 1 & 1 & 1 & 1 \end{pmatrix} , \tag{7.63}$$

which yields a matrix $M^{(2)}$ of the form

$$M^{(2)} = M_0^{(2)} + M_1^{(2)} z^{-1} + \mathcal{O}\left(\frac{1}{z^2}\right) , \qquad M_0^{(2)} = \mathrm{Diag}(-i\Omega, +i\Omega, -i\Omega, +i\Omega) \,. \tag{7.64}$$

One thus finds four modes propagating at speed $c = 1$, two ingoing and two outgoing modes. We expect them to be associated with the scalar and polar gravitational degrees of freedom.

In order to discriminate between the scalar and gravitational modes, it is useful to pursue the diagonalisation up to the next-to-leading order. This can be done by following, step by step, the algorithm of chapter 5, which leads us to introduce the successive matrices $P^{(3)}$ and $P^{(4)}$, with

$$\begin{aligned} P^{(3)} &= I_4 + \frac{i}{2\Omega z} \begin{pmatrix} 0 & 0 & -1 & -2 \\ 0 & 0 & 1 & -(1+2\Omega^2) \\ 1 & 2 & 0 & 0 \\ -1 & 1-2\Omega^2 & 0 & 0 \end{pmatrix} , \\ P^{(4)} &= \begin{pmatrix} -3a+b & 1 & 0 & 0 \\ 1 & 0 & 0 & 0 \\ 0 & 0 & -3a+\bar{b} & 1 \\ 0 & 0 & 1 & 0 \end{pmatrix} , \end{aligned} \tag{7.65}$$

with the complex coefficient $b$ defined by

$$b = -\frac{1}{2} + \frac{\varepsilon^2}{24} i\Omega (1 - 3\Omega^2 - 36 i\Omega) \,. \tag{7.66}$$

Hence, we obtain a new vector $Y^{(4)}$ whose corresponding matrix $M^{(4)}$ is given by

$$\begin{aligned} M^{(4)} &= \mathrm{Diag}(-i\Omega, -i\Omega, i\Omega, i\Omega) \\ &\quad + \frac{1}{z}\,\mathrm{Diag}\left(-1 - i\Omega, 3 - i\Omega\big(1 + \varepsilon^2/3\big), -1 + i\Omega, 3 + i\Omega\big(1 + \varepsilon^2/3\big)\right) \\ &\quad + \mathcal{O}\left(\frac{1}{z^2}\right), \end{aligned} \tag{7.67}$$

up to order $\varepsilon^2$. As a consequence, we can now easily integrate the equation for $Y^{(4)}$ up to sub-leading order when $z \gg 1$ (up to $\varepsilon^2$) and we obtain

$$Y^{(4)} \approx \left(c_-\ \mathfrak{s}_-^\infty(z)\,, \quad d_-\ \mathfrak{g}_-^\infty(z)\,, \quad c_+\ \mathfrak{s}_+^\infty(z)\,, \quad d_+\ \mathfrak{g}_+^\infty(z)\right)^\top, \tag{7.68}$$

where $c_\pm$ and $d_\pm$ are integration constants while

$$\mathfrak{g}_\pm^\infty(z) \approx e^{\pm i\Omega z} z^{3 \pm i\Omega(1+\varepsilon^2/3)} = e^{\pm i z_*}, \qquad \mathfrak{s}_\pm^\infty(z) \approx e^{\pm i\Omega z} z^{-1 \pm i\Omega}. \tag{7.69}$$

The two modes $\mathfrak{g}_\pm^\infty$ follow the same behaviour as the axial modes obtained in eq. (6.139): those can be dubbed gravitational modes, while the other two modes $\mathfrak{s}_\pm^\infty$ correspond to scalar modes.

We can then determine the behaviour of the metric perturbations $K$, $\chi$, $H_1$ and $H_0$ by combining the matrices $P^{(i)}$, with $i = 1, \dots, 4$ as

$$Y = PY^{(4)} \quad \text{with} \quad P = P^{(1)}P^{(2)}P^{(3)}P^{(4)}. \tag{7.70}$$

with the leading order terms of each coefficient of $P$ given by

$$P \approx \frac{1}{z^2}\begin{pmatrix} -1 & -\dfrac{1}{2iz\Omega} & -1 & \dfrac{1}{2iz\Omega} \\ -\dfrac{iaz^2}{\Omega} & \dfrac{az}{2\Omega^2} & \frac{iaz^2}{\Omega} & \dfrac{az}{2\Omega^2} \\ -3a + b & 1 & 3a - b & -1 \\ -3a - b & 1 & -(3a + \overline{b}) & -1 \end{pmatrix} \tag{7.71}$$

Hence, the metric and the scalar perturbations are non-trivial linear combinations of the so-called gravitational and scalar modes. This shows that the metric and the scalar variables are dynamically entangled.

### 7.4.2. Near the horizon

The asymptotic behaviour of polar perturbations near the horizon is technically more complex to analyse than the previous case because we need more steps to “diagonalise” the matrix $M$ and then to integrate asymptotically the system for the perturbations. However, the procedure is straightforward following the algorithm presented in chapter 5. For this reason, we do not give the details of the calculation here but instead present the final result. The curious reader is referred to appendix A.2 in which such details are given.

After several changes of variables, one obtains a first order differential system satisfied by a vector $\tilde{Y}$ whose corresponding matrix $\tilde{M}$ is of the form

$$\tilde{M} = \frac{1}{z-1}\tilde{M}_{-1} + \mathcal{O}(1) \ , \tag{7.72}$$

where the leading order term $M_{_{-1}}$ is, up to $\varepsilon^2$, given by

$$\tilde{M}_{-1} = \text{Diag}\left[-i\Omega\left(1-\frac{21}{10}\varepsilon^2\right), +i\Omega\left(1-\frac{21}{10}\varepsilon^2\right),\right. \\ \left. -i\Omega\left(1-\frac{21}{10}\varepsilon^2\right), +i\Omega\left(1-\frac{21}{10}\varepsilon^2\right)\right] + \mathcal{O}(\varepsilon^3) \ . \tag{7.73}$$

One recognises that the coefficients of $M_{-1}$ correspond to the leading order term in the asymptotic expansion of $\pm i\Omega z_*$ around $z = 1$, given in eq. (6.83). Indeed, we see that

$$\tilde{M} = i\Omega \, \frac{\mathrm{d}z_*}{\mathrm{d}z} \, \text{Diag}\,(-1, +1, -1, +1) + \mathcal{O}(1) \ , \tag{7.74}$$

and then integrating the equation for $\tilde{Y}$ becomes trivial as

$$\frac{\mathrm{d}Y}{\mathrm{d}z_*} \approx \text{Diag}\,(-i\Omega, +i\Omega, -i\Omega, +i\Omega)\,\tilde{Y} \ , \tag{7.75}$$

which leads to the solution

$$\tilde{Y} = \left(c_- \mathfrak{p}^{\mathrm{h}}_-(z), \quad c_+ \mathfrak{p}^{\mathrm{h}}_+(z), \quad d_- \mathfrak{p}^{\mathrm{h}}_-(z), \quad d_+ \mathfrak{p}^{\mathrm{h}}_+(z)\right)^{\top} \ , \tag{7.76}$$

where $c_\pm$ and $d_\pm$ are integration constants, and we introduced the polar modes (up to $\varepsilon^2$),

$$\mathfrak{p}^{\mathrm{h}}_\pm(z) \approx e^{\pm i\Omega z_*} = (z-1)^{\pm i\Omega(1-21\varepsilon^2/10)} \ . \tag{7.77}$$

Several remarks are in order. First, exactly as in the analysis of the asymptotics at infinity, one cannot discriminate between the gravitational mode and the scalar mode at leading order since they are equivalent at this order. Going to next-to-leading orders would be needed in order to further characterise each mode. Then, computing the behaviour of each mode at the horizon in terms of the metric perturbation functions, in a similar way to what was done at spatial infinity, is possible but not enlightening since the expressions are very involved. Finally, notice that the results above eqs. (7.69) and (7.77) are consistent with the behaviours found in [223], as one can see in their equation (17).

## 7.5. 4dEGB solution

In order to compute the asymptotical behaviour of the polar modes for the 4dEGB solution, we proceed similarly to what was done in the previous sections, writing the system as $\mathrm{d}Y/\mathrm{d}z = M(z)Y$, with $Y = \begin{pmatrix} K, & \delta\phi, & H_1, & H_0 \end{pmatrix}^\top$, and applying the algorithm given in chapter 5. The exact form of the matrix $M$ and the explicit steps of the algorithm are not given here for simplicity but one can find them in appendix A.3.

### 7.5.1. Spatial infinity

At spatial infinity, the diagonalized matrix $\tilde{M}$ is found to be

$$\begin{aligned}\tilde{M}(z) &= \mathrm{Diag}(0, 0, -i\Omega, i\Omega) \\ &\quad + \frac{1}{z}\,\mathrm{Diag}\left[-5 - i\sqrt{\lambda}, -5 + i\sqrt{\lambda}, 1 - i\Omega(1+\beta), 1 + i\Omega(1+\beta)\right] \\ &\quad + \mathcal{O}\left(\frac{1}{z^2}\right).\end{aligned}$$

This leads to an asymptotic solution where $Y$ is a combination of 4 modes, where we recognise two polar gravitational modes,

$$\mathfrak{g}^\infty_\pm(z) \approx e^{\pm i\Omega z} z^{1\pm i\Omega(1+\beta)}\,, \tag{7.78}$$

and identify the other two as scalar modes,

$$\mathfrak{s}^\infty_1(z) \approx z^{-5+i\sqrt{\lambda}}\,, \quad \mathfrak{s}^\infty_2(z) \approx z^{-5-i\sqrt{\lambda}}\,. \tag{7.79}$$

We can recover the behaviour of the metric perturbations $K$, $\delta\phi$, $H_1$ and $H_0$ which are the components of $Y$ by using the explicit expression of the matrix $\tilde{P}$. After a direct calculation, we find the following behaviour for $Y$ when $z$ goes to infinity:

$$Y \approx \begin{pmatrix} \dfrac{-i}{\Omega z} & \dfrac{i}{\Omega z} & \dfrac{2 - i\sqrt{\lambda} - \lambda}{\Omega^2} z & \dfrac{2 + i\sqrt{\lambda} - \lambda}{\Omega^2} z \\ \xi & \bar{\xi} & \dfrac{z^6}{24\beta(1+\beta)} & \dfrac{z^6}{24\beta(1+\beta)} \\ -1 & 1 & \dfrac{2 - i\sqrt{\lambda}}{i\Omega} z^2 & \dfrac{2 + i\sqrt{\lambda}}{i\Omega} z^2 \\ 1 & 1 & \dfrac{2}{3} z^3 & \dfrac{2}{3} z^3 \end{pmatrix} \begin{pmatrix} c_+ \mathfrak{g}^\infty_+ \\ c_- \mathfrak{g}^\infty_- \\ d_1 \mathfrak{s}^\infty_1 \\ d_2 \mathfrak{s}^\infty_2 \end{pmatrix}, \tag{7.80}$$

where

$$\xi = \frac{i(53\lambda - 4)}{48\beta(\beta+1)\Omega^3} + \frac{57\lambda + 466}{576\beta\Omega^2} + \frac{26327 i(\beta+1)}{2304\beta\Omega} - \frac{175\beta^2 - 1954\beta + 175}{36864\beta}\,. \tag{7.81}$$

and $c_\pm$ and $d_\pm$ are constants.

This result calls for a few comments. First, we can see from eq. (7.79) that the scalar modes are not propagating at infinity: even though it is possible to identify two branches corresponding to two sign choices, the corresponding modes do not contain exponentials, and the leading order depends on $\lambda$. This implies that there is no choice of $z_*$ such that the two scalar modes can be expressed as $\mathfrak{s}_\pm^\infty(z_*) \simeq e^{\pm i z_*/c_0}$, with $c_0$ a constant speed independant of $\lambda$. Such a behaviour for scalar modes leads to the conclusion that defining quasinormal modes of the scalar sector in the usual way (through outgoing boundary conditions at infinity) for this solution is not possible.

Second, one can compare the asymptotic behaviour of the scalar modes with what is obtained by considering only scalar perturbations onto a fixed background; this is done in section 7.6 and we see that the two behaviours are very similar, even though they slightly differ. Third, one can observe that the 4-dimensional matrix above section 7.5.1 is ill-defined in the GR limit where $\beta \to 0$. In fact, the second line of the matrix tends to infinity in this limit. This could be expected, since in that limit there is no degree of freedom associated with the scalar perturbation, which is obtained precisely from the second line of the matrix. One could solve this problem by setting $\chi = \beta\,\delta\phi$ and considering the vector $\begin{pmatrix} K, & \chi, & H_1, & H_0 \end{pmatrix}^\top$, similarly to what was done for the EsGB solution in eq. (7.56).

### 7.5.2. Near the horizon

Near the horizon, we use the variable $x = 1/\sqrt{z-1}$, as for axial modes. Using the algorithm, we find a change of vector $Y = \tilde{P}\tilde{Y}$ such that the associated matrix, that we denote $\tilde{M}_x$ exactly as in eq. (6.151), is diagonal and is explicitly given by

$$\tilde{M}_x = \frac{1}{x}\,\mathrm{Diag}(-1,0,0,2) + \mathcal{O}\left(\frac{1}{x^2}\right). \tag{7.82}$$

Solving the first order system is then immediate and the asymptotic expressions of the components of the 4-dimensional vector $\tilde{Y}$ (written as functions of $z$) are combinations of the four modes

$$\mathfrak{g}_1^{\mathrm{h}}(z) \simeq 1\,, \quad \mathfrak{g}_2^{\mathrm{h}}(z) \simeq \frac{1}{z-1}\,, \quad \mathfrak{s}_1^{\mathrm{h}}(z) \simeq 1 \quad \text{and} \quad \mathfrak{s}_2^{\mathrm{h}}(z) \simeq \sqrt{z-1}\,. \tag{7.83}$$

We have named two of these modes $\mathfrak{s}_i$ (for "scalar") because they contain a nonzero $\delta\phi$ contribution, as can be seen by expressing these modes in terms of the original perturbative quantities, using the explicit expression for the matrix $\tilde{P}$ provided by the algorithm [3]. Indeed, the relation between each of the above

3. One can also see from (7.84) that $\delta\phi$ is a combination of only these two modes at the horizon, which strengthens this denomination.

modes and the initial perturbations is given by

$$Y \simeq \begin{pmatrix} \frac{\zeta_1}{\sqrt{z-1}} & \zeta_2\sqrt{z-1} & \zeta_4\sqrt{z-1} & \zeta_6\sqrt{z-1} \\ 0 & 0 & 1 & \sqrt{z-1} \\ \frac{1}{z-1} & 1 & 0 & \zeta_7\sqrt{z-1} \\ 0 & \frac{\zeta_3}{\sqrt{z-1}} & \frac{\zeta_5}{\sqrt{z-1}} & \zeta_8 \end{pmatrix} \begin{pmatrix} c_1\, \mathfrak{g}_1^{\mathrm{h}} \\ c_2\, \mathfrak{g}_2^{\mathrm{h}} \\ d_1\, \mathfrak{s}_1^{\mathrm{h}} \\ d_2\, \mathfrak{s}_2^{\mathrm{h}} \end{pmatrix}, \tag{7.84}$$

where $c_i$ and $d_i$ are integration constants while $\zeta_i$ are constants whose expressions are given explicitly by

$$\begin{aligned}
\zeta_1 &= -\frac{4i(2\beta+1)\sqrt{-2\beta^2+\beta+1}\Omega}{4(2\beta\Omega+\Omega)^2+(\beta-1)^2}, \\
\zeta_2 &= -\frac{4i\sqrt{1-\beta}}{(2\beta+1)^{3/2}\Omega\nu}\Big[(\beta-1)^2\left(6\beta^2-2\beta-1\right)(2\beta+1)^3\Omega^2 \\
&\qquad + (\beta-1)^4\beta(\beta(2\beta(8\lambda-1)+8\lambda-5)+1) \\
&\qquad + 4(2\beta(\beta+1)-1)(2\beta+1)^5\Omega^4\Big], \\
\zeta_3 &= -\frac{2i(1-\beta)^{3/2}\beta}{\sqrt{2\beta+1}(2\beta(\beta+1)-1)\Omega}, \\
\zeta_4 &= \frac{4(1-\beta)^{3/2}}{\sqrt{2\beta+1}\nu}\Big[(\beta-1)^2(2\beta(4\beta(\beta(3\lambda-1)+\lambda-2)+2\lambda+1)+1) \\
&\qquad - 4(2\beta(4\beta(\beta\lambda+\lambda+1)-2\lambda+1)-1)(2\beta\Omega+\Omega)^2\Big], \\
\zeta_5 &= \frac{4\sqrt{1-\beta}\beta\sqrt{2\beta+1}}{2\beta(\beta+1)-1}, \\
\zeta_6 &= \frac{8(\beta-1)^2\beta}{4(2\beta+1)^3\Omega^2+(\beta-1)^2(6\beta+1)}, \\
\zeta_7 &= \frac{8i\beta\Omega\left(12(\beta+1)(2\beta+1)^3\Omega^2+(\beta-1)^2(\beta(10\beta+13)+7)\right)}{12(\beta-1)(2\beta+1)^4\Omega^2+3(\beta-1)^3(6\beta+1)(2\beta+1)}, \\
\zeta_8 &= \frac{4(\beta-1)\beta\left(4(2\beta\Omega+\Omega)^2+(\beta-1)^2\right)}{\sqrt{-2\beta^2+\beta+1}\left(4(2\beta+1)^3\Omega^2+(\beta-1)^2(6\beta+1)\right)},
\end{aligned} \tag{7.85}$$

with

$$\nu = (2\beta(\beta+1)-1)\left(4(2\beta\Omega+\Omega)^2+(\beta-1)^2\right)^2. \tag{7.86}$$

This behaviour is similar to what we have obtained for the axial perturbations. One cannot exhibit ingoing and outgoing modes: instead, the perturbations have non-oscillating behaviours at the horizon.

## 7.6. Linear perturbations of the scalar field about a fixed background

We consider in this section the perturbations of the scalar only, while the perturbations of the metric are fixed to zero: $h_{\mu\nu} = 0$. This corresponds to a decoupling limit in which the backreaction of the scalar onto the metric is neglected. One can compute the equation satisfied by $\delta\phi(r)$ for an arbitrary background but its general expression is too cumbersome to be written here. In the case $q = 0$, it can be extracted from the quadratic Lagrangian computed in [196]. Instead, we concentrate on the expression of this equation for three background solutions considered in this manuscript, namely the BCL, stealth Schwarzschild and 4dEGB solutions.

### 7.6.1. Effective potential

Let us consider the equation for the scalar field perturbation $\delta\phi(r)$ when $h_{\mu\nu} = 0$. In general, for a Horndeski theory, it is a second-order linear differential equation of the form

$$c_2(r)\delta\phi''(r) + c_1(r)\delta\phi'(r) + c_0(r)\delta\phi(r) = 0\,. \tag{7.87}$$

We aim to obtain the asymptotical behaviour of $\delta\phi$ near some value $r_0$ of $r$, for example $r_0 = +\infty$. It is not possible to take directly the limit $r \longrightarrow r_0$ for each coefficient $c_i$, since one does not know in general how the first and second derivative of $\delta\phi$ scale with respect to each other.

One therefore changes variables in order to obtain a simpler equation. Let us write

$$\delta\phi(r) = \kappa(r)\psi(r)\,, \tag{7.88}$$

which implies

$$c_2\kappa\frac{\mathrm{d}^2\psi}{\mathrm{d}r^2} + (2\kappa' c_2 + \kappa c_1)\frac{\mathrm{d}\psi}{\mathrm{d}r} + (\kappa'' c_2 + \kappa' c_1 + \kappa c_0)\psi = 0\,. \tag{7.89}$$

One can then get rid of the first derivative by imposing

$$\frac{\kappa'}{\kappa} = -\frac{c_1}{2c_2}\,. \tag{7.90}$$

The equation becomes

$$\frac{\mathrm{d}^2\psi}{\mathrm{d}r^2} + \left(\frac{\kappa''}{\kappa} + \frac{c_1}{c_2}\frac{\kappa'}{\kappa} + \frac{c_0}{c_2}\right)\psi = 0\,. \tag{7.91}$$

Using eq. (7.90), we find the relation

$$\frac{\kappa''}{\kappa} = \frac{c_2' c_1 - c_2 c_1'}{2c_2^2} - \frac{c_1}{2c_2}\frac{\kappa'}{\kappa} = \frac{c_2' c_1 - c_2 c_1'}{2c_2^2} + \left(\frac{c_1}{2c_2}\right)^2, \tag{7.92}$$

which finally leads to the equation

$$-\frac{\mathrm{d}^2\psi}{\mathrm{d}r^2} + V_\psi(r)\psi = 0 \quad \text{with} \quad V_\psi(r) = \frac{c_2 c_1' - c_2' c_1}{2c_2^2} + \left(\frac{c_1}{2c_2}\right)^2 - \frac{c_0}{c_2}\,. \tag{7.93}$$

In order to obtain the behaviour near $r_0$, one can then decompose $V_\psi(r)$ around $r_0$ and solve directly for $\psi$. The solution will be the expansion of $\psi$ around $r_0$. One must then come back to $\delta\phi$ by using eqs. (7.88) and (7.90).

### 7.6.2. BCL background

When the background is the BCL metric, one shows that the differential equation satisfied by $\chi(t,r)$ (defined from $\delta\phi(t,r)$ in eq. (7.6)) is given by

$$\frac{\partial^2\chi}{\partial t^2} + \frac{1}{2}A(r)\frac{\partial^2\chi}{\partial r^2} + \frac{1}{r}\left(1 + \frac{\mu^2\xi}{2r^2}\right)\frac{\partial\chi}{\partial r} - W(r)\chi = 0\,, \tag{7.94}$$

where $A(r)$ is the function entering into the BCL metric and

$$W(r) = \frac{1}{4r^4}\left(2r^2(3+2\lambda) - 4r(1+\lambda)\mu - 2(1+2\lambda)\mu^2\xi - \frac{1}{2}\frac{(2r-\mu)^2}{A(r)}\right).$$

As $A(r) > 0$, one immediately sees that $\chi(t,r)$ satisfies an elliptic equation and is therefore not propagating.

We now consider the Fourier component of $\chi(t,r)$, namely $\chi(r)$, and use the results of section 7.6.1 to obtain the asymptotic behaviour of $\chi$ at infinity ($\chi$ plays the role of $\delta\phi$ here). The coefficients $c_0$, $c_1$ and $c_2$ can be deduced from eq. (7.94):

$$c_0 = -W(r) - \omega^2\,, \quad c_1 = \frac{1}{r}\left(1 + \frac{\mu^2\xi}{2r^2}\right) \quad \text{and} \quad c_2 = \frac{1}{2}A(r)\,. \tag{7.95}$$

One therefore obtains

$$V_\psi(r) = \frac{2}{A}\omega^2 + \frac{1}{r^2A^2}\Big[2(\lambda+1) - 4(\lambda+1)\frac{\mu}{r} + \frac{1}{4}(7 - 14\xi + (8-12\xi)\lambda)\left(\frac{\mu}{r}\right)^2 + 3\xi(\lambda+1)\left(\frac{\mu}{r}\right)^3 + \xi^2(\lambda+1)\left(\frac{\mu}{r}\right)^4\Big]\,, \tag{7.96}$$

and

$$\kappa \propto \frac{1}{2rA}\,. \tag{7.97}$$

When $r \longrightarrow +\infty$, eq. (7.93) simplifies to

$$\tilde{\psi}'' = 2\omega^2\tilde{\psi}\,, \tag{7.98}$$

which means that the behaviour at infinity of $\chi(r)$ is given by

$$\chi(r) = \frac{1}{2r}\left(b_1 e^{\sqrt{2}\omega r} + b_2 e^{-\sqrt{2}\omega r}\right)\,, \tag{7.99}$$

where $b_1$ and $b_2$ are integration constants. This agrees with the asymptotic behaviour found for the scalar mode in eq. (7.19). Therefore, it seems that the asymptotic behaviour of the scalar perturbation when the metric is fixed coincides with the asymptotic behavior of the scalar part of the polar modes.

In order to confirm this intuition, we study eq. (7.93) when $r \longrightarrow r_+$. The resulting equation is

$$\tilde{\psi}'' + \frac{1}{4(r-r_+)^2}\tilde{\psi} = 0\,, \tag{7.100}$$

and the general solution corresponds to

$$\chi(r) = \frac{1}{\sqrt{r-r_+}}\left(b_1 + b_2 \ln(r-r_+)\right)\,, \tag{7.101}$$

where $b_1$ and $b_2$ are integration constants. We observe that this result is also fully consistent with the asymptotic analysis in eq. (7.27).

### 7.6.3. Stealth background

A similar analysis can be made when the background is the stealth Schwarzschild solution. For simplicity, we distinguish again the three cases where the only non-vanishing parameter is $\gamma \neq 0$, $\beta \neq 0$ or $\alpha \neq 0$.

When $\gamma \neq 0$, the equation for $\delta\phi$ is given by eq. (7.87) with

$$c_2 = 1\,, \quad c_1 = \frac{\mu(r-\mu) + 2i(\mu r^5)^{1/2}}{r\mu}\,,$$
$$c_0 = -\frac{\omega\left[5i(r\mu)^{3/2} - 3i(r^5\mu)^{1/2} + 2\omega r^4\right]}{2r\mu(r-\mu)^2}\,. \tag{7.102}$$

The renormalisation $\kappa$ is such that

$$\kappa(r) = \exp\left[-2i\omega\sqrt{r/\mu}(r+3\mu)\right]\left(\frac{\sqrt{r/\mu}+1}{\sqrt{r/\mu}-1}\right)^{i\omega\mu}\,, \tag{7.103}$$

and then $\psi$ is solution of the second order equation

$$4r^2\psi'' + \psi = 0\,, \tag{7.104}$$

which can be solved immediately to get

$$\psi(r) = a_1\sqrt{r} + a_2\sqrt{r}\ln r\,, \tag{7.105}$$

where $a_1$ and $a_2$ are integration constants. Going back to the original variable $\delta\phi$, we recover the asymptotic behaviours of the scalar mode obtained in eqs. (7.34) and (7.35).

The case where $\beta \neq 0$ is treated in exactly the same way. Taking now

$$\kappa(r) = \exp\left[-2i\omega\sqrt{r\mu}\right]\left(\frac{\sqrt{r/\mu}+1}{\sqrt{r/\mu}-1}\right)^{i\omega\mu}, \tag{7.106}$$

we show that the field $\psi$ satisfies the equation

$$4r^2\psi'' + (4\lambda+1)\psi = 0\,, \tag{7.107}$$

which, again, can be solved immediately:

$$\psi(r) = \sqrt{r}\left(a_+ r^{+i\sqrt{\lambda}} + a_- r^{-i\sqrt{\lambda}}\right), \tag{7.108}$$

where $a_\pm$ are constants. We find again that the perturbation is not propagating. Furthermore, these results agree with the full asymptotic analysis of the solutions of the polar system given in eq. (7.35).

Finally, in the case $\alpha \neq 0$, the equation satisfied by $\delta\phi$ at linear order disappears, since the quadratic Lagrangian for $\delta\phi$ is a total derivative.

### 7.6.4. 4dEGB background

In the case of the 4dEGB solution, one finds

$$c_0(r) = \frac{8\alpha\ell(\ell+1)}{r^2}\Big(4\sigma A + 2\sqrt{A}A - 2r\sigma A' + \sqrt{A}(2 - 2rA' + r^2A'')\Big), \tag{7.109}$$

$$c_1(r) = 8\alpha\Big(4\sqrt{A}A' - r\sigma A'^2 + \sigma A(4A' - 2rA'')\Big), \tag{7.110}$$

$$c_2(r) = -16\alpha\sqrt{A}\Big(2A + 2\sigma A\sqrt{A} - r\sigma\sqrt{A}A'\Big). \tag{7.111}$$

We observe that time does not appear in the equations, since $\omega$ is absent: this means that $\delta\phi$ satisfies an elliptic equation rather than the expected hyperbolic equation. The fact that $\delta\phi$ does not propagate could be related a strong coupling problem.

Applying the reasoning presented in section 7.6.1, one finds that the asymptotical behaviour of $\delta\phi$ is

$$\delta\phi = Ar^{-i\sqrt{1+\lambda}} + Br^{+i\sqrt{1+\lambda}}. \tag{7.112}$$

We do not recover exactly the asymptotical behaviour found in eq. (7.79) where both metric and scalar perturbations have been considered. However, the behaviours are very similar. This result differs from the previous two cases, for which the behaviour of the decoupled scalar perturbations and the scalar mode found from the full system agreed at both the horizon and infinity. It can be seen as the effect of a more important backreaction of the scalar field onto the metric. One can note that the behaviours still agree in the $\lambda \longrightarrow +\infty$ limit, implying that the coupling between the metric and the scalar perturbations becomes subdominant in that case.

## Conclusion

In this chapter, we have applied the novel approach introduced in chapter 5 to study linear polar BH perturbations in the context of DHOST theories. The method is very generic and enables one to obtain the asymptotic behaviours of the perturbations at spatial infinity and near the black hole horizon without reformulating their dynamics in terms of a Schrödinger-like equation. The knowledge of these asymptotic behaviours is essential to define and compute QNMs, characterised by outgoing conditions at spatial infinity and ingoing conditions at the horizon.

The study of polar perturbations is more challenging than the treatment of axial perturbations done in chapter 6 because the scalar field and metric perturbations are now coupled and we have not found a generalised Schrödinger-like reformulation of the system for the considered BH solutions. The only option left was thus to apply the method of chapter 5, providing the asymptotic behaviours of the solutions at spatial infinity and near the horizon for each type of BH.

For the BCL solution, we have identified two pairs of modes at the boundaries. One pair consists of an ingoing mode and an outgoing mode, which look similar to the usual gravitational modes. By contrast, the other two modes, corresponding to scalar modes, possess an asymptotic behaviour that appears pathological. Indeed, they do not propagate neither at infinity nor at the horizon.

For the stealth black hole solution, we have found that the gravitational polar modes behave asymptotically as their axial counterparts. In the stealth models with $\alpha \neq 0$, their behaviour is similar to the standard GR behaviour but in a disformed Schwarzschild metric, with a different horizon and characterised by a radially-dependent time shift. The scalar modes, however, have very different behaviours. Indeed, in the case $\alpha \neq 0$, they do not see the shift of the horizon. In the other two cases, it is not possible to extract propagating waves at infinity and at the horizon, which might be a sign of strong coupling.

In the case of the EsGB solution, we found that both gravitational and scalar modes in the polar sector behave healthily at the horizon and at infinity. Indeed, we found that they corresponded to ingoing and outgoing waves propagating at speed $c = 1$. This solution is the only one that exhibits such regular asymptotic behaviours for all modes.

Finally, we showed that the 4dEGB polar perturbations were very pathological. Indeed, at infinity, only gravitational polar modes seem to be propagating while the scalar perturbations do not yield a wave equation. Moreover, it is not possible to define ingoing and outgoing modes for both modes at the horizon.

This work opens a new window for the investigation of black hole perturbations in modified gravity. The potential of the new method presented in chapter 5 has been illustrated here with just a couple of examples. In chapter 8, we pursue the presentation of this method from the numerical point of view, going to the actual computation of QNMs.

CHAPTER 8

# NUMERICAL COMPUTATION OF QUASI-NORMAL MODES

Contents

Computations of QNMs require one to impose specific boundary conditions at both the BH horizon and at infinity, namely that perturbations be ingoing at the former and outgoing at the latter. However, when one does not have a Schrödinger-like reformulation of the perturbation equations, one cannot impose such boundary conditions in general since the combination of metric perturbations that corresponds to a given wave propagation direction is complicated to obtain.

Let us consider the case of polar perturbations of BHs in DHOST theories. As these perturbations do not have a Schrödinger-like reformulation, one could think that it is not possible to compute their QNMs. However, using the asymptotic decoupling presented in chapter 7, we are able to find the boundary conditions on the metric perturbation functions $K$, $\delta\phi$, $H_1$ and $H_0$ that correspond to QNM boundary conditions. This allows us to integrate numerically the perturbation equations from the first-order system directly. The aim of this chapter is to apply this numerical method for the BCL BH, after presenting it in the simpler context of axial Schwarzschild perturbations.

We start by reviewing different ways of computing QNMs when the perturbation equations can be cast into a Schrödinger-like equation in section 8.1. Then, in section 8.2, we show the principle of the method by applying it to the Schwarzschild BH using the asymptotic expressions obtained in chapter 5. Finally, we pave the way for a computation in the case of the BCL solution in section 8.3. The work presented in this chapter has not been published and is still ongoing.

## 8.1. Computation of quasi-normal modes from the Schrödinger equation

### 8.1.1. Mathematical problem

The reformulation of BH perturbations as Schrödinger equations of the form

$$\frac{\mathrm{d}^2 X}{\mathrm{d}r_*^2} + \left(\omega^2 - V\right) X = 0\,, \tag{8.1}$$

for the different modes, as was presented in the case of GR in chapter 4 (see eqs. (4.37) and (6.23), where $\hat{Y}_1$ has been renamed $X$ here), is very useful to gain a physical understanding of the behaviour of perturbations. Indeed, one can interpret these equations as wave propagation equations with speed $c = 1$ and scattering potential $V$. One can then impose boundary conditions relevant to wave propagation problems, noting that bound states cannot exist since the potentials are positive [200]. These boundary conditions are that perturbations be propagating outwards at infinity and inwards at the BH horizon.

As the potential $V$ goes to zero both at the horizon and at infinity, we proved in eq. (4.46) the function $X$ was such that

$$\begin{aligned} X(t,r) &\approx \mathcal{A}_{\mathrm{hor}}\, e^{-i\omega(t-r_*)} + \mathcal{B}_{\mathrm{hor}}\, e^{-i\omega(t+r_*)} \quad \text{when} \quad r \longrightarrow \mu\,, \\ X(t,r) &\approx \mathcal{A}_{\infty}\, e^{-i\omega(t-r_*)} + \mathcal{B}_{\infty}\, e^{-i\omega(t+r_*)} \quad \text{when} \quad r \longrightarrow \infty\,, \end{aligned} \tag{8.2}$$

where $\approx$ represents equality up to subleading terms. Imposing the boundary conditions then corresponds to the choice

$$\mathcal{A}_{\mathrm{hor}} = \mathcal{B}_{\infty} = 0\,. \tag{8.3}$$

The computation of QNMs becomes an eigenvalue problem: one has a second-order ordinary differential equation (ODE) with boundary conditions, and must find all the values of $\omega$ such that solutions exist. The mathematical literature on this topic is abundant, and many different methods, both numerical and analytical, were developed in the past 50 years to compute QNMs. Many of them are presented in [182, 201]. We present here a few of them that are relevant for the generalization of the computation of QNMs to modified gravity theories.

### 8.1.2. Numerical computations

#### 8.1.2.1. Shooting method

The most simple way to solve an ODE is to integrate it numerically, starting from one of its boundaries. In the present case, we do not have two boundary conditions at one point but two boundary conditions at different points: such an integration method is therefore called a “shooting method”; indeed, one can integrate eq. (8.1) from both boundaries and see what are the values of $\omega$ such that both solutions match in the bulk.

Such a procedure was first proposed in [217]. In this work, the function numerically integrated was not directly $X$ but the phase $f$ of $X$, such that

$$X = \exp\left(i \int \mathrm{d}r_* f(r_*)\right). \tag{8.4}$$

One integrates from a $r_* < 0$ such that $|V(r_*)| \ll \omega^2$; the boundary condition is $f = -\omega$ in order to have ingoing waves at the BH horizon. The resulting solution for $X$ is called $X_{\mathrm{hor}}$. One also integrates from a $r_* > 0$ such that $|V(r_*)| \ll \omega^2$ with the choice $f = +\omega$ in that case. The solution is called $X_\infty$. At some point in the middle of spacetime, one must check that both solutions match. Since eq. (8.1) is linear, one cannot simply expect $X_{\mathrm{hor}}$ and $X_\infty$ to match as they are defined up to a constant; however, the Wronskian defined by

$$W(r_*) = \begin{vmatrix} X_{\mathrm{hor}}(r_*) & X_\infty(r_*) \\ X'_{\mathrm{hor}}(r_*) & X'_\infty(r_*) \end{vmatrix} \tag{8.5}$$

must be zero. The numerical method then consists in the computation of the Wronskian at some point in spacetime as a function of $\omega$, and the search of the zeroes of this function.

The main source of error in this method comes from the fact that the equations are not integrated from $r_* = \pm\infty$ but from finite values of $r_*$, leading to small numerical errors in the boundary conditions. This means that the numerical solution computed is not simply an ingoing mode (when integrating from the horizon), but an ingoing mode to which a small outgoing perturbation is added. However, since outgoing modes behave like $e^{i\omega r_*}$ when $r_* \longrightarrow -\infty$, the perturbation will grow exponentially with $r_*$, leading to possible big errors at the point where the Wronskian is computed. This numerical difficulty was already presented in [217]. The method was improved in [224, 225]. Nowadays, this method has reached sufficient precision to be used for computations [202, 226, 227, 228].

#### 8.1.2.2. Continued fraction method

The continued fraction method was proposed by Leaver [198], and uses a result from the theory of the hydrogen atom to integrate eq. (8.1). The idea is to use an ansatz that incorporates the essential singularity behaviour of the perturbations. In the case of the Schwarzschild BH, we know for example from eqs. (4.40), (8.2) and (8.3) that at infinity, the expression of $X$ should be

$$X \approx e^{i\omega r_*} = e^{i\omega r} r^{i\mu\omega}, \tag{8.6}$$

while at the horizon

$$X \approx e^{-i\omega r_*} = (r-\mu)^{-i\mu\omega}\,. \tag{8.7}$$

Therefore, it is natural to look for a solution of the form

$$X(r) = e^{i\omega r} r^{i\mu\omega} \left(\frac{r-\mu}{r}\right)^{-i\mu\omega} f(r)\,, \tag{8.8}$$

with $f$ an arbitrary function. One can check that the ansatz (8.8) yields eqs. (8.6) and (8.7) as soon as $f$ is a constant at both boundaries. Since the perturbations should not have singularities elsewhere in spacetime, we expect $f$ to be bounded everywhere and can therefore expand it as a power of the variable $1-\mu/r$:

$$f(r) = \sum_{n=0}^{+\infty} a_n \left(\frac{r-\mu}{r}\right)^n\,. \tag{8.9}$$

Leaver then finds that the coefficients $a_n$ are linked via a three-term recurrence relation:

$$\alpha_n a_{n+1} + \beta_n a_n + \gamma_n a_{n-1} = 0 \quad \text{and} \quad \alpha_0 a_1 + \beta_0 a_0 = 0\,. \tag{8.10}$$

This relation has two different branches of solutions, which is expected since it is the case for eq. (8.1). However, only one branch will be regular at the boundaries [1]. Only specific values of $\omega$ will correspond to a solution that contains only this branch: these values are the QNMs and can be found using a mathematical result from Gautschi [229]. Indeed, they are such that a given continued fraction made of the $\alpha_n$, $\beta_n$ and $\gamma_n$ is zero. One then recovers an implicit equation that can be solved for $\omega$, yielding the Schwarzschild QNMs.

This method can be generalized to more complicated potentials, such as the potential for a Kerr or RN BH [230]. It can also be generalized to coupled systems, leading to a "matricial continued fraction" method [228]. It allows one to compute very precisely many overtones without numerical issues.

#### 8.1.2.3. Results for Schwarzschild

We present in table 8.1 the numerical values for QNMs computed in [182] for $\ell = 2$, in order to provide a reference for the QNM computations that will be presented in this chapter. The frequencies are written as $\omega = \omega_R + i\omega_I$. The frequencies computed for the axial and polar perturbations are the same: this property is known as the *isospectrality theorem* [197]. One should note that if $\omega_R + i\omega_i$ is a Schwarzschild QNM, then $-\omega_R + i\omega_I$ is also a Schwarzschild QNM. Indeed, if $X$ is a solution of eq. (8.1) for a frequency $\omega = \omega_R + i\omega_I$, with the boundary conditions (8.3), then $\overline{X}$ is also a solution of eq. (8.1) with the same boundary conditions and $\omega = -\omega_R + i\omega_I$ (see [200]).

1. For example, at infinity, one branch should go to a constant while the other one behaves like $e^{-2i\omega r} r^{-2i\mu\omega}$.

| Overtone | $\mu\omega_R$ | $-\mu\omega_I$ |
|---|---|---|
| 0 | 0.74734 | 0.17792 |
| 1 | 0.69342 | 0.54783 |
| 2 | 0.60211 | 0.95655 |
| 3 | 0.50301 | 1.4103 |
| 4 | 0.41503 | 1.8937 |
| 5 | 0.33860 | 2.3912 |
| 6 | 0.26650 | 2.8958 |
| 7 | 0.18564 | 3.4077 |
| 8 | 0.00000 | 4.0000 |
| 9 | 0.12653 | 4.6053 |
| 10 | 0.15311 | 5.1217 |

Table 8.1. – Values of quasi-normal modes for the Schwarzschild BH computed in [182] for $\ell = 2$. One should note that the values for the axial and polar modes are the same. Only the first 5 decimals are given, but the precision is much higher (around $10^{-12}$).

### 8.1.3. Analytical results

Analytical results for the values of $\omega$ solving eqs. (8.1) and (8.3) are in general very hard to obtain. Due to the Schrödinger-like formulation of the problem, this difficulty can be linked to the difficulty of solving the Schrödinger equation for a nonconstant potential. Therefore, apart from very specific examples presented in [182], QNMs cannot be computed exactly in general, and in particularly not for a Schwarzschild BH. It is however possible to find ways to estimate the QNM frequencies using approximate solutions to the Schrödinger equation.

Such a technique, initially proposed in [231, 232], relies on an adaptation of the Wentzel–Kramers–Brillouin (WKB) method used extensively in quantum mechanics. It follows from the remark made previously in [233] that given the potential for perturbations around the Schwarzschild metric (see eq. (4.41) for example), QNMs can be seen as waves traveling around the black hole, trapped in some finite region of space of maximum potential and slowly "leaking out". The idea of the method is therefore to assume that $\omega^2$ is close to the maximum of the potential $V$. Therefore, one can expand $\omega^2 - V$ as a Taylor series:

$$\omega^2 - V \approx Q_0 + \frac{1}{2}(r_* - r_*^{\mathrm{m}})^2 Q_2\,, \tag{8.11}$$

where $r_*^{\mathrm{m}}$ is the tortoise coordinate at which $V$ is maximal, and $Q_2 = -\,\mathrm{d}^2V/\mathrm{d}r_*^2$. The coefficient $Q_0$ contains $\omega^2$, while the coefficient $Q_2$ depends only on the form of the potential $V$. Equation (8.1) then becomes

$$\frac{\mathrm{d}^2X}{\mathrm{d}r_*^2} + \left(Q_0 + \frac{1}{2}(r_* - r_*^{\mathrm{m}})^2 Q_2\right)X = 0\,. \tag{8.12}$$

Such an equation can be solved analytically in terms of parabolic cylinder functions [182]. One finds that at infinity and horizon, the behaviour is not similar to eq. (8.2), which was expected given the fact that the approximation for $\omega^2 - V$ diverges for $r_* \longrightarrow \pm\infty$ instead of going to a constant. It is still possible to identify a term that diverges at both boundaries and a term that is regular; imposing that the coefficient in front of the former be zero gives

$$\frac{Q_0}{\sqrt{2Q_2}} = i\left(n + \frac{1}{2}\right). \tag{8.13}$$

This means that $Q_0$ is quantized, and therefore that $\omega$ is quantized too: an analytical approximation for the values of QNMs has been found.

The precision of the approximation is quite poor when done like it has been presented here; increasing the number of terms in eq. (8.11), as was done in [234, 235, 236, 237], yields better results. As an illustration, we reproduce in table 8.2 the results for $\ell = 2$ obtained in [237] using sixth-order WKB approximation. We observe that the results match the numerical results of table 8.1 up to 1 %.

| Overtone | $\mu\omega_R$ | $-\mu\omega_I$ |
|---|---|---|
| 0 | 0.74724 | 0.17802 |
| 1 | 0.69264 | 0.54702 |
| 2 | 0.59704 | 0.95522 |

Table 8.2. – Values of quasi-normal modes for the Schwarzschild BH computed in [237] for $\ell = 2$, using the sixth-order WKB approximation.

### 8.1.4. QNMs from the first-order system

One can note that several methods presented in sections 8.1.2 and 8.1.3 rely on the existence of a Schrödinger-like reformulation for the wave propagation equations. Indeed, the WKB formulation requires a potential, and the continued fraction method is conceived for a second-order ODE. However, as explained in section 4.5, such a formulation does not always exist. It is therefore relevant to look for a way to adapt the existing methods of QNM computation to the kind of system we study in this manuscript, meaning a first-order system of the form $\mathrm{d}Y/\mathrm{d}r = MY$.

The WKB method will not be adaptable because it depends heavily on the Schrödinger-like reformulation. In the following, we will present an adaptation of the integration method of section 8.1.2.1 in the case of first-order systems. This work is still ongoing and has not yet been published; it should be seen as a mere proof of concept for the feasibility of the computation of QNMs from a first-order system of equations.

## 8.2. QNMs from the first-order system for the Schwarzschild BH

### 8.2.1. Ansätze for the metric perturbations

Let us consider the first-order system for the axial perturbations of a Schwarzschild BH. This system is given in eq. (4.18) and involves the variable $Y = \big(h_0,\ \ h_1/\omega\big)^{\top}$. One can go back to the metric perturbations themselves by considering the system

$$\frac{\mathrm{d}Y}{\mathrm{d}r} = M(r)Y\,, \quad M(r) = \begin{pmatrix} 2/r & 2i\lambda(r-\mu)/\omega r^3 - i\omega \\ -i\omega r^2/(r-\mu)^2 & -\mu/r(r-\mu) \end{pmatrix}\,, \tag{8.14}$$

for which the asymptotic behaviours for $h_0$ and $h_1$ are given in eqs. (5.24) and (5.33).

Similarly to what was done for the continued fraction method, we factorize the essential singularities and poles at both $r = +\infty$ and $r = \mu$ by defining the ansätze

$$\begin{aligned} h_0(r) &= e^{i\omega r} r^{1+i\omega\mu}\left(\frac{r-\mu}{r}\right)^{-i\mu\omega} f_0(r)\,, \\ h_1(r) &= e^{i\omega r} r^{1+i\omega\mu}\left(\frac{r-\mu}{r}\right)^{-1-i\mu\omega} f_1(r)\,. \end{aligned} \tag{8.15}$$

The system (8.14) then becomes a set of two equations for $f_0$ and $f_1$:

$$\begin{aligned} 0 =& -\frac{1}{2}(u-1)(u+1)^2 f_0'(u) + \left[\frac{1}{2}(1-u^2) + i\Omega(u^2+2u-1)\right] f_0(u) \\ & \qquad - \frac{i}{2\Omega}\left[(u-1)(u+1)^2\lambda + 4\Omega^2\right] f_1(u)\,, \\ 0 =& -\frac{1}{2}(u-1)(u+1)^2 f_1'(u) - 2i\Omega f_0(u) \\ & \qquad + \left[\frac{1}{2}(u^2-1) + i\Omega(u^2+2u-1)\right] f_1(u)\,, \end{aligned} \tag{8.16}$$

where we defined $\Omega = \mu\omega$ and $u = \mu/r - 1$, such that $-1 \le u \le 1$, the horizon being $u = 1$ and infinity being $u = -1$.

One can notice that the system (8.16) is singular at both boundaries $u = 1$ and $u = -1$. This effect is due to the choice of ansatz that singles out one branch of the asymptotic solutions given in eqs. (5.24) and (5.33). If one requires $f_0$ and $f_1$ to be finite at each boundary, one can write eq. (8.16) for $u = \pm 1$. First, at $u = -1$, one obtains

$$\begin{aligned} -2i\Omega\left(f_0(-1) + f_1(-1)\right) &= 0\,, \\ -2i\Omega\left(f_0(-1) + f_1(-1)\right) &= 0\,. \end{aligned} \tag{8.17}$$

Then, at $u = +1$, one has

$$\begin{aligned} 2i\Omega\left(f_0(+1) - f_1(+1)\right) &= 0\,, \\ -2i\Omega\left(f_0(+1) - f_1(+1)\right) &= 0\,. \end{aligned} \tag{8.18}$$

One can observe that both systems are degenerate, leading to only one equation in both cases:

$$f_1(-1) = -f_0(-1) \quad \text{and} \quad f_1(1) = f_0(1)\,. \tag{8.19}$$

This can be understood by looking at the number of degrees of freedom system (8.16) contains. Indeed, in general, this system describes propagation in both directions, so it contains two modes both at the horizon and infinity. Imposing the ansätze of eq. (8.15) singles out outgoing modes at infinity and ingoing modes at the horizon by making them regular solutions of eq. (8.16). Therefore, imposing regularity for $f_0$ and $f_1$ is equivalent to imposing the physical boundary conditions for QNMs. In that sense, we can note that eq. (8.19) is coherent with the choice $c_- = 0$ in eq. (5.24) and $c_+ = 0$ in eq. (5.33).

One therefore sees that imposing regularity at both boundaries is enough to ensure the required boundary conditions are verified [2]. The idea is then to integrate system (8.16) from the horizon and from infinity and to find the values of $\omega$ that make both solutions match in the middle of spacetime. Instead of doing this via step-by-step integration, we use a global resolution method that relies on spectral decomposition of the functions.

## 8.2.2. Spectral method

### 8.2.2.1. Polynomial approximation

We will solve numerically the system (8.16) using a decomposition of both $f_0$ and $f_1$ on a basis of orthogonal polynomials, the Chebyshev polynomials. A review of their properties as well as the procedure of decomposition of a function on a basis of polynomials can be found in [238]; we only review the core results here.

2. The system is still singular at the boundaries, which means that usual numerical methods will fail at this point. We will need to find a way to go around this singularity, which will be described in the next section.

These polynomials are written $T_n(u)$, for $u \in [-1,1]$ and $n \in \mathbf{N}$. They verify the relation

$$\int_{-1}^{1} \mathrm{d}u \, \frac{T_n(u)T_m(u)}{\sqrt{1-u^2}} = \frac{\pi}{2}(1+\delta_{0n})\delta_{nm} \, . \tag{8.20}$$

This allows one to define an inner product on functions on the interval $[-1,1]$ by

$$\langle f, g \rangle = \int_{-1}^{1} \mathrm{d}u \, \frac{f(u)g(u)}{\sqrt{1-u^2}} \, . \tag{8.21}$$

The Chebyshev polynomials are then orthogonal with respect to this inner product, and one can prove that they form a basis of the space of such functions: any sufficiently regular function $g$ can be approximated by an infinite sum of these polynomials:

$$g = \sum_{n=0}^{+\infty} \hat{g}^n T_n \, , \quad \text{with} \quad \hat{g}_n = \frac{\langle g, T_n \rangle}{\|T_n\|^2} \, . \tag{8.22}$$

One can then define an approximation of $g$ by truncating the sum in eq. (8.22) up to some order $N$:

$$P_N(g) = \sum_{n=0}^{N} \hat{g}_n T_n \, . \tag{8.23}$$

Of course, one can prove that the approximation $P_N(g)$ converges to $g$ when $N$ increases.

Interestingly, one can also approximate $g$ by its value at a given set of points, called *collocation points*, and this representation is linked to the decomposition onto Chebyshev polynomials for the right choice of points. For example, if we define

$$u_n = \cos\left(\frac{\pi n}{N}\right) \quad \text{for} \quad n \in [\![0, N]\!] \, , \tag{8.24}$$

then the function $g$ can be approximated by its interpolant $I_N(g)$:

$$I_N(g) = \sum_{n=0}^{N} \tilde{g}_n T_n \quad \text{with} \quad \tilde{g}_n = \frac{\sum_{i=0}^{N} g(u_i) T_n(u_i) w_i}{\sum_{i=0}^{N} T_n(u_i)^2 w_i} \, , \tag{8.25}$$

where we defined the weights $w_i$ such that $w_0 = w_N = \pi/2N$ and $w_i = \pi/N$ for $i \in [\![1, N-1]\!]$. This is called the Chebyshev-Gauss-Lobatto quadrature. This approximation is very interesting since it does not require to know the function $g$ at every point in $[-1,1]$, contrary to the approximation by $P_N(g)$ that requires to compute an integral. In general, $P_N(g)$ and $I_N(g)$ are different; however, as $N$ goes to infinity, they both converge to the original function $g$.

Furthermore, the main advantage of spectral methods is that the approximation $I_N(g)$ converges towards $g$ faster than any power of $N$. In practice, this means that 30 points are enough to approximate $f$ with machine precision: the representation of functional equations involves much smaller systems than what is obtained with finite differences for example.

#### 8.2.2.2. Equations converted to a linear system

In a spectral method, we represent a function $g$ by a vector $(g(u_n))_{n\in[\![0,N]\!]}$ that we denote $\tilde{g}$. Multiplication by any function $g$ then becomes a matrix operating on the vector $\tilde{g}$:

$$\widetilde{fg} = (f(u_n)g(u_n))_{n\in[\![0,N]\!]} = \mathrm{Diag}(f(u_n))\tilde{g}\,, \tag{8.26}$$

where we left the range for $n$ implicit in the final step. However, some operations are much easier to represent if the expression of $g$ is given in terms of polynomials: for example, if one writes $\hat{g}$ the vector of coefficients $\hat{g}_n$ appearing in $P_N(g)$, then the derivative operator $\mathrm{d}/\mathrm{d}u$ is a matrix $D$ with

$$D_{ij} = \frac{2j}{1+\delta_{i0}} \quad \text{if} \quad j \geq i+1 \quad \text{and} \quad i+j \equiv 0\,[2]\,, \quad D_{ij} = 0 \quad \text{else}\,. \tag{8.27}$$

This operator could not be represented as a matrix operating on $\tilde{g}$! One must then juggle between the representation $\tilde{g}$ in the "coordinate space" and $\hat{g}$ in the "coefficient space".

In order to numerically solve the system (8.16), we will use the coefficient space representation of $f_0$ and $f_1$ as variables, since derivatives are involved; however, since multiplication by a coefficient is much easier to execute in the coordinate space, we will switch to the coordinate space for the last step. Schematically, we will therefore represent system (8.16) by a linear equation of the form

$$(C_{\mathrm{der}} \cdot P \cdot D + C_{\mathrm{fun}} \cdot P)\hat{F} = 0\,, \tag{8.28}$$

where

- $C_{\mathrm{der}}$ and $C_{\mathrm{fun}}$ represent multiplication by the coefficients of the derivatives ($f_0'(u)$ and $f_1'(u)$) and the original functions ($f_0(u)$ and $f_1(u)$) respectively;
- $P$ converts coefficient space representation into coordinate space representation;
- $\hat{F}$ is a vector of size $2(N+1)$ of the form

$$\hat{F} = \begin{pmatrix}\hat{f}_{00} & \cdots & \hat{f}_{0N} & \hat{f}_{10} & \cdots & \hat{f}_{1N}\end{pmatrix}^{\top}\,, \tag{8.29}$$

  with $\hat{f}_{0i}$ and $\hat{f}_{1i}$ are respectively the coefficients of the approximations $P_N(f_0)$ and $P_N(f_1)$.

In order to impose ingoing boundary conditions at the horizon and outgoing boundary conditions at infinity, one must verify the relations of eq. (8.19). We can notice that the value of one of the functions $f_0$ and $f_1$ is still free at each boundary; therefore, we will proceed in the following way. We will first solve the system between $u = -1$ and $u = 0$ while choosing without loss of generality $f_1(-1) = 1$, leading to solutions $f_0^{\infty}$ and $f_1^{\infty}$. We will then solve the system between $u = 0$ and $u = 1$ while fixing similarly $f_1(1) = 1$, leading to solutions $f_0^{\rm h}$ and $f_1^{\rm h}$. The last step will be to match the solutions at $u = 0$.

In order to impose the value of $f_0$ and $f_1$ at the boundaries, we observe that the first and last collocation points defined in eq. (8.24) correspond respectively to infinity and the horizon. Henceforth, imposing $f_1^{\infty}(-1) = 1$ and $f_0^{\infty}(-1) = -1$ can be done by replacing the first and $N + 2^{\rm th}$ lines of system (8.28) respectively by the equations

$$\sum_{n=0}^{N} \hat{f}_{0n} T_n(-1) = -1 \quad \text{and} \quad \sum_{n=0}^{N} \hat{f}_{1n} T_n(-1) = 1\,. \tag{8.30}$$

Moreover, replacing these lines of the equations will allow us to avoid the singularity present at $u = -1$. A similar procedure [3] involving the $N + 1^{\rm th}$ and last equations allows one to fix the values of $f_0^{\rm h}$ and $f_1^{\rm h}$ at $u = 1$.

Once the boundary conditions are enforced by the replacement of the relevant equations, eq. (8.28) becomes

$$(C_{\rm der} \cdot P \cdot D + C_{\rm fun} \cdot P)\hat{F} = B\,, \tag{8.31}$$

with $B$ a nonzero vector: the numerical computation of the solution then boils down to the resolution of a linear system. This procedure yields $\hat{F}$, the vector containing the coefficients of the decompositions of $f_0$ and $f_1$ onto Chebyshev polynomials. From this vector, we can compute both functions at every $u$ in $[-1, 1]$. On fig. 8.1, we show the result of the resolution for $N = 30$ and $\Omega = 0.5 - 0.3i$. We can observe that the coefficients decrease exponentially with $n$, which indicates that the numerical solution found is correctly represented by its polynomial interpolant. We also observe that the coefficients stop decreasing once machine precision is reached. We can see however that the functions are not continuous at $u = 0$. This could be expected: we have no indication that the $\Omega$ chosen is actually a quasinormal mode.

On fig. 8.2, we perform the same numerical resolution of eq. (8.16) but with $\Omega$ chosen to be the fundamental QNM given in table 8.1. We observe that the functions are still not continuous: this is due to the fact that the choices

3. In practice, to solve the system between $u = -1$ and $u = 0$, we work with a rescaled coordinate $x = 2u + 1$ that varies between $-1$ and 1: this allows us to avoid the singularities present at $u = 1$. We proceed similarly to solve between $u = 0$ and $u = 1$.

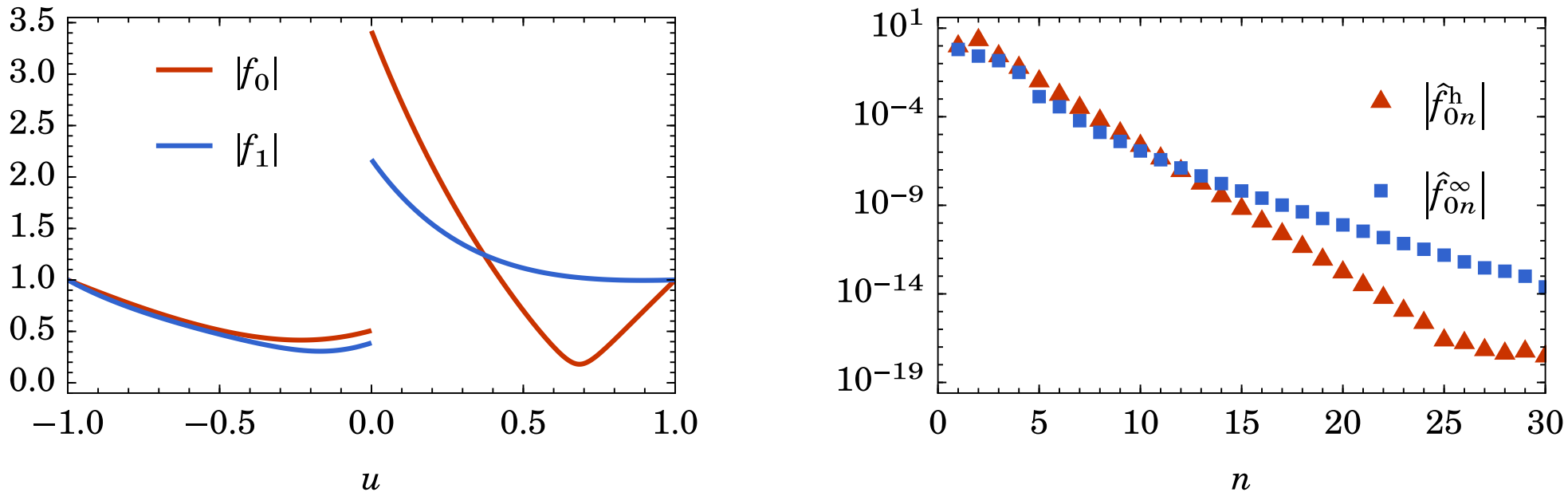

(a) Plot of the functions (b) Plot of the coefficients

Figure 8.1. – Result of the integration of eq. (8.16) for $\Omega = 0.5 - 0.3i$ and $\ell = 2$ ($\lambda = 2$). The number of grid points was taken to be $N = 30$. Only the coefficients of the function $f_0$ integrated from the horizon ($\hat{f}^{\mathrm{h}}_{0n}$) and from infinity ($\hat{f}^{\infty}_{0n}$) are shown in order to avoid cluttering the plot, but the coefficients of $f_1$ behave similarly.

$f^{\infty}_1(-1) = 1$ and $f^{\mathrm{h}}_1(1) = 1$ do not necessarily lead to the same overall factor for the solutions $f^{\infty}_1$ and $f^{\mathrm{h}}_1$. Therefore, the relevant quantity to quantify the matching at $u = 0$ is not the discontinuity but the linear dependance between $\left(f_0, \ f_1\right)^{\mathsf{T}}$ integrated from $u = -1$ and from $u = +1$. Therefore, finding the QNMs boils down to looking for the zeroes of the determinant $d$ defined by

$$d(\Omega) = \begin{vmatrix} f^{\mathrm{h}}_0 & f^{\infty}_0 \\ f^{\mathrm{h}}_1 & f^{\infty}_1 \end{vmatrix} . \tag{8.32}$$

This determinant can be seen as a function of $\Omega$. In the next section, we investigate the behaviour of $d$ in the complex plane.

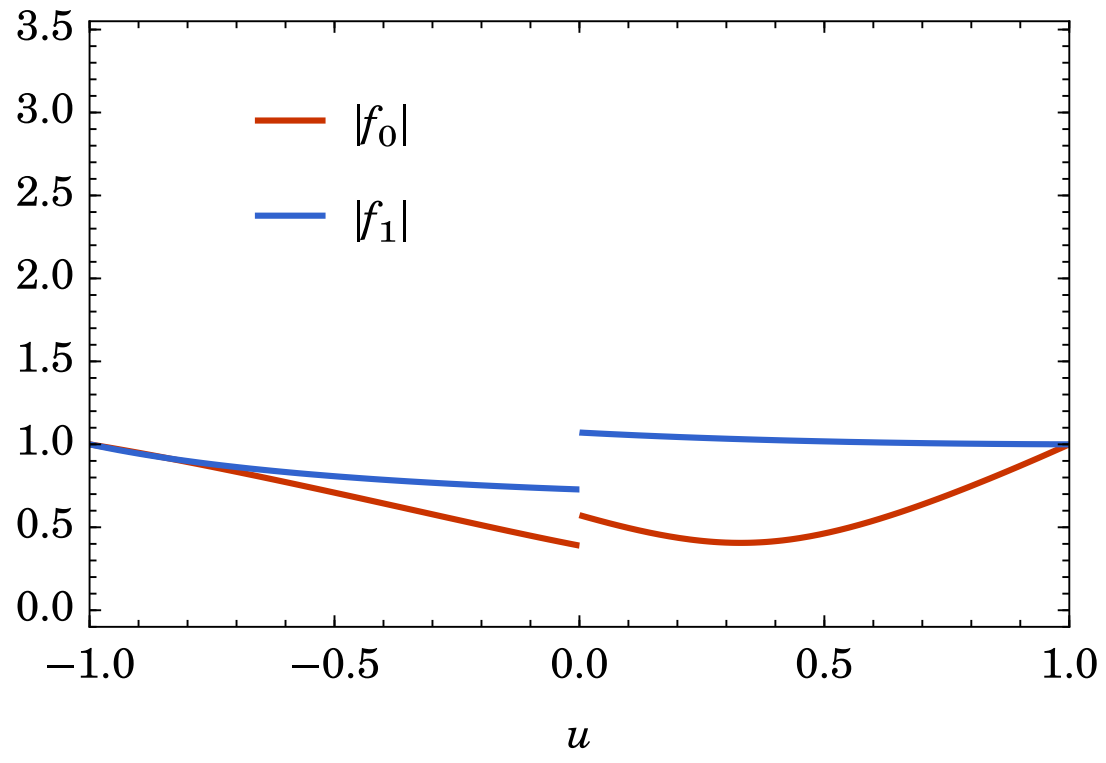

Figure 8.2. – Result of the integration of eq. (8.16) for the fundamental QNM of a Schwarzschild BH with $\ell = 2$ (see table 8.1). We took $N = 30$.

### 8.2.3. Numerical results

We study the zeroes of the function $d(\Omega)$ in the complex plane. We restrict ourselves to $\Re(\Omega) > 0$. We first plot $\ln|d|$ in the complex plane, which allows us to guess the positions of the zeroes, and we then use a numerical minimization procedure on $\ln|d|$ to find the corresponding value of $\Omega$. One numerical constraint appears when one wants to compute overtones: inverting system (8.31) becomes hard due to the operator of the left-hand side having a bad condition number. The solution is to use arbitrary precision numbers, up to $10^{-60}$ for the highest computed overtones.

In table 8.3 we give the values of the QNMs obtained via our method. We also provide in fig. 8.3 a proof that the value obtained for a QNM via our method converges when $N$ is increased [4].

| Overtone | $\Omega_R$ | $-\Omega_I$ |
|---|---|---|
| 0 | 0.74734 | 0.17793 |
| 1 | 0.69342 | 0.54783 |
| 2 | 0.60211 | 0.95655 |
| 3 | 0.50300 | 1.4103 |
| 4 | 0.41461 | 1.8938 |

Table 8.3. – Values of quasi-normal modes for the Schwarzschild BH obtained via our method, for $\ell = 2$, written as $\Omega = \Omega_R + i\Omega_I$. The results match the values obtained in [182] and presented in table 8.1.

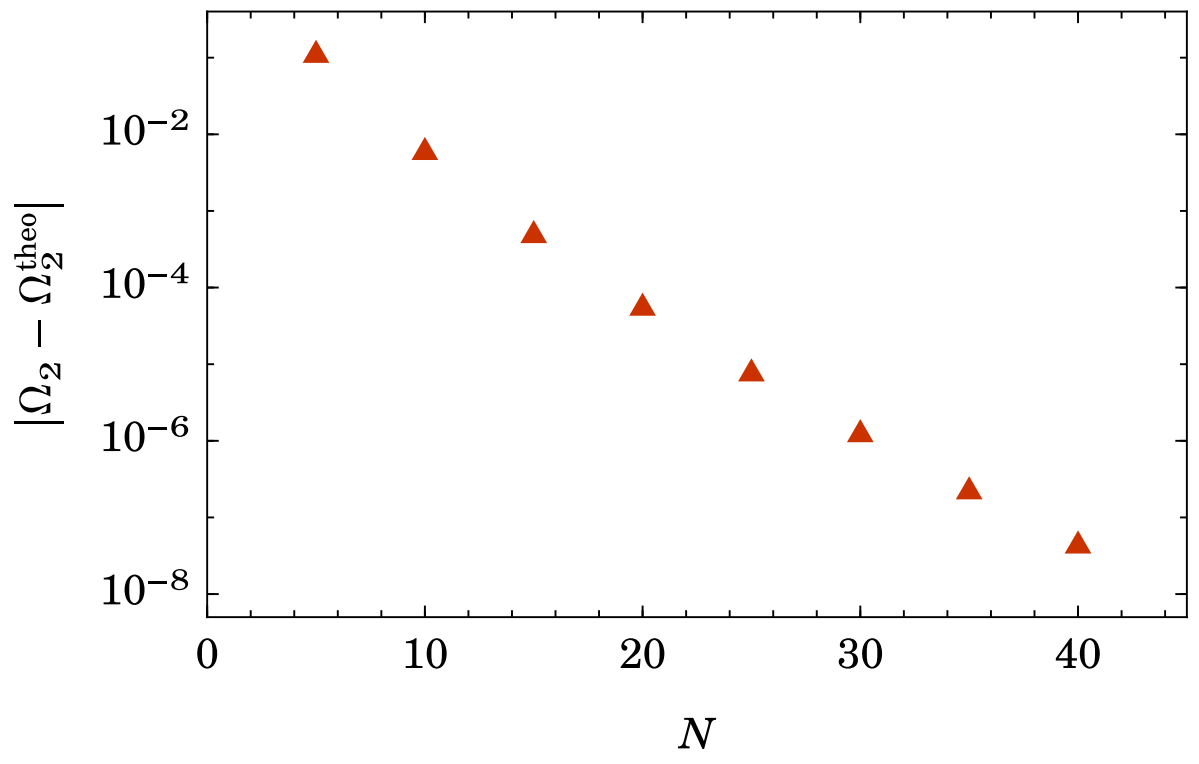

Figure 8.3. – Difference between the $n = 2$ QNM for $\ell = 2$ computed with our method (noted $\Omega_2$) and the one given in [182] (noted $\Omega_2^{\rm theo}$), for different values of $N$.

4. The arbitrary precision is increased accordingly in order to ensure that numerical errors do not blow up when the system is inverted.

Another method, proposed in [239], can be applied: one can also write system (8.28) in the coefficient space only, and consider the change of functions $f_R = f_1/\Omega$. The system will then take the very simple form

$$(A_0 + \Omega A_1 + \Omega^2 A_2)\hat{F} = 0\,, \tag{8.33}$$

with $\hat{F}$ a vector containing the coefficients of $f_0$ and $f_R$. Finding the values of $\Omega$ such that $\hat{F}$ is nonzero is called a polynomial eigenvalue problem, and has been studied in the mathematical literature. A possible resolution is given in [239]. One should note that in that case, it is not necessary to impose boundary conditions: as a polynomial approximation can only be regular, regularity at the boundaries does not need to be imposed. This is called a "behavioural" boundary condition [240]. Such a method has been generalized to the first-order system in [181, 211] in order to compute the QNMs of several BH solutions; however, it is not very efficient and leads to many numerical errors in that case.

## 8.3. Results for the BCL solution

The application to the case of polar BCL perturbations is very similar to what was done for the Schwarzschild BH in section 8.2: we will use an ansatz for the perturbations that singles out the asymptotic behaviour we want to impose, then get the new equations and integrate them from both infinity and the horizon. Using eqs. (7.18), (7.22), (7.28) and (7.29), we can use the following ansätze for the perturbations of the metric in the polar sector:

$$\begin{aligned}
K(r) &= e^{i\omega r} r^{i\omega\mu}\left(\frac{r-r_+}{r}\right)^{-i\omega r_0} f_K(r)\,,\\
\chi(r) &= e^{i\omega r} r^{-1+i\omega\mu}\left(\frac{r-r_+}{r}\right)^{-1-i\omega r_0} f_\chi(r)\\
H_1(r) &= e^{i\omega r} r^{1+i\omega\mu}\left(\frac{r-r_+}{r}\right)^{-1-i\omega r_0} f_1(r)\,,\\
H_0(r) &= e^{i\omega r} r^{1+i\omega\mu}\left(\frac{r-r_+}{r}\right)^{-1-i\omega r_0} f_0(r)\,.
\end{aligned} \tag{8.34}$$

One can note that we impose the gravitational behaviour at each boundary: this way, we expect to excite only the gravitational mode.

We can then change variables by defining $u = 2r_+/r - 1$. The resulting equations are singular both when $u = 1$ and $u = -1$, as was the case for Schwarzschild. Imposing regularity at $u = 1$ yields the equations

$$\begin{aligned}
f_K(1) &= \frac{-2r_+(\lambda+1) + 2i\omega r_0(r_+ + r_-)}{-i\omega r_+(1+2i\omega r_0)} f_1(1)\,, \quad f_\chi(1) = -\frac{2r_0 r_- r_+}{1+2i\omega r_0} f_1(1)\,,\\
f_0(1) &= \frac{r_0(r_+ + r_-)^2 + 2i\omega r_+^4}{r_+^2(r_+ + r_-)(1+2i\omega r_0)} f_1(1)\,.
\end{aligned} \tag{8.35}$$

Imposing regularity at $u = -1$ gives

$$f_K(-1) = -\frac{1}{i\omega} f_1(-1) \quad \text{and} \quad f_0(-1) = -f_1(-1)\,. \tag{8.36}$$

The three relations present in eq. (8.35) are coherent with eq. (7.29). They assure us that only one mode is left in the system: we have selected the ingoing gravitational mode at the horizon. However, there are only two relations in eq. (8.36), meaning that we have not selected only one mode. Since we know this equation should be coherent with eq. (7.22), we can find the last relation we need using eq. (7.22). If we consider only the outgoing gravitational mode at infinity $\mathfrak{g}_{+}^{\infty}(r)$ and go back to the original variables, we see that the decompositions of $H_1$ and $\chi$ are linked by

$$f_\chi(-1) = -\frac{r_+ r_-}{i\omega} f_1(-1)\,. \tag{8.37}$$

This last relation assures us that we get only outgoing gravitational perturbations at infinity.

At the moment when this manuscript is written, the numerical procedure has not been successfully conveyed, due to numerical errors in the integration from infinity when $\xi \neq 0$. Nevertheless, when $\xi = 0$, one recovers the Schwarzschild QNMs, confirming that imposing the gravitational boundary conditions on the full polar first-order system is enough to perform a numerical resolution. On fig. 8.4, we show the result of numerical integration in that case. We observe that the scalar perturbation is found to be zero everywhere, which is consistent with the decoupling limit $\xi = 0$.

## Conclusion

In this chapter, we have implemented a numerical method that allows us to compute QNMs from the first-order system describing the equations of perturbations, without resorting to a Schrödinger-like formulation like presented in chapter 4. This is very interesting for the study of polar perturbations in the context of scalar-tensor theories. Indeed, these theories contain two degrees of freedom, the usual one found in GR and a scalar. These two degrees of freedom cannot be decoupled in general, meaning that independent Schrödinger-like equations governing the propagation of each mode separately cannot be found: this means that most existing numerical techniques cannot be applied.

We used the asymptotic behaviour found in chapter 7 using the algorithm presented in chapter 5 in order to build ansätze for the functions entering the perturbations, singling out the relevant asymptotic behaviours at infinity and the horizon: outoing waves in the former case and ingoing waves in the latter.

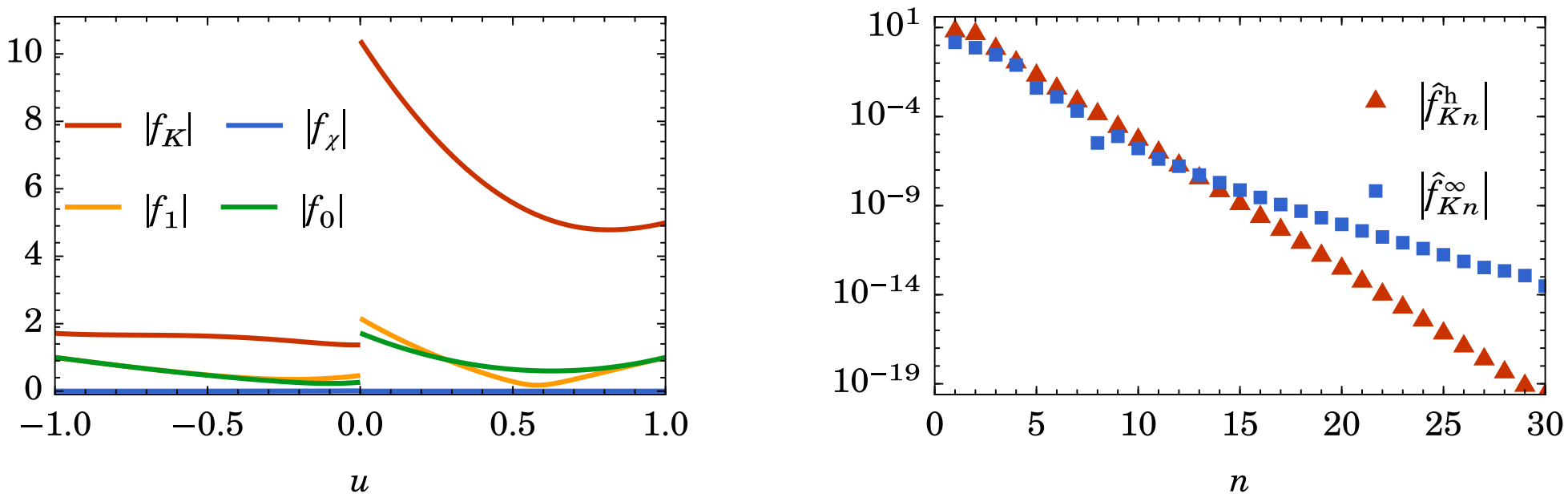

(a) Plot of the functions

(b) Plot of the coefficients

Figure 8.4. – Result of the integration of the BCL equations of polar perturbations for $\Omega = 0.5 - 0.3i$, $\xi = 0$ and $\ell = 2$ ($\lambda = 2$). The number of grid points was taken to be $N = 30$. Only the coefficients of the function $f_K$ integrated from the horizon ($\hat{f}^{\mathrm{h}}_{Kn}$) and infinity ($\hat{f}^{\infty}_{Kn}$) are shown in order to avoid cluttering the plot, but the coefficients of the other functions behave similarly.

We then integrated the resulting equations using a spectral method relying on approximation by Chebyshev polynomials.

The equations were integrated from infinity and from the horizon and matched in the bulk of spacetime for various values of $\Omega$. If the metric perturbations integrated from the horizon and infinity were linearly dependent at the matching point, we declared that the corresponding value of $\Omega$ was a QNM. Using this procedure, we recovered the QNMs of a Schwarzschild BH computed in [182] with precision $10^{-5}$. We proved that increasing the number of polynomials for the approximation makes the results more precise, and noted that doing this requires using numbers with precision higher than machine precision, due to numerical errors.

We finally applied this method to the BCL black hole. While the results are not complete yet, we prove that one can impose gravitational behaviour at both boundaries and recover the spectrum of a Schwarzschild BH when $\xi = 0$. This constitutes a proof of concept that the method developed in this chapter could allow to compute QNMs in both gravitational and scalar sectors using the asymptotic behaviours provided by the algorithm.

# CONCLUSION

This thesis focused on the study of GW propagation around a hairy BH in DHOST theories. This study is of paramount importance in the present context of gravitational physics for several reasons. First, a very diverse set of modified gravity theories and BH solutions have been proposed in the past 60 years, and it is important to check that existing solutions have appropriate GW propagation behaviours. Indeed, some solutions might contain apparent pathologies: studying BH perturbations could allow one to rule them out. Second, QNM measurements from ringdown data are becoming more and more precise, which means it is important to compute QNMs numerically for modified gravity solutions, as their comparison with experiments will lead to meaningful constraints on these theories. Third, the presence of behaviours quantitatively different from GR could be a guide towards the development of new experimental measurements.

In this context, we developed throughout this thesis a new method of computation of QNMs of BH solutions in DHOST theories. After a review of the construction of modified gravity theories in chapter 1, in which we described the DHOST action and its link with other theories such as the ones that include a coupling with an EGB term, we described in chapter 2 several nonrotating BH solutions that have been proposed in the literature in the context of DHOST theories. More precisely, we studied four different solutions. The first one, called BCL after the names of the authors of the paper describing it, is similar to a Reissner-Nordström BH with imaginary charge. The second one, called "stealth", reproduces exactly the metric sector of Schwarzschild while featuring a nonzero scalar field. The third one, called EsGB, is known as a small modification of the Schwarzschild metric, the deviation being controlled by a coupling with the Gauss-Bonnet parameter. Finally, the last BH solution comes from the compactification of a higher-dimensional EGB theory. While all studied BHs exhibit healthy metric sectors, we proved in the rest of the manuscript that all of them — except the EsGB solution — exhibit pathologies when one considers their perturbations.

In chapter 3, we paused from the description of BH dynamics in DHOST to describe a reformulation of these theories in a frame where the form of the action can be easily interpreted from a geometrical point of view. This reformulation allows one to better emphasize the fact that the degeneracy on which the construction of DHOST theories relies so much is not a coincidence but on the contrary comes from a deeper property of scalar-tensor theories.

In order to study the perturbations of the BHs introduced in chapter 2, we needed to define a perturbation framework. This was done in chapter 4, in which we reviewed the classical computations made by Regge, Wheeler and Zerilli for the Schwarzschild solution in GR that led to the derivation of Schrödinger-like equations for the perturbations. We also showed how the original perturbed Einstein's equations could be expressed as two decoupled first-order systems, through which we recovered the asymptotic behaviour of the perturbations without resorting to the Schrödinger-like reformulation in chapter 5. This is the idea behind the main result of this thesis.

Indeed, the most important idea that results from this thesis is that it is possible to study the dynamics of perturbations without resorting to a Schrödinger-like reformulation. While this is not a game-changing statement in the case of GR, it allowed us to decouple the degrees of freedom asymptotically and study the propagation of GW in the case of DHOST theories as we described in chapters 6 and 7. In order to do this, we used the first-order formulation of the perturbation equations that stems directly from the perturbed Einstein's equations. Then, we decoupled the system asymptotically by using a specific procedure that we described in chapter 5.

This procedure allowed us to discriminate between theories that have well-behaved metric and scalar sectors and theories that contain pathologies. As an example, we found that the asymptotic propagation properties in each sector were fine in the case of the EsGB solution, while we proved that it was not possible to describe directions of propagation in all cases for most of the other solutions. This could be seen in the case of odd-parity perturbations, as proven in chapter 6, as well as for even-parity perturbations, as proven in chapter 7. Therefore, we were able to use our study to rule out solutions, as we initially set out to do. Although the results were obtained only for DHOST solutions, we can expect our method to hold for many other modified theories of gravity.

Meanwhile, we obtained in chapter 6 the effective metric in which odd-parity perturbations propagate. This allowed us to study the horizons and singularities effectively seen by odd-parity perturbations on several BH backgrounds. The results confirmed the pathologies obtained using the asymptotic study we presented previously.

Furthermore, obtaining the behaviour of perturbations on both spacetime boundaries allowed us to find the proper conditions to set on perturbations at both the horizon and infinity in order to compute QNMs numerically, as we presented in chapter 8. Although the results obtained were not complete, the method used was promising as a proof of concept since it allowed us to recover the QNMs of the Schwarzschild solution in GR using only the first-order system and those of the BCL solution in the limit where scalar and gravitation are decoupled.

The results obtained in this manuscript pave the way for several new projects that would allow us to deepen our understanding of BH perturbations in modified gravity theories. As a first example, it would be interesting to apply the method we developed to new solutions in modified theories of gravity that are not DHOST. Indeed, the only requirements for our asymptotic analysis to work are the existence of a first-order system of equations, which is in general linked to the structure of Einstein's equations, and the possibility to find an analytic expression for the BH horizon position.

Comparisons with experiments would also require an improvement on the numerical part of the computation presented in chapter 8. This could be done by using existing solvers for eigenvalue problems. As a further improvement, the computation of QNMs for rotating solutions would also be required. Although one can expect this generalisation to be quite complicated due to the addition of a new angular variable which would make the Regge-Wheeler and Zerilli gauges obsolete, it could still be possible to study the slow rotation limit in which such tools are still available. A complete study would also contain a processing pipeline for LIGO/Virgo data, allowing one to finally compute constraints on existing theories from GW signals.

A more complete study of polar perturbations could be also performed, using the results of chapter 7 as a guide. Indeed, we observe that the asymptotic behaviours of axial and polar gravitational perturbations are very similar: this is a hint that the effective metrics might be the same, a statement that we could prove by looking at the dispersion relations in both cases.

Finally, one could study the stability of solutions in a more precise way by looking at the time evolution of a perturbation with initially compact support outside the horizon. This is already studied in the literature using the Green function associated with the Schrödinger-like equation; the possibility to generalise such results to a Green function associated with the first-order system should be studied and could lead to strong results concerning the stability of solutions along with new ways of differentiating between GR and modified gravity beyond a measurement of QNMs.

# Appendices

# PERTURBATION EQUATIONS AND ALGORITHM STEPS

In this section, we give the explicit perturbation equations for the even sector of perturbations of BHs described in chapter 2 solution, using the form

$$M_A \frac{\mathrm{d}Y}{\mathrm{d}r} = M_B Y \,. \tag{A.1}$$

This corresponds to a first-order system of the form (5.1), with

$$M = M_A^{-1} M_B \,. \tag{A.2}$$

We then describe the different steps of the algorithm of chapter 5 leading to a asymptotically diagonalized system both at the horizon and at infinity:

$$Y = \tilde{P}\tilde{Y} \,, \quad \frac{\mathrm{d}\tilde{X}}{\mathrm{d}r} = \tilde{M}\tilde{Y} \,. \tag{A.3}$$

We also give the asymptotic expression of the matrix $\tilde{M}$ in each case.

## A.1. Stealth solutions

Let us first turn to the stealth solutions, designed to mimc the metric sector of GR. The background is given in section 2.3.

### A.1.1. Case $\alpha \neq 0$

We consider the case where only $\alpha \neq 0$ (while $\beta$, $\gamma$ and $\delta$ are zero). In that case, the matrices $M_A$ and $M_B$ are decomposed as

$$M_A = M_{A[0]} + M_{A[1]}\omega \qquad M_B = M_{B[0]} + M_{B[1]}\omega + M_{B[2]}\omega^2 \,, \tag{A.4}$$

with

$$\left(M_{A[0]}\right)_{11} = \frac{2\alpha q^2 (2r-\mu)\sqrt{\mu r}}{r-\mu} \,, \quad \left(M_{A[0]}\right)_{12} = 0 \,,$$
$$\left(M_{A[0]}\right)_{13} = 0 \,, \quad \left(M_{A[0]}\right)_{14} = -4\alpha q^2 \sqrt{\mu r} \,,$$

$$
\begin{aligned}
\left(M_{A[0]}\right)_{21} &= -\frac{r\left(\mu^2\left(4\alpha q^2+1\right)-3r\left(\mu+2\alpha\mu q^2\right)+2r^2\right)}{(r-\mu)^2}\,,\\
\left(M_{A[0]}\right)_{22} &= \frac{8\alpha q\sqrt{\mu r}}{r-\mu}\,,\\
\left(M_{A[0]}\right)_{23} &= 0\,,\quad \left(M_{A[0]}\right)_{24} = 2r\left(\frac{2\alpha q^2(r-2\mu)}{r-\mu}+1\right)\,,\\
\left(M_{A[0]}\right)_{31} &= -2\alpha q^2\sqrt{\mu r}\,,\quad \left(M_{A[0]}\right)_{32} = -\frac{2\alpha\mu q}{r}\,,\\
\left(M_{A[0]}\right)_{33} &= -\left(\left(2\alpha q^2+1\right)(r-\mu)\right)\,,\quad \left(M_{A[0]}\right)_{34} = 0\,,\\
\left(M_{A[0]}\right)_{41} &= \frac{r\left(r-\mu\left(2\alpha q^2+1\right)\right)}{r-\mu}\,,\quad \left(M_{A[0]}\right)_{42} = \frac{2\alpha q\sqrt{\frac{\mu}{r}}(\mu-2r)}{r-\mu}\,,\\
\left(M_{A[0]}\right)_{43} &= 0\,,\quad \left(M_{A[0]}\right)_{44} = -r\left(2\alpha q^2+1\right)\,,\\
\left(M_{A[1]}\right)_{11} &= -2ir^2\left(2\alpha q^2+1\right)\,,\quad \left(M_{A[1]}\right)_{12} = 8i\alpha q\sqrt{\mu r}\,,\\
\left(M_{A[1]}\right)_{13} &= 0\,,\quad \left(M_{A[1]}\right)_{14} = 0\,,\\
\left(M_{A[1]}\right)_{21} &= 0\,,\quad \left(M_{A[1]}\right)_{22} = 0\,,\\
\left(M_{A[1]}\right)_{23} &= 0\,,\quad \left(M_{A[1]}\right)_{24} = 0\,,\\
\left(M_{A[1]}\right)_{31} &= 0\,,\quad \left(M_{A[1]}\right)_{32} = 0\,,\\
\left(M_{A[1]}\right)_{33} &= 0\,,\quad \left(M_{A[1]}\right)_{34} = 0\,,\\
\left(M_{A[1]}\right)_{41} &= 0\,,\quad \left(M_{A[1]}\right)_{42} = 0\,,\\
\left(M_{A[1]}\right)_{43} &= 0\,,\quad \left(M_{A[1]}\right)_{44} = 0\,,
\end{aligned} \tag{A.5}
$$

and

$$
\begin{aligned}
\left(M_{B[0]}\right)_{11} &= -\frac{4\alpha\lambda q^2\sqrt{\mu r}}{r-\mu}\,,\\
\left(M_{B[0]}\right)_{12} &= \frac{4\alpha\mu q\left(\mu\left(\lambda+2\alpha\lambda q^2+1\right)+r\left(-\lambda+2\alpha q^2-1\right)\right)}{r(r-\mu)\left(r-\mu\left(2\alpha q^2+1\right)\right)}\,,\\
\left(M_{B[0]}\right)_{13} &= 4\alpha q^2\left(-\lambda+2r\left(\frac{1}{-\mu-2\alpha\mu q^2+r}+\frac{1}{\mu-r}\right)-1\right)-2(\lambda+1)\,,\\
\left(M_{B[0]}\right)_{14} &= -\frac{4\alpha q^2\sqrt{\mu r}\left(-\mu+2\alpha q^2 r+r\right)}{(r-\mu)\left(r-\mu\left(2\alpha q^2+1\right)\right)}\,,\\
\left(M_{B[0]}\right)_{21} &= \frac{2\lambda r\left(r-\mu\left(2\alpha q^2+1\right)\right)}{(r-\mu)^2}\,,\\
\left(M_{B[0]}\right)_{22} &= -\frac{4\alpha q\sqrt{\frac{\mu}{r}}}{(r-\mu)^2\left(r-\mu\left(2\alpha q^2+1\right)\right)}\left(\mu^2\left(\lambda+2\alpha(\lambda+3)q^2+2\right)\right.\\
&\qquad \left.-\mu r\left(3\lambda+4\alpha(\lambda+2)q^2+5\right)+(2\lambda+3)r^2\right)\,,
\end{aligned}
$$

$$\left(M_{B[0]}\right)_{23} = \frac{16\alpha^2 q^4 (\mu r)^{3/2}}{(r-\mu)^2 \left(r-\mu\left(2\alpha q^2+1\right)\right)}\,,$$

$$\left(M_{B[0]}\right)_{24} = 2r\left(\frac{\lambda+2\alpha\lambda q^2-2}{\mu-r}+\frac{2\alpha q^2 r}{(r-\mu)^2}+\frac{2\left(\alpha q^2+1\right)}{\mu+2\alpha\mu q^2-r}\right)\,,$$

$$\left(M_{B[0]}\right)_{31} = 0\,,$$

$$\left(M_{B[0]}\right)_{32} = -\frac{2\alpha\mu q\left(\mu\left(3\alpha q^2+2\right)+2r\left(\alpha q^2-1\right)\right)}{r^2\left(r-\mu\left(2\alpha q^2+1\right)\right)}\,,$$

$$\left(M_{B[0]}\right)_{33} = \frac{\mu\left(\mu\left(2\alpha q^2+1\right)^2-r\left(8\alpha^2 q^4+4\alpha q^2+1\right)\right)}{r\left(r-\mu\left(2\alpha q^2+1\right)\right)}\,,$$

$$\left(M_{B[0]}\right)_{34} = \frac{\alpha q^2\left(4\alpha q^2+1\right)\sqrt{\frac{\mu}{r}}(\mu+r)}{r-\mu\left(2\alpha q^2+1\right)}\,,$$

$$\left(M_{B[0]}\right)_{41} = 0\,,$$

$$\left(M_{B[0]}\right)_{42} = \frac{\alpha\sqrt{\mu}q\left(\mu^2\left(4\alpha q^2+1\right)-7r\left(\mu+2\alpha\mu q^2\right)+6r^2\right)}{r^{3/2}(r-\mu)\left(r-\mu\left(2\alpha q^2+1\right)\right)}\,,$$

$$\left(M_{B[0]}\right)_{43} = -\frac{4\alpha^2 q^4\sqrt{\mu^3 r}}{(r-\mu)\left(r-\mu\left(2\alpha q^2+1\right)\right)}\,,$$

$$\left(M_{B[0]}\right)_{44} = \frac{\mu\left(\mu+2\alpha\mu q^2\left(3\alpha q^2+2\right)+r\left(2\alpha q^2\left(\alpha q^2-1\right)-1\right)\right)}{(r-\mu)\left(r-\mu\left(2\alpha q^2+1\right)\right)}\,,$$

$$\left(M_{B[1]}\right)_{11} = -ir\left(2\alpha q^2\left(3-\frac{r^2}{(r-\mu)^2}\right)+\frac{r}{\mu-r}+3\right)\,,$$

$$\left(M_{B[1]}\right)_{12} = \frac{4i\alpha q\sqrt{\frac{\mu}{r}}\left(\mu+4\alpha\mu q^2-r\left(2\alpha q^2+1\right)\right)}{r-\mu\left(2\alpha q^2+1\right)}\,,$$

$$\left(M_{B[1]}\right)_{13} = -\frac{8i\alpha\sqrt{\mu}q^2 r^{3/2}\left(-\mu\left(4\alpha q^2+1\right)+2\alpha q^2 r+r\right)}{(r-\mu)\left(r-\mu\left(2\alpha q^2+1\right)\right)}\,,$$

$$\left(M_{B[1]}\right)_{14} = \frac{2ir\left(-\mu+2\alpha q^2 r+r\right)\left(-\mu\left(4\alpha q^2+1\right)+2\alpha q^2 r+r\right)}{(r-\mu)\left(r-\mu\left(2\alpha q^2+1\right)\right)}\,,$$

$$\left(M_{B[1]}\right)_{21} = \frac{2i\alpha\sqrt{\mu}q^2 r^{5/2}(2r-\mu)}{(r-\mu)^3}\,,$$

$$\left(M_{B[1]}\right)_{22} = \frac{8i\alpha\mu q r}{(r-\mu)^2}\,,$$

$$\left(M_{B[1]}\right)_{23} = \frac{4ir^2\left(-\mu\left(4\alpha q^2+1\right)+2\alpha q^2 r+r\right)}{(r-\mu)^2}\,,$$

$$\left(M_{B[1]}\right)_{24} = \frac{4i\alpha q^2\sqrt{\mu r^5}}{(r-\mu)^2}\,,$$

$$
\begin{aligned}
\left(M_{B[1]}\right)_{31} &= -ir\left(\frac{2\alpha q^2 r}{r-\mu}+1\right), \\
\left(M_{B[1]}\right)_{32} &= \frac{2i\alpha q\sqrt{\frac{\mu}{r}}\left(3\mu^2\left(2\alpha q^2+1\right)+2r^2\left(\alpha q^2+1\right)-\mu r\left(6\alpha q^2+5\right)\right)}{(r-\mu)\left(r-\mu\left(2\alpha q^2+1\right)\right)}, \\
\left(M_{B[1]}\right)_{33} &= -\frac{4i\alpha\sqrt{\mu}q^2 r^{3/2}\left(2\alpha q^2+1\right)}{\mu+2\alpha\mu q^2-r}, \\
\left(M_{B[1]}\right)_{34} &= -\frac{ir\left(2\alpha q^2+1\right)\left(-\mu+2\alpha q^2 r+r\right)}{r-\mu\left(2\alpha q^2+1\right)}, \\
\left(M_{B[1]}\right)_{41} &= -\frac{2i\alpha q^2\sqrt{\mu r^5}}{(r-\mu)^2}, \quad \left(M_{B[1]}\right)_{42} = -\frac{2i\alpha\mu q r}{(r-\mu)^2}, \\
\left(M_{B[1]}\right)_{43} &= \frac{ir^2\left(2\alpha q^2+1\right)}{\mu-r}, \quad \left(M_{B[1]}\right)_{44} = 0, \\
\left(M_{B[2]}\right)_{11} &= 0, \quad \left(M_{B[2]}\right)_{12} = 0, \\
\left(M_{B[2]}\right)_{13} &= 0, \quad \left(M_{B[2]}\right)_{14} = 0, \\
\left(M_{B[2]}\right)_{21} &= -\frac{2r^4\left(2\alpha q^2+1\right)}{(r-\mu)^2}, \quad \left(M_{B[2]}\right)_{22} = \frac{8\alpha q\sqrt{\mu r^5}}{(r-\mu)^2}, \\
\left(M_{B[2]}\right)_{23} &= 0, \quad \left(M_{B[2]}\right)_{24} = 0, \\
\left(M_{B[2]}\right)_{31} &= 0, \quad \left(M_{B[2]}\right)_{32} = 0, \\
\left(M_{B[2]}\right)_{33} &= 0, \quad \left(M_{B[2]}\right)_{34} = 0, \\
\left(M_{B[2]}\right)_{41} &= 0, \quad \left(M_{B[2]}\right)_{42} = 0, \\
\left(M_{B[2]}\right)_{43} &= 0, \quad \left(M_{B[2]}\right)_{44} = 0.
\end{aligned}
\tag{A.6}
$$

Let us give the expression of the diagonalized system at infinity. One poses $x = \sqrt{r}$ and $\alpha = \zeta/2q^2$. In order to speed up the process of using the algorithm, one starts with a change of variables of the form $\mathrm{Diag}(1, qx^3, x^2, 1)$. Then, the algorithm steps one uses are the following:

1. putting the system in block Jordan form;
2. reducing the Poincaré rank with a $\mathrm{Diag}(1, x, 1, 1)$ transformation;
3. normalising the subleading order;
4. reducing the Poincaré rank with a $\mathrm{Diag}(1, x, 1, 1)$ transformation;
5. putting the system in block Jordan form;
6. normalising the subleading order;
7. reducing the Poincaré rank with a $\mathrm{Diag}(1, x, 1, x)$ transformation;
8. putting the system in block Jordan form;
9. normalising the sub-subleading order;

10. reducing the Poincaré rank with a $\mathrm{Diag}(1, x, x^2, x^3)$ transformation;
11. putting the system in block Jordan form;
12. splitting the system into decoupled subsystems: the gravitational part is decoupled at this stage and one keeps working with the scalar part only;
13. putting the system in block Jordan form;
14. normalising the subleading order;
15. reducing the Poincaré rank with a $\mathrm{Diag}(1, x)$ transformation;
16. putting the system in block Jordan form;
17. using an exponential shift $\mathrm{Diag}(\exp(-2i\sqrt{\mu}\omega x), \exp(-2i\sqrt{\mu}\omega x))$ to go to a nilpotent leading order;
18. normalising the subleading order;
19. reducing the Poincaré rank with a $\mathrm{Diag}(1, x)$ transformation;
20. putting the system in block Jordan form.

One is finally left with a matrix of the form

$$\tilde{M}_{\mathrm{grav}} = \mathrm{Diag}(-2i\sqrt{1+\zeta}\omega, +2i\sqrt{1+\zeta}\omega)x + \mathrm{Diag}(-2i\zeta\mu\omega, +2i\zeta\mu\omega)\frac{1}{x} + \mathcal{O}\left(\frac{1}{x^2}\right) \tag{A.7}$$

for the gravitational part, and a matrix of the form

$$\tilde{M}_{\mathrm{scal}} = \mathrm{Diag}(-7-2i\sqrt{\lambda}, -7+2i\sqrt{\lambda})\frac{1}{x} + \mathcal{O}\left(\frac{1}{x^2}\right) \tag{A.8}$$

for the scalar part. Going back to the original variables $Y$ and $r$, one recovers the asymptotic behaviour of eq. (7.37).

The study at the horizon can be tricky, since there are singularities both at $r = \mu$ and $r = \mu(1+\zeta)$. We concentrate on the latter case, which is the horizon seen by axial gravitons (see chapter 6). At this radius, the system is diagonalizable and no algorithm step is needed except a Jordan reduction.

### A.1.2. Case $\beta \neq 0$

We now consider the case where only $\beta \neq 0$ (while $\alpha$, $\gamma$ and $\delta$ are zero). In that case, the matrices $M_A$ and $M_B$ are decomposed as

$$M_A = M_{A[0]} + M_{A[1]}\omega \qquad M_B = M_{B[0]} + M_{B[1]}\omega + M_{B[2]}\omega^2\,, \tag{A.9}$$

with

$$\left(M_{A[0]}\right)_{11} = \frac{4\beta q^4\sqrt{\mu^3 r}}{r-\mu}\,, \quad \left(M_{A[0]}\right)_{12} = \frac{16\beta\mu q^3}{r-\mu}\,,$$

$$
\begin{aligned}
&\left(M_{A[0]}\right)_{13}=0\,,\quad \left(M_{A[0]}\right)_{14}=8\beta q^4\sqrt{\mu r}\,,\\
&\left(M_{A[0]}\right)_{21}=\frac{r\left(\mu^2\left(4\beta q^4-1\right)-2r^2+3\mu r\right)}{(r-\mu)^2}\,,\quad \left(M_{A[0]}\right)_{22}=\frac{16\beta q^3\sqrt{\mu^3 r}}{(r-\mu)^2}\,,\\
&\left(M_{A[0]}\right)_{23}=0\,,\quad \left(M_{A[0]}\right)_{24}=\frac{2r\left(\mu\left(4\beta q^4-1\right)+r\right)}{r-\mu}\,,\\
&\left(M_{A[0]}\right)_{31}=0\,,\quad \left(M_{A[0]}\right)_{32}=-\frac{4\beta\mu q^3}{r}\,,\\
&\left(M_{A[0]}\right)_{33}=\mu-r\,,\quad \left(M_{A[0]}\right)_{34}=0\,,\\
&\left(M_{A[0]}\right)_{41}=r\,,\quad \left(M_{A[0]}\right)_{42}=\frac{4\beta\mu^{3/2}q^3}{\sqrt{r}(\mu-r)}\,,\\
&\left(M_{A[0]}\right)_{43}=0\,,\quad \left(M_{A[0]}\right)_{44}=-r\,,\\
&\left(M_{A[1]}\right)_{11}=-2ir^2\,,\quad \left(M_{A[1]}\right)_{12}=-16i\beta q^3\sqrt{\mu r}\,,\\
&\left(M_{A[1]}\right)_{13}=0\,,\quad \left(M_{A[1]}\right)_{14}=0\,,\\
&\left(M_{A[1]}\right)_{21}=0\,,\quad \left(M_{A[1]}\right)_{22}=0\,,\\
&\left(M_{A[1]}\right)_{23}=0\,,\quad \left(M_{A[1]}\right)_{24}=0\,,\\
&\left(M_{A[1]}\right)_{31}=0\,,\quad \left(M_{A[1]}\right)_{32}=0\,,\\
&\left(M_{A[1]}\right)_{33}=0\,,\quad \left(M_{A[1]}\right)_{34}=0\,,\\
&\left(M_{A[1]}\right)_{41}=0\,,\quad \left(M_{A[1]}\right)_{42}=0\,,\\
&\left(M_{A[1]}\right)_{43}=0\,,\quad \left(M_{A[1]}\right)_{44}=0\,,
\end{aligned}
\tag{A.10}
$$

and

$$
\begin{aligned}
&\left(M_{B[0]}\right)_{11}=0\,,\quad \left(M_{B[0]}\right)_{12}=-\frac{8\beta(\lambda+1)\mu q^3}{r(r-\mu)}\,,\\
&\left(M_{B[0]}\right)_{13}=-\frac{2\left((\lambda+1)\mu^2-2\mu r\left(\lambda+4\beta q^4+1\right)+(\lambda+1)r^2\right)}{(r-\mu)^2}\,,\\
&\left(M_{B[0]}\right)_{14}=-\frac{16\beta q^4\sqrt{\mu^3 r}}{(r-\mu)^2}\,,\\
&\left(M_{B[0]}\right)_{21}=\frac{2\lambda r}{r-\mu}\,,\quad \left(M_{B[0]}\right)_{22}=-\frac{8\beta(\lambda+1)\mu^{3/2}q^3}{\sqrt{r}(r-\mu)^2}\,,\\
&\left(M_{B[0]}\right)_{23}=\frac{16\beta q^4(\mu r)^{3/2}}{(r-\mu)^3}\,,\quad \left(M_{B[0]}\right)_{24}=-\frac{2r\left(\mu^2\left(\lambda+8\beta q^4\right)+\lambda r^2-2\lambda\mu r\right)}{(r-\mu)^3}\,,\\
&\left(M_{B[0]}\right)_{31}=0\,,\quad \left(M_{B[0]}\right)_{32}=0\,,\\
&\left(M_{B[0]}\right)_{33}=\frac{\mu\left(\mu-r\left(4\beta q^4+1\right)\right)}{r(r-\mu)}\,,\quad \left(M_{B[0]}\right)_{34}=\frac{2\beta q^4\sqrt{\frac{\mu}{r}}(\mu+r)}{r-\mu}\,,\\
&\left(M_{B[0]}\right)_{41}=0\,,\quad \left(M_{B[0]}\right)_{42}=0\,,
\end{aligned}
$$

$$
\begin{aligned}
&\left(M_{B[0]}\right)_{43} = -\frac{4\beta q^4\sqrt{\mu^3 r}}{(r-\mu)^2}\,, \quad \left(M_{B[0]}\right)_{44} = \frac{\mu\left(\mu + 2\beta\mu q^4 + r\left(2\beta q^4 - 1\right)\right)}{(r-\mu)^2}\,,\\
&\left(M_{B[1]}\right)_{11} = -\frac{ir\left(3\mu^2 - \mu r\left(4\beta q^4 + 5\right) + 2r^2\right)}{(r-\mu)^2}\,, \quad \left(M_{B[1]}\right)_{12} = \frac{16 i\beta q^3\sqrt{\mu^3 r}}{(r-\mu)^2}\,,\\
&\left(M_{B[1]}\right)_{13} = 0\,, \quad \left(M_{B[1]}\right)_{14} = \frac{2ir\left(\mu\left(4\beta q^4 - 1\right) + r\right)}{r-\mu}\,,\\
&\left(M_{B[1]}\right)_{21} = \frac{4i\beta\mu^{3/2}q^4 r^{5/2}}{(r-\mu)^3}\,, \quad \left(M_{B[1]}\right)_{22} = \frac{16 i\beta\mu^2 q^3 r}{(r-\mu)^3}\,,\\
&\left(M_{B[1]}\right)_{23} = \frac{4ir^2\left(\mu\left(4\beta q^4 - 1\right) + r\right)}{(r-\mu)^2}\,, \quad \left(M_{B[1]}\right)_{24} = -\frac{8i\beta q^4\sqrt{\mu r^5}}{(r-\mu)^2}\,,\\
&\left(M_{B[1]}\right)_{31} = -ir\,, \quad \left(M_{B[1]}\right)_{32} = -\frac{4i\beta q^3\sqrt{\mu r}}{r-\mu}\,,\\
&\left(M_{B[1]}\right)_{33} = 0\,, \quad \left(M_{B[1]}\right)_{34} = -ir\,,\\
&\left(M_{B[1]}\right)_{41} = 0\,, \quad \left(M_{B[1]}\right)_{42} = -\frac{4i\beta\mu q^3 r}{(r-\mu)^2}\,,\\
&\left(M_{B[1]}\right)_{43} = \frac{ir^2}{\mu - r}\,, \quad \left(M_{B[1]}\right)_{44} = 0\,,\\
&\left(M_{B[2]}\right)_{11} = 0\,, \quad \left(M_{B[2]}\right)_{12} = 0\,,\\
&\left(M_{B[2]}\right)_{13} = 0\,, \quad \left(M_{B[2]}\right)_{14} = 0\,,\\
&\left(M_{B[2]}\right)_{21} = -\frac{2r^4}{(r-\mu)^2}\,, \quad \left(M_{B[2]}\right)_{22} = -\frac{16\beta q^3\sqrt{\mu r^5}}{(r-\mu)^2}\,,\\
&\left(M_{B[2]}\right)_{23} = 0\,, \quad \left(M_{B[2]}\right)_{24} = 0\,,\\
&\left(M_{B[2]}\right)_{31} = 0\,, \quad \left(M_{B[2]}\right)_{32} = 0\,,\\
&\left(M_{B[2]}\right)_{33} = 0\,, \quad \left(M_{B[2]}\right)_{34} = 0\,,\\
&\left(M_{B[2]}\right)_{41} = 0\,, \quad \left(M_{B[2]}\right)_{42} = 0\,,\\
&\left(M_{B[2]}\right)_{43} = 0\,, \quad \left(M_{B[2]}\right)_{44} = 0\,.
\end{aligned}
\tag{A.11}
$$

Let us give the expression of the diagonalized system at infinity. One poses $x = \sqrt{r}$ and $\beta = \xi/4q^4$. In order to speed up the process of using the algorithm, one starts with a change of variables of the form $\mathrm{Diag}(1, qx^3, x^2, 1)$. Then, the algorithm steps one uses are the following:

1. putting the system in block Jordan form;
2. reducing the Poincaré rank with a $\mathrm{Diag}(1, x, 1, 1)$ transformation;
3. normalising the subleading order;
4. reducing the Poincaré rank with a $\mathrm{Diag}(1, x, 1, 1)$ transformation;
5. putting the system in block Jordan form;

6. normalising the subleading order;
7. reducing the Poincaré rank with a $\mathrm{Diag}(1,x,1,x)$ transformation;
8. putting the system in block Jordan form;
9. normalising the sub-subleading order;
10. reducing the Poincaré rank with a $\mathrm{Diag}(1,x,x^2,x^3)$ transformation;
11. putting the system in block Jordan form;
12. splitting the system into decoupled subsystems: the gravitational part is decoupled at this stage and one keeps working with the scalar part only;
13. putting the system in block Jordan form;
14. using an exponential shift $\mathrm{Diag}(\exp(-2i\sqrt{\mu}\omega x),\exp(-2i\sqrt{\mu}\omega x))$ to go to a nilpotent leading order;
15. normalising the subleading order;
16. reducing the Poincaré rank with a $\mathrm{Diag}(1,x)$ transformation;
17. putting the system in block Jordan form.

One is finally left with a matrix of the form

$$\tilde{M}_{\text{grav}} = \mathrm{Diag}(-2i\omega, +2i\omega)x + \mathrm{Diag}(-2-2i\mu\omega, -2+2i\mu\omega)\frac{1}{x} + \mathcal{O}\left(\frac{1}{x^2}\right) \quad \text{(A.12)}$$

for the gravitational part, and a matrix of the form

$$\tilde{M}_{\text{scal}} = \mathrm{Diag}(-7-2i\sqrt{\lambda}, -7+2i\sqrt{\lambda})\frac{1}{x} + \mathcal{O}\left(\frac{1}{x^2}\right) \quad \text{(A.13)}$$

for the scalar part. Going back to the original variables $Y$ and $r$, one recovers the asymptotic behaviour of eqs. (7.33) and (7.35). One can note that the diagonalisation procedure in that case is very similar to the one presented in the case $\alpha \neq 0$.

At the horizon, the system is diagonalizable: no algorithm step is needed and the behaviours of eq. (7.34) can be obtained immediately.

### A.1.3. Case $\gamma \neq 0$

We finally consider the case where only $\gamma \neq 0$ (while $\alpha$, $\beta$ and $\delta$ are zero). In that case, the matrices $M_A$ and $M_B$ are decomposed as

$$M_A = M_{A[0]} + M_{A[1]}\omega \qquad M_B = M_{B[0]} + M_{B[1]}\omega + M_{B[2]}\omega^2\,, \quad \text{(A.14)}$$

with

$$\left(M_{A[0]}\right)_{11} = 0\,, \quad \left(M_{A[0]}\right)_{12} = -\frac{4\gamma\mu q^3 r^2}{r-\mu}\,, \quad \left(M_{A[0]}\right)_{13} = 0\,,$$

$$
\begin{aligned}
&\left(M_{A[0]}\right)_{14}=0\,,\quad \left(M_{A[0]}\right)_{21}=\frac{r(\mu-2r)}{r-\mu}\,,\quad \left(M_{A[0]}\right)_{22}=-\frac{4\gamma\mu^{3/2}q^3r^{5/2}}{(r-\mu)^2}\,,\\
&\left(M_{A[0]}\right)_{23}=0\,,\quad \left(M_{A[0]}\right)_{24}=2r\,,\quad \left(M_{A[0]}\right)_{31}=0\,,\\
&\left(M_{A[0]}\right)_{32}=0\,,\quad \left(M_{A[0]}\right)_{33}=\mu-r\,,\quad \left(M_{A[0]}\right)_{34}=0\,,\\
&\left(M_{A[0]}\right)_{41}=r\,,\quad \left(M_{A[0]}\right)_{42}=0\,,\quad \left(M_{A[0]}\right)_{43}=0\,,\\
&\left(M_{A[0]}\right)_{44}=-r\,,\quad \left(M_{A[1]}\right)_{11}=-2ir^2\,,\quad \left(M_{A[1]}\right)_{12}=0\,,\\
&\left(M_{A[1]}\right)_{13}=0\,,\quad \left(M_{A[1]}\right)_{14}=0\,,\quad \left(M_{A[1]}\right)_{21}=0\,,\\
&\left(M_{A[1]}\right)_{22}=0\,,\quad \left(M_{A[1]}\right)_{23}=0\,,\quad \left(M_{A[1]}\right)_{24}=0\,,\\
&\left(M_{A[1]}\right)_{31}=0\,,\quad \left(M_{A[1]}\right)_{32}=0\,,\quad \left(M_{A[1]}\right)_{33}=0\,,\\
&\left(M_{A[1]}\right)_{34}=0\,,\quad \left(M_{A[1]}\right)_{41}=0\,,\quad \left(M_{A[1]}\right)_{42}=0\,,\\
&\left(M_{A[1]}\right)_{43}=0\,,\quad \left(M_{A[1]}\right)_{44}=0\,,
\end{aligned}
\tag{A.15}
$$

and

$$
\begin{aligned}
&\left(M_{B[0]}\right)_{11}=0\,,\quad \left(M_{B[0]}\right)_{12}=0\,,\\
&\left(M_{B[0]}\right)_{13}=-\frac{2\left((\lambda+1)\mu^2+2\gamma\mu q^4r^3+(\lambda+1)r^2-2(\lambda+1)\mu r\right)}{(r-\mu)^2}\,,\\
&\left(M_{B[0]}\right)_{14}=\frac{2\gamma\sqrt{\mu}q^4r^{5/2}(\mu+r)}{(r-\mu)^2}\,,\\
&\left(M_{B[0]}\right)_{21}=\frac{2\lambda r}{r-\mu}\,,\quad \left(M_{B[0]}\right)_{22}=0\,,\quad \left(M_{B[0]}\right)_{23}=-\frac{4\gamma\mu^{3/2}q^4r^{7/2}}{(r-\mu)^3}\,,\\
&\left(M_{B[0]}\right)_{24}=\frac{2r\left(-\lambda\mu^2+\gamma\mu q^4r^3+r^2\left(\gamma\mu^2q^4-\lambda\right)+2\lambda\mu r\right)}{(r-\mu)^3}\,,\\
&\left(M_{B[0]}\right)_{31}=0\,,\quad \left(M_{B[0]}\right)_{32}=0\,,\\
&\left(M_{B[0]}\right)_{33}=-\frac{\mu}{r}\,,\quad \left(M_{B[0]}\right)_{34}=0\,,\\
&\left(M_{B[0]}\right)_{41}=0\,,\quad \left(M_{B[0]}\right)_{42}=0\,,\\
&\left(M_{B[0]}\right)_{43}=0\,,\quad \left(M_{B[0]}\right)_{44}=-\frac{\mu}{r-\mu}\,,\\
&\left(M_{B[1]}\right)_{11}=-\frac{ir(2r-3\mu)}{r-\mu}\,,\quad \left(M_{B[1]}\right)_{12}=-\frac{4i\gamma q^3\sqrt{\mu r^7}}{(r-\mu)^2}\,,\\
&\left(M_{B[1]}\right)_{13}=0\,,\quad \left(M_{B[1]}\right)_{14}=2ir\,,\\
&\left(M_{B[1]}\right)_{21}=0\,,\quad \left(M_{B[1]}\right)_{22}=-\frac{4i\gamma\mu q^3r^4}{(r-\mu)^3}\,,\\
&\left(M_{B[1]}\right)_{23}=\frac{4ir^2}{r-\mu}\,,\quad \left(M_{B[1]}\right)_{24}=0\,,\\
&\left(M_{B[1]}\right)_{31}=-ir\,,\quad \left(M_{B[1]}\right)_{32}=0\,,
\end{aligned}
$$

$$
\begin{aligned}
&(M_{B[1]})_{33} = 0\,, \quad (M_{B[1]})_{34} = -ir\,,\\
&(M_{B[1]})_{41} = 0\,, \quad (M_{B[1]})_{42} = 0\,,\\
&(M_{B[1]})_{43} = \frac{ir^2}{\mu - r}\,, \quad (M_{B[1]})_{44} = 0\,,\\
&(M_{B[2]})_{11} = 0\,, \quad (M_{B[2]})_{12} = 0\,,\\
&(M_{B[2]})_{13} = 0\,, \quad (M_{B[2]})_{14} = 0\,,\\
&(M_{B[2]})_{21} = -\frac{2r^4}{(r-\mu)^2}\,, \quad (M_{B[2]})_{22} = 0\,,\\
&(M_{B[2]})_{23} = 0\,, \quad (M_{B[2]})_{24} = 0\,,\\
&(M_{B[2]})_{31} = 0\,, \quad (M_{B[2]})_{32} = 0\,,\\
&(M_{B[2]})_{33} = 0\,, \quad (M_{B[2]})_{34} = 0\,,\\
&(M_{B[2]})_{41} = 0\,, \quad (M_{B[2]})_{42} = 0\,,\\
&(M_{B[2]})_{43} = 0\,, \quad (M_{B[2]})_{44} = 0\,.
\end{aligned} \tag{A.16}
$$

Let us give the expression of the diagonalized system at infinity written in the variable $x$ such that $x = \sqrt{r}$. In order to speed up the process of using the algorithm, one starts with a change of variables of the form $\mathrm{Diag}(1, x^2, x^2, x)$. Then, the algorithm steps one uses are the following:

1. putting the system in block Jordan form;
2. splitting the system into decoupled subsystems;
3. using an exponential shift $\mathrm{Diag}(\exp\left(-2i\omega x^3/3\sqrt{\mu}\right), \exp\left(-2i\omega x^3/3\sqrt{\mu}, 1, 1\right)$ to go to a nilpotent leading order;
4. reducing the Poincaré rank with a $\mathrm{Diag}(1, x, 1, x)$ transformation;
5. putting the system in block Jordan form;
6. at this stage the gravitational part is decoupled; one pursues with the scalar part by normalizing the subleading order;
7. reducing the Poincaré rank with a $\mathrm{Diag}(1, x)$ transformation;
8. putting the system in block Jordan form;
9. using an exponential shift of the form $\mathrm{Diag}(\exp\left(-2i\sqrt{\mu}\omega x\right), \exp\left(-2i\sqrt{\mu}\omega x\right))$ to go to a nilpotent leading order;
10. normalizing the subleading order;
11. reducing the Poincaré rank with a $\mathrm{Diag}(1, x)$ transformation;
12. putting the system in block Jordan form.

One is finally left with a matrix of the form

$$
\tilde{M}_{\mathrm{grav}} = \mathrm{Diag}(-2i\omega, +2i\omega)x + \mathrm{Diag}(-2i\mu\omega, +2i\mu\omega)\frac{1}{x} + \mathcal{O}\left(\frac{1}{x^2}\right) \tag{A.17}
$$

for the gravitational part, and a matrix of the form

$$\tilde{M}_{\text{scal}} = \begin{pmatrix} -5 & 0 \\ 1 & -5 \end{pmatrix} \frac{1}{x} + \mathcal{O}\left(\frac{1}{x^2}\right). \tag{A.18}$$

for the scalar part. Going back to the original variables $Y$ and $r$, one recovers the leading asymptotic behaviour of eqs. (7.33) and (7.35).

At the horizon, the system is diagonalizable: no algorithm step is needed and the behaviours of eq. (7.34) can be obtained immediately, similarly to the case $\beta \neq 0$.

## A.2. EsGB solution

We consider the EsGB solution given in section 2.4. We write the matrices $M_A$ and $M_B$ as series in $\varepsilon$:

$$\begin{aligned} M_A &= M_{A,0} + M_{A,1}\,\varepsilon + M_{A,2}\,\varepsilon^2 + \mathcal{O}(\varepsilon^3)\,, \\ M_B &= M_{B,0} + M_{B,1}\,\varepsilon + M_{B,2}\,\varepsilon^2 + \mathcal{O}(\varepsilon^3)\,. \end{aligned} \tag{A.19}$$

The coefficients of the matrices $M_{A,i}$ and $M_{B,i}$ for $i \in \{0,1,2\}$ are given by

$$\left(M_{A,0}\right)_{11} = -2i\Omega z^2\,, \quad \left(M_{A,0}\right)_{12} = -\frac{8i\Omega}{z}\,, \quad \left(M_{A,0}\right)_{13} = 0\,,$$
$$\left(M_{A,0}\right)_{14} = 0\,, \quad \left(M_{A,0}\right)_{21} = \frac{z}{z-1} - \frac{2z^2}{z-1}\,,$$
$$\left(M_{A,0}\right)_{22} = -\frac{8}{(z-1)z^2} - \frac{4z}{z-1} + \frac{8}{(z-1)z}\,,$$
$$\left(M_{A,0}\right)_{23} = 0\,, \quad \left(M_{A,0}\right)_{24} = 2z\,, \quad \left(M_{A,0}\right)_{31} = 0\,, \quad \left(M_{A,0}\right)_{32} = 0\,,$$
$$\left(M_{A,0}\right)_{33} = 1-z\,, \quad \left(M_{A,0}\right)_{34} = 0\,, \quad \left(M_{A,0}\right)_{41} = z\,, \quad \left(M_{A,0}\right)_{42} = -\frac{4}{z^2}\,,$$
$$\left(M_{A,0}\right)_{43} = 0\,,$$
$$\left(M_{A,0}\right)_{44} = -z\,, \quad \left(M_{A,1}\right)_{11} = 0\,,$$
$$\left(M_{A,1}\right)_{12} = -\frac{8i\rho_2\Omega}{3z^4} - \frac{4i\rho_2\Omega}{z^3} - \frac{8i\rho_2\Omega}{z^2}\,,$$
$$\left(M_{A,1}\right)_{13} = 0\,, \quad \left(M_{A,1}\right)_{14} = 0\,, \quad \left(M_{A,1}\right)_{21} = 0\,,$$
$$\left(M_{A,1}\right)_{22} = -\frac{10\rho_2}{3(z-1)z^5} - \frac{32\rho_2}{15(z-1)z^4} - \frac{5\rho_2}{(z-1)z^3} + \frac{8\rho_2}{(z-1)z^2} - \frac{73\rho_2 z}{15(z-1)}\,,$$
$$\left(M_{A,1}\right)_{23} = 0\,, \quad \left(M_{A,1}\right)_{24} = 0\,, \quad \left(M_{A,1}\right)_{31} = 0\,, \quad \left(M_{A,1}\right)_{32} = 0\,,$$
$$\left(M_{A,1}\right)_{33} = 0\,, \quad \left(M_{A,1}\right)_{34} = 0\,, \quad \left(M_{A,1}\right)_{41} = 0\,,$$
$$\left(M_{A,1}\right)_{42} = -\frac{4\rho_2}{3z^5} - \frac{2\rho_2}{z^4} - \frac{4\rho_2}{z^3}\,,$$

$$\left(M_{A,1}\right)_{43} = 0\,, \quad \left(M_{A,1}\right)_{44} = 0\,, \quad \left(M_{A,2}\right)_{11} = \frac{8i\Omega}{z^4} - \frac{8i\Omega}{z}\,,$$

$$\begin{aligned}\left(M_{A,2}\right)_{12} = & -\frac{2i\rho_2^2\Omega}{9z^7} - \frac{4i\rho_3\Omega}{9z^7} - \frac{496i\Omega}{15z^7} - \frac{56i\rho_2^2\Omega}{75z^6} - \frac{4i\rho_3\Omega}{3z^6} - \frac{8i\Omega}{5z^6} \\ & - \frac{73i\rho_2^2\Omega}{30z^5} - \frac{11i\rho_3\Omega}{3z^5} - \frac{8i\Omega}{3z^5} - \frac{146i\rho_2^2\Omega}{45z^4} - \frac{4i\rho_3\Omega}{z^4} + \frac{32i\Omega}{z^4} \\ & - \frac{73i\rho_2^2\Omega}{15z^3} - \frac{4i\rho_3\Omega}{z^3} - \frac{146i\rho_2^2\Omega}{15z^2} + \frac{8i\Omega}{z^2} - \frac{8i\Omega}{z}\,,\end{aligned}$$

$$\left(M_{A,2}\right)_{13} = 0\,, \quad \left(M_{A,2}\right)_{14} = 0\,,$$

$$\left(M_{A,2}\right)_{21} = -\frac{44}{5z^5} - \frac{31}{3z^4} - \frac{12}{z^3} - \frac{4}{z^2} - \frac{8}{3z} - \frac{1}{3}\,,$$

$$\begin{aligned}\left(M_{A,2}\right)_{22} = & -\frac{8\rho_2^2}{27(z-1)z^8} - \frac{16\rho_3}{27(z-1)z^8} - \frac{1312}{15(z-1)z^8} - \frac{341\rho_2^2}{450(z-1)z^7} \\ & - \frac{47\rho_3}{36(z-1)z^7} + \frac{928}{15(z-1)z^7} - \frac{2501\rho_2^2}{1050(z-1)z^6} - \frac{71\rho_3}{21(z-1)z^6} \\ & - \frac{56}{15(z-1)z^6} - \frac{73\rho_2^2}{45(z-1)z^5} - \frac{4\rho_3}{3(z-1)z^5} + \frac{1228}{15(z-1)z^5} \\ & - \frac{584\rho_2^2}{225(z-1)z^4} - \frac{4\rho_3}{5(z-1)z^4} - \frac{892}{15(z-1)z^4} - \frac{73\rho_2^2}{12(z-1)z^3} \\ & + \frac{4\rho_3}{(z-1)z^3} + \frac{62}{3(z-1)z^3} + \frac{146\rho_2^2}{15(z-1)z^2} - \frac{32}{3(z-1)z^2} \\ & - \frac{12511\rho_2^2 z}{1890(z-1)} - \frac{12511\rho_3 z}{3780(z-1)} - \frac{166z}{15(z-1)} \\ & - \frac{4}{3(z-1)} + \frac{20}{3(z-1)z}\,,\end{aligned}$$

$$\left(M_{A,2}\right)_{23} = 0\,,$$

$$\left(M_{A,2}\right)_{24} = -\frac{48}{5z^5} - \frac{8}{5z^4} - \frac{5}{3z^3} + \frac{26}{3z^2}\,,$$

$$\left(M_{A,2}\right)_{31} = 0\,, \quad \left(M_{A,2}\right)_{32} = 0\,,$$

$$\left(M_{A,2}\right)_{33} = -\frac{52}{15z^6} + \frac{19}{5z^5} - \frac{1}{3z^4} + \frac{8}{3z^3} - \frac{4}{z^2} + \frac{1}{z} + \frac{1}{3}\,,$$

$$\left(M_{A,2}\right)_{34} = 0\,,$$

$$\left(M_{A,2}\right)_{41} = \frac{2}{z^5} + \frac{2}{z^4} + \frac{2}{z^3}\,,$$

$$\begin{aligned}\left(M_{A,2}\right)_{42} = & -\frac{\rho_2^2}{9z^8} - \frac{2\rho_3}{9z^8} - \frac{548}{15z^8} - \frac{28\rho_2^2}{75z^7} - \frac{2\rho_3}{3z^7} - \frac{24}{5z^7} - \frac{73\rho_2^2}{60z^6} \\ & - \frac{11\rho_3}{6z^6} - \frac{16}{3z^6} - \frac{73\rho_2^2}{45z^5} - \frac{2\rho_3}{z^5} + \frac{28}{z^5} - \frac{73\rho_2^2}{30z^4} \\ & - \frac{2\rho_3}{z^4} - \frac{4}{z^4} - \frac{73\rho_2^2}{15z^3} - \frac{4}{z^2}\,,\end{aligned}$$

$$\left(M_{A,2}\right)_{43} = 0\,,$$
$$\left(M_{A,2}\right)_{44} = \frac{23}{5z^5} + \frac{4}{5z^4} + \frac{7}{6z^3} - \frac{10}{3z^2} + \frac{1}{z}\,, \tag{A.20}$$

and

$$\left(M_{B,0}\right)_{11} = \frac{3i\Omega z}{z-1} - \frac{2i\Omega z^2}{z-1}\,, \quad \left(M_{B,0}\right)_{12} = \frac{16i\Omega}{(z-1)z^2} + \frac{4i\Omega z}{z-1} - \frac{16i\Omega}{(z-1)z}\,,$$
$$\left(M_{B,0}\right)_{13} = -2\lambda - 2\,, \quad \left(M_{B,0}\right)_{14} = 2i\Omega z\,,$$
$$\left(M_{B,0}\right)_{21} = -\frac{2\Omega^2 z^4}{(z-1)^2} + \frac{2\lambda z^2}{(z-1)^2} - \frac{2\lambda z}{(z-1)^2}\,,$$
$$\left(M_{B,0}\right)_{22} = \frac{8\lambda}{(z-1)^2 z^2} + \frac{24}{(z-1)^2 z^2} - \frac{8\lambda}{(z-1)^2 z} - \frac{8\Omega^2 z}{(z-1)^2} - \frac{24}{(z-1)^2 z}\,,$$
$$\left(M_{B,0}\right)_{23} = \frac{4i\Omega z^2}{z-1}\,, \quad \left(M_{B,0}\right)_{24} = -\frac{2\lambda z}{z-1}\,,$$
$$\left(M_{B,0}\right)_{31} = -i\Omega z\,, \quad \left(M_{B,0}\right)_{32} = \frac{12i\Omega}{z^2}\,,$$
$$\left(M_{B,0}\right)_{33} = -\frac{1}{z}\,, \quad \left(M_{B,0}\right)_{34} = -i\Omega z\,, \quad \left(M_{B,0}\right)_{41} = 0\,,$$
$$\left(M_{B,0}\right)_{42} = -\frac{4}{z^3(z-1)} + \frac{12}{z^2(z-1)} - \frac{4}{z-1}\,,$$
$$\left(M_{B,0}\right)_{43} = -\frac{i\Omega z^2}{z-1}\,,$$
$$\left(M_{B,0}\right)_{44} = \frac{1}{1-z}\,, \quad \left(M_{B,1}\right)_{11} = 0\,,$$
$$\left(M_{B,1}\right)_{12} = -\frac{2i\rho_2\Omega}{(z-1)z^5} + \frac{52i\rho_2\Omega}{15(z-1)z^4} + \frac{9i\rho_2\Omega}{(z-1)z^3} - \frac{8i\rho_2\Omega}{(z-1)z^2} + \frac{73i\rho_2\Omega z}{15(z-1)}\,,$$
$$\left(M_{B,1}\right)_{13} = 0\,, \quad \left(M_{B,1}\right)_{14} = 0\,, \quad \left(M_{B,1}\right)_{21} = 0\,,$$
$$\begin{aligned}\left(M_{B,1}\right)_{22} = &-\frac{12\rho_2}{(z-1)^2 z^6} + \frac{8\lambda\rho_2}{3(z-1)^2 z^5} + \frac{16\rho_2}{(z-1)^2 z^5} + \frac{4\lambda\rho_2}{3(z-1)^2 z^4} \\ &+ \frac{4\rho_2}{(z-1)^2 z^4} + \frac{4\lambda\rho_2}{(z-1)^2 z^3} + \frac{24\rho_2}{(z-1)^2 z^3} - \frac{8\lambda\rho_2}{(z-1)^2 z^2} \\ &- \frac{8\rho_2\Omega^2}{3(z-1)^2 z^2} - \frac{32\rho_2}{(z-1)^2 z^2} - \frac{4\rho_2\Omega^2}{(z-1)^2 z} - \frac{8\rho_2\Omega^2}{(z-1)^2}\,,\end{aligned}$$
$$\left(M_{B,1}\right)_{23} = 0\,, \quad \left(M_{B,1}\right)_{24} = 0\,, \quad \left(M_{B,1}\right)_{31} = 0\,,$$
$$\left(M_{B,1}\right)_{32} = \frac{4i\rho_2\Omega}{z^5} + \frac{6i\rho_2\Omega}{z^4} + \frac{12i\rho_2\Omega}{z^3}\,,$$
$$\left(M_{B,1}\right)_{33} = 0\,, \quad \left(M_{B,1}\right)_{34} = 0\,, \quad \left(M_{B,1}\right)_{41} = 0\,,$$
$$\left(M_{B,1}\right)_{42} = -\frac{6\rho_2}{(z-1)z^6} + \frac{6\rho_2}{5(z-1)z^5} + \frac{\rho_2}{(z-1)z^4} + \frac{16\rho_2}{(z-1)z^3} - \frac{73\rho_2}{15(z-1)}\,,$$
$$\left(M_{B,1}\right)_{43} = 0\,, \quad \left(M_{B,1}\right)_{44} = 0\,,$$

$$
\begin{aligned}
\left(M_{B,2}\right)_{11} &= \frac{i\Omega}{3} + \frac{44i\Omega}{5z^5} + \frac{7i\Omega}{3z^4} + \frac{4i\Omega}{z^3} - \frac{4i\Omega}{z^2} + \frac{8i\Omega}{3z}, \\
\left(M_{B,2}\right)_{12} &= -\frac{22i\rho_2^2\Omega}{27(z-1)z^8} - \frac{44i\rho_3\Omega}{27(z-1)z^8} + \frac{5984i\Omega}{15(z-1)z^8} - \frac{503i\rho_2^2\Omega}{450(z-1)z^7} \\
&\quad - \frac{65i\rho_3\Omega}{36(z-1)z^7} - \frac{6656i\Omega}{15(z-1)z^7} - \frac{338i\rho_2^2\Omega}{175(z-1)z^6} - \frac{16i\rho_3\Omega}{7(z-1)z^6} \\
&\quad + \frac{112i\Omega}{15(z-1)z^6} + \frac{73i\rho_2^2\Omega}{30(z-1)z^5} + \frac{13i\rho_3\Omega}{3(z-1)z^5} - \frac{956i\Omega}{5(z-1)z^5} \\
&\quad + \frac{949i\rho_2^2\Omega}{225(z-1)z^4} + \frac{24i\rho_3\Omega}{5(z-1)z^4} + \frac{3772i\Omega}{15(z-1)z^4} + \frac{219i\rho_2^2\Omega}{20(z-1)z^3} \\
&\quad - \frac{110i\Omega}{3(z-1)z^3} - \frac{146i\rho_2^2\Omega}{15(z-1)z^2} + \frac{56i\Omega}{3(z-1)z^2} + \frac{12511i\rho_2^2\Omega z}{1890(z-1)} \\
&\qquad + \frac{12511i\rho_3\Omega z}{3780(z-1)} + \frac{166i\Omega z}{15(z-1)} + \frac{4i\Omega}{3(z-1)} - \frac{44i\Omega}{3(z-1)z}, \\
\left(M_{B,2}\right)_{13} &= \frac{8\lambda}{z^6} + \frac{8}{z^6} - \frac{8\lambda}{z^3} - \frac{8}{z^3}, \\
\left(M_{B,2}\right)_{14} &= \frac{112i\Omega}{5z^5} - \frac{48i\Omega}{5z^4} - \frac{29i\Omega}{3z^3} - \frac{22i\Omega}{3z^2}, \\
\left(M_{B,2}\right)_{21} &= \frac{18}{5(z-1)^2z^6} - \frac{44\lambda}{15(z-1)^2z^5} - \frac{34}{5(z-1)^2z^5} - \frac{2\lambda}{5(z-1)^2z^4} \\
&\quad - \frac{2}{15(z-1)^2z^4} - \frac{2\lambda}{3(z-1)^2z^3} - \frac{4\Omega^2z^3}{3(z-1)^2} - \frac{4}{(z-1)^2z^3} \\
&\quad + \frac{4\lambda}{3(z-1)^2z^2} - \frac{4\Omega^2z^2}{3(z-1)^2} + \frac{188\Omega^2}{15(z-1)^2z^2} + \frac{16}{3(z-1)^2z^2} \\
&\quad + \frac{2\lambda z}{3(z-1)^2} + \frac{2\lambda}{(z-1)^2} - \frac{32\Omega^2 z}{3(z-1)^2} + \frac{14\Omega^2}{3(z-1)^2} \\
&\qquad + \frac{68\Omega^2}{15(z-1)^2z} + \frac{4}{(z-1)^2} - \frac{2}{(z-1)^2z},
\end{aligned}
$$

$$
\begin{aligned}
\left(M_{B,2}\right)_{22} = &-\frac{73\rho_2^2\Omega^2}{15(z-1)^2 z}-\frac{146\rho_2^2\Omega^2}{15(z-1)^2}-\frac{146\rho_2^2\Omega^2}{45(z-1)^2 z^2}-\frac{73\rho_2^2\Omega^2}{30(z-1)^2 z^3}\\
&-\frac{56\rho_2^2\Omega^2}{75(z-1)^2 z^4}-\frac{2\rho_2^2\Omega^2}{9(z-1)^2 z^5}-\frac{8z\Omega^2}{(z-1)^2}-\frac{4\rho_3\Omega^2}{(z-1)^2 z}\\
&-\frac{4\rho_3\Omega^2}{(z-1)^2 z^2}-\frac{11\rho_3\Omega^2}{3(z-1)^2 z^3}-\frac{4\rho_3\Omega^2}{3(z-1)^2 z^4}-\frac{4\rho_3\Omega^2}{9(z-1)^2 z^5}\\
&-\frac{16\Omega^2}{3(z-1)^2 z}+\frac{8\Omega^2}{3(z-1)^2}+\frac{64\Omega^2}{3(z-1)^2 z^2}+\frac{16\Omega^2}{(z-1)^2 z^3}\\
&+\frac{248\Omega^2}{15(z-1)^2 z^4}-\frac{224\Omega^2}{15(z-1)^2 z^5}-\frac{146\lambda\rho_2^2}{15(z-1)^2 z^2}+\frac{73\lambda\rho_2^2}{15(z-1)^2 z^3}\\
&+\frac{73\lambda\rho_2^2}{45(z-1)^2 z^4}+\frac{73\lambda\rho_2^2}{90(z-1)^2 z^5}+\frac{253\lambda\rho_2^2}{150(z-1)^2 z^6}+\frac{118\lambda\rho_2^2}{225(z-1)^2 z^7}\\
&+\frac{2\lambda\rho_2^2}{9(z-1)^2 z^8}-\frac{584\rho_2^2}{15(z-1)^2 z^2}+\frac{146\rho_2^2}{5(z-1)^2 z^3}+\frac{73\rho_2^2}{15(z-1)^2 z^4}\\
&+\frac{73\rho_2^2}{30(z-1)^2 z^5}+\frac{553\rho_2^2}{50(z-1)^2 z^6}-\frac{377\rho_2^2}{75(z-1)^2 z^7}-\frac{8\rho_2^2}{5(z-1)^2 z^8}\\
&-\frac{2\rho_2^2}{(z-1)^2 z^9}-\frac{8\lambda}{(z-1)^2 z}+\frac{16\lambda}{3(z-1)^2 z^2}-\frac{16\lambda}{(z-1)^2 z^3}\\
&+\frac{64\lambda}{(z-1)^2 z^4}-\frac{56\lambda}{(z-1)^2 z^5}+\frac{56\lambda}{15(z-1)^2 z^6}-\frac{928\lambda}{15(z-1)^2 z^7}\\
&+\frac{344\lambda}{5(z-1)^2 z^8}-\frac{4\lambda\rho_3}{(z-1)^2 z^3}+\frac{\lambda\rho_3}{3(z-1)^2 z^5}+\frac{7\lambda\rho_3}{3(z-1)^2 z^6}\\
&+\frac{8\lambda\rho_3}{9(z-1)^2 z^7}+\frac{4\lambda\rho_3}{9(z-1)^2 z^8}-\frac{20\rho_3}{(z-1)^2 z^3}+\frac{8\rho_3}{(z-1)^2 z^4}\\
&+\frac{13\rho_3}{3(z-1)^2 z^5}+\frac{19\rho_3}{(z-1)^2 z^6}-\frac{16\rho_3}{3(z-1)^2 z^7}-\frac{2\rho_3}{(z-1)^2 z^8}\\
&-\frac{4\rho_3}{(z-1)^2 z^9}-\frac{24}{(z-1)^2 z}+\frac{16}{(z-1)^2 z^2}-\frac{48}{(z-1)^2 z^3}\\
&+\frac{384}{(z-1)^2 z^4}-\frac{392}{(z-1)^2 z^5}+\frac{536}{5(z-1)^2 z^6}\\
&-\frac{2848}{5(z-1)^2 z^7}+\frac{3112}{5(z-1)^2 z^8}-\frac{96}{(z-1)^2 z^9},\\
\left(M_{B,2}\right)_{23} = &-\frac{356i\Omega}{15(z-1)z^4}-\frac{116i\Omega}{15(z-1)z^3}-\frac{8i\Omega}{(z-1)z^2}+\frac{4i\Omega z}{3(z-1)}\\
&+\frac{4i\Omega}{3(z-1)}+\frac{20i\Omega}{(z-1)z},
\end{aligned}
$$

$$
\begin{aligned}
(M_{B,2})_{24} = &-\frac{12}{(z-1)z^6}+\frac{154\lambda}{15(z-1)z^5}+\frac{40}{(z-1)z^5}+\frac{34\lambda}{15(z-1)z^4} \quad ,\\
&-\frac{8}{(z-1)z^4}+\frac{7\lambda}{3(z-1)z^3}+\frac{4}{(z-1)z^3}-\frac{28\lambda}{3(z-1)z^2}-\frac{24}{(z-1)z^2}\\
&-\frac{2\lambda}{3(z-1)}-\frac{2\lambda}{3(z-1)z}\\
(M_{B,2})_{31} = &-\frac{18i\Omega}{z^5}+\frac{2i\Omega}{z^4}+\frac{2i\Omega}{z^3}+\frac{8i\Omega}{z^2},\\
(M_{B,2})_{32} = &\frac{i\rho_2^2\Omega}{3z^8}+\frac{2i\rho_3\Omega}{3z^8}+\frac{1548i\Omega}{5z^8}+\frac{28i\rho_2^2\Omega}{25z^7}+\frac{2i\rho_3\Omega}{z^7}-\frac{128i\Omega}{5z^7}\\
&+\frac{73i\rho_2^2\Omega}{20z^6}+\frac{11i\rho_3\Omega}{2z^6}-\frac{24i\Omega}{z^6}+\frac{73i\rho_2^2\Omega}{15z^5}+\frac{6i\rho_3\Omega}{z^5}-\frac{188i\Omega}{z^5}\\
&+\frac{73i\rho_2^2\Omega}{10z^4}+\frac{6i\rho_3\Omega}{z^4}+\frac{4i\Omega}{z^4}+\frac{73i\rho_2^2\Omega}{5z^3}-\frac{8i\Omega}{z^3}+\frac{12i\Omega}{z^2},\\
(M_{B,2})_{33} = &\frac{62}{3z^7}-\frac{116}{5z^6}+\frac{1}{z^5}-\frac{8}{z^4}+\frac{12}{z^3}-\frac{1}{3z},\\
(M_{B,2})_{34} = &-\frac{57i\Omega}{5z^5}+\frac{24i\Omega}{5z^4}+\frac{31i\Omega}{6z^3}+\frac{14i\Omega}{3z^2}+\frac{i\Omega}{z},\\
(M_{B,2})_{41} = &-\frac{18}{5z^6}-\frac{4}{z^5}-\frac{14}{3z^4}-\frac{2}{z^3}-\frac{2}{z^2},\\
(M_{B,2})_{42} = &-\frac{23\rho_2^2}{27(z-1)z^9}-\frac{46\rho_3}{27(z-1)z^9}-\frac{2276}{15(z-1)z^9}-\frac{221\rho_2^2}{150(z-1)z^8}\\
&-\frac{29\rho_3}{12(z-1)z^8}+\frac{796}{3(z-1)z^8}-\frac{7963\rho_2^2}{2100(z-1)z^7}-\frac{205\rho_3}{42(z-1)z^7}\\
&-\frac{344}{15(z-1)z^7}+\frac{73\rho_2^2}{60(z-1)z^6}+\frac{23\rho_3}{6(z-1)z^6}+\frac{436}{5(z-1)z^6}\\
&+\frac{73\rho_2^2}{50(z-1)z^5}+\frac{26\rho_3}{5(z-1)z^5}-\frac{864}{5(z-1)z^5}+\frac{73\rho_2^2}{60(z-1)z^4}\\
&+\frac{10\rho_3}{(z-1)z^4}+\frac{82}{3(z-1)z^4}+\frac{292\rho_2^2}{15(z-1)z^3}-\frac{4}{(z-1)z^3}+\frac{32}{3(z-1)z^2}\\
&-\frac{12511\rho_2^2}{1890(z-1)}-\frac{12511\rho_3}{3780(z-1)}-\frac{166}{15(z-1)}-\frac{4}{3(z-1)z},\\
(M_{B,2})_{43} = &\frac{86i\Omega}{15(z-1)z^4}+\frac{29i\Omega}{15(z-1)z^3}+\frac{7i\Omega}{3(z-1)z^2}\\
&-\frac{i\Omega z}{3(z-1)}+\frac{2i\Omega}{3(z-1)}-\frac{4i\Omega}{(z-1)z},\\
(M_{B,2})_{44} = &\frac{69}{5z^6(z-1)}-\frac{301}{15z^5(z-1)}+\frac{25}{6z^4(z-1)}-\frac{14}{3z^3(z-1)}\\
&+\frac{35}{3z^2(z-1)}-\frac{1}{3z(z-1)}-\frac{1}{3(z-1)}.
\end{aligned}
\tag{A.21}
$$

The diagonalised system at infinity was already obtained in section 7.4. Here, we describe the diagonalisation procedure at the horizon. One poses $x = 1/(z-1)$ and writes the system using this new variable. Then, the algorithm steps one uses are the following:

1. putting the system in block Jordan form;
2. normalising the subleading order;
3. using a $\mathrm{Diag}(1,1,1,x)$ transformation which helps put subleading orders in block diagonal form;
4. putting the system in block Jordan form;
5. normalising the subleading order;
6. using a $\mathrm{Diag}(1,1,x,x)$ transformation which helps put subleading orders in block diagonal form;
7. putting the system in block Jordan form;
8. normalising the two first subleading orders;
9. using a $\mathrm{Diag}(1,1,1,x^2)$ transformation which helps put subleading orders in block diagonal form;
10. putting the system in block Jordan form;
11. normalising the subleading order;
12. reducing the Poincaré rank with a $\mathrm{Diag}(1,x,x^2,x^3)$ transformation;
13. putting the system in block Jordan form.

At this point, one is left with a matrix of the form

$$\begin{aligned}\tilde{M} = \mathrm{Diag}(&-2 - i\Omega\left(1 - \frac{21}{10}\varepsilon^2\right), -2 + i\Omega\left(1 - \frac{21}{10}\varepsilon^2\right), \\ &-3 - i\Omega\left(1 - \frac{21}{10}\varepsilon^2\right), -3 + i\Omega\left(1 - \frac{21}{10}\varepsilon^2\right)) + \mathcal{O}\left(\frac{1}{x^2}\right). \quad \text{(A.22)}\end{aligned}$$

By integrating the (A.3) using eq. (A.22) and going back to the variable $z$, we see that we recover, up to unimportant powers of $z-1$, the modes given in eq. (7.77).

## A.3. 4dEGB solution

We now give the perturbation equations for the 4dEGB solution described in section 2.5. The coefficients of the matrices $M_A$ and $M_B$ are given in terms of the function $f$ defined in eq. (2.63) by

$$(M_A)_{11} = -2i\Omega\left(2\beta - 2\beta f(z)^2 + z^2\right),$$
$$(M_A)_{12} = 0, \quad (M_A)_{13} = 0, \quad (M_A)_{14} = 0,$$

$$
\begin{aligned}
(M_A)_{21} = &-\frac{2}{zf(z)}\Big[-4\beta\sigma+\left(z^3+6\beta z\right)f'(z)\\
&-6\beta zf(z)^2f'(z)+f(z)\left(z^2-6\beta\right)+2\beta f(z)^3\Big]\,,\\
(M_A)_{22} =& -\frac{16\beta\left(-zf'(z)+f(z)+\sigma\right)}{zf(z)}\,,\\
(M_A)_{23} =& \,0\,,\\
(M_A)_{24} =& \frac{2\left(2\beta-2\beta f(z)^2+z^2\right)}{z}\,,\\
(M_A)_{31} =& \,0\,,\quad (M_A)_{32}=0\,,\\
(M_A)_{33} =& -\frac{f(z)^2\left(2\beta-2\beta f(z)^2+z^2\right)}{z}\,,\\
(M_A)_{34} =& \,0\,,\\
(M_A)_{41} =& \frac{2\beta-4\beta\sigma zf'(z)+4\beta f(z)\left(\sigma-zf'(z)\right)+2\beta f(z)^2+z^2}{z}\,,\\
(M_A)_{42} =& -\frac{8\beta\sigma\left(-zf'(z)+f(z)+\sigma\right)}{z}\,,\\
(M_A)_{43} =& \,0\,,\\
(M_A)_{44} =& -\frac{2\beta-2\beta f(z)^2+z^2}{z}\,,
\end{aligned}
\tag{A.23}
$$

and

$$
\begin{aligned}
(M_B)_{11} =& \frac{2i\Omega\left(2\beta-2\beta f(z)^2+z^2\right)\left(zf'(z)-f(z)\right)}{zf(z)}\,,\\
(M_B)_{12} =& \frac{16i\beta\Omega\left(2\beta-2\beta f(z)^2+z^2\right)}{z\left(2\beta-4\beta z(f(z)+\sigma)f'(z)+2\beta f(z)(f(z)+2\sigma)+z^2\right)}\Big[z^2f(z)f''(z)\\
&+\left(f(z)-zf'(z)\right)\left(-zf'(z)+f(z)+\sigma\right)\Big]\,,\\
(M_B)_{13} =& \frac{4f(z)\left(2\beta+z^2\right)f'(z)}{z}-\frac{8\beta f(z)^3f'(z)}{z}+f(z)^2\left(\frac{4\beta\lambda}{z^2}+2\right)\\
&+\frac{2\beta f(z)^4}{z^2}-\frac{2\left(2\beta\lambda+\beta+(\lambda+2)z^2\right)}{z^2}\,,\\
(M_B)_{14} =& \frac{2i\Omega\left(2\beta-2\beta f(z)^2+z^2\right)\left(-2\beta+2\beta f(z)\left(f(z)-2zf'(z)\right)+z^2\right)}{z\left(2\beta-4\beta z(f(z)+\sigma)f'(z)+2\beta f(z)(f(z)+2\sigma)+z^2\right)}\,,\\
(M_B)_{21} =& \frac{2\left(2\beta\Omega^2-\frac{4\beta\lambda\sigma f'(z)}{z}+\lambda+\frac{\beta(2\lambda-1)}{z^2}+1\right)}{f(z)^2}\\
&+\frac{8\beta\lambda\sigma-4f'(z)\left(z^3+2\beta(\lambda+1)z\right)}{z^2f(z)}+\frac{8\beta f(z)f'(z)}{z}\\
&-\frac{2\Omega^2\left(2\beta+z^2\right)}{f(z)^4}-\frac{2\beta f(z)^2}{z^2}+\frac{4\beta(\lambda+1)}{z^2}-2\,,
\end{aligned}
$$

$$
\begin{aligned}
(M_B)_{23} &= \frac{4i\Omega\left(2\beta - 2\beta f(z)^2 + z^2\right)}{zf(z)^2}\,,\\
(M_B)_{31} &= -\frac{i\Omega\left(-2\beta + 2\beta f(z)\left(f(z) - 2zf'(z)\right) + z^2\right)}{z}\,,\\
(M_B)_{32} &= -\frac{8i\beta\Omega\left(2\beta - 2\beta f(z)^2 + z^2\right)}{z\left(2\beta - 4\beta z(f(z)+\sigma)f'(z) + 2\beta f(z)(f(z)+2\sigma) + z^2\right)}\\
&\quad \Big[z^2 f(z)f''(z) + \left(f(z) - zf'(z)\right)\left(-zf'(z) + f(z) + \sigma\right)\Big]\,,\\
(M_B)_{33} &= \frac{2f(z)\left(-z\left(2\beta - 4\beta f(z)^2 + z^2\right)f'(z) - 2\beta f(z)\left(f(z)^2 - 1\right)\right)}{z^2}\,,\\
(M_B)_{34} &= -\frac{i\Omega\left(2\beta - 2\beta f(z)^2 + z^2\right)\left(-2\beta + 2\beta f(z)\left(f(z) - 2zf'(z)\right) + z^2\right)}{z\left(2\beta - 4\beta z(f(z)+\sigma)f'(z) + 2\beta f(z)(f(z)+2\sigma) + z^2\right)}\,,\\
(M_B)_{41} &= 0\,,\\
(M_B)_{43} &= -\frac{i\Omega\left(2\beta - 2\beta f(z)^2 + z^2\right)}{zf(z)^2}\,.
\end{aligned}
\tag{A.24}
$$

The remaining coefficients are

$$
\begin{aligned}
(M_B)_{22} =& \frac{8\beta}{z^2 f(z)^2\left(2\beta - 4\beta\sigma z f'(z) - 4\beta z f(z)f'(z) + 4\beta\sigma f(z) + 2\beta f(z)^2 + z^2\right)}\\
&\Big[-4\beta\lambda\sigma^2 - 4\beta\sigma^2 + 3z^4 f(z)^3 f''(z) - z^4 f(z)f''(z) + 8\beta\sigma z^2 f(z)^2 f''(z)\\
&+ 5\beta z^2 f(z)^5 f''(z) - 6\beta z^2 f(z)^3 f''(z) + 9\beta z^2 f(z)f''(z)\\
&+ 6z^5 f(z)f'(z)^3 - 6\sigma z^4 f(z)f'(z)^2 - 9z^4 f(z)^2 f'(z)^2 - z^4 f'(z)^2\\
&- 8\beta\sigma z^3 f'(z)^3 - 20\beta z^3 f(z)^3 f'(z)^3 + 12\beta z^3 f(z)f'(z)^3\\
&+ 2\lambda\sigma z^3 f'(z) + 3\sigma z^3 f(z)^2 f'(z) + 3\sigma z^3 f'(z)\\
&+ 2z^3 f(z)f'(z) - 8\beta\lambda\sigma^2 z^2 f'(z)^2 - 8\beta\lambda\sigma z^2 f(z)f'(z)^2\\
&+ 20\beta\sigma z^2 f(z)^3 f'(z)^2 + 4\beta\sigma z^2 f(z)f'(z)^2 + 5\beta\sigma f(z)^5\\
&+ 45\beta z^2 f(z)^4 f'(z)^2 - 30\beta z^2 f(z)^2 f'(z)^2 + 9\beta z^2 f'(z)^2\\
&+ 8\beta\lambda\sigma^3 z f'(z) + 24\beta\lambda\sigma^2 z f(z)f'(z) - 4\beta\sigma^2 f(z)^2\\
&+ 12\beta\lambda\sigma z f(z)^2 f'(z) + 4\beta\lambda\sigma z f'(z) - 8\beta\sigma^3 f(z)\\
&+ 8\beta\sigma^3 z f'(z) + 8\beta\sigma^2 z f(z)f'(z) - 25\beta\sigma z f(z)^4 f'(z)\\
&+ 6\beta\sigma z f(z)^2 f'(z) - 5\beta\sigma z f'(z) - 30\beta z f(z)^5 f'(z)\\
&+ 24\beta z f(z)^3 f'(z) - 18\beta z f(z)f'(z) + 6z^5 f(z)^2 f'(z)f''(z)\\
&- 8\beta\sigma z^3 f(z)f'(z)f''(z) - 20\beta z^3 f(z)^4 f'(z)f''(z)\\
&+ 12\beta z^3 f(z)^2 f'(z)f''(z) - 2\lambda\sigma z^2 f(z) + 3\sigma z^2 f(z)^3\\
&- 3\sigma z^2 f(z) + 3z^2 f(z)^4 - z^2 f(z)^2 - 8\beta\lambda\sigma^3 f(z)\\
&- 12\beta\lambda\sigma^2 f(z)^2 - 4\beta\lambda\sigma f(z)^3 - 4\beta\lambda\sigma f(z)\\
&- 2\beta\sigma f(z)^3 + 5\beta\sigma f(z) + 5\beta f(z)^6 - 6\beta f(z)^4
\end{aligned}
$$

$$+ 9\beta f(z)^2 - 2\lambda\sigma^2 z^2 - 2\sigma^2 z^2\Big], \tag{A.25}$$

$$\begin{aligned}
(M_B)_{24} =& \frac{2}{z^2 f(z)^2 \left(2\beta - 4\beta\sigma z f'(z) - 4\beta z f(z) f'(z) + 4\beta\sigma f(z) + 2\beta f(z)^2 + z^2\right)} \\
&\Big[ - 4\beta^2\lambda - 12\beta^2 + 4z^5 f(z) f'(z) - 4\beta\sigma z^4 f(z) f'(z)^2 \\
&- 16\beta z^4 f(z)^2 f'(z)^2 + 4\beta\lambda\sigma z^3 f'(z) + 4\beta\lambda z^3 f(z) f'(z) \\
&+ 2\beta\sigma z^3 f(z)^2 f'(z) + 2\beta\sigma z^3 f'(z) - 12\beta z^3 f(z)^3 f'(z) + 12\beta z^3 f(z) f'(z) \\
&+ 8\beta^2\sigma z^2 f(z)^3 f'(z)^2 + 8\beta^2\sigma z^2 f(z) f'(z)^2 + 48\beta^2 z^2 f(z)^4 f'(z)^2 \\
&- 32\beta^2 z^2 f(z)^2 f'(z)^2 - 8\beta^2\lambda\sigma z f(z)^2 f'(z) + 8\beta^2\lambda\sigma z f'(z) \\
&- 8\beta^2\lambda z f(z)^3 f'(z) + 8\beta^2\lambda z f(z) f'(z) - 10\beta^2\sigma z f(z)^4 f'(z) \\
&- 20\beta^2\sigma z f(z)^2 f'(z) + 14\beta^2\sigma z f'(z) - 36\beta^2 z f(z)^5 f'(z) \\
&- 20\beta^2 z f(z) f'(z) + 2z^4 f(z)^2 - 4\beta\lambda\sigma z^2 f(z) + 2\beta\sigma z^2 f(z)^3 \\
&- 2\beta\sigma z^2 f(z) + 7\beta z^2 f(z)^4 - 8\beta z^2 f(z)^2 + 8\beta^2\lambda\sigma f(z)^3 \\
&- 8\beta^2\lambda\sigma f(z) + 4\beta^2\lambda f(z)^4 + 2\beta^2\sigma f(z)^5 + 12\beta^2\sigma f(z)^3 \\
&+ 40\beta^2 z f(z)^3 f'(z) - 14\beta^2\sigma f(z) + 6\beta^2 f(z)^6 - 8\beta^2 f(z)^4 + 14\beta^2 f(z)^2 \\
&- \lambda z^4 - 2z^4 - 4\beta\lambda z^2 + \beta z^2\Big],
\end{aligned} \tag{A.26}$$

$$\begin{aligned}
(M_B)_{42} =& - \frac{8\beta}{z^2 f(z) \left(2\beta - 4\beta\sigma z f'(z) - 4\beta z f(z) f'(z) + 4\beta\sigma f(z) + 2\beta f(z)^2 + z^2\right)} \\
&\Big[2\beta\sigma + z^4 f(z)^2 f''(z) - 4\beta\sigma z^2 f(z) f''(z) + 2\beta z^2 f(z)^4 f''(z) \\
&- 6\beta z^2 f(z)^2 f''(z) + z^5 f'(z)^3 - \sigma z^4 f'(z)^2 - z^4 f(z) f'(z)^2 \\
&+ 6\beta z^3 f'(z)^3 - z^3 f(z)^2 f'(z) - z^3 f'(z) + 6\beta\sigma z^2 f(z)^2 f'(z)^2 \\
&+ 14\beta z^2 f(z)^3 f'(z)^2 - 14\beta z^2 f(z) f'(z)^2 - 8\beta\sigma z f(z)^3 f'(z) \\
&- 10\beta z f(z)^4 f'(z) + 12\beta z f(z)^2 f'(z) - 2\beta z f'(z) + z^5 f(z) f'(z) f''(z) \\
&+ 6\beta z^3 f(z) f'(z) f''(z) + \sigma z^2 f(z)^2 + z^2 f(z)^3 + z^2 f(z) + 2\beta\sigma f(z)^4 \\
&+ 2\beta f(z)^5 - 4\beta f(z)^3 + 2\beta f(z) + \sigma z^2 - 6\beta z^3 f(z)^2 f'(z)^3 - 4\beta\sigma f(z)^2 \\
&- 6\beta z^3 f(z)^3 f'(z) f''(z) + 8\beta\sigma z f(z) f'(z) - 6\beta\sigma z^2 f'(z)^2\Big],
\end{aligned} \tag{A.27}$$

and

$$\begin{aligned}
(M_B)_{44} =& - \frac{2}{z^2 f(z) \left(2\beta - 4\beta\sigma z f'(z) - 4\beta z f(z) f'(z) + 4\beta\sigma f(z) + 2\beta f(z)^2 + z^2\right)} \\
&\Big[4\beta^2\sigma + z^5 f'(z) - 2\beta\sigma z^4 f'(z)^2 - 4\beta z^4 f(z) f'(z)^2 + 4\beta\sigma z^3 f(z) f'(z) \\
&- 2\beta z^3 f(z)^2 f'(z) + 4\beta z^3 f'(z) + 4\beta^2\sigma z^2 f(z)^2 f'(z)^2 - 4\beta^2\sigma z^2 f'(z)^2 \\
&+ 16\beta^2 z^2 f(z)^3 f'(z)^2 - 16\beta^2 z^2 f(z) f'(z)^2 - 8\beta^2\sigma z f(z)^3 f'(z) \\
&+ 16\beta^2\sigma z f(z) f'(z) - 16\beta^2 z f(z)^4 f'(z) - 4\beta^2 z f'(z) + 28\beta^2 z f(z)^2 f'(z)
\end{aligned}$$

$$- 2\beta\sigma z^2 f(z)^2 + 2\beta z^2 f(z)^3 - 6\beta z^2 f(z) + 4\beta^2\sigma f(z)^4 \\ - 8\beta^2\sigma f(z)^2 + 4\beta^2 f(z)^5 - 8\beta^2 f(z)^3 + 4\beta^2 f(z) - 2\beta\sigma z^2 \Big] . \tag{A.28}$$

Let us now describe the algorithm steps needed to decouple the system at infinity. We consider only the case $\sigma = -1$ here. The algorithm is executed as follows (renaming the variable $z$ to $x$):

1. putting the system in block Jordan form;
2. normalising the subleading order;
3. reducing the Poincaré rank with a $\mathrm{Diag}(1,x,1,1)$ transformation;
4. reducing the Poincaré rank with a $\mathrm{Diag}(1,x,1,1)$ transformation;
5. reducing the Poincaré rank with a $\mathrm{Diag}(1,x,1,1)$ transformation;
6. putting the system in block Jordan form;
7. reducing the Poincaré rank with a $\mathrm{Diag}(1,x,1,1)$ transformation;
8. putting the system in block Jordan form;
9. normalising the subleading order;
10. using a $\mathrm{Diag}(1,1,1,x)$ transformation which helps put subleading orders in block diagonal form;
11. normalising the two subleading orders;
12. using a $\mathrm{Diag}(1,1,1,x)$ transformation which helps put subleading orders in block diagonal form;
13. normalising the two subleading orders;
14. reducing the Poincaré rank with a $\mathrm{Diag}(1,x,x^2,1)$ transformation;
15. putting the system in block Jordan form;
16. at this stage the gravitational part is decoupled; one goes on with the scalar part by normalising the subleading order;
17. reducing the Poincaré rank with a $\mathrm{Diag}(1,x)$ transformation;
18. putting the system in block Jordan form.

At the end of the procedure, one is left with a matrix of the form

$$\tilde{M}_{\mathrm{grav}} = \mathrm{Diag}(-i\Omega, i\Omega) + \mathrm{Diag}(-1 - i(1+\beta)\Omega, -1 + i(1+\beta)\Omega)\frac{1}{x} \tag{A.29}$$

for the gravitational part, and

$$\tilde{M}_{\mathrm{scal}} = \mathrm{Diag}(-5 - i\sqrt{\lambda}, -5 + i\sqrt{\lambda})\frac{1}{x} \tag{A.30}$$

for the scalar part. One recovers the modes given in eqs. (7.78) and (7.79).

At the horizon, the system can be quickly diagonalised by normalising the subleading order and using a transformation of the form $\mathrm{Diag}(1,x,1,1)$, with $x = 1/\sqrt{z-1}$.